\documentclass{report}
\usepackage[utf8]{inputenc}

\usepackage{amsmath}
\usepackage{xspace}
\usepackage{tikz}
\usepackage{cleveref}
\usepackage{amsthm}
\usepackage{subfigure}
\usepackage{amsfonts}
\usepackage{enumitem}
\usepackage{cite}
\usepackage{setspace}
\usepackage{enumitem}
\usepackage{ulem}

\usepackage[subfigure]{tocloft}

\usepackage{slantsc}

\usepackage[scaled=.7]{beramono}

\usepackage[T1]{fontenc}

\usepackage{nomencl}
\makenomenclature

\usepackage[margin=1in]{geometry}

\definecolor{contrastcolour}{rgb}{0,0,0}
\definecolor{backgroundcolour}{rgb}{1,1,1}

\title{}
\date{}

\definecolor{darkgreen}{rgb}{0.0,0.5,0.0}

\renewcommand{\dag}{DAG\xspace}
\newcommand{\smp}{SMP\xspace}
\newcommand{\imped}{impedensable\xspace}
\newcommand{\Imped}{Impedensable\xspace}
\newcommand{\mds}{MDS\xspace}
\newcommand{\gc}{GC\xspace}
\newcommand{\vc}{VC\xspace}
\newcommand{\ds}{DS\xspace}
\newcommand{\sdmds}{SDMDS\xspace}
\newcommand{\mvc}{MVC\xspace}
\newcommand{\mis}{MIS\xspace}
\newcommand{\is}{IS\xspace}
\newcommand{\tds}{2DS\xspace}
\newcommand{\dags}{DAGs\xspace}
\newcommand{\ldag}{$\prec$-DAG\xspace}
\newcommand{\dc}{DC\xspace}
\newcommand{\mm}{MM\xspace}
\renewcommand{\sp}{SP\xspace}

\newcommand{\gmrit}{GMRIT\xspace}
\newcommand{\po}{PO\xspace}

\newtheorem{definition}{Definition}[chapter]
\newtheorem{example}{Example}[chapter]

\newtheorem{examplesmpcont}{Example \thechapter.\smp: continuation}
\newtheorem{exampledscont}{Example \thechapter.\mds: continuation}

\newtheorem{algorithm}{Algorithm}[chapter]
\newtheorem{theorem}{Theorem}[chapter]
\newtheorem{corollary}{Corollary}[chapter]
\newtheorem{lemma}{Lemma}[chapter]
\newtheorem{observation}{Observation}[chapter]

\newtheorem{examplegccont}{Example \thechapter.Colouring: Continuation}
\newtheorem{examplemaxcont}{Example \thechapter.Max: Continuation}

\setlist{nosep}

\usepackage{titlesec}

\titleformat{\chapter}[display]
{\vspace{-.275in}\centering\normalfont\bfseries}
{\MakeUppercase\chaptertitlename\ \thechapter}
{1ex}
{\normalsize\MakeUppercase}

\titleformat*{\section}{\normalsize\bfseries}
\titleformat*{\subsection}{\normalsize\bfseries}
\titleformat*{\subsubsection}{\normalsize\bfseries}

\titlespacing{\chapter}{0pt}{0pt}{0pt}[0pt]
\titlespacing{\section}{0pt}{0pt}{0pt}[0pt]
\titlespacing{\subsection}{0pt}{0pt}{0pt}[0pt]
\titlespacing{\subsubsection}{0pt}{0pt}{0pt}[0pt]

\begin{document}

\thispagestyle{empty}
\begin{center}
~

\vspace{.75in}
CONVERGENCE SANS SYNCHRONIZATION

\vspace{1.33in}
By

\vspace{.33in}
Arya Tanmay Gupta

\vspace{2.83in}
A DISSERTATION

\vspace{.33in}
Submitted to\\
Michigan State University\\
in partial fulfillment of the requirements\\
for the degree of

\vspace{.33in}
Computer Science -- Doctor of Philosophy

\vspace{.33in}
2024    
\end{center}

\newpage 

\doublespacing
\chapter*{
    Abstract
}
\mark{ABSTRACT}

We currently see a steady rise in the usage and size of multiprocessor systems, and so the community is evermore interested in developing fast parallel processing algorithms. However, most algorithms require a synchronization mechanism, which is costly in terms of computational resources and time.

If an algorithm can be executed in asynchrony, then it can use all the available computation power, and the nodes can execute without being scheduled or locked. 
However, to show that an algorithm guarantees convergence in asynchrony, we need to generate the entire global state transition graph and check for the absence of cycles. 
This takes time exponential in the size of the global state space.

In this dissertation, we present a theory that explains the necessary and sufficient properties of a multiprocessor algorithm that guarantees convergence even without synchronization. We develop algorithms for various problems that do not require synchronization. Additionally, we show for several existing algorithms that they can be executed without any synchronization mechanism.

A significant theoretical benefit of our work is in proving that an algorithm can converge even in asynchrony. Our theory implies that we can make such conclusions about an algorithm, by only showing that the local state transition graph of a computing node forms a partial order, rather than generating the entire global state space and determining the absence of cycles in it. Thus, the complexity of rendering such proofs, formal or social, is phenomenally reduced.

Experiments show a significant reduction in time taken to converge, when we compare the execution time of algorithms in the literature versus the algorithms that we design. We get similar results when we run an algorithm, that guarantees convergence in asynchrony, under a scheduler versus in asynchrony. These results include some important practical benefits of our work.
\thispagestyle{empty}
\newpage

\singlespacing
\thispagestyle{empty}
\vspace*{\fill}
\noindent 
Copyright by\\
ARYA TANMAY GUPTA\\
2024 
\vspace*{\fill}

\newpage

\pagenumbering{roman}
\setcounter{page}{4}

\vspace*{\fill}
\begin{center}
    To\\ my grandfather, Giriraj Nandan,\\
    for his heritage that I am proud to share,\\
    and the rest of my family,\\
    grandma Madhuri Devi,\\
    parents Gopal and Sonali,\\
    brother and comrade Veer Akku Gaurang,\\
    for their love.
\end{center}
\vspace*{\fill}
\newpage

\doublespacing
\chapter*{Acknowledgements}
\mark{ACKNOWLEDGEMENTS}

    The experiments present in this dissertation were supported through computational resources and services provided by the Institute for Cyber-Enabled Research, Michigan State University.
    ~\\
    - - - - - - - - - - - - - - - - 
    ~ ~ ~ ~ ~ ~ ~ ~ ~ ~ ~ ~ ~ ~ ~ ~ ~ ~ ~ ~ ~ ~ ~ ~ ~ ~ ~ ~ ~ ~ ~ ~ ~ ~ ~ ~ ~ ~ ~ ~ ~ ~ 
    - - - - - - - - - - - - - - - - \\
    - - - - - - - - - - - - - - - - - - - - - - - - - - - - - - - - - - - - - - - - - - - - - - - - - - - - - - - - - - - - - - - - - - - - - - - - 

This dissertation would not have been concluded without the sincere efforts of my advisor, Professor Sandeep S Kulkarni. He motivates me to be thirsty for knowledge and mends me as I go along; I am always enlightened by our discussions. I always have a chance to play with Mathematics and to experiment, in my mind, and on a computer. I can enjoy my work only because of the freedom that he has given to me, and the support that he has shown in all these years. His advice is enlightening, not only with respect to my research, but also to circumstances in my personal life and my academic career. His supervision is outstanding; it is my sole ambition in my academic career to provide a `Kulkarni mentorship' to my students. This is the only way, methinks, through which I will be able to pay forward what I got from him.
In addition to Prof Kulkarni, I owe thanks to several other folks. I note some of them in the following paragraphs. 

The other members of my doctoral program committee, Drs Abdol-Hossein Esfahanian, Eric Torng and Shaunak D Bopardikar, are some of the most supportive faculty who I worked with during my doctoral program. Their questions and comments about my work have led this dissertation to include some nontrivial lemmas, insights and details.
I am thankful to Dr Esfahanian, whose kind supervision and trust have boosted me outstandingly; I am thankful for his hand on my back in moments of immense self-doubt.
The support and insightful suggestions that I got from Dr Bopardikar and Dr Torng are unparalleled; they helped me see some interesting nontrivial properties of systems that I present in this dissertation.

Archimedes of Syracuse, Srinivasa Ramanujan, John Forbes Nash Jr, and other extraordinary scientists have taught us that innovation and discovery can strike in the most unusual of circumstances, and so can great enlightening discussions -- while teaching a class, at a conference or breakfast, while going to sleep, or even when one is swimming in a pool but cannot help ponder over some mathematical model or problem. I am thankful to Vijay Garg, Maurice Herlihy, Shlomi Dolev, Borzoo Bonakdarpour, Chase Bruggeman, Aljoscha Meyer, for the most valuable discussions and deliberations. Their valuable suggestions and queries have led to shaping this dissertation to a great extent.

I wish to thank my groupmates: Duong Ngoc Nguyen, Gabe Appleton, Jesus Garcia, Raaghav Ravishankar, William Schultze, Ishaan Lagvankar, Luke Sperling,  Joel DeBoer, Eliezer Amponsah, for the rigorous discussions that we had in our group meetings. Their fellowship has not only helped me in my work but also kept my spirits high. 

A dissertation is not only an extract of concluding intellectual thoughts and efforts, it is also a consequence of besting non-intellectual battles. I wish to mark some acknowledgements in this realm, next.

The emotion of sadness gives us strength through sustaining (self-)compassion in us (paraphrased: Susan Cain (2022). Bittersweet. \cite{Cain2022}). I owe thanks to all who mourned with me at my losses and stood by me when I exploited an opportunity or made miscalculations. I am not very social, but even so, I was blessed with some great friends and mentors, whose fellowship provided me with strength and spirit in the most despairing of circumstances.

Before joining MSU, the Graduate School trained us (the incoming TAs) over several scenarios related to pedagogy, research, and life at MSU.
This was during the time when the first wave of COVID-19 shook the world. I am grateful for the support that the Graduate School has shown to me, especially Dr Stefanie Baier. My first non-work, non-business, informal, in person interaction post COVID-19 with people happened at OISS Coffee Hour which was resumed in the Fall of 2021 after COVID. I am thankful to the support that Office of International Students and Scholars (OISS) has shown to me, especially Dr Krista Beatty. 

We are not so different chemically, biologically, however, historically, we are unique; our thoughts, our actions, our differences, have reasons hidden deep in our stories, our biographies (paraphrased: Oliver Sacks (1985). The Man who Mistook his Wife for a Hat. \cite{Sacks1998}). Friends, colleagues and mentors such as Stefanie Baier, Hima Rawal, Chase Bruggeman, Frost Peanut, Samara Chamoun, Gloria Ashaolu, Tianyi(Titi) Kou-Herrema, Sunia Tanweer, Subhaprad Ash, Abby Hack, Ellen Searle, Nkolay Ivanov, Saviour Kitcher, Sevan Chanakian, DR Mossman, Bryce Carlton, Christian Yanez, Natasha George, have (basically) kept me emotionally and spiritually alive. I cannot thank them enough, and I am not sure whether I will be able to repay them or pay forward the love and completeness that I experienced in their company. There are some friendships where I did not get to interact much during my time at MSU, but they are as fierce as any; I am thankful to Shahrul Akhtar, Anubhav Gupta, Shikhar Goel, Harshita Trikha, Purushottam Verma, for being unconditional.

I am immensely thankful to my family, to whom I dedicate this dissertation, for the care and love that they have shown to me. I am immensely indebted to the patience that my family had to render, when I was not available for them, but they were available for me; I can only hope to repay them. I am thankful to the almighty for keeping my family healthy and safe.

One continuously keeps learning and, even unknowingly, keeps diffusing what he has learned throughout life. I wish to thank my students who I got the opportunity to teach and mentor during my service for 12 semesters, under the title of Graduate Teaching Assistant, with various capacities, at MSU. I have come in contact with some brilliant and heartwarming souls who not only carved the way that I teach but also the way that I think. 

Finally, in current times, it is difficult to convince people about the beauty of theoretical work. The attractive shine of some areas that are rigorously applicable to be a technology has overshadowed the fervor for conducting research just for the sake of the development of science. This includes funding agencies, fellow colleagues, and even a layman. This seems true for theoretical work in any field. Sometimes, we even start to doubt ourselves as to whether our work will be appreciated in the community, just because it is `too theoretical'. I acknowledge the austerity, perseverance, integrity, judgement, that I manifested which led me to conclude this dissertation in its current form. I am thankful that I sustained myself in all kinds of circumstances. I am thankful to all the persons, especially those who I have mentioned in this Acknowledgement, who consistently inspirited me. I am thankful to Michigan State University for assisting and providing me with every kind of resource that I need to conduct my research.

This Acknowledgement was edited recursively over a period of several months. I sincerely apologize if I forgot to mention someone significant.
\begin{flushright}
- \=Ary\`a T\`anm\`ay G\`upt\=a
\end{flushright}
\newpage

\singlespacing 
\setcounter{tocdepth}{0}
\renewcommand{\contentsname}{\vspace{-.85in}\normalsize\centerline{TABLE OF CONTENTS}\vspace{-.6in}}
\tableofcontents
\newpage

\doublespacing 
\addcontentsline{toc}{chapter}{LIST OF SYMBOLS}

\setlength{\nomlabelwidth}{1.5cm}

\nomenclature{$\exists$}{there exists at least one; there exist some}
\nomenclature{$\exists^^21$}{there exists a unique}
\nomenclature{$\forall$}{for all; for each; for every; for any}
\nomenclature{$:$}{such that; where}
\nomenclature{$Z \setminus Z'$}{(setminus) a set containing elements of $Z$, excluding elements of $Z'$}
\nomenclature{$Adj_i$}{set of nodes within distance $1$ from $i$, excluding $i$}
\nomenclature{$Adj^x_i$}{set of nodes within distance $x$ from $i$, excluding $i$}
\nomenclature{$dist(x, y)$}{length of shortest path between nodes $x$ and $y$}
\nomenclature{$V(G)$}{vertex set of graph $G$}
\nomenclature{$E(G)$}{edge set of graph $G$}
\nomenclature{$\lceil x \rceil$}{ceiling($x$); smallest integer greater than or equal to a real number $x$}
\nomenclature{$\lfloor x \rfloor$}{floor($x$); largest integer smaller than or equal to a real number $x$}
\nomenclature{$\cup$}{union}
\nomenclature{$N_i$}{$Adj_i\cup \{i\}$}
\nomenclature{$N_i^x$}{$Adj_i^x\cup \{i\}$}
\nomenclature{~$\mid$~$Z\mid$}{size of the set $Z$; number of elements in the set $Z$}

\renewcommand{\nomname}{List of Symbols}
\printnomenclature
\newpage


\setcounter{page}{1}
\pagenumbering{arabic}

\chapter{INTRODUCTION}

Given the available technology, increasing the number of computing processors is significantly cheaper than making a single chip more powerful.
The development and demand of computing systems are, presently, profoundly affected by this fact, and we see a continuous rise in the usage and size of multiprocessor systems.
A multiprocessor system contains multiple computing nodes performing executions to solve one problem. Each node can run one or more processes. 

In this dissertation, we assume that each computing node runs only one process to allow maximum parallelization. For this reason, throughout this dissertation, we use the term node for processes and processors alike. A node can be viewed as an independent computing machine, and a multiprocessor system consists of multiple nodes.


In a multiprocessor system, each node may store some data. 
When the system executes, all nodes read the data that they need 
through a shared memory or by passing messages.

The goal for the computing nodes is to collectively reach a state where the problem is deemed solved. 
%
To achieve this,
these nodes need to read the data in a controlled manner so that they synchronize with each other. 
%
To illustrate this, consider the problem of graph colouring.
An algorithm for this problem can be developed as follows: if node $i$ reads that it has a conflicting colour with at least one of its neighbours, then it changes its colour to the minimum possible available colour.
If this algorithm is run in an interleaving fashion (where one node executes at a time), then it will converge to a state where all nodes have non-conflicting colours. 

In a uniprocessor execution (or a multiprocessor execution where only one node executes at a time), 
this algorithm runs correctly because the global states form a directed acyclic graph (\dag), and all the sink nodes of this \dag are optimal states.
This property is also necessary and sufficient for the correctness of an 
algorithm that must reach an optimal state and then terminate/stutter.
However, if the above algorithm is run on a multiprocessor system in asynchrony, it may never converge.
This can be explained by the following example. Consider a graph with only two nodes, $i$ and $j$, connected to each other by an edge. Consider that both $i$ and $j$ are initialized with colour value 1. Suppose that $i$ reads the value of $j$ and decides to change its colour to 2. However, if $j$ reads the state of $i$ asynchronously before the changes made by $i$ are reflected, it will also decide to change its colour to 2. Similarly, in the next step, both nodes may decide to change their colour to 1. This can continue forever and the system may never converge.

Convergence cannot be guaranteed, from an arbitrary state, in the above system because of the presence of a cycle in the global state transition graph, which, in turn, is induced due to the race conditions arising among neighbouring nodes. 
Deployment of proper synchronization mechanisms (in this case, local mutual exclusion) eliminates such behaviour.

From above, we observe: (1) under the assumption of synchronization, analysis of system behaviour and design of algorithms becomes easier, and (2) if such assumptions are removed, then we may observe cyclic behaviour in state transitions, making it tedious to analyze the system, and in some cases, even preventing the system from convergence. Thus, designing an algorithm that guarantees convergence without synchronization is not trivial.

There are various synchronization primitives considered in the literature. A common synchronization primitive is to use a \textit{scheduler/daemon} that identifies how/when nodes can execute.
The synchronization model used in the above-discussed algorithm for graph colouring uses the \textit{central scheduler}, which chooses only one node per time step to execute. Other than this, there are other scheduler-based synchronization models, e.g., a \textit{distributed scheduler} chooses one or more nodes, possibly arbitrarily, per time step, and a \textit{synchronous scheduler} chooses all the nodes in each time step.
All schedulers implicitly assume a barrier requirement that one step has to complete before the next step can begin. This means that if multiple nodes are executing a step then it is necessary to wait until all executing nodes finish their step before starting the next one. 


Apart from enforcing the above-discussed scheduler-based synchronization primitives, there are other synchronization primitives like local mutual exclusion, token ring and semaphores. In all such models, the idea is to prohibit certain processes to execute, depending on the availability of resources or semantics of the subject algorithm. This assumption restricts the usage of resources and flow of data, and thus, makes the design of algorithms easy, ensuring correctness.

Enforcing synchronization introduces an overhead, which can be very costly in terms of computational resources and time. For this reason, the community is interested in developing algorithms that require minimum possible synchronization. If an algorithm can be executed in asynchrony, then it can use all the available computation power, and the nodes can execute without being scheduled or locked. 
If we understand the underlying behaviour of such algorithms, then we can easily analyze if an algorithm possesses this quality, and if not, we can make minimal changes to it and transform it into an algorithm that does not require synchronization.


While there have been instances of algorithms that are correct under asynchrony (e.g., \cite{Bhagat2015,Fang2005,Assran2020}), we do not have a model that captures and explains the behaviour of an arbitrary algorithm that convergences in asynchrony. Hence, in this dissertation, we focus on developing a theory for multiprocessor algorithms that guarantees convergence under asynchrony. Specifically, we validate the following thesis statement:
\begin{center}
    \begin{tabular}{@{}l@{}}
        \texttt{
        \textbf{
        Synchronization is not required in a multiprocessor system which can identify, in every
        }}
        \\
        \texttt{
        \textbf{
        suboptimal global state, at least one node 
        whose local state will indefinitely prevent
        }}
        \\
        \texttt{
        \textbf{
        convergence.
        }
        }
    \end{tabular}
\end{center}

\subsubsection*{Organization of the Chapter}

\noindent 
We validate the above thesis statement by building on a recent work by 
Garg \cite{Garg2020}. We briefly discuss Garg's model of lattice-linearity in \Cref{section:ll-intro}. 
In \Cref{section:contributions}, we enumerate the contributions of this dissertation. In \Cref{section:applications}, we discuss the applications of the problems that we study in this dissertation. 
\Cref{section:organization} outlines the organization of the following chapters in this dissertation.

\section{Lattice-Linearity}\label{section:ll-intro}

The recent introduction of lattice-linearity (by Garg, SPAA 2020) \cite{Garg2020} has shown that algorithms can be developed in such a way that the nodes can execute asynchronously.

A critical observation from \cite{Garg2020} is that if the global states form a lattice, then, under some additional constraints, an algorithm traversing that lattice is fully tolerant to asynchrony.
The key idea of lattice-linearity is that in such systems, a node can determine that its local state is not feasible in any reachable optimal global state, so, it has to change its state to reach an optimal state.
Thus, if node $i$ changes its state from $i[st]$ to $i[st']$ it never revisits state $i[st]$ again. 

Since it is guaranteed for a node in a violating state that its current state has to be rejected permanently, and that no optimal global state can be reached in its current local state, it can change its state even if it is relying on the old values of its neighbours. 
This allows the nodes to run without synchronization and the system is yet guaranteed to reach an optimal state.
There are some additional constraints regarding what the next chosen local state $i[st']$ will be, once $i$ rejects its current state $i[st]$. These constraints are problem-dependent; we discuss these constraints, in detail, in the following chapters, according to their relevance to the subject problem and the algorithm being studied.

As a consequence of the property that a local state once rejected is never visited again, the local states visited by each individual node form a total order. The lattice structure induced among the global states is a consequence of this total order induction in the local state transition graph.

In \cite{Garg2020}, Garg studies a restricted class of problems, which we call \textit{lattice-linear problems}. In such problems, the nodes that are in a violating local state can be distinctly determined. In addition, the problems studied in \cite{Garg2020} do not allow \textit{self-stabilization}.

\section{Contributions of the Dissertation}\label{section:contributions}

We first study whether there exist any lattice-linear problems that allow self-stabilization. To this end, we show that the parallel processing version developed for Karatsuba's multiplication algorithm (cf. \cite{Karatsuba1962}) by Cesari and Maeder \cite{Cesari1996} guarantees convergence without synchronization. This algorithm is lattice-linear and is self-stabilizing.
We study multiple algorithms for multiplication and modulo operations that are lattice-linear, and effectively, they guarantee convergence in asynchrony. This study is presented in \Cref{chapter:mulmod}; a preliminary version of this chapter appeared in SSS 2023.

The problem that we study next is whether we can
develop algorithms for non-lattice-linear problems (problems in which violating nodes cannot be determined in a suboptimal global state) that converge in asynchrony. To this end, we observe that, under problem-specific constraints, convergence can be guaranteed in non-lattice-linear problems algorithmically even if one or more lattices are induced only in a subset of the state space. This leads to the introduction of eventually lattice-linear algorithms, and we develop eventually lattice-linear algorithms for service-demand based minimal dominating set, minimal vertex cover, maximal independent set, graph colouring and 2-dominating set problems.
This study is presented in \Cref{chapter:ella}; a preliminary version of this chapter appeared in SSS 2021 and the full chapter was published in JPDC 2024.
We also present some experimental results, comparing the runtime of our algorithm for maximal independent set with other algorithms in the literature. Experimental results show that our algorithm converges much faster as compared to other algorithms.

In eventually lattice-linear algorithms, lattices are induced only among a subset of the global states, so a developer must guarantee that the system, initialized in an arbitrary or specified state, is (1) guaranteed to reach a state in one of the lattices, and then (2) guaranteed to reach an optimal state.
Thus, the problem that we study next is whether it is possible to induce lattices in the entire state space if the underlying problem is a non-lattice-linear problem. To this end, we observe that lattices can be induced algorithmically in the entire state space in non-lattice-linear problems. 
Thus, we introduce fully lattice-linear algorithms, and develop algorithms for minimal dominating set, graph colouring, minimal vertex cover and maximal independent set problems.
We also present a parallel-processing lattice-linear 2-approximation algorithm for the vertex cover problem.
This study is presented in \Cref{chapter:flla}; a preliminary version of this chapter appeared in SRDS 2023. We transform the lattice-linear algorithm for minimal dominating set (which is originally a distance-4 algorithm) to a distance-1 algorithm.
We present some experimental results, comparing the runtime of our algorithms for minimal dominating set with other algorithms in the literature. Experimental results show that the distance-1 algorithm that we develop for minimal dominating set converges much faster as compared to other algorithms.

We also show for an algorithm developed by Goswami et. al (SSS, 2022) that it is lattice-linear, and is thus, tolerant to asynchrony. This algorithm was developed to solve the gathering problem on a finite number of robots on an infinite triangular grid. The authors of this algorithm originally assumed a distributed scheduler. We also show that this algorithm converges in $2n$ rounds, which is less than the time complexity originally showed (i.e., $2.5n$ rounds). Apart from this, we find that some guards used in the original algorithm are redundant, so we remove them and present a new updated algorithm, which uses only a subset of guards from the original algorithm. This study is presented in Chapter \ref{chapter:flla} (\Cref{section:gsgs}); this study appeared in EDCC 2024.

We have that lattice induction is sufficient to allow asynchrony, but it is not a necessary condition. In other words, all algorithms that induce a lattice structure in the state space (that results from the induction of total order in the local state transition graph) allow asynchrony, but given an arbitrary algorithm that can tolerate asynchrony, it is possible that it is not lattice-linear. Thus, we investigate to find the necessary and sufficient conditions under which an algorithm guarantees convergence in asynchrony. To this end, we find that asynchrony can be allowed if and only if the local states visited by individual nodes form a partial order. Due to the partial order induced among the local states, the global states form a directed acyclic graph (DAG). Effectively, we introduce the classes of \dag-inducing problems and \dag-inducing algorithms.
We study the dominant clique problem and the shortest path problems as \dag-inducing problems, and the maximal matching problem as a non-\dag-inducing problem. We provide algorithms for all these problems. This study is presented in \Cref{chapter:dag}.

It is noteworthy from above that problems such as dominant clique and shortest path cannot be modelled within the class of lattice-linear problems, and maximal matching cannot be modelled within the class of non-lattice-linear problems. The algorithms that we present for these problems cannot be modelled within the class of lattice-linear algorithms. This is because here, the local state transition graph of each individual node forms a partial order, and this behaviour cannot be modelled within a discrete structure such as the total order.


For several problems we study in this dissertation, we develop multiple algorithms. In such cases, we analyze those algorithms in such a way that the differences in their behaviour become clear.

The algorithms that we develop theoretically converge faster as compared to the other algorithms in the literature -- in terms of the number of moves, or the order of time. We have conducted several experiments to investigate the benefit of asynchrony. Our algorithms that were put to this test are observed to run faster than other algorithms in the literature. We also observe that algorithms that guarantee convergence in asynchrony run faster in asynchrony as compared to the same algorithm running under a scheduler.

\section{Applicability of Our Work}\label{section:applications}

The major focus and benefit of this dissertation is in developing algorithms that guarantee convergence in asynchrony, and in developing a theory that explains the properties of such algorithms.

A significant theoretical benefit of our work is in proving that an algorithm can converge even in asynchrony. Our theory implies that we can show that an algorithm can tolerate asynchrony, by only showing that the local state transition graph of an arbitrary computing node forms a partial order, rather than generating the entire global state space and determining the absence of cycles in it. Thus, the complexity of rendering such proofs, formal or social, is phenomenally reduced. We also study the time complexity of an arbitrary asynchrony-tolerant algorithm, and how it is tied to the complexity class of the subject problem.

Experiments show a significant reduction in time taken to converge, when we compare the execution time of algorithms in the literature versus the algorithms that we design. We get the same results when we run an algorithm, that guarantees convergence in asynchrony, under a scheduler versus in asynchrony. These results include some remarkable practical benefits of our work.

Some applications of the specific problems studied in this dissertation are listed as follows.
Dominating set is applied in communication and wireless networks where it is used to compute the virtual backbone of a network.
Vertex cover is applicable in (1) computational biology, where it is used to eliminate repetitive DNA sequences -- providing a set covering all desired sequences, and (2) economics, where it is used in camera instalments -- it provides a set of locations covering all hallways of a building.
Independent set is applied in computational biology, where it is used in discovering stable genetic components for designing engineered genetic systems.
Graph colouring is applicable in (1) chemistry, where it is used to design storage of chemicals -- a pair of reacting chemicals are not stored together,
and (2) communication networks, where it is used in wireless networks to compute radio frequency assignment.

Matching has applications in numerous areas including social networks. Shortest path problem has applications in network routing and geographical route navigation. Dominant clique problem has applications in social networks and ecology (cf. \cite{Cohen1978}).

The applications of integer multiplication include the computation of power, matrix products which has applications in a plethora of fields including artificial intelligence and game theory, the sum of fractions and coprime base.
Modular arithmetic has applications in theoretical mathematics, where it is heavily used in number theory and various topics (e.g., groups, rings, fields, knots) in abstract algebra. 
Modular arithmetic also has applications in applied mathematics, where it is used in computer algebra and cryptography. It has applications also in chemistry and the visual and musical arts. In many of these applications, the value of the divisor is fixed.
In addition to these applications, it is needless to note that multiplication and modulo are among the fundamental mathematical operations.

\section{Organization of the Dissertation}\label{section:organization}

The following chapters are organized as follows. 
In \Cref{chapter:preliminaries}, we discuss preliminary symbols and definitions that we utilize in the chapters that follow.
In \Cref{chapter:mulmod}, we study the properties of lattice-linearity in the parallel processing version for Karatsuba's multiplication algorithm (cf. \cite{Karatsuba1962}) present in \cite{Cesari1996}. We study some more algorithms for multiplication and modulo operations that are lattice-linear.
In \Cref{chapter:ella}, we introduce the class of \textit{eventually lattice-linear algorithms}, and present example algorithms for several problems, along with some experimental results.
In \Cref{chapter:flla}, 
we introduce the class of \textit{fully lattice-linear algorithms}, and present example algorithms for several problems, along with some experimental results. 
%
%
\Cref{chapter:dag} investigates the conditions that are necessary and sufficient, under which an algorithm can guarantee convergence even in asynchrony. Like the previous chapters, in this chapter also, we present asynchrony-tolerant example algorithms for several problems. 
We discuss the related work in \Cref{chapter:literature}. We conclude in \Cref{chapter:conclusion}. 

This dissertation is written such that we assume that a reader is aware of some basic results in graph theory and distributed systems. For a reader who is not aware of these topics, we provide a preliminary overview of these subjects in \Cref{chapter:introductory-concepts}.
We provide a list of peer-reviewed publications that emerged from this dissertation in \Cref{chapter:publications}.
\chapter{PRELIMINARIES}\label{chapter:preliminaries}

Most algorithms that we study in this dissertation are graph algorithms where the input is a graph $G$, $V(G)$ is the set of its vertices, $E(G)$ is the set of its edges.
In our algorithms, each computation node will simulate a distinct vertex of a graph, so we use the term nodes for vertices; in the rest of this dissertation, $V(G)$ will be called a set of the nodes of graph $G$; this can also be visualized as each vertex of $G$ acting as a computation node.

For the graph $G$, $n=|V(G)|$, and $m=|E(G)|$. For a node $i\in V(G)$, $Adj_i$ is the set of nodes adjacent to $i$, and $Adj^x_i$ is the set of nodes within distance-$x$ from $i$, excluding $i$. Neighbourhood of $i$ is adjacency of $i$, including $i$, i.e., $N_i=Adj_i\cup\{i\}$ and $N_i^x=Adj_i^x\cup\{i\}$.
Degree of a node $i$ is the number of nodes connected to $i$ by an edge, i.e., $deg(i)=|Adj_i|$. The length of shortest path from $i$ to node $j$ is denoted by $dis(i,j)$.

For a finite natural number $x$, $[1:x]$ denotes a sequence of all natural numbers from 1 to $x$.

\section{Modeling Distributed Systems}\label{section:preliminaries-distributed-systems}

A parallel/distributed algorithm consists of nodes where each node is associated with a set of variables. \textit{Local state} of a node $i$ is a sequence of values of all its variables. A \textit{global state}, say $s$, is a sequence of local states of all nodes. Effectively, we represent $s$ as a vector, where $s[i]$ itself is a vector of the variables of node $i$.
We denote the \textit{state space} by $S$, which is the set of all global states that a given system can obtain. 

Each node in $V(G)$ is associated with rules. Each rule at node $i$ checks values of the variables of the nodes in $N_i^x$ (where the value of $x$ depends on the subject problem and acting algorithm) and updates the variables of $i$. A \textit{rule} at a node $i$ is of the form $g \longrightarrow a_c$, where the \textit{guard} $g$ is a Boolean expression over variables in $N_i^x$ and the \textit{action} $a_c$ is a set of instructions that updates the variables of $i$ if $g$ is true. 
A node is \textit{enabled} iff at least one of its guards is true, otherwise it is \textit{disabled}.

A \textit{move} is an event in which an enabled node updates its variables.
A \textit{round} is a sequence of events in which every node evaluates its guards at least once, and makes a move iff it is enabled.

In many algorithms presented in this dissertation, at a time, atmost one guard per node holds true. In the case that more than one guard is true and more than one action is to be executed, the actions will be executed in the order in which they are written, and in the order in which their respective guards are written. If a node executes more than one actions in a single move, then all those actions will take into consideration the updations made by the actions that were executed before them.

The \textit{state transition system} $\mathcal{S}$ on the state space $S$ is a discrete structure that defines all the possible transitions that can take place among the states of $S$. Under a given algorithm $A$, $\mathcal{S}$ is a directed graph such that $V(\mathcal{S})=S$, and $E(\mathcal{S})=\{\langle s, s'\rangle | \langle s, s'\rangle$ is a state transition under $A\}$.

We assume $S_o$ to be the set of \textit{optimal} global states: the system is deemed converged once it reaches a state in $S_o$. All other global states are \textit{suboptimal}.
An algorithm $A$ is \textit{self-stabilizing} with respect to the subset $S_o$ of $S$ iff  it satisfies the following properties: (1) \textit{convergence}: starting from an  arbitrary state, any sequence of computations of $A$ reaches a state in $S_o$, and (2) \textit{closure}: any computation of $A$ starting from $S_o$ always stays in $S_o$. 
$A$ is a \textit{silent} self-stabilizing algorithm if no node makes a move once a state in $S_o$ is reached.

We use the \textit{work complexity} of an algorithm with respect to time consumed; it is the summation of the time taken by all nodes to execute whenever they were enabled.
The term \textit{time complexity} is also used with respect to time consumed, however, this term, on the other hand, describes the order of time taken for an algorithm to reach an optimal state; it does not provide any information to the reader/observer about the work complexity of an algorithm, but it tells how much time a distributed system as a whole will take to converge.
As an example, if an algorithm executes actions 1, 2, and 3, each taking 1 time unit, such that actions 1 and 2 are executed concurrently and action 3 is executed only after they finish, then the time complexity is 2 units and the work complexity is 3 units. 

\subsection{Execution Without Synchronization. }\label{section:asynchrony}

Typically, we view the \textit{computation sequence} of an algorithm as a sequence of global states $\langle s_0, s_1, \cdots\rangle$, where $s_{t+1}~(t\geq 0)$ is obtained by executing some action by one or more nodes (as decided by the scheduler) in $s_t$.  
For the sake of discussion, assume that only node $i$ executes in state $s_t$, and it has only one variable. 
The computation prefix uptil $s_{t}$ is $\langle s_0, s_1, \cdots, s_t\rangle$. The state that the system traverses to after $s_t$ is $s_{t+1}$.
Under proper synchronization, $i$ would evaluate its guards on the \textit{current} local states of its distance-$x$ ($x\geq 1$) neighbours in $s_t$, resulting in the system reaching $s_{t+1}$.

We assume that an observer oracle is able to take a consistent snapshot of the system instantly without having to stop the executions of the nodes. However, the computing nodes may not know the fresh local states of other nodes. To understand how the execution works in asynchrony, let $s[j]$ be the local state of node $j$ in state $s$. 
If $i$ executes in asynchrony, then since the reads performed by the nodes, as well as their movements, are not coordinated with each other, the read operation performed by $i$ may return older values.
As a result, $i$ views the global state that it is in to be $s'$ 
where, for an arbitrary node $j\neq i$, $s'[j]\in\{s_0[j], s_1[j], \cdots, s_t[j]\}$. This means that $i$ can read older local states of other nodes from arbitrarily older global states.

In the case that $i$ may read older values, $s_{t+1}$ is evaluated as follows.
If all guards of $i$ evaluate to false in global state $s'$, then the system will continue to remain in state $s_t$, i.e., $s_{t+1} = s_{t}$.
If some guard $g$ evaluates to true in $s'$ then $i$ will execute its corresponding action $a_c$.
Here, we have the following observations:
(1) $s_{t+1}[i]$ is the state that $i$ obtains after executing an action in $s'$, and (2) $\forall j\neq i$, $s_{t+1}[j] = s_t[j]$.

\subsection{Variations of Asynchrony}

In this dissertation, we are interested in two models: arbitrary asynchrony (AA) and asynchrony with monotonous read (AMR). In the AA model, as described above, a node can read old values of other nodes arbitrarily; here, we only assume that if some information is sent from a node, it eventually reaches the target node.
Similar to AA, in AMR, the nodes execute asynchronously.
However, the AMR model adds another restriction: the values of variables of other nodes are read/received in the order in which they were updated/sent.

Algorithms present in this dissertation that require AMR model are also guaranteed to converge when there is a much more relaxed requirement that given a pair of arbitrary nodes $i$ and $j$, node $i$ can read arbitrarily old values of $j$ (as allowed by AA), but $i$ will eventually stop reading/receiving the values that $j$ obtained but rejected. However, in this more relaxed model, our proofs that describe the time complexity and other characteristics of such algorithms are, clearly, not valid.

In both AA and AMR, node $i$ reads the most recent state of itself. 

Our algorithms are independent of, and allow, both the message-passing model and the shared memory model, and our correctness and time-complexity proofs remain correct in both these models. We do not assume any node failures or byzantine behaviours, other than the nodes running without synchronization.

\subsection{Simple Examples}\label{subsection:colouring-and-max}

We discuss two example problems and one algorithm, each, to solve them. Our intent here is to show the difference between the working of algorithms that guarantee convergence in asynchrony, versus the algorithms that cannot make this guarantee. Specifically, we show how the naive algorithm for graph colouring that we discuss in the Introduction does not guarantee convergence in asynchrony. We introduce the max problem, and we discuss an algorithm for it that converges in asynchrony.

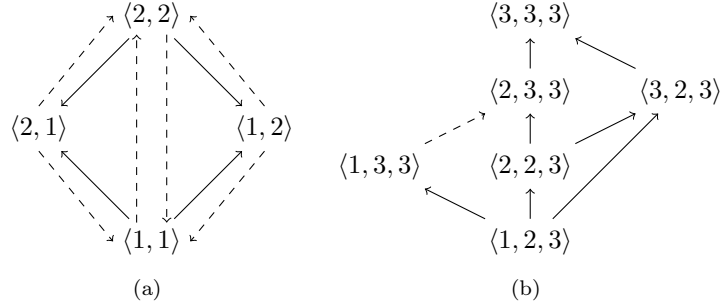
\begin{figure}[h]
    \centering
    \subfigure[]{
        \begin{tikzpicture}
            \node (a1) at (0,-3) {$\langle 1,1\rangle$};
            \node (a2) at (1.5,-1.5) {$\langle 1,2\rangle$};
            \node (a3) at (-1.5,-1.5) {$\langle 2,1\rangle$};
            \node (a4) at (0,0) {$\langle 2,2\rangle$};
            
            \draw[->] (a1) -- (a2); \draw[->] (a1) -- (a3);
            \draw[dashed, ->] (a2.south) -- (a1.east); \draw[dashed, ->] (a3.south) -- (a1.west);
            \draw[->] (a4) -- (a2); \draw[->] (a4) -- (a3);
            \draw[dashed, ->] (a2.north) -- (a4.east); \draw[dashed, ->] (a3.north) -- (a4.west);
            \draw[dashed, ->] (.2,-.25) -- (.2,-2.75); \draw[dashed, ->] (-.2,-2.75) -- (-.2,-.25);
        \end{tikzpicture}
    }
    \subfigure[]{
        \begin{tikzpicture}
            \node (a123) at (0,0) {$\langle 1,2,3\rangle$};
            \node (a133) at (-2,1) {$\langle 1,3,3\rangle$};
            \node (a223) at (0,1) {$\langle 2,2,3\rangle$};
            \node (a323) at (2,2) {$\langle 3,2,3\rangle$};
            \node (a233) at (0,2) {$\langle 2,3,3\rangle$};
            \node (a333) at (0,3) {$\langle 3,3,3\rangle$};
            
            \draw[->] (a123) -- (a133); \draw[->] (a123) -- (a223); \draw[->] (a123) -- (a323);
            \draw[->] (a223) -- (a233);
            \draw[->] (a233) -- (a333);
            \draw[->] (a223) -- (a323);
            \draw[->] (a323) -- (a333);
            
            \draw[dashed, ->] (a133) -- (a233);
        \end{tikzpicture}
    }
    \caption{(a) State transitions of a naive 2-node algorithm for graph colouring. (b) State transitions under a naive algorithm for the max problem.}
    \label{figure:colouring-and-max}
\end{figure}

\begin{example}
    \textbf{Colouring.}
    Consider the example algorithm for the graph colouring problem discussed in the Introduction. This algorithm, if run under a central scheduler, guarantees convergence.
    If the initial state is $\langle 1,1\rangle$, then depending on which node executes first, it is guaranteed to converge to an optimal state (see \Cref{figure:colouring-and-max} (a), solid lines going out from $\langle 1,1\rangle$ and from $\langle 2,2\rangle$).
    However, if this algorithm is run in AA or AMR model, then it does not guarantee convergence. In AA model, e.g., all possible global state transitions are possible, depending on the oldness of the values that the nodes are reading from each other (see \Cref{figure:colouring-and-max} (a), all dotted and solid lines).
    \qed 
\end{example}

\begin{example}\label{example:max}
    \textbf{Max.}
    Now consider the max problem on 3 nodes, and let the initial state be $\langle 1,2,3\rangle$. Consider the following naive algorithm: if a node $i$ reads that there is another node $j$ whose value is greater than that of $i$, then $i$ changes its own value to be equal to the value of $j$.
    If this algorithm is run under a central scheduler, then convergence is guaranteed (possible transitions are represented by solid edges in \Cref{figure:colouring-and-max} (b)).
    Notice that if this algorithm is run in AMR or AA model, then also convergence is guaranteed.
    However, AA or AMR model result in additional transitions. Specifically, state $\langle 1,3,3\rangle$ has a predecessor $\langle 1,2,3\rangle$, so in state $\langle 1,3,3\rangle$, node 1 can read an old local state of node 2 (i.e., 2) and not the current state of node 2 (i.e., 3). Thus, another transition $\langle 1,3,3\rangle \longrightarrow \langle 2,3,3\rangle$ is allowed in this case (presented as a dashed edge in \Cref{figure:colouring-and-max} (b)).
    \qed 
\end{example}

\section{Embedding a $\prec$-lattice in Global States}\label{section:<-lattice}

In this section, we discuss the structure of a lattice in the state space which, under proper constraints, allows an algorithm to converge to an optimal state.
To describe the embedding,
we define a total order $\prec_l$; all local states of a node $i$ are totally ordered under $\prec_l$. 
Using $\prec_l$, we define a partial order $\prec_g$ among global states as follows.

We say that $s \prec_g s^\prime$ iff $(\forall i: s[i]=s'[i]\lor s[i]\prec_l s'[i]) \land (\exists i: s[i]\prec_ls'[i])$.
Also, $s=s'$ iff $\forall i: s[i] = s'[i]$. 
For brevity, we use $\prec$ to denote $\prec_l$ and $\prec_g$: $\prec$ corresponds to $\prec_l$ while comparing local states, and $\prec$ corresponds to $\prec_g$ while comparing global states. 
We also use the symbol `$\succ$' which is such that $s\succ s'$ iff $s' \prec s$.
Similarly, we use symbols `$\preceq$' and `$\succeq$'; e.g., $s\preceq s'$ iff  $s=s' \lor s \prec s'$.
We call the lattice, formed from such partial order, a \textit{$\prec$-lattice}.

\begin{definition}\label{definition:<-lattice}
    \textit{\boldmath $\prec$-\textbf{\textit{lattice}}}. 
    Given a total relation $\prec_l$ that orders the values of $s[i]$ (the local state of node $i$ in state $s$), for each $i$, the $\prec$-lattice corresponding to $\prec_l$ is defined by the following partial order:
    $s \prec s'$ iff $(\forall i: s[i] \preceq_l s'[i]) \land (\exists i: s[i] \prec_l s'[i])$.
\end{definition}

A $\prec$-lattice constraints how global states can transition among one another: a global state $s$ can transition to state $s'$ only if $s\prec s'$.

In the $\prec$-lattice discussed above, we can define the meet and join of two states in the standard way: the meet (respectively, join), of two states $s_1$ and $s_2$ is a state $s_3$ where $\forall i: s_3[i]$ is equal to $min(s_1[i], s_2[i])$ (respectively, $max(s_1[i], s_2[i])$), 
where for a pair of local states $x$ and $y$, if $x\prec_l y$, then $\min(x, y) = \min(y, x)=x$ and 
$\max(x, y) = \max(y, x)=y$.
We are interested in the $\prec$-lattices where a join can be found for any pair of global states, however, a meet may not be found for some but not all the pairs of global states, the examples of which we study, in this dissertation, in the following chapters. This makes a $\prec$-lattice an incomplete lattice.

By varying $\prec_l$ that identifies a total order among the states of a node, one can obtain different lattices. A $\prec$-lattice, embedded in the state space, is useful for permitting the algorithm to execute asynchronously.
Under proper constraints on how the lattice is formed, convergence is ensured. We discuss these constraints in the next section.

\section{Introduction to Lattice-Linearity}\label{section:lattice-linearity}

Lattice-linearity has been shown to allow asynchrony among the computing nodes whenever it is induced in problems. The key idea of lattice linearity is that if a node can determine if its current state is not feasible in any optimal global state, then it can perform executions even based on old and inconsistent values, i.e., without needing any synchronization. We call such nodes \textit{\imped}
\footnote{The term `\imped' is similar to the notion of a node being \textit{forbidden} introduced in \cite{Garg2020}. This word itself comes from predicate detection background \cite{Chase1995}. 
We changed the notation to avoid the misinterpretation of the English meaning of the word `forbidden'.}
nodes (an \textit{impediment} to progress if does not execute, \textit{indispensable} to execute for progress).

\begin{definition}\label{definition:impedensable-node}\cite{Garg2020} \textbf{\textit{\Imped node}}. $\textsc{\Imped}$ $(i$, $s$, $\mathcal{P})\equiv \lnot \mathcal{P}(s)$ $\land$ $(\forall s'\succ s:s'[i]=s[i]\Rightarrow\lnot \mathcal{P}(s'))$. \end{definition}

\begin{examplemaxcont}
    Under the algorithm for the max problem present in \Cref{subsection:colouring-and-max}, we notice that in any suboptimal global state $s$, an impedensable node $i$ is a node that does not have the highest value stored in it. For example in state $\langle 2,2,3\rangle$ node 1 and node 2 are \imped. If any of these nodes retains its state, then it will prevent the system from convergence.
    \qed 
\end{examplemaxcont}

\begin{examplegccont}
    Notice that in the graph colouring problem, an \imped node cannot be determined. This is because, in any given global state and any chosen node, convergence can be achieved without changing the local state of that node.
    \qed 
\end{examplegccont}

The problems in which nodes can make such decisions are called \textit{lattice linear problems}. In such problems, a node once discards its local state, by definition, discards it forever. This can form a total order among the states of a node, resulting in the induction of a lattice among the global states. In such a system, because a local state that is discarded is never revisited, computing nodes can run without synchronization and the system is guaranteed to converge correctly even when the nodes read old values of each other.

Now, because convergence is guaranteed in such systems and nodes do not revisit the local states that they reject, in any suboptimal global state, there must be at least one node that must change its state, i.e., there must be at least one \imped node in every suboptimal global state.

\begin{definition}\textbf{\textit{Impedensable global state}}.\label{definition:imped-state}
    $\textsc{\Imped}(s,\mathcal{P})\equiv \exists i:\textsc{\Imped}(i,s,\mathcal{P})$.
\end{definition}

\begin{examplemaxcont}
    Under the algorithm for the max problem present in \Cref{subsection:colouring-and-max}, we notice that in any suboptimal global state, there is at least one \imped node. All suboptimal global states are \imped global states.
    \qed 
\end{examplemaxcont}

\begin{examplemaxcont}
    Under the algorithm for the max problem presented in \Cref{subsection:colouring-and-max}, the local states form a total order. E.g., for node 1, this order is $1\rightarrow 2 \rightarrow 3$. Since node 2 is initialized in local state $2$, the total order induced among its local states is $2 \rightarrow 3$.
    
    Since all nodes follow the same algorithm, the partial order formed among the local states is essentially the same; its starting point only depends on the local state of initialization for each individual node.
    
    Due to the total order formed among the local states of each individual node, the global states form a $\prec$-lattice, which is shown in \Cref{figure:colouring-and-max} (b).
    \qed 
\end{examplemaxcont}

In this section, we discuss \textit{lattice-linear problems}, i.e., the problems where the description of the problem statement creates the lattice structure. Such problems can be represented by a predicate under which the states in $S$ form a lattice.
Such problems have been discussed in \cite{Garg2020, Garg2021, Garg2022}. 

A \textit{lattice-linear problem} $P$ can be represented by a predicate $\mathcal{P}$ such that if any node $i$ is violating $\mathcal{P}$ in a state $s$, then it must change its state. Otherwise, the system will not satisfy $\mathcal{P}$.
Let $\mathcal{P}(s)$ be true iff the global state $s$ satisfies $\mathcal{P}$. A node violating $\mathcal{P}$ in $s$ is called an \textit{\imped} node.

If a node $i$ is \imped in state $s$, then in any state $s'$ such that $s'\succ s$, if the state of $i$ remains the same, then the algorithm will not converge.
Thus, predicate $\mathcal{P}$ induces a total order among the local states visited by a node, for all nodes. Consequently, the discrete structure that gets induced among the global states is a $\prec$-lattice, as described in \Cref{definition:<-lattice}. 
We say that $\mathcal{P}$, satisfying \Cref{definition:impedensable-node}, is \textit{lattice-linear} with respect to that $\prec$-lattice.
$\mathcal{P}$ is used by the nodes to determine if they are \imped, using \Cref{definition:impedensable-node}.

\begin{definition}\label{definition:ll-predicate}\cite{Garg2020}\textbf{\textit{Lattice-linear predicate}}.
    $\mathcal{P}$ is an LLP with respect to a $\prec$-lattice induced among the global states iff $\forall s\in S: \lnot\mathcal{P}(s) \Rightarrow \exists i:\textsc{\Imped}(i,s,\mathcal{P})$.
\end{definition}

\begin{examplemaxcont}
    \label{example:max-predicate}
    The predicate, governing the algorithm for the max problem presented in \Cref{subsection:colouring-and-max}, can be noted as follows.
    \begin{center}
        $\textsc{\Imped-Max}(i)\equiv \exists j: j[val]>i[val]$.
    \end{center}
    When this predicate is combined with the actions imposed the algorithm, the state transition graph, as presented in \Cref{figure:colouring-and-max} (b), is formed. This transition graph is a $\prec$-lattice.
    \qed 
\end{examplemaxcont}

Now we complete the definition of lattice-linear problems. In a lattice-linear problem $P$, given any suboptimal global state $s$, $P$ specifies all and the only nodes which cannot retain their local states. 
$\mathcal{P}$ is thus designed conserving this nature of the subject problem $P$, following Definitions \ref{definition:impedensable-node} and \ref{definition:ll-predicate}.

\begin{definition}\label{definition:ll-problem}
    \textbf{\textit{Lattice-linear problems}}.
    Problem $P$ is lattice-linear 
    iff there exists a predicate $\mathcal{P}$ and a $\prec$-lattice such that
    
    \begin{itemize}
        \item $P$ is deemed solved iff the system reaches a state where $\mathcal{P}$ is true,
        \item $\mathcal{P}$ is lattice-linear with respect to the $\prec$-lattice induced among the states in $S$, i.e., $\forall s: \neg \mathcal{P}(s) \Rightarrow \exists i:\textsc{\Imped}(i,s,\mathcal{P})$, and
        \item $\forall s:(\forall i:\textsc{\Imped}(i,s,\mathcal{P})\Rightarrow (\forall s':\mathcal{P}(s')\Rightarrow s'[i]\neq s[i]))$.
    \end{itemize}
\end{definition}

\noindent\textbf{\textit{Remark}}: A $\prec$-lattice, induced under $\mathcal{P}$, allows asynchrony because if a node, reading old values, reads the current state $s$ as $s'$, then $s'\prec s$. So $\lnot\mathcal{P}(s')\Rightarrow \lnot\mathcal{P}(s)$ because $\textsc{\Imped}(i,s',\mathcal{P})$ and $s'[i]=s[i]$.

\begin{examplemaxcont}
    The max problem, as discussed in \Cref{subsection:colouring-and-max} is a lattice-linear problem. This is because, in any suboptimal state, we can determine all the nodes that are \imped.
    \qed 
\end{examplemaxcont}

\begin{examplegccont}
    Since an \imped node cannot be determined in the graph colouring problem, we put the graph colouring problem in the class of non-lattice-linear problems. We discuss more on such problems later in this dissertation (in \Cref{section:nllp}, \Cref{chapter:ella} and \Cref{chapter:flla}).
    \qed 
\end{examplegccont}

\begin{definition}\label{definition:ssll-predicate}
    \textbf{\textit{Self-stabilizing lattice-linear predicate}}.
    Continuing from \Cref{definition:ll-problem},
    $\mathcal{P}$ is a self-stabilizing lattice-linear predicate if and only if the supremum of the lattice, that $\mathcal{P}$ induces, is an optimal state.
\end{definition}

\noindent Note that a self-stabilizing lattice-linear predicate $\mathcal{P}$ can also be true in states other than the supremum of the $\prec$-lattice. 

\begin{example}\label{example:mom-definition}\textbf{\textit{SMP}}.
    We describe a lattice-linear problem, the \textit{stable (man-optimal) marriage problem} (\smp) from \cite{Garg2020}. In \smp, all men (respectively, women) rank women (respectively men) in terms of their preference (lower rank is preferred more). A man proposes to one woman at a time based on his preference list, and the proposal may be accepted or rejected.
    
    A global state is represented as a vector $s$ where the vector $s[i]$ contains a scalar that represents the rank of the woman, according to the preference of man $i$, whom $i$ proposes.
    
    SMP can be defined by the predicate
    \begin{center}
        $\mathcal{P}_{\smp}\equiv \forall m,m':m\neq m'\Rightarrow s[m]\neq s[m']$.
    \end{center}
    $\mathcal{P}_{\smp}$ is true iff no two men are proposing to the same woman. A man $m$ is \imped iff there exists $m'$ such that $m$ and $m'$ are proposing to the same woman $w$ and $w$ prefers $m'$ over $m$. Thus,
    \begin{center}
        $\textsc{\Imped-\smp}(m,s,\mathcal{P}_{\smp})$ $\equiv$ $\exists m'$ $:$ $s[m]=s[m']\land rank(s[m],m')< rank(s[m],m)$.
    \end{center}
    If $m$ is \imped, he increments $s[m]$ by 1 until all his choices are exhausted. Following this algorithm, an optimal state, i.e., a state where the sum of regret of men is minimized, is reached.
    \qed 
\end{example}

A key observation from the stable marriage problem (\smp) and other problems from \cite{Garg2020} is that 
the states in $S$ form \textit{one} lattice, which contains a global infimum $\ell$ and \textit{possibly} a global supremum $u$ i.e., $\ell$  and $u$ are the states such that $\forall s\in S, \ell\preceq s$ and $u \succeq s$.

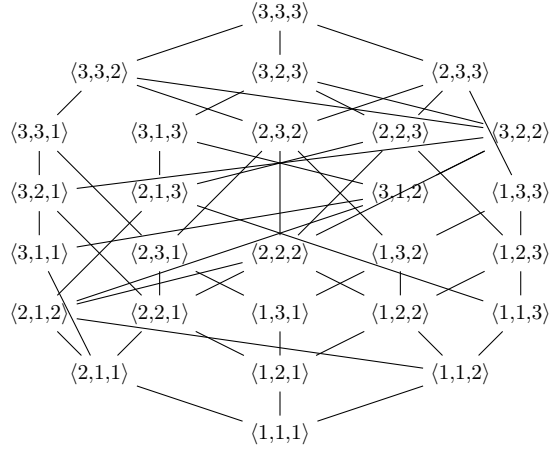
\begin{figure}[ht]
    \centering 
    \begin{tikzpicture}[scale=.8,every node/.style={scale=.8}]
        \node at (0,0) (a) {$\langle$1,1,1$\rangle$};
        
        \node at (-3,1) (b) {$\langle$2,1,1$\rangle$};
        \node at (0,1) (c) {$\langle$1,2,1$\rangle$};
        \node at (3,1) (d) {$\langle$1,1,2$\rangle$};
        
        \node at (4,2) (e) {$\langle$1,1,3$\rangle$};
        \node at (2,2) (f) {$\langle$1,2,2$\rangle$};
        \node at (4,3) (g) {$\langle$1,2,3$\rangle$};
        
        \node at (0,2) (h) {$\langle$1,3,1$\rangle$};
        \node at (2,3) (i) {$\langle$1,3,2$\rangle$};
        \node at (4,4) (j) {$\langle$1,3,3$\rangle$};
        
        \node at (-4,2) (k) {$\langle$2,1,2$\rangle$};
        \node at (-2,4) (l) {$\langle$2,1,3$\rangle$};
        
        \node at (-2,2) (m) {$\langle$2,2,1$\rangle$};
        \node at (0,3) (n) {$\langle$2,2,2$\rangle$};
        \node at (2,5) (o) {$\langle$2,2,3$\rangle$};
        
        \node at (-2,3) (p) {$\langle$2,3,1$\rangle$};
        \node at (0,5) (q) {$\langle$2,3,2$\rangle$};
        \node at (3,6) (r) {$\langle$2,3,3$\rangle$};
        
        \node at (-4,3) (s) {$\langle$3,1,1$\rangle$};
        \node at (2,4) (t) {$\langle$3,1,2$\rangle$};
        \node at (-2,5) (u) {$\langle$3,1,3$\rangle$};
        
        \node at (-4,4) (v) {$\langle$3,2,1$\rangle$};
        \node at (4,5) (w) {$\langle$3,2,2$\rangle$};
        \node at (0,6) (x) {$\langle$3,2,3$\rangle$};
        
        \node at (-4,5) (y) {$\langle$3,3,1$\rangle$};
        \node at (-3,6) (z) {$\langle$3,3,2$\rangle$};
        \node at (0,7) (parent) {$\langle$3,3,3$\rangle$};
        
        \draw (a) -- (b);
        \draw (a) -- (c);
        \draw (a) -- (d);
        \draw (d) -- (e);
        \draw (d) -- (f);
        \draw (c) -- (f);
        \draw (e) -- (g);
        \draw (f) -- (g);
        \draw (c) -- (h);
        \draw (f) -- (i);
        \draw (h) -- (i);
        \draw (g) -- (j);
        \draw (i) -- (j);
        \draw (b) -- (k);
        \draw (d) -- (k);
        \draw (e) -- (l);
        \draw (k) -- (l);
        \draw (b) -- (m);
        \draw (c) -- (m);
        \draw (m) -- (n);
        \draw (n) -- (o);
        \draw (g) -- (o);
        \draw (l) -- (o);
        \draw (f) -- (n);
        \draw (k) -- (n);
        \draw (h) -- (p);
        \draw (m) -- (p);
        \draw (p) -- (q);
        \draw (i) -- (q);
        \draw (n) -- (q);
        \draw (o) -- (r);
        \draw (j) -- (r);
        \draw (q) -- (r);
        \draw (b) -- (s);
        \draw (k) -- (t);
        \draw (s) -- (t);
        \draw (t) -- (u);
        \draw (l) -- (u);
        \draw (s) -- (v);
        \draw (m) -- (v);
        \draw (n) -- (w);
        \draw (t) -- (w);
        \draw (v) -- (w);
        \draw (w) -- (x);
        \draw (u) -- (x);
        \draw (o) -- (x);
        \draw (p) -- (y);
        \draw (v) -- (y);
        \draw (y) -- (z);
        \draw (w) -- (z);
        \draw (q) -- (z);
        \draw (z) -- (parent);
        \draw (r) -- (parent);
        \draw (x) -- (parent);
    \end{tikzpicture}
    \caption{Lattice for \smp with 3 men and 3 women; $\ell=\langle 1,1,1\rangle$ and $u=\langle 3,3,3\rangle$. Transitive edges are not shown for brevity.}
    \label{figure:smplattice}
\end{figure}

\begin{examplesmpcont}
    As an illustration of \smp, consider the case where we have 3 men and 3 women. 
    The lattice induced in this case is shown in \Cref{figure:smplattice}. In this figure, every vector represents the global state $s$ such that $s[i]$ represents the rank of the woman, according to the preference of man $i$, whom $i$ proposes. 
    The algorithm begins in the state $\langle 1, 1, 1\rangle$ (i.e., each man starts with his first choice) and continues its execution in this lattice. 
    The algorithm terminates in the lowest state in the lattice where no node is \imped.
    \qed
\end{examplesmpcont}

In \smp and other problems in \cite{Garg2020}, the algorithm needs to be initialized to $\ell$ to reach an optimal solution.
If we start from a state $s$ such that $s \neq \ell$, then the algorithm can only traverse the lattice from $s$. Hence, upon termination, it is possible that the optimal solution is not reached. 
In other words, such algorithms cannot be self-stabilizing unless $u$ is optimal. 

\begin{examplesmpcont}\label{examplesmpcont:smp-preferences}
    Consider that men and women are $M=\langle A,J,T\rangle$ and $W=\langle K,Z,M\rangle$ indexed in that sequence respectively. Let that proposal preferences of men are $A=\langle Z,K,M\rangle$, $J=\langle Z,K,M\rangle$ and $T=\langle K,M,Z\rangle$, and women have ranked men as $Z=\langle A,J,T\rangle$, $K=\langle J,T,A\rangle$ and $M=\langle T,J,A\rangle$. The optimal state (starting from $\langle 1,1,1\rangle$) is $\langle 1,2,2\rangle$.
    Starting from $\langle 1,2,3\rangle$, the algorithm terminates at $\langle 1,2,3\rangle$ which is not optimal.
    Starting from $\langle 3,1,2\rangle$, the algorithm terminates declaring that no solution is available.
    \qed 
\end{examplesmpcont}

\chapter{PARALLELIZING MULTIPLICATION AND MODULO}\label{chapter:mulmod}

We observe that the community is evermore interested in designing algorithms for problems with least possible synchronization assumptions. Asynchrony in algorithms has been earnestly a desired property for algorithms to possess. Recently, Garg (SPAA, 2020) \cite{Garg2020} showed that if lattices are induced among the state space, then an algorithm that traverses those lattices, under some additional constraints, guarantees convergence in asynchrony.
In \Cref{section:<-lattice}, we formally described the structure of such lattices. However, the problems studied in \cite{Garg2020} are constrained to a class, which we call lattice-linear problems, and do not allow self-stabilization. In this chapter, we study whether there exist lattice-linear problems that allow self-stabilization.

Specifically, in this chapter, we study some parallel processing algorithms for multiplication and modulo operations. We demonstrate that the state transitions that are formed under these algorithms satisfy lattice-linearity, and these algorithms induce a lattice among the global states. 
Lattice-linearity implies that these algorithms can be implemented in asynchronous environments, where the nodes are allowed to read old information from each other. It means that these algorithms are guaranteed to converge correctly without any synchronization overhead. 
These algorithms also exhibit snap-stabilizing properties, i.e., starting from an arbitrary state, the sequence of state transitions made by the system strictly follows its specification.

The algorithms present in this chapter tolerate asynchrony in AMR model (cf. \Cref{section:asynchrony}).

\subsubsection*{Organization of the Chapter}

\noindent This chapter is organized as follows.
In \Cref{section:mul-mod-preleminaries}, we describe the definitions and notations that we specifically use in this chapter.
We study lattice linearity of the multiplication operation in \Cref{section:mul-parallel}.
Then, in \Cref{section:mod-parallel}, we study the lattice linearity of the modulo operation.
Finally, we summarize the chapter in \Cref{section:mul-mod-summary}. We discuss the nature of the problems that are not naturally lattice-linear in \Cref{section:nllp}; this section lays the foundation of the next chapter in which we study such problems and present algorithms for them.

\section{Some Specific Preliminaries}\label{section:mul-mod-preleminaries}

This chapter focuses on multiplication and modulo operations, where the operands are $n$ and $m$. In the computation $n\times m$ or $n\mod m$, $n$ and $m$ are the values of these numbers respectively, and $|n|$ and $|m|$ are the length of the bitstrings required to represent $n$ and $m$ respectively.
\footnote{{Since $n$ and $m$ are sequence of bit-values and if $x$ is a sequence or a set, $|x|$ is used to denote the number of elements in $x$, $|n|$ and $|m|$ are the lengths of these bitstrings respectively. This notation does not represent the magnitude of their values.}}
If $n$ is a bitstring, then $n[k]$ is the $k^{th}$ bit of $n$ (indices start from 1). For a bitstring $n$, $n[1]$ is the most significant bit of $n$ and $n[|n|]$ is the least significant bit of $n$. We use $n[j:k]$ to denote the bitstring from $j^{th}$ bit to $k^{th}$ bit of $n$; this includes $n[i]$ and $n[j]$. For simplicity, we stipulate that $n$ and $m$ are of lengths in some power of $2$.
Since size of $n$ and $m$ may be substantially different, we provide complexity results that are of the form $O(f(n,m))$ in all cases, where $f$ is a function of $n$ and $m$.

\subsection{Additional Operations}

We use the following string operations: (1) $append(a$, $b)$, appends $b$ to the end of $a$ in $O(1)$ time, (2) $rshift(a, k)$, deletes rightmost $k$ bits of $a$ in $O(k)$ time, and (3)  $lshift(a, k)$, appends $k$ zeros to the right of $a$ in $O(k)$ time. 

$n \times m$ or $n \mod m$ are typically thought of as arithmetic operations. However,  when $n$ and $m$ are large, we view them as algorithms. In this chapter, we view them as parallel/distributed algorithms where the nodes collectively perform computations to converge to the final output. 

In several places, we have used the functions $\textsc{Mod}(x,y)$, $\textsc{Mul}(x,y)$, and $\textsc{Sum}(x,y)$. These functions, respectively, compute $x\mod y, x\times y$ and $x+y$.

\subsection{Modulo: Some Classic Sequential Models}\label{section:mod-sequential}

In this subsection, we discuss some sequential algorithms for computing $n \mod m$. We will utilize these preliminary algorithms to analyze the effective time complexity of parallelized modulo operation. We consider two instances, one where both $n$ and $m$ are inputs and another where $n$ is an input but $m$ is hardcoded. We utilize these algorithms in \Cref{section:parallel-modulo}.

The latter algorithm is motivated by algorithms such as RSA \cite{Rivest1978} where the value of $n$ changes based on the message to be encrypted/decrypted, but the value of $m$ is fixed once the keys are determined. Thus, some pre-processing can potentially improve the performance of the modulo operation;
we observe that certain optimizations are possible. While the time and space complexities required for preprocessing in this algorithm are high, thereby making it impractical, it demonstrates a gap between the lower and upper bound in the complexity. 

\subsubsection*{Modulo by Long Division}\label{subsection:long-division}

\begin{figure}[ht]
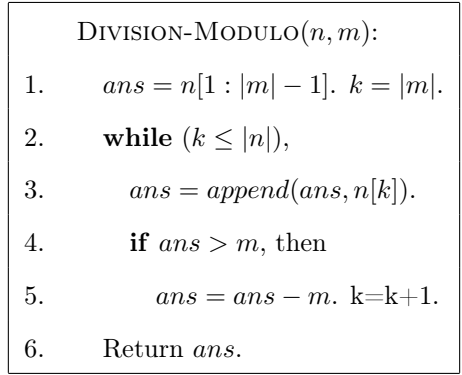

    \centering 
    \doublespacing 
    \begin{tabular}{|l l|}
        \hline 
         & \textsc{Division-Modulo}($n,m$):\\
        1. & \quad $ans = n[1:|m|-1]$. $k = |m|$.\\
        2. & \quad \textbf{while} $(k \leq |n|)$, \\
        3. & \quad \quad $ans=append(ans,n[k])$.\\
        4. & \quad \quad \textbf{if} $ans>m$, then\\
        5. & \quad \quad \quad $ans=ans-m$. k=k+1.\\
        6. & \quad Return $ans$.\\
        \hline 
    \end{tabular}
    \caption{The standard long-division algorithm to compute $n$ mod $m$.}
    \label{figure:standard-long-div}
\end{figure}

The standard long-division algorithm to compute $n$ mod $m$ is shown in \Cref{figure:standard-long-div}. 

Clearly, the number of iterations in this algorithm is bounded by $|n|$. In each of these iterations, the worst case complexity is $O(|m|)$ to perform the subtraction operation. Thus, the time complexity of standard long division is $O(|n|\times |m|)$ when $m$ and $n$ are both inputs to the algorithm. 

\subsubsection*{Modulo by Constructing DFA }\label{subsection:dfa-modulo}

If the value of $m$ is hardcoded in the algorithm, we use it to reduce the cost of the modulo operation by creating a deterministic finite automaton (DFA) $M=\langle Q$, $\Sigma$, $\delta$, $q_0\rangle$, where (1) $Q=\{q_0..q_{m-1}\}$ is the set of all possible states of $M$, (2) $\Sigma=\{0,1\}$ is the alphabet set, (3) $\delta$ is the transition function, the details of which we study in this section, and (4) $q_0$ is the initial state. If $M$ has read the first $k$ digits of $n$ then its state would be $n[1:k] \mod m$. When $M$ reads the next digit of $n$, $n[1:(k+1)] \mod m$ is evaluated depending on the next digit. If the next digit is 0, the next state will be $(2\times (n[1:k] \mod m)) \mod m$, otherwise, it will be  $(2\times (n[1:k] \mod m)+1) \mod m$.
Since the value of $m$ is hardcoded, this DFA is assumed to be pre-computed. 
As an example, for $m=3$, the corresponding DFA is provided in \Cref{example:mod3}.

\begin{figure}[ht]
    \centering
    \begin{tikzpicture}
        \node [circle, draw=black] at (0,0) (q0) {$q_0$};
        \node [circle, draw=black] at (2,0) (q1) {$q_1$};
        \node [circle, draw=black] at (4,0) (q2) {$q_2$};
        
        \draw[->] (q0.north west) to [out=135,in=225,looseness=5] (q0.south west);
        
        \draw[->,thick] (0,1) -- (0,.5);
        
        \draw[->] (q0.north east) to [out=45, in=135] (q1.north west);
        \draw[->] (q1.south west) to [out=225, in=315] (q0.south east);
        
        
        \draw[->] (q1.north east) to [out=45, in=135] (q2.north west);
        \draw[->] (q2.south west) to [out=225, in=315] (q1.south east);
        
        \draw[->] (q2.north east) to [out=45,in=315,looseness=5] (q2.south east);
        
        \node at (1,1) {1};
        \node at (1,-1) {1};
        \node at (-1,0) {0};
        
        \node at (3,1) {0};
        \node at (3,-1) {0};
        
        \node at (5,0) {1};
    \end{tikzpicture}
    \caption{A finite automaton $M_3$ computing $n\mod 3$ for any $n\in \mathbb{N}$.}
    \label{figure:mod3-dfa}
\end{figure}
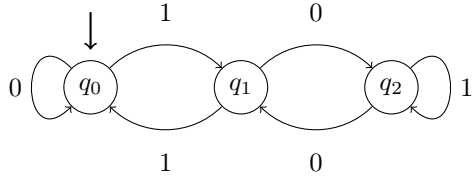

\begin{example}\label{example:mod3}
    A finite automaton $M_3$ computing $n\mod 3$ ($m=3$ is fixed) for any $n\in \mathbb{N}$ is shown in \Cref{figure:mod3-dfa}.
    \qed 
\end{example}

With this DFA, the cost of computing the modulo operation corresponds to one DFA transition for each digit of $n$. Hence, the complexity of the corresponding operation is $O(|n|)$.
$M$ does not have any accepting states, which is unlike a usual finite automaton; the final state of $M$ only tells us the value of the remainder which we would obtain after the computation of the expression $n\mod m$.

We define the construction of the transition function $\delta$ of $M$ in \Cref{figure:dfa-modulo}.
$\delta$ is constructed using the magnitude of $m$, and it is capable of reading a bitsting $n$ starting from $n[1]$ and traversing through $n[|n|]$, reading every bit, sequentially, in each step. The problem is as follows: let that $M$ has read the first $z$ bits of $n$ that evaluates to the value $K$ (here, $K=n[1:z]$), and let $K\mod m=k$, so $M$ is in state $q_k$. From here, we have to determine the next state based on whether the next bit $M$ reads is 0 or 1, which would mean that the total value read will be $2\times K+0$ or $2\times K+1$.

We start by assigning edges from $q_0$. $(2\times0+0) \mod m$ is 0 and $(2\times 0+1)\mod m$ is 1, for example. So $\delta(q_0,0)=0$ and $\delta(q_0,1)=1$. So we assign $q_0$ to transition to $q_0$ on input 0, and to $q_1$ on input 1. This method induces labelled edges between the states of $M$, such that those labels define which edge should be traversed based on what bit is read. After reading the first bit, assuming that $M$ has read the value $K$ so far and $M$ is in state $q_k$, we assign the next state to be $(2\times k+0)\mod m$ and $(2\times k+1)\mod m$ for inputs 0 and 1 respectively. This is because $(2\times K+0)\mod m$ and $(2\times K+1)\mod m$ will be equal to $(2\times k+0)\mod m$ and $(2\times k+1)\mod m$ respectively. Note that if $2\times k+0$ (resp., $2\times k+1$) is greater than $m$, then we assign $(2\times k+0)\mod m$ (resp., $(2\times k+1)\mod m$) to be $(2\times k+0)-m$ (resp., $(2\times k+1)-m$) as $2\times k+0$ (and $2\times k+1$) cannot exceed $2\times m$.

\begin{figure}[ht]
    \centering
    \doublespacing 
    \begin{tabular}{|l l|}
        \hline 
         & Elaborated definition of $\delta$.\\
         & \\
        1. & $v=0$. $m_{rs}=rshift(m, 1)$.\\
         & \quad $b=$ last bit of $m$. $i=0$.\\
        2. & \textbf{for}( ; $i< m_{rs}$; $i=i+1$),\\
        3. & \quad $\delta(q_i,0)=q_v$. $v= v+1$.\\
        4. & \quad $\delta(q_i,1)=q_v$. $v= v+1$.\\
        5. & \textbf{if} $b=1$, then\\
        6. & \quad {$\delta(q_i,0)=q_v$. $\delta(q_i,1)=q_0$. $v=1$.}\\
        7. & \textbf{else}, then\\
        8. & \quad {$\delta(q_i,0)=q_0$. $\delta(q_i,1)=q_1$. $v=2$.}\\
        9. & $i=i+1$.\\
        10. & \textbf{for}( ; $i< m$; $i=i+1$),\\
        11. & \quad $\delta(q_i,0)=q_v$. $v= v+1$.\\
        12. & \quad $\delta(q_i,1)=q_v$. $v= v+1$.\\
        13. & {$\delta(q_i,0)=q_0$. $\delta(q_i,1)=q_1$.}\\
        \hline 
    \end{tabular}
    \caption{Definition of the transition function $\delta$.}
    \label{figure:dfa-modulo}
\end{figure}

It can be clearly observed that while this approach takes $|n|$ steps, each step taking a constant amount of time, the time complexity (as well as the space complexity) of the required preprocessing is $O(m\times |m|)$, which is very high. Therefore, this approach is not practical when $m$ is large. However, we consider it to observe that we obtain a time complexity of $O(|n|)$ to run this automaton to evaluate for the modulo operation if $m$ can be hardcoded. 

\section{Parallelized Multiplication Operation}\label{section:mul-parallel}

In this section, we demonstrate that the parallelized version of the standard multiplication algorithm as well as a parallelized version of Karatsuba's \cite{Karatsuba1962} algorithm presented in \cite{Cesari1996} meet the requirements of lattice-linearity, i.e. a system of nodes traverses a lattice of global states and provide the final output. 
We consider the problem where we want to compute $n\times m$.

\subsection{Parallelizing Standard Multiplication}\label{subsection:multiplicaiton-standard}

In this subsection, we present the parallelization of the standard multiplication algorithm. First, we discuss the key idea of the sequential algorithm, then we elaborate on the lattice-linearity of its parallelization.

\subsubsection*{Key Idea}

In the standard multiplication, we multiply $m$ with one digit of $n$ at a time (for each digit of $n$), and then add all these multiplications, after left shifting those resultant strings appropriately. Suppose that we have two strings $a=n[1:\lfloor |n|/2\rfloor]\times m$ and $b=n[\lfloor |n|/2\rfloor+1:|n|]\times m$. Then the resultant multiplication $n\times m$ will be equal to $lshift(a,|n|-\lceil |n|/2\rceil)+b$.

\subsubsection*{Parallelization}

This algorithm requires $2\times |n|-1$ nodes, and induces a binary tree among them. The root of the tree is marked as node $1$ and any node $i$ ($1\leq i\leq |n|-1$) has two children: node $2i$ and node $2i+1$.

In this algorithm, every node stores two variables: $shift$ and $ans$. We demonstrate that the computation of each of these variables is lattice-linear. \\~

\noindent \textbf{Computation of {\boldmath $i[shift]$}}: At the lowest level (level 1), the value of $shift$ is set to $0$. 
Consequently, at the next level (level 2), $shift$ is set to $1$. At all higher levels, $shift$ of any node is computed to be twice the value of $shift$ of its children, i.e., $i[shift]$ is set to $2\times (2i)[shift]$. This can be viewed as a lattice-linear computation where a node is \imped iff the following condition is satisfied. A \imped node updates its value to be either $0$ (at level 1), $1$ (at level 2), or $2\times (2i)[shift]$ (at level 3 and higher).  
{
$$\begin{array}{c}
\textsc{\Imped-Multiplication-Standard-Shift}(i)\equiv\\
    \begin{cases}
        i[shift]\neq 0 & \text{if $i\geq|n|$}\\
        i[shift]\neq 1 & \text{if $(2i)[shift]=(2i+1)[shift]=0$}\\
        i[shift]\neq 2\times (2i)[shift] & \text{if $(2i)[shift]=(2i+1)[shift]\geq 1$}
    \end{cases}
\end{array}
$$
}

\noindent\textbf{Computation of {\boldmath$i[ans]$}}: The value of $ans$ at the lowest level (level 1) is set to be the corresponding bit of $n$. At level 2, $i[ans]$ is computed to be equal to the bitstring stored in $ans$ of left child (left-shifted by $i[shift]$ bits (by 1 bit)) multiplied with $m$, added to the bitstring stored in $ans$ of right child multiplied with $m$. At every level above level 2, $i[ans]$ is set by left shifting $(2i)[ans]$ by the $i[shift]$, and then adding $(2i+1)[ans]$ to that value. Thus, to propagate the value of $ans$ among the nodes correctly, we declare them to be \imped as follows. 
{
\noindent$$\begin{array}{c}
\textsc{\Imped-Multiplication-Standard-Ans}(i)\equiv\\
    \begin{cases}
        i[ans]\neq n[i-|n|+1] & \quad \quad \text{if $i\geq|n|$.}\\
        i[ans]\neq lshift(((2i)[ans]\times m)$, $i[shift])\\ \quad \quad +((2i+1)[ans]\times m) & \quad \quad \text{if $(2i)[shift]=(2i+1)[shift]=0$.}\\
        i[ans]\neq lshift((2i)[ans]$, $i[shift])+(2i+1)[ans] & \quad \quad \text{if $(2i)[shift]=(2i+1)[shift]\geq 1$.}
    \end{cases}
\end{array}
$$
}

We observe that determining $\textsc{\Imped-Multiplication-Standard-Ans}(i)$ is more complex than determining $\textsc{\Imped-Multiplication-Standard-Shift}(i)$. However, we can eliminate it by observing that it suffices to update $ans$ when $shift$ is updated. This requires that $i[shift]$ and $i[ans]$ are updated at a node $i$ atomically in a single step. In that case, we can view the algorithm as \Cref{algorithm:mul-normal-tree}.

\begin{algorithm}Parallelized standard multiplication algorithm.\label{algorithm:mul-normal-tree}
\end{algorithm}
$$
\begin{array}{|l|}
    \hline 
    \text{\textit{Rules for node $i$}}.\\~\\
    \textsc{\Imped-Multiplication-Standard-Ans}(i)\longrightarrow\\
    \begin{array}{l}
        \begin{cases} 
            i[shift]=0,i[ans] = n[i-|n|+1]. & \text{if $i\geq|n|$.}\\
            i[shift]=1,\\ \quad \quad i[ans]=lshift(((2i)[ans]\times m)$, $1)+((2i+1)[ans]\times m). & \text{if $(2i)[shift]=0$.}\\
            i[shift]=2\times (2i)[shift],\\ \quad \quad i[ans]=lshift((2i)[ans]$, $i[shift])+(2i+1)[ans]. &  \text{otherwise.}
        \end{cases}
    \end{array}~\\
    \hline
\end{array}
$$

From the above description, we see that the standard multiplication algorithm satisfies the constraints of lattice-linearity, {which we prove in the following part of this subsection}. This algorithm executes in $O(|m|\times\lg |n|)$ time. Its work complexity is $O(|n|\times |m|)$, which is same as the time complexity of the standard multiplication algorithm. The example below demonstrates the working of \Cref{algorithm:mul-normal-tree}.

\begin{figure}[ht]
    \centering
    \begin{tikzpicture}[scale=1.3,every node/.style={scale=.7}]
        \node [draw=black,label=below:node 1] (a1) at (1.5,3) {1010000+110111};
        
        \node [draw=black,label=below:node 2] (a2) at (.5,2) {0+101};
        \node [draw=black,label=below:node 3] (a3) at (4.5,2) {101000+1111};
       
        \node [draw=black,label=below:node 4] (a4) at (-.5,1) {0+0};
        \node [draw=black,label=below:node 5] (a5) at (1.5,1) {0+101};
        \node [draw=black,label=below:node 6] (a6) at (3.5,1) {1010+0};
        \node [draw=black,label=below:node 7] (a7) at (5.5,1) {1010+101};
        
        \node [draw=black,label=below:node 8] (a8) at (-1,0) {0};
        \node [draw=black,label=below:node 9] (a9) at (0,0) {0};
        \node [draw=black,label=below:node 10] (a10) at (1,0) {0};
        \node [draw=black,label=below:node 11] (a11) at (2,0) {1};
        \node [draw=black,label=below:node 12] (a12) at (3,0) {1};
        \node [draw=black,label=below:node 13] (a13) at (4,0) {0};
        \node [draw=black,label=below:node 14] (a14) at (5,0) {1};
        \node [draw=black,label=below:node 15] (a15) at (6,0) {1};
        
        \draw[->] (a1) -- (a2); \draw[->] (a1) -- (a3);
        \draw[->] (a2) -- (a4); \draw[->] (a2) -- (a5); \draw[->] (a3) -- (a6); \draw[->] (a3) -- (a7);
        \draw[->] (a4) -- (a8); \draw[->] (a4) -- (a9); \draw[->] (a5) -- (a10); \draw[->] (a5) -- (a11); \draw[->] (a6) -- (a12); \draw[->] (a6) -- (a13); \draw[->] (a7) -- (a14); \draw[->] (a7) -- (a15);
        
        \node at (-2,0) {$shift = 0$}; \node at (-2,1) {$shift = 1$}; \node at (-2,2) {$shift = 2~(10)$}; \node at (-2,3) {$shift = 4~(100)$};
        
        \node at (4,3) {(=10000111)};
    \end{tikzpicture}
    \caption{Multiplication of 00011011 and 0101 in base 2. }
    \label{figure:11011times101}
\end{figure}
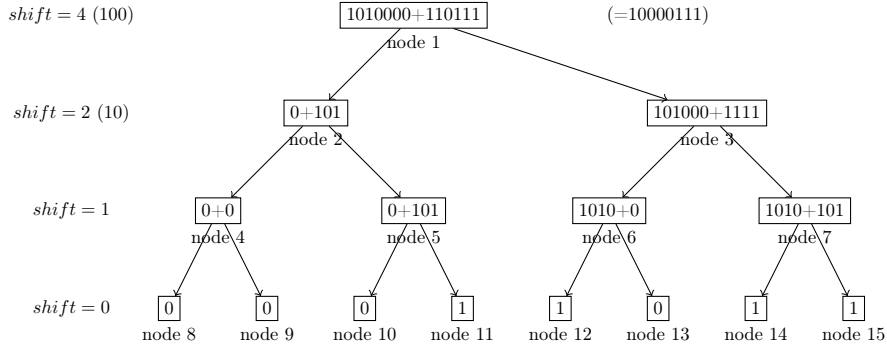

\begin{example}
    In \Cref{figure:11011times101}, we demonstrate the multiplication of the bitstrings 00011011 (value of $n$) and 0101 (value of $m$) under  \Cref{algorithm:mul-normal-tree}. 
    \qed 
\end{example}

\subsubsection*{Lattice-Linearity}\label{subsubsection:ll-mul-normal}

\begin{lemma}\label{lemma:mul-normal-predicate-ll}
    Given the input bitstrings $n$ and $m$, the predicate 
    \begin{center}
        $\forall i:\lnot\textsc{\Imped-Multiplication-Standard-Ans}(i)$
    \end{center}
    is lattice-linear on $2|n|-1$ computing nodes.
\end{lemma}

\begin{proof}
    Let us assume that node 1 does not have the correct value of $1[ans]=n\times m$. This implies that (1) node 1 has a non-updated value in $1[ans]$ or $1[shift]$, in which case node 1 is \imped, or (2) node 2 does not have the correct values $2[ans]=n[1:|n|/2]\times m$ or $2[shift]$ or node 3 does not have the correct values $3[ans]=n[|n|/2+1:|n|]\times m$ or $3[shift]$.
    
    Recursively, this can be extended to any node $i$. Let that node $i$ has stored an incorrect value in $i[ans]$ or $i[shift]$. If $i\leq |n|-1$, then this means that (1) node $i$ has a non-updated value in $i[ans]$ or $i[shift]$, in which case node $i$ is \imped, or (2) node $2i$ or node $2i+1$ do not have the correct values in $(2i)[ans]=n[|n|-2(|n|-2i)+1:|n|-2(|n|-2i)+2]$ or $(2i+1)[ans]=n[|n|-2(|n|-2i-1)+1:|n|-2(|n|-2i-1)+2]$ respectively. If $i\geq |n|$, then this implies that node $i$ has not read the correct value $n[i-|n|+1]$, in which case, again, node $i$ is \imped.
    
    From these cases, we have that given a global state $s$, where $s=\langle\langle 1[ans]$, $1[shift]\rangle,$ $\langle 2[ans]$, $2[shift]\rangle$, $...$, $\langle (2|n|-1)[ans]$, $(2|n|-1)[shift]\rangle\rangle$, if $s$ is \imped, there is at least one node which is \imped.

    
    Next, we show that if some node is \imped, then node 1 will not store the correct answer. $\forall i:i\in[1:2|n|-1]$ node $i$ is \imped if it has a non-updated value in $i[ans]$ ($i[ans]=n[|n|-2(|n|-i)+1:|n|-2(|n|-i)+2]\times m$ if $i\leq |n|-1$ and $i[ans]=n[i-|n|+1]$ if $i\geq |n|$) or $i[shift]$. This implies that the parent of node $i$ will also store incorrect value in its $ans$ or $shift$ variable. Recursively, we have that node 1 stores an incorrect value in $1[ans]$, and thus the global state is \imped.
\end{proof}

{
\begin{lemma}\label{lemma:mul-normal-predicate-self-stabilizing}
   The predicate
    \begin{center}
        $\forall i:\lnot\textsc{\Imped-Multiplication-Standard-Ans}(i)$
    \end{center}
    is a lattice-linear self-stabilizing predicate.
\end{lemma}
\begin{proof}
Since the lattice linearity was shown in \Cref{lemma:mul-normal-predicate-ll}, we only focus on the self-stabilizing aspect here. To show this, we need to show that this predicate is true in the supremum state. 

    We note that if $s$ is the supremum of the induced lattice, then there is no outgoing edge from $s$ to any other global state in the state transition graph. It means that in $s$, no node is enabled, and so, no node is \imped. Thus, we have that the predicate 
    \begin{center}
        $\forall i:\lnot\textsc{\Imped-Multiplication-Standard-Ans}(i)$
    \end{center}
    holds true in $s$.
    Since $s$ is an arbitrary supremum, this predicate 
    is a self-stabilizing predicate.
\end{proof}
}

{
\begin{theorem}\label{theorem:algo-mul-normal-algo-ss}
    \Cref{algorithm:mul-normal-tree} is silent and self-stabilizing.
\end{theorem}
\begin{proof}
    
    The nodes that have ID $\geq n$ (leaf nodes) read the bit-values directly from the input (cf. first rule of \Cref{algorithm:mul-normal-tree}), so their value is fixed immediately when they make their first move. 
    Then, these nodes will not change their states. 
    After that, recursively, all other nodes will update their state with respect to the state of their children (cf. second and third rules of \Cref{algorithm:mul-normal-tree}). If the nodes are arbitrarily initialized, then several nodes may need to update their state more than once.
    
    This process will continue for all nodes in the tree, and the nodes will converge to a stable state bottom-to-top, in an acyclic fashion. Therefore, eventually, the root node will correct its own state. At this point, no node is enabled and the value of $ans$ in the root node provides the answer. Thus, \Cref{algorithm:mul-normal-tree} is silent and self-stabilizing.
\end{proof}
}



\subsection{Parallelized Karatsuba's Multiplication Operation}\label{subsection:karatsuba}

In this section, we study the lattice-linearity of the parallelization (of Karatsuba's \cite{Karatsuba1962} algorithm) that was presented in \cite{Cesari1996}. First we discuss the idea behind the sequential Karatsuba's algorithm, and then we elaborate on its lattice-linearity.

\subsubsection*{Key Idea of Sequential Karatsuba's Algorithm}\label{sububsection:idea-karatsuba}

The input is a pair of bitstrings $n$ and $m$.
This algorithm is recursive in nature. As the base case, when the length of $n$ and $m$ equals 1 then, the multiplication result is trivial. 
When the length is greater than 1, we let $n=append(a, b)$ and $m=append(c, d)$, where  $a$ and $b$ are half the length of $n$, and $c$ and $d$ are half the length of $m$. Here, $append(a, b)$, for example, represents concatenation of $a$ and $b$, which equals $n$.

Let $z=2^{|b|}$. 
$n\times m$ can be computed as $a\times c\times z^2+(a\times d+b\times c)\times z+b\times d$. $a\times d+b\times c$ can be computed as $(a+b)\times (c+d)-a\times c-b\times d$.
Thus, to compute $n \times m$, it suffices to compute 3 multiplications $a\times c$, $b\times d$ and $(a+b)\times (c+d)$.
Hence, we can eliminate one of the multiplications.
In the following section, we analyse the lattice-linearity of the parallelization of this algorithm as described in \cite{Cesari1996}.

\subsubsection*{The CM Parallelization \cite{Cesari1996} for Karatsuba's Algorithm}

The Karatsuba multiplication algorithm involves dividing the input string into substrings and use them to evaluate the multiplication recursively. In the parallel version of this algorithm, the recursive call is replaced by utilizing other (\textit{children}) nodes to treat those substrings. We elaborate more on this in the following paragraphs. Consequently, 
this algorithm induces a tree among the computing nodes, where every non-leaf node has three children. 
This algorithm works in two phases, top-down and bottom-up. This algorithm uses four variables to represent the state of each node $i$: $i[n]$, $i[m]$, $i[ans]$ and $i[shift]$ respectively.

In the sequential Karatsuba's algorithm, both of the input strings $n$ and $m$ are divided into two substrings each, and the algorithm then runs recurively on three different input pairs computed from those excerpt bitstrings. In the parallel version, those recursive calls are replaced by \textit{activating} three children nodes \cite{Cesari1996}. As a result of such parallelization, if there is no carry-forwarding due to addition, we require $\lg |n|$ levels, for which a total of $|n|^{\lg 3}$ nodes are required. However, if there is carry-forwarding due to additions, then we require $2 \lg |n|$ levels, for which a total of $|n|^{2\lg 3}$ nodes are required.

In the top-down phase, if $|i[m]| > 1$ or $|i[n]|>1$, then $i$ writes (1) $a$ and $c$ to its left child, node $3i-1$ ($(3i-1)[m]=a$ and $(3i-1)[n]=c$), (2) $b$ and $d$ to its middle child, node $3i$ ($(3i)[m]=b$ and $(3i)[n]=d$), and (3) $a+b$ and $c+d$ to its right child, node $3i+1$ ($(3i+1)[m]=a+b$ and $(3i+1)[n]=c+d$). 
If $|i[m]|=|i[n]|=1$, i.e., in the base case, the bottom-up phase begins and node $i$ sets $i[ans]=i[m]\times i[n]$ that can be computed trivially since $|i[m]|=|i[n]|=1$.

In the bottom-up phase, node $i$ sets 
$i[ans] = (3i-1)[ans]\times z^2 + ((3i+1)[ans]-((3i-1)[ans]+(3i)[and]))\times z + (3i)[ans]$.
Notice that multiplication by $z$ and $z^2$ corresponds to bit shifts and does not need an actual multiplication. 
Consequently, the product of $m\times n$ for node $i$ is computed by this algorithm.

With some book-keeping (storing the place values of most significant bits of $a+b$ and $c+d$), a node $i$ only needs to write the rightmost $\frac{|i[m]|}{2}$ and $\frac{|i[n]|}{2}$ bits to its children. Thus, we can safely assume that when a node writes $m$ and $n$ to any of its children, then $m$ and $n$ of that child are of equal length and are of length in some power of 2. (If we do not do the book-keeping, the required number of nodes increases, this number is upper bounded by $|n|^{2\lg 3}$ as the number of levels is upper bounded by $2\lg |n|$; this observation was not made in \cite{Cesari1996}.) However, we do not show such book-keeping in the algorithm for brevity. Thus this algorithm would require $2\lg |n|$ levels, i.e., $|n|^{2\lg 3}$ nodes.\\~

\noindent\textbf{\textit{Computation of \textit{i[shift]}}}:
This algorithm utilizes $shift$ to compute $z$. A node $i$ updates $i[shift]$ by doubling the value of $shift$ from its children. A node $i$ evaluates that it is \imped because of an incorrect value of $i[shift]$ by evaluating the following macro. 
{
$$
\begin{array}{c}
    \textsc{\Imped-Multiplication-Karatsuba-Shift}(i)\equiv\\
    \begin{array}{l}
        \begin{cases}
            |i[m]|=1\land |i[n]|=1\land i[shift]\neq 0~~OR\\
            (3i)[shift]=(3i-1)[shift]=0\leq (3i+1)[shift]\land i[shift]\neq 1~~OR\\
            0<(3i)[shift]=(3i-1)[shift]\leq (3i+1)[shift]\land i[shift]\neq (3i)[shift]\times 2.\\
        \end{cases}
    \end{array}
\end{array}
$$
}

\noindent\textbf{\textit{Computation of \textit{i[m]} and \textit{i[n]}}}:
To ensure that the data flows down correctly, we declare a node $i$ to be \imped as follows.
{
$$\begin{array}{c}
    \textsc{\Imped-Multiplication-Karatsuba-TopDown}(i)\equiv\\
    \begin{array}{l}
        \begin{cases}
            i=1\land (i[m]\neq m\lor i[n]\neq n)~~OR\\
            ((|i[m]|>1\land |i[n]|>1)\land\\
            ((3i-1)[m] \neq i[m]\Big[1:\frac{|i[m]|}{2}\Big]~~OR\\
            (3i-1)[n] \neq i[n]\Big[1:\frac{|i[n]|}{2}\Big]~~OR\\
            (3i)[m] \neq i[m]\Big[\frac{|i[m]|}{2}+1:|i[m]|\Big]~~OR\\
            (3i)[n] \neq i[n]\Big[\frac{|i[n]|}{2}+1:|i[n]|\Big]~~OR\\
            (3i+1)[m]\neq i[m]\Big[1:\frac{|i[m]|}{2}\Big]+i[m]\Big[\frac{|i[m]|}{2}+1:|i[m]|\Big]~~OR\\
            (3i+1)[n]\neq i[n]\Big[1:\frac{|i[n]|}{2}\Big]+i[n]\Big[\frac{|i[n]|}{2}+1:|i[n]|\Big])).\\
        \end{cases}
    \end{array}
\end{array}
$$
}

\noindent\textbf{\textit{Computation of \textit{i[ans]}}}:
To determine if a node $i$ has stored $i[ans]$ incorrectly, it evaluates to be \imped as follows.
{
$$\begin{array}{c}
    \textsc{\Imped-Multiplication-Karatsuba-BottomUp}(i)\equiv\\
    \begin{array}{l}
        \begin{cases}
            |i[m]|=1 \land |i[n]|=1\land i[ans]\neq i[m]\times i[n]~~OR\\
            |i[m]|>1\land |i[n]|>1\land (i[ans]\neq lshift((3i-1)[ans], i[shift])\\
            \quad \quad +lshift((3i+1)[ans]-(3i-1)[ans]-(3i)[ans],(3i)[shift])\\
            \quad \quad +(3i+1)[ans])
        \end{cases}
    \end{array}
\end{array}
$$
}

Thus, \Cref{algorithm:parallel-karatsuba}
 is described as follows:
\newpage 
 
\begin{algorithm}\label{algorithm:parallel-karatsuba}Parallel processing version of Karatsuba's algorithm.
\end{algorithm}
{
$$
\begin{array}{|l|}
    \hline
    \text{\textit{Rules for node $i$}}.\\~\\
    \textsc{\Imped-Multiplication-Karatsuba-Shift}(i)\longrightarrow\\
    \begin{array}{l}
        \begin{cases}
            i[shift]=0 & \text{if $|i[m]|=1\land |i[n]|=1\land i[shift]\neq 0$.}\\
            i[shift]=1 & \text{if $(3i)[shift]=(3i-1)[shift]=0$}\\
             & \quad \quad \text{$\leq (3i+1)[shift]\land i[shift]\neq 1$}\\
            i[shift]=(3i)[shift]\times 2 & \text{otherwise}
        \end{cases}
    \end{array}
    ~\\
    \textsc{\Imped-Multiplication-Karatsuba-TopDown}(i)\longrightarrow\\
    \begin{array}{l}
        \begin{cases}
            i[m]=m,i[n]=n & \text{if $i=1\land (i[m]\neq m\lor i[n]\neq n)$.}\\
            (3i-1)[m] = i[m]\Big[1:\frac{|i[m]|}{2}\Big] & \text{if $(3i-1)[m] \neq i[m]\Big[1:\frac{|i[m]|}{2}\Big]$.}\\
            (3i-1)[n] = i[n]\Big[1:\frac{|i[n]|}{2}\Big] & \text{if $(3i-1)[n] \neq i[n]\Big[1:\frac{|i[n]|}{2}\Big]$.}\\
            (3i)[m] = i[m]\Big[\frac{|i[m]|}{2}+1:|i[m]|\Big] & \text{if $(3i)[m] \neq i[m]\Big[\frac{|i[m]|}{2}+1:|i[m]|\Big]$.}\\
            (3i)[n] = i[n]\Big[\frac{|i[n]|}{2}+1:|i[n]|\Big] & \text{if $(3i)[n] \neq i[n]\Big[\frac{|i[n]|}{2}+1:|i[n]|\Big]$.}\\
            (3i+1)[m]=i[m]\Big[1:\frac{|i[m]|}{2}\Big]\\ \quad +i[m]\Big[\frac{|i[m]|}{2}+1:|i[m]|\Big] & \text{if $(3i+1)[m]\neq i[m]\Big[1:\frac{|i[m]|}{2}\Big]$.}\\
             & \text{\quad \quad $+i[m]\Big[\frac{|i[m]|}{2}+1:|i[m]|\Big]$.}\\
            (3i+1)[n]=i[n]\Big[1:\frac{|i[n]|}{2}\Big]\\ \quad +i[n]\Big[\frac{|i[n]|}{2}+1:|i[n]|\Big] & \text{otherwise}\\
        \end{cases}
    \end{array}
    ~\\
    \textsc{\Imped-Multiplication-Karatsuba-BottomUp}(i)\longrightarrow\\
    \begin{array}{l}
        \begin{cases}
            i[ans]=i[m]\times i[n] & \text{if $|i[m]|=1 \land |i[n]|=1$}\\
            i[ans]=lshift((3i-1)[ans], i[shift])+\\
            \quad \quad lshift((3i+1)[ans]-(3i-1)[ans]\\
            \quad \quad \quad -(3i)[ans],(3i)[shift])+(3i+1)[ans]) & \text{otherwise.}
        \end{cases}
    \end{array}
    ~\\
    \hline
\end{array}
$$
}

\Cref{algorithm:parallel-karatsuba} converges in $O(|n|)$ time \cite{Cesari1996}, and its work complexity is $O(n^{\lg 3})$, which is the time complexity of the sequential Karatsuba's algorithm \cite{Cesari1996}. 

\begin{example}\label{example:1111times1111}
    \Cref{figure:100times100} evaluates $100\times 100$ following \Cref{algorithm:parallel-karatsuba}.  
    \qed 
\end{example}
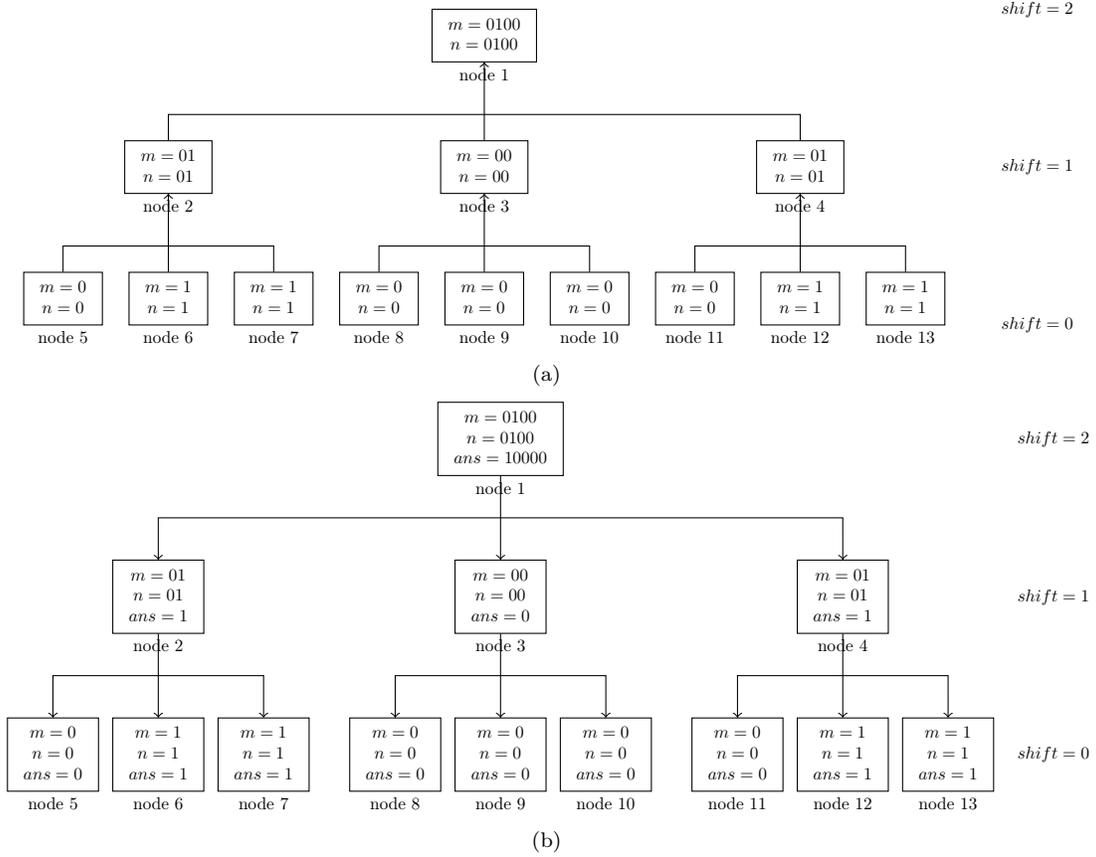
\begin{figure}[ht]
    \centering
    \subfigure[]{
        \begin{tikzpicture}[scale=.7,every node/.style={scale=0.65}]
            \node [draw=black,label=below:node 1] (a1) at (0,0) {\begin{tabular}{c}$m=0100$\\$n=0100$\end{tabular}};
            
            \node [draw=black,,label=below:node 2] (a2) at (-6,-2.5) {\begin{tabular}{c}$m=01$\\$n=01$\end{tabular}};
            \node [draw=black,,label=below:node 3] (a3) at (0,-2.5) {\begin{tabular}{c}$m=00$\\$n=00$\end{tabular}};
            \node [draw=black,,label=below:node 4] (a4) at (6,-2.5) {\begin{tabular}{c}$m=01$\\$n=01$\end{tabular}};
            
            \node [draw=black,,label=below:node 5] (a5) at (-8,-5) {\begin{tabular}{c}$m=0$\\$n=0$\end{tabular}};
            \node [draw=black,,label=below:node 6] (a6) at (-6,-5) {\begin{tabular}{c}$m=1$\\$n=1$\end{tabular}};
            \node [draw=black,,label=below:node 7] (a7) at (-4,-5) {\begin{tabular}{c}$m=1$\\$n=1$\end{tabular}};
            \node [draw=black,,label=below:node 8] (a8) at (-2,-5) {\begin{tabular}{c}$m=0$\\$n=0$\end{tabular}};
            \node [draw=black,,label=below:node 9] (a9) at (0,-5) {\begin{tabular}{c}$m=0$\\$n=0$\end{tabular}};
            \node [draw=black,,label=below:node 10] (a10) at (2,-5) {\begin{tabular}{c}$m=0$\\$n=0$\end{tabular}};
            \node [draw=black,,label=below:node 11] (a11) at (4,-5) {\begin{tabular}{c}$m=0$\\$n=0$\end{tabular}};
            \node [draw=black,,label=below:node 12] (a12) at (6,-5) {\begin{tabular}{c}$m=1$\\$n=1$\end{tabular}};
            \node [draw=black,,label=below:node 13] (a13) at (8,-5) {\begin{tabular}{c}$m=1$\\$n=1$\end{tabular}};
            
            \node at (10.5,.5) {$shift=2$};
            \node at (10.5,-2.5) {$shift=1$};
            \node at (10.5,-5.5) {$shift=0$};
            
            \draw (a2) -- (-6,-1.5) -- (0,-1.5); \draw[->] (a3) -- (a1); \draw (a4) -- (6,-1.5) -- (0,-1.5);
            
            \draw (a5) -- (-8,-4) -- (-6,-4); \draw[->] (a6) -- (a2); \draw (a7) -- (-4,-4) -- (-6,-4);
            \draw (a8) -- (-2,-4) -- (0,-4); \draw[->] (a9) -- (a3); \draw (a10) -- (2,-4) -- (0,-4);
            \draw (a11) -- (4,-4) -- (6,-4); \draw[->] (a12) -- (a4); \draw (a13) -- (8,-4) -- (6,-4);
        \end{tikzpicture}
    }
    \subfigure[]{
        \begin{tikzpicture}[scale=.7,every node/.style={scale=.65}]
            \node [draw=black,label=below:node 1] (a1) at (0,.5) {\begin{tabular}{c}$m=0100$\\$n=0100$\\$ans=10000$\end{tabular}};
            
            \node [draw=black,,label=below:node 2] (a2) at (-6.5,-2.5) {\begin{tabular}{c}$m=01$\\$n=01$\\$ans=1$\end{tabular}};
            \node [draw=black,,label=below:node 3] (a3) at (0,-2.5) {\begin{tabular}{c}$m=00$\\$n=00$\\$ans=0$\end{tabular}};
            \node [draw=black,,label=below:node 4] (a4) at (6.5,-2.5) {\begin{tabular}{c}$m=01$\\$n=01$\\$ans=1$\end{tabular}};
            
            \node [draw=black,,label=below:node 5] (a5) at (-8.5,-5.5) {\begin{tabular}{c}$m=0$\\$n=0$\\$ans=0$\end{tabular}};
            \node [draw=black,,label=below:node 6] (a6) at (-6.5,-5.5) {\begin{tabular}{c}$m=1$\\$n=1$\\$ans=1$\end{tabular}};
            \node [draw=black,,label=below:node 7] (a7) at (-4.5,-5.5) {\begin{tabular}{c}$m=1$\\$n=1$\\$ans=1$\end{tabular}};
            \node [draw=black,,label=below:node 8] (a8) at (-2,-5.5) {\begin{tabular}{c}$m=0$\\$n=0$\\$ans=0$\end{tabular}};
            \node [draw=black,,label=below:node 9] (a9) at (0,-5.5) {\begin{tabular}{c}$m=0$\\$n=0$\\$ans=0$\end{tabular}};
            \node [draw=black,,label=below:node 10] (a10) at (2,-5.5) {\begin{tabular}{c}$m=0$\\$n=0$\\$ans=0$\end{tabular}};
            \node [draw=black,,label=below:node 11] (a11) at (4.5,-5.5) {\begin{tabular}{c}$m=0$\\$n=0$\\$ans=0$\end{tabular}};
            \node [draw=black,,label=below:node 12] (a12) at (6.5,-5.5) {\begin{tabular}{c}$m=1$\\$n=1$\\$ans=1$\end{tabular}};
            \node [draw=black,,label=below:node 13] (a13) at (8.5,-5.5) {\begin{tabular}{c}$m=1$\\$n=1$\\$ans=1$\end{tabular}};
            
            \node at (10.5,.5) {$shift=2$};
            \node at (10.5,-2.5) {$shift=1$};
            \node at (10.5,-5.5) {$shift=0$};
            
            \draw[<-] (a2) -- (-6.5,-1) -- (0,-1); \draw[<-] (a3) -- (a1); \draw[<-] (a4) -- (6.5,-1) -- (0,-1);
            
            \draw[<-] (a5) -- (-8.5,-4) -- (-6.5,-4); \draw[<-] (a6) -- (a2); \draw[<-] (a7) -- (-4.5,-4) -- (-6.5,-4);
            \draw[<-] (a8) -- (-2,-4) -- (0,-4); \draw[<-] (a9) -- (a3); \draw[<-] (a10) -- (2,-4) -- (0,-4);
            \draw[<-] (a11) -- (4.5,-4) -- (6.5,-4); \draw[<-] (a12) -- (a4); \draw[<-] (a13) -- (8.5,-4) -- (6.5,-4);
        \end{tikzpicture}
    }
    \caption{Demonstration of multiplication of 100 and 100 in base 2: (a) top down (b) bottom up.
    }
    \label{figure:100times100}
\end{figure}

\subsubsection*{Lattice-Linearity}\label{subsubsection:ll-karatsuba}

\begin{lemma}\label{lemma:mul-karatsuba}
    Given the input bitstrings $n$ and $m$, the predicate 
    \begin{center}
        $\forall i:\lnot(\textsc{\Imped-Multiplication-Karatsuba-Shift}(i)\lor$\\
        $\textsc{\Imped-Multiplication-Karatsuba-TopDown}(i) \lor$\\
        $\textsc{\Imped-Multiplication-Karatsuba-BottomUp}(i))$
    \end{center}
    is lattice-linear on $|n|^{2\lg 3}$ computing nodes.
\end{lemma}

\begin{proof}
    For the global state to be optimal, in this problem, we require node 1 to store the correct multiplication result in $1[ans]$. To achieve this, each node $i$ must have the correct value stored in $i[n]$ and $i[m]$, and their children must store correct values of $n$, $m$ and $ans$ according to the values of $i[n]$ and $i[m]$. This in turn requires all nodes to store the correct $i[shift]$ values.

    Let us assume 
    for contradiction
    that node 1 does not store the correct value in $1[ans]$ as $n\times m$. This implies that (1) node 1 does not have an updated value in $1[n]$ or $1[m]$, or (2) node 1 has a non-updated value of $1[ans]$, (3) node 1 has not written the updated values to $2[n]$ \& $2[m]$, $3[n]$ \& $3[m]$ or $4[n]$ \& $4[m]$, (4) node 1 has a non-updated value in $1[shift]$, or (5) nodes 2, 3 or 4 have incorrect values in their respective $n$, $m$, $ans$ or $shift$ variables. In cases (1),...,(4), node 1 is \imped.
    
    Recursively, this can be extended to any node $i$. Let node $i$ has stored an incorrect value in $i[ans]$ or $i[shift]$. Let $i>1$. Then (1) node $i$ has a non-updated value in $i[shift]$, $i[ans]$, $i[n]$ or $i[m]$, or (2) if $|i[m]|>1$ or $|i[n]|>1$, node $i$ has not written updated values to $(3i-1)[n]$ \& $(3i-1)[m]$ or $(3i)[n]$ \& $(3i)[m]$ or $(3i+1)[n]$ \& $(3i+1)[m]$, in which case node $i$ is \imped. In both these cases, node $i$ is \imped. It is also possible that at least one of the children of node $i$ has incorrect values in its respective $n$, $m$, $ans$ or $shift$ variables.
    
    From these cases, we have that given a global state $s$, where $s=\langle\langle 1[n],$ $1[m]$, $1[ans]\rangle$, $\langle 2[n]$, $2[m]$, $2[ans]\rangle$, $...$
    $\rangle$, if $s$ is \imped, there is at least one node which is \imped. This shows that if the global state is \imped, then there exists some node $i$ which is \imped.
    
    Next, we show that if some node is \imped, then node 1 will not store the correct answer. Node 1 is \imped if it has not read the correct value $1[m]$ and $1[n]$. Additionally, $\forall i:i\in[1:n^{2\lg 3}]$ node $i$ is \imped if (1) it has non-updated values in $i[ans]$ or $i[shift]$, (2) $i$ has not written the correct values to $(3i-1)[n]$ \& $(3i-1)[m]$ or $(3i)[n]$ \& $(3i)[m]$ or $(3i+1)[n]$ \& $(3i+1)[m]$. This implies that the parent of $i$ will also store incorrect value in its $ans$ or $shift$ variable. Recursively, we have that node 1 stores an incorrect value in $1[ans]$. Thus, the global state is \imped.
    %
\end{proof}

{
With the arguments similar to those made in the proof of \Cref{lemma:mul-normal-predicate-self-stabilizing} and \Cref{theorem:algo-mul-normal-algo-ss}, we have the following.
\begin{lemma}
    The predicate
    \begin{center}
        $\forall i:\lnot(\textsc{\Imped-Multiplication-Karatsuba-Shift}(i)\lor$\\
        $\textsc{\Imped-Multiplication-Karatsuba-TopDown}(i) \lor$\\
        $\textsc{\Imped-Multiplication-Karatsuba-BottomUp}(i))$
    \end{center}
    is a lattice-linear self-stabilizing predicate.
\end{lemma}

\begin{theorem}
    \Cref{algorithm:parallel-karatsuba} is silent and self-stabilizing.
\end{theorem}
}

\noindent\textbf{\textit{Remark}}: 
{In \Cref{algorithm:parallel-karatsuba}, an \imped node updates the state of its children. We have done so for the brevity of the presentation of the algorithm. In practice, each child will notice that its local state does not tally with the state of its parent, and will then update its own state. 
}


\section{Parallel Processing Modulo Operation}\label{section:parallel-modulo}\label{section:mod-parallel}

In this section, we demonstrate parallel processing systems which can be used to compute a given modulo operation $n \mod m$.

\subsection{Using $|$\textit{n}$|$ Processors}\label{subsection:mod-n-processors}

In this subsection, we discuss a lattice-linear method to compute modulo operation which requires $|n|$ processors.
First, we discuss the key idea of the sequential algorithm, then we elaborate on the lattice-linearity of its parallelization.

\subsubsection*{Key Idea}

This algorithm is based on the standard modulo operation. Suppose that we have computed $a=n[1:|n|-1]\mod m$ and $b=n[|n|]$. Then we have that the resultant value of $n\mod m$ is $(a\times 2+b)\mod m$, which is also equal to $(lshift(a,1)+b)\mod m$.

\subsubsection*{Parallelization}

Every node, sequentially, reads a distinct bit of the input dividend $n$. Every node $i$ will eventually store the value of $n[1:i]$ under modulo $m$. The last node, indexed as node $|n|$, will store the final value, i.e. $n\mod m$. We demonstrate two ways of executing this algorithm. One way is with using the machine $M$ that we constructed in \Cref{subsection:dfa-modulo}. Another way is to perform the computation without $M$ where we use $\textsc{Division-Modulo}()$ that we defined in \Cref{subsection:long-division}. We demonstrate these methods in the following.

\subsubsection*{Using \textit{M}}
In this part, we will utilize $M$ to compute $n\mod m$. Since every node $i$ must store the value of $n[1:i]\mod m$, the \imped node $i$ can be defined as follows.
{
\begin{equation*}
    \textsc{\Imped-Linear-Modulo}(i)\equiv
    \begin{cases}
        i[ans] \neq n[1] & \text{if $i=1$}\\
        (i[ans]\neq M((i-1)[ans],n[i]) & \text{otherwise}
    \end{cases}
\end{equation*}
}
In the definition of $\textsc{\Imped-Linear-Modulo}(i)$, $M(i,j)$ means that $M$ is being invoked with an initial state $q_i$ and an input $j\in\{0,1\}$, i.e. $M(i,j)=\delta(q_i,j)$. If $\delta(q_i,j)=q_k$, then the execution of $M(i,j)$ will give $k$ as output. The algorithm to compute $n\mod m$ is demonstrated in \Cref{algorithm:linear-modulo-with-M}.

\begin{algorithm}\label{algorithm:linear-modulo-with-M} Computing modulo on $|n|$ processors using $M$.
\end{algorithm}
    $$
    \begin{array}{|l|}
        \hline
        \text{Rules for node $i$}.\\~\\
        \begin{array}{c}
            \textsc{\Imped-Linear-Modulo}(i)\longrightarrow\\
            \begin{cases}
                i[ans] = n[1] & \text{if $i=1$}.\\
                i[ans]=M((i-1)[ans],n[i]) & \text{otherwise}.
            \end{cases}
        \end{array}
        ~\\
        \hline
    \end{array}
    $$

The time complexity of this algorithm is $O(|n|)$. 
However, this method needs a preprocessing of $O(m\times |m|)$, which is quite high and impractical, especially if $m$ is large. We present this result only to demonstrate that some pre-processing can reduce the complexity of the modulo operation substantially.

\begin{theorem}
    Given the input bitstrings $n$ and $m$, the predicate 
    \begin{center}
        $\forall i:\lnot\textsc{\Imped-Linear-Modulo}(i)$
    \end{center}
    is lattice-linear on $|n|$ computing nodes.
\end{theorem}

\begin{proof}
    Let us assume that node $|n|$ has incorrect value in $|n|[ans]$. This implies that (1) node $|n|$ does not have an updated value in $(|n|)[ans]$, in which case node $|n|$ is \imped, or (2) node $|n|-1$ has an incorrect value in $(|n|-1)[ans]$.
    
    Recursively, this can be extended to any node $i$. Let that node $i$ has stored an incorrect value in $i[ans]$, then (1) node $i$ has a non-updated value in $i[ans]$, in which case, node $i$ is imedensable, or (2) node $i-1$ has an incorrect value in $(i-1)[ans]$. 
    From these cases, we have that given a global state $s$, where $s=\langle 1[ans], 2[ans],..., |n|[ans]\rangle$, if $s$ is \imped, there is at least one node which is \imped.
    
    This shows that if the global state is \imped, then there exists some node $i$ which is \imped.
    
    Next, we show that if some node is \imped, then node 1 will not store the correct answer. If node $i$ is \imped, then node $i$ has a non-updated value in $i[ans]$. 
    This implies that node $i+1$ will also store incorrect value in $(i+1)[ans]$. Recursively, we have that node $|n|$ stores an incorrect value in $|n|[ans]$, and thus the global state is \imped.
    %
\end{proof}

{
With the arguments similar to those made in the proof of \Cref{lemma:mul-normal-predicate-self-stabilizing} and \Cref{theorem:algo-mul-normal-algo-ss}, we have the following.
\begin{lemma}
    The predicate
    \begin{center}
        $\forall i:\lnot\textsc{\Imped-Linear-Modulo}(i)$
    \end{center}
    is a lattice-linear self-stabilizing predicate.
\end{lemma}

\begin{theorem}
    \Cref{algorithm:linear-modulo-with-M} is silent and self-stabilizing.
\end{theorem}
}

\subsubsection*{Using Long Division}

If we utilize $\textsc{Division-Modulo}()$ instead of $M$ in \Cref{algorithm:linear-modulo-with-M} (and, subtraction in place of $M$ in the definition of $\textsc{\Imped-Linear-Modulo}(i)$), then
every node takes $O(|m|)$ time because of the subtraction in \textsc{Division-Modulo}(), which implies that the total work complexity is $O(|n|\times|m|)$. Node $i$ will compute the value of $n[1:i]\mod m$ by the end of time-step $t$.
Therefore, the time complexity of this algorithm is $O(|n|\times|m|)$ to compute $n\mod m$, which is the same as the work complexity of this algorithm.

\subsubsection*{Discussion}

The behaviour of these methods is lattice-linear, but similar to a uniprocessor computation, in the sense that if we ran these methods on a uniprocessor machine, then it will take the same order of time. In \Cref{subsection:mod->n-processors}, we present algorithms which exploit the power of a distributed system better. 

\subsection{Using 4 $|$\textit{n}$|$/$|$\textit{m}$|$ $-$ 1 Processors}\label{subsection:mod->n-processors}

In this section, we present a parallel processing algorithm to compute $n\mod m$ using $4\times |n|/|m|-1$ computing nodes. 
First, we discuss the key idea of the sequential algorithm, then we elaborate on the lattice-linearity of its parallelization.

\subsubsection*{Key Idea}

This algorithm better parallizes the idea discussed in \Cref{subsection:mod-n-processors}. Suppose that we have computed $a=n[1:\lfloor |n|/2\rfloor]\mod m$ and $b=n[\lfloor |n|/2\rfloor+1:|n|]$. Then the resultant value of $n\mod m$ is $(a\times 2^{\lceil |n|/2\rceil}+b)\mod m$, which is also equal to $(lshift(a,\lceil |n|/2\rceil)+b)\mod m$.

\subsubsection*{Parallelization}

The algorithm induces a binary tree among the nodes based on their ids; there are $2\times|n|/|m|$ nodes in the lowest level (level 1).
This algorithm starts from the leaves where all leaves compute and store, in sequence, a substring of $n$ of length $|m|/2$ under modulo $m$. In the induced binary tree, the computed modulo result by sibling nodes at level $\ell$ is sent to the parent. Consecutively, those parents at level $\ell+1$, contiguously, store a larger substring of $n$ (double the bits that each of their children covers) under modulo $m$. We elaborate this procedure in this subsection.
This algorithm uses three variables to represent the state of each node $i$: $i[shift]$, $i[pow]$ and $i[ans]$.\\~

\noindent\textbf{Computation of \textit{i[shift]}: } 
The variable $shift$ stores the required power of 2. At any node at level 1, $shift$ is $0$. At level 2, the value of $shift$ at any node is $|m|/2$. At any higher level, the value of $shilft$ is twice the value of shift of its children. 
\textsc{\Imped-Log-Modulo-Shift}, in this context, is defined below. 

{
$$\begin{array}{c}
\textsc{\Imped-Log-Modulo-Shift}(i)\equiv\\
    \begin{cases}
        i[shift]\neq 0 & \text{if $i\geq 2\times|n|/|m|$}\\
        i[shift]\neq |m|/2 & \text{if $(2i)[shift]=(2i+1)[shift]=0$}\\
        i[shift]\neq2\times (2i)[shift] & \text{if $(2i)[shift]=(2i+1)[shift]\geq |m|/2$}
    \end{cases}
\end{array}
$$
}

\noindent\textbf{Computation of \textit{i[pow]}: } 
The goal of this computation is to set $i[pow]$ to be  $2^{i[shift]} \mod m$, whenever the level of $i$ is greater than 1. This can be implemented using the following definition for \textsc{\Imped-Log-Modulo-Pow}. 

$$
\begin{array}{l}
\textsc{\Imped-Log-Modulo-Pow}(i)\equiv\\
    \begin{cases}
        i[pow]\neq 1 & \text{if $i[shift]=0$}\\
        i[pow]\neq 2^{\frac{|m|}{2}} & \text{if $i[shift]=|m|/2$}\\
        i[pow]\neq((2i)[pow])^2\mod m & \text{otherwise}
    \end{cases}
\end{array}
$$

By definition, $i[pow]$ is less than $m$. Also, computation of $pow$ requires multiplication of two numbers that are upper-bounded by $|m|$. Hence, this computation can benefit from parallelization of \Cref{algorithm:parallel-karatsuba}. However, as we will see later, the complexity of this algorithm (for modulo) is dominated by the modulo operation happening in individual nodes which is $O(|m|^2)$, we can use the sequential version of Karatsuba's algorithm for multiplication, without affecting the order of the time complexity of this algorithm. 

\noindent\textbf{Computation of \textit{i[ans]}: }
We split $n$ into strings of size $\frac{|m|}{2}$, the number representing this substring is less than $m$. At the lowest level (level 1), $i[ans]$ is set to the corresponding substring. 
At higher levels, $i[ans]$ is set to $(i[pow]\times (2i)[ans] + (2i+1)[ans]) \mod m$. This computation also involves multiplication of two numbers whose size is upper bounded by $|m|$. An \imped node $i$ from a non-updated $i[ans]$ can be evaluated using \textsc{\Imped-Log-Modulo-Ans}$(i)$.

{
$$\begin{array}{c}
\textsc{\Imped-Log-Modulo-Ans}(i)\equiv\\
    \begin{cases}
        i[ans]\neq n[(i-2\times\dfrac{|n|}{|m|})\times \dfrac{|m|}{2}+1:(i-2\times\dfrac{|n|}{|m|}+1)\times \dfrac{|m|}{2}] & \text{if $i[shift]=0$}\\
        i[ans]\neq\textsc{Mod}(\textsc{Sum}(\textsc{Mul}((2i)[ans],i[pow]),(2i+1)[ans]),m) & \text{otherwise}
    \end{cases}
\end{array}
$$
}

We describe the algorithm as \Cref{algorithm:log-modulo}.

\begin{algorithm}\label{algorithm:log-modulo} Modulo computation by inducing a tree among the nodes.
\end{algorithm}
$$\begin{array}{|l|}
        \hline 
        \text{\textit{Rules for node} $i$.}\\~\\
        \textsc{\Imped-Log-Modulo-Shift}(i)\longrightarrow\\
        \begin{array}{ll}
            \begin{cases}
                    i[shift]= 0 & \text{if $i\geq 2\times|n|/|m|$}\\
                i[shift]= \dfrac{|m|}{2} & \text{if $(2i)[shift]=(2i+1)[shift]=0$}\\
                i[shift]= 2\times (2i)[shift] & \text{if $(2i)[shift]=(2i+1)[shift]\geq |m|/2$}
            \end{cases}
        \end{array}
        ~\\
        \textsc{\Imped-Log-Modulo-Pow}(i)\longrightarrow\\
        \begin{array}{ll}
            \begin{cases}
                i[pow]= 1 & \text{if $i[shift]=0$}\\
                i[pow]= 2^{\frac{|m|}{2}} & \text{if $i[shift]=|m|/2$}\\
                i[pow]= \textsc{Mod}(\textsc{Mul}(i[pow],i[pow]),m) & \text{otherwise}
            \end{cases}
        \end{array}
        ~\\
        \textsc{\Imped-Log-Modulo-Ans}(i)\longrightarrow\\
        \begin{array}{ll}
            \begin{cases}
                i[ans]= n[(i-2\times\dfrac{|n|}{|m|})\times \dfrac{|m|}{2}+1:(i-2\times\dfrac{|n|}{|m|}+1)\times \dfrac{|m|}{2}] & \text{if $i[shift]=0$}\\
                i[ans]= \textsc{Mod}(\textsc{Sum}(\textsc{Mul}((2i)[ans],i[pow]),(2i+1)[ans]),m) & \text{otherwise}
            \end{cases}
        \end{array}
        ~\\
        \hline 
    \end{array}
$$

\begin{example}
    \Cref{figure:11011mod11} shows the computation of $11011\mod 11$ as performed by \Cref{algorithm:log-modulo}.
    \qed 
\end{example}
\begin{figure}[ht]
    \centering
    \begin{tikzpicture}[scale=1.1,every node/.style={scale=.65}]
        \node [draw=black,label=below:{node 1}] (a1) at (3.5,3) {1+10 mod 11};
        
        \node [draw=black,label=below:{node 2}] (a2) at (1.5,2) {1 mod 11};
        \node [draw=black,label=below:{node 3}] (a3) at (5.5,2) {10+0 mod 11};
        
        \node [draw=black,label=below:{node 4}] (a4) at (.5,1) {00 mod 11};
        \node [draw=black,label=below:{node 5}] (a5) at (2.5,1) {01 mod 11};
        \node [draw=black,label=below:{node 6}] (a6) at (4.5,1) {10 mod 11};
        \node [draw=black,label=below:{node 7}] (a7) at (6.5,1) {11 mod 11};
        
        \node [draw=black,label=below:{node 8}] (a8) at (0,0) {0};
        \node [draw=black,label=below:{node 9}] (a9) at (1,0) {0};
        \node [draw=black,label=below:{node 10}] (a10) at (2,0) {0};
        \node [draw=black,label=below:{node 11}] (a11) at (3,0) {1};
        \node [draw=black,label=below:{node 12}] (a12) at (4,0) {1};
        \node [draw=black,label=below:{node 13}] (a13) at (5,0) {0};
        \node [draw=black,label=below:{node 14}] (a14) at (6,0) {1};
        \node [draw=black,label=below:{node 15}] (a15) at (7,0) {1};
        
        \draw[->] (a1) -- (a2); \draw[->] (a1) -- (a3);
        \draw[->] (a2) -- (a4); \draw[->] (a2) -- (a5); \draw[->] (a3) -- (a6); \draw[->] (a3) -- (a7);
        \draw[->] (a4) -- (a8); \draw[->] (a4) -- (a9); \draw[->] (a5) -- (a10); \draw[->] (a5) -- (a11); \draw[->] (a6) -- (a12); \draw[->] (a6) -- (a13); \draw[->] (a7) -- (a14); \draw[->] (a7) -- (a15);
        
        \node at (-2.5,0) {$pow = 1$};
        \node at (-2.5,1) {$pow = 10$};
        \node at (-2.5,2) {$pow = 100\mod 11~(=1)$}; \node at (-2.5,3) {$pow = 10000\mod 11~(=1)$};
        
        \node at (6,3) {(=0)};
    \end{tikzpicture}
    \caption{Processing $11011 \mod 11$ following \Cref{algorithm:log-modulo}.}
    \label{figure:11011mod11}
\end{figure}

\subsubsection*{Lattice-Linearity}

\begin{theorem}\label{theorem:mod-log}
    Given the input bitstrings $n$ and $m$, the predicate 
    \begin{center}
        $\forall i:\lnot(\textsc{\Imped-Log-Modulo-Shift}(i)\lor$\\
        $\textsc{\Imped-Log-Modulo-Pow}(i)\lor$\\
        $\textsc{\Imped-Log-Modulo-Ans}(i))$
    \end{center}
    is lattice-linear on $4|n|/|m|-1$ computing nodes.
\end{theorem}

\begin{proof}
    For the global state to be optimal, under this algorithm, we require node 1 to store the correct modulo result in $1[ans]$. To achieve this, each node $i$ must have the correct value of $i[ans]$. This in turn requires all nodes to store the correct $i[shift]$ and $i[pow]$ values.

    Let us assume
    for contradiction
    that node 1 has incorrect value in $1[ans]$. This implies that (1) node 1 has a non-updated value in $i[shift]$ or $i[pow]$, or (2) node 1 does not have an updated value in $i[ans]$. In both these cases, node 1 is \imped. It is also possible that node 2 or node 3 have an incorrect value in their variables.
    
    Recursively, this can be extended to any node $i$. Let node $i$ has stored an incorrect value in $i[ans]$. If $i<2(|n|/|m|)$, then (1) node $i$ has a non-updated value in $i[ans]$, $i[pow]$ or $i[shift]$, in which case node $i$ is \imped or (2) node $2i$ or node $2i+1$ have incorrect value in their respective $shift$, $pow$ or $ans$ variables. 
    If $i\geq 2(|n|/|m|)$, then $i$ does not have values $i[shift]=0$, $i[pow]=1$ or a correct $i[ans]$ value, in which case $i$ is \imped.
    From these cases, we have that given a global state $s$, where $s=\langle\langle 1[shift]$, $1[pow]$, $1[ans]\rangle$, $\langle 2[shift]$, $2[pow]$, $2[ans]\rangle$, $...$, $\langle (4|n|/|m|-1)[shift]$, $(4|n|/|m|-1)[pow]$, $(4|n|/|m|-1)[ans]\rangle\rangle$, if $s$ is \imped, there is at least one node which is \imped.
    
    This shows that if the global state is \imped, then there exists some node $i$ which is \imped.
    
    Next, we show that if some node is \imped, then node 1 will not store the correct answer.
    $\forall i:i\in[1:4|n|/|m|-1]$ node $i$ is \imped if it has non-updated values in $i[ans]$, $i[pow]$ or $i[shift]$. This implies that the parent of node $i$ also stores incorrect value in its $ans$ variable. Recursively, we have that node 1 stores an incorrect value in $1[ans]$, and thus the global state is \imped.
    %
\end{proof}

{
With the arguments similar to those made in the proof of \Cref{lemma:mul-normal-predicate-self-stabilizing} and \Cref{theorem:algo-mul-normal-algo-ss}, we have the following.
\begin{lemma}
    The predicate
    \begin{center}
        $\forall i:\lnot(\textsc{\Imped-Log-Modulo-Shift}(i)\lor$\\
        $\textsc{\Imped-Log-Modulo-Pow}(i)\lor$\\
        $\textsc{\Imped-Log-Modulo-Ans}(i))$
    \end{center}
    is a lattice-linear self-stabilizing predicate.
\end{lemma}

\begin{theorem}
    \Cref{algorithm:log-modulo} is silent and self-stabilizing.
\end{theorem}
}


\subsubsection*{Time Complexity Analysis}

\Cref{algorithm:log-modulo} is a general algorithm that uses the \textsc{Mod}( \textsc{Mul}($\cdots$)) and \textsc{Mod}( \textsc{Sum}($\cdots$)).
For some given $x,y$ and $z$ values, \textsc{Mod}(\textsc{Mul}($x,y$),$z$) (resp., \textsc{Mod} (\textsc{Sum}($x,y$),$z$)) involves first the multiplication (resp., addition) of two input values $x$ and $y$ and then evaluating the resulting value under modulo $z$.
These functions can be implemented in different ways. Choices for these implementations affect the time complexity. 
We consider the following approaches.

\subsubsection*{Modulo via Long Division}

First, we consider the standard approach for computing \textsc{Mod}(\textsc{Mul}($\cdots$)) and \textsc{Mod}(\textsc{Sum}($\cdots$)). Observe that in \Cref{algorithm:log-modulo}, if we compute  \textsc{Mod}(\textsc{Mul}$(x,y)$) then $x,y < m$. Hence, we can use Karatsuba's parallelized algorithm from \Cref{section:mul-parallel}, where both the input numbers are less than $m$. Using the analysis from \Cref{section:mul-parallel}, we have that each multiplication operation has a time complexity of $O(|m|)$.

Subsequently, to compute the mod operation, we need to compute $xy \mod m$ where $xy$ is upto $2|m|$ digits long. Using the standard approach of long division, we will need $|m|$ iterations where in each iteration, we need to do a subtraction operation with numbers that are $|m|$ digits long. Hence, the complexity of this approach is $O(|m|^2)$ per modulo operation. Since this complexity is higher than the cost of multiplication, the overall time complexity is $O(|m|^2\times \lg \dfrac{|n|}{|m|})$. 

\subsubsection*{Modulo by Using $M$}
The previous approach used $m$ and $n$ as inputs. Next, we consider the case where $m$ is hardcoded in the algorithm. As discussed in \Cref{section:mod-sequential}, we observe that these problems occur in practice. Our analysis is intended to provide lower bounds on the complexity of the modulo operation when $m$ is hardcoded. Similar to \Cref{subsection:dfa-modulo}, the pre-processing required in these algorithms makes them impractical in practice. However, we present them to show that there is a potential to reduce the complexity by some pre-processing. 

We can use \Cref{algorithm:parallel-karatsuba} for multiplication; each multiplication operation will have a time complexity of $O(|m|)$.
Subsequently, to compute the mod operation, we need to compute $xy \mod m$ where $xy$ is upto $2|m|$ digits long.  Using $M$, we will need $2m$ iterations; each iteration takes a constant amount of time. Hence, the complexity of this approach is $O(|m|)$ per modulo operation. Since this complexity is higher than the cost of multiplication, the overall time complexity is $O(|m|\times \lg \dfrac{|n|}{|m|})$. 

\subsubsection*{Modulo by Constructing Transition Functions}
In this part, we again consider the case where $m$ is hardcoded.
%
%
If $m$ is fixed, we can create a table $\delta_{sum}$ of size $m\times m$ where an entry at location $(i,j)$ represents $i+j \mod m$ in $O(m^2)$ time. Using $\delta_{sum}$, we can create another transition function $\delta_{mul}$ of size $m\times m$ where an entry at location $(i,j)$ represents $i\times j \mod m$ in $O(m^2)$ time. 

Using a preprocessed $\delta_{mul}$, the time complexity of a \textsc{Mod}(\textsc{Mul}($\cdots$)) operation becomes $O(1)$. As a result, the overall complexity of the modulo operation becomes $O(\lg \dfrac{|n|}{|m|})$. The pre-processing required in this method also is high. However, the effective time complexity of the modulo operation is reduced even more, as compared to the method that uses $M$, which is discussed above.

\section{Discussion on Common Properties of These Algorithms}\label{section:discussion}

In this section, we look at some common properties that are present in the problems and algorithms discussed in the preceding sections. Effectively, we also provide an alternate visualization to the abstraction of the lattices induced by the algorithms present in this chapter.

\subsection{Data Dependency Among Nodes}

In \Cref{subsubsection:ll-karatsuba}, for example, we showed how \Cref{algorithm:parallel-karatsuba} is lattice-linear by showing that given any suboptimal global state, we can point out specific nodes that are \imped. Any \imped node $i$ has only one choice of action, which implies that a total order is induced among all the local states that $i$ can visit. Such a total order, induced among the local states of every node, gives rise to the induction of a lattice among the global states.

Let that \textit{source} of the variable $i[var]$ in a node $i$ be the node that $i$ depends on to evaluate $i[var]$. For example, under \Cref{algorithm:parallel-karatsuba}, node $i$ depends on node $3i-1$, node $3i$ and node $3i+1$ to evaluate $i[ans]$. Thus the source nodes for node $i$ with respect to the evaluation of $i[ans]$ are node $3i-1$, node $3i$ and node $3i+1$. Similarly the source node of $i$ with respect to $i[m]$ or $i[n]$ is node $\Big\lfloor\dfrac{i+1}{3}\Big\rfloor$.

Let that $\textsc{Source}(i,var)$ is the set of nodes that are the source of $i$ with respect to $var$. Thus, under \Cref{algorithm:parallel-karatsuba}, for example, $\textsc{Source}(i,ans)=\{3i-1,3i,3i+1\}$. $\textsc{Source}(i,m)=\Big\{\Big\lfloor\dfrac{i+1}{3}\Big\rfloor\Big\}$. However, $\textsc{Source}(1,m)=\phi$ because $i$ is receiving $m$ as part of the input. Similarly, for all other algorithms, we can define the source nodes for all the nodes with respect to any given variable.

Let $\textsc{Variables}(i)$ be the set of the names of all variables of node $i$. We use a macro $\textsc{Depends}(i)$; a recursive definition for this macro is present in \Cref{figure:depends}.
\begin{figure}[ht]
    \centering
    \doublespacing 
    \begin{tabular}{|l|}
        \hline 
        $\textsc{Depends}(i)\supseteq \bigcup\limits_{var\in \textsc{Variables}(i)}\textsc{Source}(i,var)$.\\
        $\textsc{Depends}(i)\supseteq \bigcup\limits_{j\in\textsc{Depends}(i), var\in \textsc{Variables}(j)}\textsc{Source}(j,var)$.\\
        \hline 
    \end{tabular}
    \caption{Definition of the macro \texttt{Depends}.}
    \label{figure:depends}
\end{figure}

\subsection{Induction of $\prec$-lattice}

In \Cref{algorithm:parallel-karatsuba}, for $m$ and $n$, the information of $m$ and $n$ for $i$ is set based on the values of the parent of $i$. Hence, the $\textsc{Depends}(i)$ will contain all the ancestor nodes of $i$ in the tree. In addition, the $ans$ variable of $i$ is based on the children of $i$. Hence $\textsc{Depends}(i)$ will (also) contain the descendants of $i$ in the tree. For example, in \Cref{figure:100times100}, $\textsc{Depends}(3)=\{1,8,9,10\}$ and $\textsc{Depends}(8)=\{1,3\}$.

Let $\textsc{Is-Bad}(i,var)$ be true if and only if $i$ is impedensable with respect to some variable $var$, i.e., based on the values of the variables in $\textsc{Source}(i,var)$, $i$ has not computed $var$ correctly yet. We define state value of a node $i$ in a global state $s$ as follows. All the macros are also computed in the same global state $s$.
\begin{center}
    $\begin{array}{l}
        \textsc{State-Value}(i,s)=\\
        |\{var|var\in\textsc{Variables}(i):\textsc{Is-Bad}(i,var)\}|\\
        + |\{var|var\in \textsc{Variables}(j): j\in\textsc{Depends}(i,var): \textsc{Is-Bad}(j,var)\}|
    \end{array}$
\end{center}

We define the rank of a global state $s$ as follows.

\begin{equation*}
    \textsc{Rank}(s)=\sum\limits_{\text{each node $i$}}\textsc{State-Value}(i,s).
\end{equation*}


From the perspective of, for example, \Cref{algorithm:parallel-karatsuba}, a total order is induced among the local state visited by a node; $\textsc{State-Value}(i)$ describes the badness of the local state of a node $i$, which decreases monotonously as the nodes execute under \Cref{algorithm:parallel-karatsuba}.
Similarly, for all other algorithms, a total order is defined similarly using $\textsc{State-Value}(i)$.

As a consequence of the total order that is defined by $\textsc{State-Value}(i)$, a lattice among the global states can be observed with respect to the rank of the system; if the rank of a state $s$ is nonzero, then there is some node $i$ that is \imped in $s$. Let that only node $i$ changes its state and as a consequence, $s$ transitions to state $s'$. Then, we have that $s\prec s'$ where $s[i]\prec s'[i]$. This forms a $\prec$-lattice among the global states where $s[i]\prec s'[i]$ iff $\textsc{State-Value}(i,s)>\textsc{State-Value}(i,s')$ and $s\prec s'$ iff $\textsc{Rank}(s)>\textsc{Rank}(s')$. 
Rank is 0 at the supremum of the lattice, which is the optimal state.

From the above observation, we have that the system is able to converge from an arbitrary state to the required state within the expected number of time steps. This allows providing new inputs to a parallel processing system without needing to refresh variables of the nodes.

A problem is lattice-linear if it can be modelled in such a way that an \imped node must change its state in order for the system to reach the optimal state \cite{Garg2020}. 
From the above discussion, we have the following theorem about multiplication and modulo operations.


\section{Summary of the Chapter}\label{section:mul-mod-summary}

The contribution of this chapter is two-fold, one is applicative and the other is mathematical. First, we show that the parallelization of multiplication and modulo is lattice-linear. Due to lattice-linearity, we have that the algorithms we study in this chapter are tolerant to asynchrony. Second, we show two different distributive lattice structures, for both multiplication and modulo, which guarantee convergence in asynchronous environments. This chapter is the first work that shows that a lattice-linear problem can be solved under two different lattice structures. Specifically, considering (any) one of these problems in both the lattice structures, (1) the numbers of nodes are different, so the size of the global states in both the lattice structures is different, and (2) the numbers of children that a node has are different.

Multiplication and Modulo are among the fundamental mathematical operations. Fast parallel processing algorithms for such operations reduce the execution time of the applications which they are employed in. In this chapter, we showed that these problems are lattice-linear.
In this context, we studied parallelization of the standard multiplication and a parallelization of Karatsuba's algorithm. In addition, we studied parallel processing algorithms for the modulo operation. 

The presence of lattice-linearity in problems and algorithms allows nodes to execute asynchronously. This is specifically valuable in parallel algorithms where synchronization can be removed as is. 
These algorithms are snap-stabilizing, which means that the state transitions of the system strictly follow its specification.
They are also self-stabilizing, i.e., the supremum states in the lattices induced under the respective predicates are the optimal states. 

Utilizing these algorithms, the available cluster or GPU power can be used to compute the multiplication and modulo operations on big-number inputs. In this case, a synchronization primitive also does not need to be deployed. Also, the circuit does not need to be refreshed before providing it with a new input. This is also very fruitful, for example, in Karatsuba's multiplication the time that it would take to refresh the circuit is $O(|n|^{2\lg 3})$, but even without refreshing, we obtain the final answer in $O(n)$ time. This shows the gravity of the utility of the self-stabilizing property of these algorithms. Thus a plethora of applications will benefit from the observations presented in this chapter.

\section{Non-Lattice-Linear Problems}\label{section:nllp}

Unlike the problems that we studied in this chapter, certain problems are \textit{non-lattice-linear problems}. In those problems, given a suboptimal global state, the problem does not stipulate, from a specific set of nodes, to change their state. In such problems, there are instances in which the \imped nodes cannot be determined naturally, i.e., in those instances
$\exists s :\lnot\mathcal{P}(s) \wedge   (\forall i : \exists s' : \mathcal{P}(s')\land s[i]=s'[i]$).
For such problems, $\prec$-lattices may be induced algorithmically, through \textit{lattice-linear algorithms}.

In \Cref{chapter:ella} (and \ref{chapter:flla}), we study non-lattice linear problems. Minimal dominating set, minimal vertex cover and maximal independent set are examples of non-lattice linear problems: in such problems for any subject node $i$, an optimal global state can be reached without changing the local state of $i$.
\chapter{EVENTUALLY LATTICE-LINEAR ALGORITHMS}\label{chapter:ella}

In \Cref{chapter:mulmod}, we study examples of lattice-linear problems and present self-stabilizing algorithms that guarantee convergence in asynchrony. These algorithms are capable of converging in asynchrony because they exploit the property of lattice-linearity of the problems that they are developed for.

In this chapter, we study whether lattice-linearity can be extended for problems that are not lattice-linear.
This is one of the issues that were pointed out in \cite{Garg2020}: whether non-lattice-linear problems can be solved under the model of lattice-linearity. The behaviour of non-lattice-linear problems makes it seem impossible since the nature of these problems does not provide the definition of \imped nodes. However, we find that a total order can be induced among the local states of nodes algorithmically, even if the problem does not does not define how \imped nodes can be identified.

In this chapter, we introduce the class of \textit{eventually lattice-linear algorithms}.
%
We present eventually lattice-linear self-stabilizing algorithms for service demand based minimal dominating set (\sdmds), minimal vertex cover (\mvc), maximal independent set (\mis), graph colouring (\gc) and 2-dominating set (\tds) problems.

Eventually lattice-linear algorithms induce lattices in a subset of the state space.
These algorithms first (1) guarantee that from any arbitrary state, the system reaches a state in one of the induced lattices, and then (2) these algorithms behave lattice-linearly, and make the system traverse that lattice and reach an optimal state.

We proceed as follows. We begin with the SDMDS problem which is a generalization of the minimal dominating set. We devise a self-stabilizing algorithm for SDMDS. We scrutinize this algorithm and decompose it into two parts, the second of which satisfies the lattice-linearity property of \cite{Garg2020} if it begins in a \textit{feasible} state.  Furthermore, the first part of the algorithm ensures that the algorithm reaches a \textit{feasible} state. We show that the resulting algorithm is self-stabilizing, and the algorithm 
has 
\textit{limited-interference} property (discussed in \Cref{subsection:ds-eventual}) due to which it is tolerant to the nodes reading old values of other nodes. 
We also demonstrate that this approach is generic. It applies to various other problems including \mvc \mis, \gc and \tds. 

The algorithms for \sdmds, \mvc and \mis converge in 1 round plus $n$ moves, the algorithm for \gc converges in $n+4m$ moves, and the algorithm for \tds converges in 1 round plus $2n$ moves. Adding to the fact that these algorithms do not require a synchronous environment to execute, these results are an improvement over the algorithms present in the literature.

We also present some experimental results that show the efficacy of eventually lattice-linear algorithms in real-time shared memory systems. Specifically, we compare our algorithm for \mis (\Cref{algorithm:rules-mis}) with algorithms presented in \cite{Hedetniemi2003} and \cite{Turau2007}. The experiments are conducted in \texttt{cuda} environment, which is built on shared memory model.

The algorithms present in this chapter tolerate asynchrony in AMR model (cf. \Cref{section:asynchrony}).

\subsubsection*{Organization of the Chapter}

\noindent This chapter is organized as follows. 
In \Cref{section:sdmds-algorithm}, we describe the algorithm for the service demand based minimal dominating set problem. In \Cref{section:sdmds-lattice-linear}, we analyze the characteristics of that algorithm and show that it is eventually lattice-linear. We use the structure of eventually lattice-linear self-stabilizing algorithms to develop algorithms for minimal vertex cover, maximal independent set, graph colouring and 2-dominating set problems, respectively, in Sections \ref{section:mvc}, \ref{section:mis}, \ref{section:gc} and \ref{section:2ds}.
Then, in \Cref{section:experiments}, we compare the convergence speed of the algorithm presented in \Cref{section:mis} with other algorithms (for the maximal independent set problem) in the literature (specifically, \cite{Hedetniemi2003} and \cite{Turau2007}).
Finally, we summarize the chapter in \Cref{section:ellss-summary}.

\section{Service Demand based Minimal Dominating Set}\label{section:sdmds-algorithm}

In this section, 
we introduce a generalization of the minimal dominating set (\mds) problem (\Cref{subsection:pd:ds}),
the service demand based minimal dominating set (\sdmds) problem, and describe an algorithm to solve it (\Cref{subsection:ds-general-algorithm}). 

\subsection{Problem Description}\label{subsection:pd:ds}

The \sdmds problem, a generalization of \mds, is a simulation, on an arbitrary graph $G$, in which all nodes have some demands to be fulfilled and they offer some services. If a node $i$ is in the dominating set then it can not only serve all its own demands $D_i$, but also offer services from, its set of services $S_i$, to its neighbours. If $i$ is not in the dominating set, then it is considered dominated only if each of its demands in $D_i$ is being served by at least one of its neighbours that is in the dominating set.

\begin{definition}\textbf{Service demand based minimal dominating set problem (SDMDS)}.
    In the \textit{service demand based minimal dominating set} problem, the input is a graph $G$ and a set of services $S_i$ and a set of demands $D_i$ for each node $i$ in $G$; the task is to compute a minimal set $\mathcal{D}$ such that for each node $i$,
    \begin{enumerate}
        \item either $i\in \mathcal{D}$, or
        \item for each demand $d$ in $D_i$, there exists at least one node $j$ in $Adj_i$ such that $d\in S_j$ and $j\in \mathcal{D}$.
    \end{enumerate}
\end{definition}

In the above generalization of the \mds problem, if all nodes have same set $X$ as their services and demands, i.e., $\forall i: S_i=X$ and $D_i=X$, then it is equivalent to \mds.

In the following subsection, we present a self-stabilizing algorithm for the minimal SDMDS problem.
Each node $i$ is associated with variable $i[st]$ with domain $\{IN, OUT\}$. $i[st]$ defines the state of $i$. We define $\mathcal{D}$ to be the set $\{i\in V(G): i[st]=IN\}$. 

\subsection{Algorithm for SDMDS Problem}\label{subsection:ds-general-algorithm}

The list of constants, provided with the input, is in \Cref{table:constants-sdmds}.

\begin{table}[ht]
    \centering 
    \doublespacing 
    \begin{tabular}{|l|l|}
        \hline
        Constant & What it stands for\\
        \hline
        $D_i$ & the set of demands of node $i$.\\
        $S_i$ & the set of services provided by node $i$.\\
        \hline
    \end{tabular}
    \caption{Constants provided with the input.}
    \label{table:constants-sdmds}
\end{table}

The macros that we utilize are described in \Cref{table:macros-sdmds-ell}.
Recall that $\mathcal{D}$ is the set of nodes which currently have the state as $IN$. A node $i$ is \textit{addable} if there is at least one demand of $i$ that is not being serviced by any neighbour of $i$ that is in $\mathcal{D}$.
A node $i$ is \textit{removable} if $\mathcal{D}\setminus\{i\}$ is also a dominating set given that $\mathcal{D}$ is a dominating set.
The \textit{dominators} of $i$ are the nodes that are (possibly) dominating node $i$: if some node $j$ is in \textsc{Dominators-Of}($i$), then there is at least one demand $d\in D_i$ such that $d\in S_j$.
$i$ is \textit{\imped} if $i$ is removable and there is no node $k$ that is removable and is of an ID higher than $i$, such that $k$ and $i$ are able to serve for some common node $j$.



\begin{table}[ht]
    \centering 
    \doublespacing 
    \begin{tabular}{|l|}
        \hline
        $\mathcal{D}\equiv \{i\in V(G): i[st] =$ $IN\}$.\\
        \textsc{Addable-SDMDS-ELL}($i)\equiv i[st]=OUT\land$\\
        \quad\quad\quad\quad $(\exists d\in D_i, \forall j\in Adj_i: d\not\in S_j\lor j[st]=OUT)$.\\
        \textsc{Removable-SDMDS-ELL}$ (i)\equiv (\forall d \in D_i : (\exists j \in Adj_i: d \in S_j \land j[st]=$ $IN))\land$\\
        \quad\quad\quad\quad $(\forall j \in Adj_i,\forall~d \in D_j:d\in S_i\Rightarrow$\\
        \quad\quad\quad\quad $(\exists k \in Adj_j, k\neq i:(d \in S_k \land k[st] =$ $IN)))$.\\
        \textsc{Dominators-Of}($i)\equiv$\\
        \quad\quad\quad\quad $\{j\in Adj_i, j[st]=IN:\exists d\in D_i:d\in S_j\}\cup\{i\}$\quad if $i[st]=IN$\\
        \quad\quad\quad\quad $\{j\in Adj_i, j[st]=IN:\exists d\in D_i:d\in S_j\}$\quad \quad \quad \quad otherwise.\\
        \textsc{Impedensable-SDMDS-ELL}$(i)\equiv i[st]=IN\land$ \textsc{Removable-SDMDS-ELL}$(i)\land$\\
        \quad\quad\quad\quad $(\forall j \in Adj_i,\forall~d \in D_j:d\in S_i\Rightarrow$\\
        \quad\quad\quad\quad $((\forall k \in$ \textsc{Dominators-Of}$(j)$, $k\neq i:(d \in S_k \land k[st] =$ $IN))\Rightarrow$\\
        \quad\quad\quad\quad $(k[id]<i[id]\lor\lnot$\textsc{Removable-SDMDS-ELL}$(k))))$.\\
        \hline
    \end{tabular}
    \caption{Macros used in the algorithm for SDMDS problem.}
    \label{table:macros-sdmds-ell}
\end{table}

The general idea our algorithm is as follows. 
\begin{enumerate}
    \item A node enters the dominating set unconditionally if it is addable.
    This ensures that $G$ enters a state where the set of nodes in $\mathcal{D}$ form a (possibly non-minimal) dominating set. 
    If $\mathcal{D}$ is a dominating set, we say that the corresponding state is a \textit{feasible} state. 
  
    \item While entering the dominating set is not lattice-linear, the instruction governing the leaving of the dominating set is lattice-linear. 
    Node $i$ leaves the dominating set iff it is \imped. 
    Specifically, if $i$ serves for a demand $d$ in $D_j$ where $j \in Adj_i$ and the same demand is also served by another node $k$ ($k\in Adj_j$) then $i$ leaves only if (1) $k[id] < i[id]$ or (2) $k$ is not removable. 
    This ensures that if some demand $d$ of $D_j$ is satisfied by both $i$ and $k$ both of them cannot leave the dominating set simultaneously. 
    This ensures that $j$ will remain dominated. 
\end{enumerate} 
Thus, the rules for \Cref{algorithm:rules-ds} are as follows: 

\begin{algorithm}\label{algorithm:rules-ds}Rules for node $i$.
    \begin{center}
        \begin{tabular}{|l|}
            \hline
            \textsc{Addable-SDMDS-ELL}$(i)\longrightarrow i[st]=IN$.\\
            \textsc{Impedensable-SDMDS-ELL}$(i)\longrightarrow i[st]=OUT$.\\
            \hline
        \end{tabular}
    \end{center}
\end{algorithm}


We decompose  \Cref{algorithm:rules-ds} into two parts: (1) \Cref{algorithm:rules-ds}.1, that only consists of first guard and action of \Cref{algorithm:rules-ds} and (2) \Cref{algorithm:rules-ds}.2, that only consists of the second guard and action of \Cref{algorithm:rules-ds}. We use this decomposition in the following section of this chapter to relate the algorithm to eventual lattice-linearity.


    
            


\section{Lattice-Linear Characteristics of the Algorithm for \sdmds}\label{section:sdmds-lattice-linear}

In this section, we analyze the characteristics of \Cref{algorithm:rules-ds} to demonstrate that it is eventually lattice-linear. We proceed as follows.
In \Cref{subsection:ds-propositions}, we state the propositions which define the feasible and optimal states of the SDMDS problem, along with some other definitions. In \Cref{subsection:guarantee-feasible}, we show that $G$ reaches a state where it manifests a (possibly non-minimal) dominating set.
In \Cref{subsection:ds-action2}, we show that after when $G$ reaches a feasible state, \Cref{algorithm:rules-ds} behaves like a lattice-linear algorithm. 
In \Cref{subsection:termination}, we show that when $\mathcal{D}$ is a minimal dominating set, no nodes are enabled. 
In \Cref{subsection:ds-eventual}, we argue that because there is a bound on interference between \Cref{algorithm:rules-ds}.1 and \Cref{algorithm:rules-ds}.2 even when the nodes read old values, \Cref{algorithm:rules-ds} is an eventually lattice-linear self-stabilizing (ELLSS) algorithm.
In \Cref{subsection:time-space-complexity-analysis}, we study the time and space complexity attributes of \Cref{algorithm:rules-ds}.

\subsection{Propositions Stipulated by the SDMDS Problem}\label{subsection:ds-propositions}

The SDMDS problem stipulates that the nodes whose state is $IN$ must collectively form a dominating set. We represent this proposition as $\mathcal{P}_{sdmds}^\prime$.
\begin{center}
    $\mathcal{P}_{sdmds}^\prime(\mathcal{D}) \equiv \forall i\in V(G):(i\in \mathcal{D}\lor (\forall d\in D_i,\exists j\in Adj_i: (d\in S_j\land j\in \mathcal{D})))$.
\end{center}
The SDMDS problem stipulates an additional condition that $\mathcal{D}$ should be a minimal dominating set. We represent this proposition as $\mathcal{P}_{sdmds}$.

\begin{center}
    $\mathcal{P}_{sdmds}(\mathcal{D})\equiv \mathcal{P}^\prime_{sdmds}(\mathcal{D})\land(\forall i\in \mathcal{D}, \lnot\mathcal{P}_{sdmds}^\prime(\mathcal{D}\setminus\{i\}))$.
\end{center}

If $\mathcal{P}_{sdmds}^\prime(\mathcal{D})$ is true, then $G$ is in a \textit{feasible} state. And, if $\mathcal{P}_{sdmds}(\mathcal{D})$ is true, then $G$ is in an \textit{optimal} state. 

Based on the above definitions, we define two scores with respect to the global state, \textsc{Rank} and \textsc{Badness}.
\textsc{Rank} determines the number of nodes needed to be added to $\mathcal{D}$ to change $\mathcal{D}$ to a dominating set. 
\textsc{Badness} determines the number of nodes that are needed to be removed from $\mathcal{D}$ to make it a minimal dominating set, given that $\mathcal{D}$ is a (possibly non-minimal) dominating set.

\begin{definition}
    $\textsc{Rank}(\mathcal{D})\equiv\min\{|\mathcal{D}^\prime|-|\mathcal{D}|:\mathcal{P}^\prime_d(\mathcal{D}^\prime)\land \mathcal{D}\subseteq \mathcal{D}^\prime\}$.
\end{definition}

\begin{definition}
    $\textsc{Badness}(\mathcal{D})\equiv\max\{|\mathcal{D}|-|\mathcal{D}^\prime|:\mathcal{P}^\prime_d(\mathcal{D}^\prime)\land \mathcal{D}^\prime\subseteq \mathcal{D}\}$.
\end{definition}

\subsection{Guarantee to Reach a Feasible State by \Cref{algorithm:rules-ds}.1}\label{subsection:guarantee-feasible}

We show that under \Cref{algorithm:rules-ds}.1, $G$ is guaranteed to reach a feasible state. 

\begin{lemma}\label{lemma:d-not-ds}
    Let $t.\mathcal{D}$ be the value of $\mathcal{D}$ at the beginning of round $t$. 
If $t.\mathcal{D}$ is not a dominating set then $(t+1).\mathcal{D}$ is a dominating set.
\end{lemma}

\begin{proof}
    Let $i$ be a node such that $i\in t.\mathcal{D}$ and $i\not\in(t+1).\mathcal{D}$, {i.e., $i$ leaves the dominating set in round $t$}. This means that $i$ remains dominated and all nodes in $Adj_i$ remain dominated, even when $i$ is removed. This implies that $i$ will not reduce the feasibility of $t.D$; it will not increase the value of \textsc{Rank}.
    
    Now let $\ell$ be a node such that $\ell\not \in t.\mathcal{D}$ which is addable when it evaluates its guards in round $t$. This implies that $\exists~d\in D_\ell$ such that $d$ is not present in $S_j$ for any $j\in Adj_\ell$ that is in the dominating set. According to the algorithm, the guard of the second action is true for $\ell$. This implies that $\ell[st]$ will be set to $IN$. 
    
    It can also be possible for the node $\ell$ that it is not addable when it evaluates its guards in round $t$. This may happen if some other nodes around $\ell$ already decided to move to $\mathcal{D}$, and as a result $\ell$ is now dominated. Hence $\ell\not\in(t+1).\mathcal{D}$ and we have that $\ell$ is dominated at round $t+1$.
    
    Therefore, we have that $(t+1).\mathcal{D}$ is a dominating set, which may or may not be minimal.
\end{proof}

From \Cref{lemma:d-not-ds}, we have that if at the beginning of some round, $G$ is in a state where $\textsc{Rank} >0$, then by the end of that round, \textsc{Rank} will be $0$.

\subsection{Lattice-Linearity of \Cref{algorithm:rules-ds}.2}\label{subsection:ds-action2}

In the following lemma, we show that \Cref{algorithm:rules-ds}.2 is lattice-linear.

\begin{lemma}\label{lemma:ds-addition}
    If $t.\mathcal{D}$ is a non-minimal dominating set then under \Cref{algorithm:rules-ds} (more specifically, \Cref{algorithm:rules-ds}.2),
    there exists at least one node such that $G$ cannot reach a minimal dominating set until that node is removed from the dominating set.
\end{lemma}

\begin{proof}
Since $\mathcal{D}$ is a dominating set, the first guard is false for all nodes in $G$.

Since $\mathcal{D}$ is not minimal, there exists at least one node that must be removed in order to make $\mathcal{D}$ minimal. Let $S^\prime$ be the set of nodes which are removable. 
Let $x$ be some node in $S^\prime$. If $x$ is not serving any node, then \textsc{Impedensable-SDMDS-ELL}($x$) is trivially true. Otherwise there exists at least one node $j$ which is served by $x$, that is, $\exists d\in D_j:d\in S_x$. We study two cases which are as follows:
(1) for some
node $j$ served by $x$, there does not exist another node $b \in S^\prime$ which serves $j$, and (2) for any node $b \in S^\prime$ such that $x$ and $b$ serve some common node $j$, $b[id]<M[id]$. 

In the first case, $x$ cannot be removed because \textsc{Impedensable-SDMDS-ELL}($x$) is false and, hence, $x$ cannot be in $S^\prime$, thereby leading to a contradiction.
In the second case, \textsc{Impedensable-SDMDS-ELL}($x$) is true and \textsc{Impedensable-SDMDS-ELL}($b$) is false since $b[id] < M[id]$. Thus, node $b$ cannot leave the dominating set until $x$ leaves. In both the cases, we have that $j$ stays dominated.

Since ID of every node is distinct, we have that there exists at least one node $x$ for which \textsc{Impedensable-SDMDS-ELL}($x$) is true. For example, \textsc{Impedensable-SDMDS-ELL} is true for the node with the highest ID in $S^\prime$; $G$ cannot reach a minimal dominating set until $x$ is removed from the dominating set.
\end{proof}

From \Cref{lemma:ds-addition}, it follows that \Cref{algorithm:rules-ds}.2 satisfies the condition of lattice-linearity as described in \Cref{section:lattice-linearity}. It follows that if we start from a state where $\mathcal{D}$ is a (possibly non-minimal) dominating set and execute \Cref{algorithm:rules-ds}.2 then it will reach  a state where $\mathcal{D}$ is a minimal dominating set even if nodes are executing with old information about others. Next, we have the following result which follows from \Cref{lemma:ds-addition}.

\begin{lemma}\label{lemma:ds-removal}
    Let $t.\mathcal{D}$ be the value of $\mathcal{D}$ at the beginning of round $t$. 
    If $t.\mathcal{D}$ is a non-minimal dominating set then $|(t+1).\mathcal{D}| \leq |t.\mathcal{D}|-1$, and $(t+1).\mathcal{D}$ is a dominating set.
\end{lemma}

\begin{proof}
From \Cref{lemma:ds-addition}, {at least one node $x$ (including the maximum ID node in $S^\prime$} from the proof of \Cref{lemma:ds-addition}) would be removed  in round $t$. Furthermore, since $\mathcal{D}$ is a dominating set, \textsc{Addable-SDMDS-ELL}($i$) is false at every node $i$. Thus, no node is added to $\mathcal{D}$ in round $t$. Thus, the $|(t+1).\mathcal{D}|\leq |t.\mathcal{D}|-1$. 
    
    For any node $x$ that is removable, \textsc{Impedensable-SDMDS-ELL}($i$) is true only if any node $j$ which is (possibly) served by $x$ has other neighbours (of a lower ID) which serve the demands which $x$ is serving to it. This guarantees that $j$ stays dominated and hence $(t+1).\mathcal{D}$ is a dominating set.
\end{proof}

\subsection{Termination of \Cref{algorithm:rules-ds}}\label{subsection:termination}

The following lemma studies the action of \Cref{algorithm:rules-ds} when $\mathcal{D}$ is a minimal dominating set.

\begin{lemma}\label{lemma:ds-minimal}
Let $t.\mathcal{D}$ be the value of $\mathcal{D}$ at the beginning of round $t$. If $\mathcal{D}$ is a minimal dominating set, then $(t+1).\mathcal{D}=t.\mathcal{D}$.

\end{lemma}

\begin{proof}
    Since $\mathcal{D}$ is a dominating set, \textsc{Addable-\sdmds-ELL}($i$) is false for every node in $V(G)$, i.e., the first action is disabled for every node in $V(G)$.
    Since $\mathcal{D}$ is minimal, \textsc{Impedensable-SDMDS-ELL}($i$) is false for every node $i$ in $\mathcal{D}$. Hence, the second action is disabled at every node $i$ in $\mathcal{D}$. 
    Thus, $\mathcal{D}$ remains unchanged.
\end{proof}

\subsection{Eventual Lattice-Linearity of \Cref{algorithm:rules-ds}}\label{subsection:ds-eventual}

\Cref{lemma:ds-addition} showed that \Cref{algorithm:rules-ds}.2 is lattice-linear. 
In this subsection, we make additional observations about \Cref{algorithm:rules-ds} to generalize the notion of lattice-linearity to eventually lattice-linear algorithms. 
We have the following observations.
\begin{enumerate}
    \item From \Cref{lemma:d-not-ds}, starting from any state, \Cref{algorithm:rules-ds} will reach a feasible state even if a node reads old information about the neighbours. This is due to the fact that \Cref{algorithm:rules-ds}.1 only adds nodes to $\mathcal{D}$. 
    If incorrect information about the state of neighbours causes $i$ not to be added to $\mathcal{D}$, this will be corrected when $i$ executes again and obtains recent information about neighbours. If incorrect information causes $i$ to be added to $\mathcal{D}$ unnecessarily, it does not affect this claim. 
    \item From \Cref{lemma:ds-addition}, if we start $G$ in a feasible state where no node has incorrect information about the neighbours in the initial state then \Cref{algorithm:rules-ds}.2 reaches a minimal dominating set. Note that this claim remains valid even if the nodes execute actions of \Cref{algorithm:rules-ds}.2 with old information about the neighbours as long as the initial information they use is correct. 
    \item We observe that \Cref{algorithm:rules-ds}.1 and \Cref{algorithm:rules-ds}.2 have very limited interference with each other, and so an arbitrary graph $G$ will reach an optimal state even if nodes are using old information. 
    
    
\end{enumerate}

From the above observations, if we allow the nodes to read old values, then the nodes can violate the feasibility of $G$ finitely many times and so $G$ will eventually reach a feasible state and stay there forever. 
We introduce the class of eventually lattice-linear algorithms (ELLA). \Cref{algorithm:rules-ds} is an ELLA.


\begin{definition}\label{definition:ella} \textbf{Eventually Lattice-Linear Algorithms (ELLA).}
    An algorithm $A$ is ELLA for a problem $P$, represented by a predicate $\mathcal{P}$, if its rules can be split into two sets of rules $F_1$ and $F_2$ and there exists a subset $S_f$ of the state space $S$, such that 
    
    \begin{enumerate}[label=(\alph*)]
        \item Any computation of $A$ (from its permitted initial states)
        eventually reaches a state
        where $S_f$ is stable in $A$,
        i.e., $S_f$ is true and remains true subsequently. 
        \item Rules in $F_1$ are disabled in a state in $S_f$.
        \item $F_2$ is a lattice-linear algorithm, i.e., it induces a total order among the local states visited by the nodes, given that the system initializes in a state in $S_f$.
    \end{enumerate}
\end{definition}

\begin{definition}\label{definition:ellssa} \textbf{Eventually Lattice-Linear Self-Stabilizing (ELLSS) Algorithms.}
    Continuing from \Cref{definition:ella}, $A$ is an ELLSS algorithm iff $F_1$ takes the system to a state in $S_f$ from an arbitrary state, and $F_2$ is capable of taking the system from any state in $S_f$ to an optimal state.
\end{definition}

\noindent\textbf{\textit{Remark}}: The algorithms that we study in this chapter are ELLSS algorithms, i.e., they follow \Cref{definition:ellssa}. Notice that \Cref{algorithm:rules-ds} is an ELLSS algorithm.

In \Cref{algorithm:rules-ds}, $F_1$ corresponds to \Cref{algorithm:rules-ds}.1 and $F_2$ corresponds to \Cref{algorithm:rules-ds}.2.
This algorithm satisfies the properties of \Cref{definition:ellssa}. 


\begin{example}\label{example:4-nodes.ella}
    We illustrate the eventual lattice-linear structure of \Cref{algorithm:rules-ds} where we consider the special case where all nodes have the same single service and demand. Effectively, it becomes a case of minimal dominating set.

    
    In \Cref{figure:half-lattices-ds.ella}, we consider an example of graph $G_4$ containing four nodes connected in such a way that they form two disjoint edges, i.e., $V(G_4)=\{v_1,v_2,v_3,v_4\}$ and $E(G_4)=\{\{v_1,v_2\},\{v_3,v_4\}\}$. 

    We write a state of this graph as $(v_1[st], v_2[st], v_3[st], v_4[st])$.
    As shown in this figure, of the 16 states in the state space, 9 are part of 4 disjoint lattices. These are feasible states, i.e., states where nodes with $st$ equals $IN$ form a (possibly non-minimal) dominating set. And, the remaining 7 are not part of any lattice. These are infeasible states, i.e., states where nodes with $st$ equals $IN$ do not form a dominating set. The states not taking part in any lattice structure (the infeasible states) are not shown in \Cref{figure:half-lattices-ds.ella}.

    In a non-feasible state, some node will be addable. The instruction executed by addable nodes is not lattice-linear: an addable node moves in the dominating set unconditionally. After this, when no node is addable, then the global state $s$ becomes feasible state, i.e., $s$ manifests a valid dominating set. In $s$, however, some nodes may be removable. Only the removable nodes can be \imped. The instruction executed by an \imped node is lattice-linear.
    
    E.g., notice in \Cref{figure:half-lattices-ds.ella} (a), assuming that the initial state is $(IN$, $IN$, $IN$, $IN)$, that $v_2$ and $v_4$ are \imped. Since they execute asynchronously, a lattice is induced among all possible global states that $G_4$ transitions through. If only $v_2$ (respectively, $v_4$) executes, the global state we obtain is $(IN$, $OUT$, $IN$, $IN)$ (respectively, $(IN$, $IN$, $IN$, $OUT)$). Since eventually both the nodes change their local states, we obtain the global state $(IN$, $OUT$, $IN$, $OUT)$.
    \qed
\end{example}

\begin{figure}[ht]
    \centering
    \subfigure[]{
        \begin{tikzpicture}[every node/.style={scale=.7}]
            \node at (0,0) (a1) {(IN,OUT,IN,OUT)};
            \node at (-1.5,-1) (a2) {(IN,OUT,IN,IN)};
            \node at (1.5,-1) (a3) {(IN,IN,IN,OUT)};
            \node at (0,-2) (a4) {(IN,IN,IN,IN)};
            \draw (a1) -- (a2);
            \draw (a1) -- (a3);
            \draw (a2) -- (a4);
            \draw (a3) -- (a4);
        \end{tikzpicture}
    }\quad\quad 
    \subfigure[]{
        \begin{tikzpicture}[every node/.style={scale=.7}]
            \node at (0,0) (a1) {(OUT,IN,OUT,IN)};
            \node at (0,-1) (a2) {//only 1 state};
        \end{tikzpicture}
    }
    \subfigure[]{
        \begin{tikzpicture}[every node/.style={scale=.7}]
            \node at (0,0) (a1) {(OUT,IN,IN,OUT)};
            \node at (0,-1) (a2) {(OUT,IN,IN,IN)};
            \draw (a1) -- (a2);
        \end{tikzpicture}
    }\quad\quad 
    \subfigure[]{
        \begin{tikzpicture}[every node/.style={scale=.7}]
            \node at (0,0) (a1) {(IN,OUT,OUT,IN)};
            \node at (0,-1) (a2) {(IN,IN,OUT,IN)};
            \draw (a1) -- (a2);
        \end{tikzpicture}
    }
    \caption{Example lattice induced by \Cref{algorithm:rules-ds}.1 in $G_4$ ($G_4$ is described in \Cref{example:4-nodes.ella}).}
    \label{figure:half-lattices-ds.ella}
\end{figure}



\subsection{Analysis  of \Cref{algorithm:rules-ds}: Time and Space complexity}\label{subsection:time-space-complexity-analysis}

\begin{theorem}\label{theorem:ds-convergence-time}

Starting from an arbitrary state, \Cref{algorithm:rules-ds} reaches an optimal state within $2n$ moves (or more precisely 1 round plus $n$ moves).



\end{theorem}


\begin{proof}
    From \Cref{lemma:d-not-ds}, we have that starting from an arbitrary state, \Cref{algorithm:rules-ds} will reach a feasible state within one round (or within $n$ moves).
    
    After that, if the input graph $G$ is not in an optimal state, then at least one node moves out such that $G$ stays in a feasible state (\Cref{lemma:ds-removal}). Thus, $G$ manifests an optimal state within $n$ additional moves.
\end{proof}


\begin{corollary}\label{corollary:algo-stabilizing-silent}
    \Cref{algorithm:rules-ds} is self-stabilizing and silent.
\end{corollary}


\begin{observation}
    At any time-step, a node will take $O((\Delta)^4\times (max_d)^2)$ time, where
    (1) $\Delta$ is the maximum degree of any node in $V(G)$, and (2) $max_d$ is the total number of distinct demands made by all the nodes in $V(G)$.
\end{observation}


\section{Applying ELLSS in Minimal Vertex Cover}\label{section:mvc}


The execution of \Cref{algorithm:rules-ds} was divided in two phases, (1) where the system reaches a feasible state (reduction of \textsc{Rank} to $0$), and (2) where the system reaches an optimal state (reduction of \textsc{Badness} to $0$).


Such design defines the concept of ELLSS algorithms. This design can be extended to numerous other problems 
where the optimal global state can be defined in terms of a minimal (or maximal) set $\mathcal{S}$ of nodes.
This includes the minimal vertex cover (\mvc) problem, maximal independent set problem and their variants. In this section, we discuss the extension to \mvc.




\begin{definition}\label{definition:mvc}\textbf{Minimal Vertex Cover.}
    In the MVC problem, the input is an arbitrary graph $G$, and the task is to compute a minimal set $\mathcal{V}$ such that for any edge $\{i,j\}\in E(G)$, $(i\in \mathcal{V})\lor (j\in \mathcal{V})$. If a node $i$ is in $\mathcal{V}$, then $i[st]=IN$, otherwise $i[st]=OUT$.
\end{definition}

The proposition $\mathcal{P}^\prime_v$ defining a feasible state and the proposition $\mathcal{P}_v$ defining the optimal state can be defined as follows.
\begin{center}
    $\mathcal{P}_v^\prime(\mathcal{V})\equiv \forall i\in V(G):((i\in \mathcal{V})\lor (\forall j\in Adj_i, j\in \mathcal{V}))$.\\
    $\mathcal{P}_v(\mathcal{V})\equiv \mathcal{P}_v^\prime(\mathcal{V}) \land (\forall i\in \mathcal{V}, \lnot\mathcal{P}_v^\prime(\mathcal{V}\setminus \{i\})).$
\end{center}

To develop an algorithm for \mvc, we utilize the macros in \Cref{table:macros-mvc-ell}. A node $i$ is \textit{removable} if all the nodes in its neighbourhood are in the vertex cover (\vc). $i$ is \textit{addable} if $i$ is not in the \vc and there is some node adjacent to it that is not in the \vc. $i$ is \textit{\imped} if  $i$ is in the \vc, and $i$ is the highest ID node that is removable in its distance-1 neighbourhood.

\begin{table}[ht]
    \centering 
    \doublespacing 
    \begin{tabular}{|l|}
        \hline
        \textsc{Removable-\mvc-ELL}$(i)\equiv \forall j \in Adj_i, j[st]=IN$.\\
        \textsc{Addable-\mvc-ELL}$(i)\equiv i[st]=OUT\land(\exists j\in Adj_i:j[st]=OUT)$.\\
        \textsc{Impedensable-\mvc-ELL}$(i)\equiv i[st]=IN$ $\land$ \textsc{Removable-\mvc-ELL}$(i)\land$\\
        \quad\quad\quad\quad $(\forall j\in Adj_i: j[id]<i[id]\lor \lnot$\textsc{Removable-\mvc-ELL}$(j))$. \\
        \hline
    \end{tabular}
    \caption{Macros used in the algorithm for MVC.}
    \label{table:macros-mvc-ell}
\end{table}

Based on the definitions above, the algorithm for \mvc is described as follows. If a node is addable, then it moves into the VC. If a node is \imped, then it moves out of the VC.
\newpage 
\begin{algorithm}\label{algorithm:rules-mvc}Rules for node $i$.
    \begin{center}
        \begin{tabular}{|l|}
            \hline
            \textsc{Addable-\mvc-ELL}$(i)\longrightarrow i[st]=IN$.\\
            \textsc{Impedensable-\mvc-ELL}$(i)\longrightarrow i[st]=OUT$.\\
            \hline
        \end{tabular}
    \end{center}
\end{algorithm}

\Cref{algorithm:rules-mvc} is an ELLSS algorithm in that it satisfies the conditions in \Cref{definition:ellssa}, where $F_1$ corresponds to the first action of \Cref{algorithm:rules-mvc}, $F_2$ corresponds to its second action, and $S_f$ is the set of the states for which $\mathcal{P}_v^\prime$ holds true.
Thus, starting from any arbitrary state, the algorithm eventually reaches a state where $\mathcal{V}$ is a minimal vertex cover.

\begin{lemma}
    \Cref{algorithm:rules-mvc} is a silent eventually lattice-linear self-stabilizing algorithm for minimal vertex cover.
\end{lemma}

\begin{proof}

    In an arbitrary non-feasible state (where the input graph $G$ does not manifest a valid \vc), there is at least one node that is addable. An addable node immediately executes the first instruction of \Cref{algorithm:rules-mvc} and moves in the \vc. This implies that by the end of the first round, we obtain a valid (possibly non-minimal) \vc.

    If the input graph $G$ is in a feasible, but not optimal, state (where $G$ manifests a non-minimal \vc), then there is at least one removable node. This implies that there is at least one \imped node $i$ in that state (e.g., the removable node with the highest ID).
    Under \Cref{algorithm:rules-mvc}, any node in $Adj_i$ will not execute until $i$ changes its state. $i$ is removable because all nodes in $Adj_i$, along with $i$, are in the vertex cover. Thus $i$ must execute so that it becomes non-removable. This shows that  the second rule in \Cref{algorithm:rules-mvc} is lattice-linear.

    In a non-minimal, but valid, \vc, there is at least one node that is \imped, thus, with every move, the size of the vertex cover, manifested by $G$, reduces by 1. Also, notice that when an \imped node $i$ changes its state, no node in $Adj_i$ changes its state simultaneously. Thus, the validity of the vertex cover is not impacted when $i$ moves. Therefore, \Cref{algorithm:rules-mvc} is self-stabilizing.

    When $G$ manifests a minimal vertex cover, no node is addable or removable. This shows that \Cref{algorithm:rules-mvc} is silent.
\end{proof}

Observe that in \Cref{algorithm:rules-mvc}, the definition of \textsc{Impedensable} relies only on the information about distance-2 neighbours. Hence, the evaluation of guards take $O(\Delta^2)$ time. In contrast, (the standard) minimal dominating set problem would require the information of distance-4 neighbours to evaluate \textsc{Impedensable}. Hence, the evaluation of guards in that would take $O(\Delta^4)$ time. This algorithm converges in $2n$ moves (or more precisely 1 round plus $n$ moves).

\section{Applying ELLSS in Maximal Independent Set}\label{section:mis}

In this section, we consider the application of ELLSS in the problem of maximal independent set (\mis). Unlike \mvc and SDMDS problems where we tried to reach a minimal set, here,  we have to obtain a maximal set. 

\begin{definition}\label{definition:mis}\textbf{Maximal Independent Set.}
    In the \textit{maximal independent set} (\mis) problem, the input is an arbitrary graph $G$, and the task is to compute a maximal set $\mathcal{I}$ such that for any two nodes $i\in\mathcal{I}$ and $j\in\mathcal{I}$, if $i\neq j$, then $\{i,j\}\neq E(G)$.
\end{definition}

The proposition $\mathcal{P}^\prime_i$ defining a feasible state and the proposition $\mathcal{P}_i$ defining the optimal state can be defined as follows.
\begin{center}
    $\mathcal{P}_i^\prime(\mathcal{I})\equiv \forall i\in V(G):((i\not\in\mathcal{I})\lor (\forall j\in Adj_i: j\not\in \mathcal{I}))$.\\
    $\mathcal{P}_i(\mathcal{I})\equiv \mathcal{P}_i^\prime(\mathcal{I})\land(\forall i\in V(G): \lnot\mathcal{P}_i^\prime(\mathcal{I}\cup\{i\}))$.
\end{center}

To develop the algorithm for \mis, we define the macros in \Cref{table:macros-mis-ell}. A node $i$ is \textit{addable} if all the neighbours of $i$ are out of the independent set(\is). A node is \textit{removable} if $i$ is in the \is and there is some neighbour of $i$ that is also in \is. $i$ is \textit{\imped} if $i$ is out of the IS, and $i$ is the highest ID node in its distance-1 neighbourhood that is addable.

\begin{table}[ht]
    \centering 
    \doublespacing 
    \begin{tabular}{|l|}
        \hline
        \textsc{Addable-MIS-ELL}($i)\equiv \forall j \in Adj_i, j[st]=OUT$.\\
        \textsc{Removable-\mis-ELL}$(i)\equiv i[st]=IN\land(\exists j\in Adj_i:j[st]=IN$).\\
        \textsc{Impedensable-\mis-ELL}$(i)\equiv i[st]=OUT\land$ \textsc{Addable-MIS-ELL}$(i)\land$\\
        \quad\quad\quad\quad $(\forall j\in Adj_i:j[id]<i[id]\lor\lnot$\textsc{Addable-MIS-ELL}$(j))$.\\
        \hline
    \end{tabular}
    \caption{Macros used in the algorithm for \mis.}
    \label{table:macros-mis-ell}
\end{table}

Based on the definitions above, the algorithm for \mis is described as follows. If a node $i$ is \imped, then it moves into the \is. If $i$ is removable, then it moves out of the \is.

\newpage 
\begin{algorithm}\label{algorithm:rules-mis}Rules for node $i$.
    \begin{center}
        \begin{tabular}{|l|}
            \hline
            \textsc{Removable-\mis-ELL}$(i)\longrightarrow i[st]=OUT$.\\
            \textsc{Impedensable-\mis-ELL}$(i)\longrightarrow i[st]=IN$.\\
            \hline
        \end{tabular}
    \end{center}
\end{algorithm}

This algorithm is an ELLSS algorithm as well: as per \Cref{definition:ellssa}, $F_1$ corresponds to the first action of \Cref{algorithm:rules-mvc}, $F_2$ corresponds to its second action, and $S_f$ is the set of the states for which $\mathcal{P}_i^\prime$ holds true.
Thus, starting from any arbitrary state, the algorithm eventually reaches a state where $\mathcal{I}$ is a maximal independent set.

\begin{lemma}
    \Cref{algorithm:rules-mis} is a silent eventually lattice-linear self-stabilizing algorithm for maximal independent set.
\end{lemma}

\begin{proof}
    In an arbitrary non-feasible state (where the input graph $G$ does not manifest a valid \is), there is at least one node that is removable. A removable node immediately executes the first instruction of \Cref{algorithm:rules-mis} and moves out of the \is. This implies that by the end of the first round, we obtain a valid (possibly non-minimal) \is.

    If the input graph $G$ is in a feasible, but not optimal, state (where $G$ manifests a non-minimal \is), then there is at least one addable node. This implies that there is at least one \imped node $i$ in that state (e.g., the addable node with the highest ID).
    Under \Cref{algorithm:rules-mvc}, any node in $Adj_i$ will not execute until $i$ changes its state. $i$ is addable because all nodes in $Adj_i$, along with $i$, are out of the independent set. Thus $i$ must execute so that it becomes non-addable. This shows that  the second rule in \Cref{algorithm:rules-mis} is lattice-linear.

    Since in a non-minimal, but valid, independent set, there is at least one node that is \imped, we have that with every move, the size of the independent set, manifested by $G$, reduces by 1. Also, notice that when an \imped node $i$ changes its state, no node in $Adj_i$ changes its state simultaneously. Thus, the validity of the independent set is not impacted when $i$ moves. Therefore, we have that \Cref{algorithm:rules-mis} is self-stabilizing.

    When $G$ manifests a maximal independent set, no node is removable or addable. This shows that \Cref{algorithm:rules-mis} is silent.
\end{proof}

In \Cref{algorithm:rules-mis}, the definition of \textsc{Addable} relies only on the information about distance-2 neighbours. Hence, the evaluation of guards take $O(\Delta^2)$ time. This algorithm converges in $2n$ moves (or more precisely 1 round plus $n$ moves).




\section{Applying ELLSS in Colouring}\label{section:gc}

In this section, we extend ELLSS algorithms to graph colouring. 
In the \textit{graph colouring} (GC) problem, the input is a graph $G$ and the task is to (re-)assign colours to all the nodes such that no two adjacent nodes have the same colour.

\begin{definition}\label{definition:gc}\textbf{Graph colouring}. 
    In the \gc problem, the input is an arbitrary graph $G$ with some initial colouring assignment $\forall i\in V(G): i[colour]\in \mathbb{N}$. The task is to (re)assign the colour values of the nodes such that any adjacent nodes should not have a conflict (i.e., should not have the same colour), and there should not be a node whose colour can be reduced without conflict.
\end{definition}

Unlike \mvc, \mds or \mis, colouring does not have a binary domain. Instead, we correspond the equivalence of changing the state to $IN$ to the case where a node sets its colour to $i[id]+n$. And, the equivalence of changing the state to $OUT$ corresponds to the case where a node decreases its colour.

The proposition $\mathcal{P}^\prime_c$ defining a feasible state and the proposition $\mathcal{P}_c$ defining an optimal state is defined below. $\mathcal{P}_c$ is true when all the nodes have lowest available colour, that is, for any node $i$ and for all colours $c$ in $[1:i[colour]-1]$, $c$ equals the colour of one of the neighbours $j$ of $i$.

\begin{center}
    $\mathcal{P}_c^\prime(G)\equiv \forall i\in V(G),\forall j\in Adj_i:i[colour]\neq j[colour]$.\\
    $\mathcal{P}_c(G)\equiv \mathcal{P}_c^\prime\land (\forall i\in V(G):(\forall c\in [1:i[colour]-1]:(\exists j\in Adj_i: j[colour]=c)))$.
\end{center}

We define the macros as shown in \Cref{table:macros-gc-ell}. A node $i$ is \textit{conflicted} if it has a conflicting colour with at least one of its neighbours. $i$ is subtractable if there is a colour value less than $i[colour]$ that $i$ can change to without a conflict with any of its neighbours. $i$ is \textit{\imped} if $i$ is not conflicted, and it is the highest ID node that is subtractable.
\begin{table}[ht]
    \centering 
    \doublespacing 
    \begin{tabular}{|l|}
        \hline
        \textsc{Conflicted-GC-ELL}($i)\equiv \exists j \in Adj_i:j[colour]=i[colour]$.\\
        \textsc{Subtractable-GC-ELL}($i)\equiv \exists c\in [1:i[colour]-1]: \forall j\in Adj_i: j[colour]\neq c$.\\
        \textsc{Impedensable-GC-ELL}$(i)\equiv\lnot$\textsc{Conflicted-GC-ELL}($i$) $\land$ \textsc{Subtractable-GC-ELL}$(i)\land$\\
        \quad\quad $(\forall j\in V(G):\lnot\textsc{Conflicted-GC-ELL}(j)\land(j[id]<i[id]\lor \lnot$\textsc{Subtractable-GC-ELL}$(j)))$.\\
        \hline
    \end{tabular} 
    \caption{Macros used in the algorithm for \gc.}
    \label{table:macros-gc-ell}
\end{table}

Unlike SDMDS, \mvc and \mis, in graph colouring (GC), each node is associated with a variable $colour$ that can take several possible values (the domain can be as large as the set of natural numbers). 
As mentioned above, the action of setting a colour value to $i[colour]+i[id]$ is done whenever a conflict is detected. 
Effectively, this is like setting the colour to an error value such that the error value of every node is distinct in order to avoid a conflict. This error value will be reduced when node $i$ becomes \imped and decreases its colour.

The actions of the algorithm are shown in \Cref{algorithm:rules-c}. If a node $i$ is \imped, then it changes its colour to the minimum possible colour value. If $i$ is conflicted, then it changes its colour value to $i[colour]+i[id]$.

\begin{algorithm}\label{algorithm:rules-c}Rules for node $i$.
    \begin{center}
        \begin{tabular}{|l|}
            \hline
            \textsc{Conflicted-GC-ELL}$(i)$ $\longrightarrow i[colour]=i[colour]+i[id]$.\\
            \textsc{Impedensable-GC-ELL}$(i)\longrightarrow$\\
            \quad\quad $i[colour]=\min\limits_{c}\{c\in [1:i[colour]-1]:(\forall j\in Adj_i: j[colour]\neq c)\}$.\\
            \hline
        \end{tabular}
    \end{center}
\end{algorithm}

\Cref{algorithm:rules-c} is an ELLSS algorithm: according to \Cref{definition:ellssa}, $F_1$ corresponds to the first action of \Cref{algorithm:rules-mvc}, $F_2$ corresponds to its second action, and and $S_f$ is the set of the states for which $\mathcal{P}_c^\prime$ holds true.
Thus, starting from any arbitrary state, the algorithm eventually reaches a state where no two adjacent nodes have the same colour and no node can reduce its colour.

\begin{lemma}
    \Cref{algorithm:rules-c} is a silent eventually lattice-linear self-stabilizing algorithm for graph colouring.
\end{lemma}

\begin{proof}
    In an arbitrary non-feasible state (where the input graph $G$ does not manifest a valid colouring), there is at least one node that is conflicted. A conflicted node immediately executes the first instruction of \Cref{algorithm:rules-c} and makes its colour equal to its $ID$ plus its colour value. Since the value $i[colour]+i[id]$ by which a node updates its colour value will resolve such conflict with one adjacent node in 1 move, $i$ will become non-conflicted in almost $deg(i)$ moves.

    If the input graph $G$ is in a feasible, but not optimal, state (where $G$ manifests a valid colouring but some nodes can reduce their colour), then there is at least one subtractable node. This implies that there is an \imped node $i$ in that state (the subtractable node with the highest ID).
    Under \Cref{algorithm:rules-c}, any node will not execute until $i$ changes its state. $i$ is subtractable because there is a colour value $c$ less that $i[colour]$ such that no node in $Adj_i$ has that colour value. Thus $i$ must execute to become non-subtractable. This shows that the second rule in \Cref{algorithm:rules-c} is lattice-linear.

    Since in a non-minimal, but valid, colouring, there is at least one node $i$ that is \imped, we have that a node will become non-subtractable in atmost $deg(i)$ moves. Notice that when an \imped node $i$ changes its state, no node changes its state simultaneously. Also, the reduced colour will not have a conflict with any other node. Thus, no conflicts arise. Therefore, we have that \Cref{algorithm:rules-c} is self-stabilizing.

    When $G$ manifests a valid non-subtractable colouring, no node is removable or addable. This shows that \Cref{algorithm:rules-c} is silent.
\end{proof}

In \Cref{algorithm:rules-c}, the definition of \textsc{\Imped} relies only on the information about distance-2 neighbours. Hence, the evaluation of guards take $O(n)$ time. This algorithm converges in $2m + (n+2m)=n+4m$ moves.

\section{Applying ELLSS in 2-Dominating Set Problem}\label{section:2ds}

The 2-dominating set (2DS) problem 
provides a stronger form of dominating set (DS), as compared to the usual \mds problem. In the \textit{2-dominating set} problem, the input is a graph $G$ with nodes having domain $\{IN,OUT\}$. The task is to compute a set $\mathcal{D}$ where some node $i\in \mathcal{D}$ iff $i[st]=IN$; $\mathcal{D}$ must be computed such that there are no two nodes $j,k\in V(G)$ that are in $\mathcal{D}$, and a node $i\in V(G)$ that is not in $\mathcal{D}$, such that $\mathcal{D}\cup\{i\}\setminus \{j,k\}$ is a valid \ds.

Unlike the SDMDS, \mvc, \mis or GC problems that simply study the condition of their immediate neighbours before they change their state, and after they would change their state, the 2-DS problem looks one step further.
Specifically, the usual \mds or \mvc problems investigate the computation of any minimal DS or VC respectively, whereas the 2DS problem requires the computation of such a DS where it must not be the case that another valid DS can be computed while removing two nodes from it and adding one node to it.

The propositions $\mathcal{P}^\prime_{d}$ defines a \ds, $\mathcal{P}_d$ defines an \mds and $\mathcal{P}_{2d}$ defines an optimal state, obtaining a \tds. These propositions are defined below.

\begin{center}
    $\mathcal{P}_{d}^\prime(\mathcal{D})$
    $\equiv \forall i\in V(G):i\in\mathcal{D}\lor (\exists j\in Adj_i:j\in\mathcal{D})$.\\
    $\mathcal{P}_{d}(\mathcal{D})$
    $\equiv\mathcal{P}_{d}'(\mathcal{D})\land(\forall(i\in V(G):\lnot\mathcal{P}_d(D\setminus \{i\})))$.\\
    $\mathcal{P}_{2d}(\mathcal{D})$
    $\equiv \mathcal{P}_{d}(\mathcal{D})  
    \land \neg (\exists i\in V(G),i\not\in\mathcal{D}:$\\$(\exists j, k\in Adj_i, j\in\mathcal{D},k\in\mathcal{D} : \mathcal{P}'_{d}(\mathcal{D} \cup \{i\}\setminus\{j, k\})))$ 
\end{center}

Our algorithm is based on the following intuition: Let $\mathcal{D}$ be an \mds. If there exists nodes $i, j$ and $k$ such that $j, k \in \mathcal{D}$ and $i \not \in \mathcal{D}$, and $\mathcal{D} \cup \{i\} - \{j, k\}$ is also a \ds, then $j$ and $k$ must be neighbours of $i$.

The macros that we utilize are in \Cref{table:macros-2ds-ell}. A node $i$ is \textit{addable} if $i[st]=OUT$ and all the neighbours of $i$ are also out of the DS. $i$ is \textit{removable} if $i[st]=IN$ and there exists at least one neighbour of $i$ that is also in the DS. A node $i$ is \textit{2-addable} if $i[st]=OUT$ there exist nodes $j$ and $k$ in the distance-2 neighbourhood of $i$ where $j[st]=IN$ and $k[st]=IN$ such that $j$ and $k$ can be removed and $i$ can be added to the DS such that $j,k$ and their neighbours stay dominated. A node is \textit{unsatisfied} if it is removable or 2-addable. A node is \textit{\imped} if it is the highest id node in its distance-4 neighbourhood that is unsatisfied.

\begin{table}[ht]
    \centering 
    \doublespacing 
    \begin{tabular}{|l|}
        \hline 
        $\textsc{Addable-2DS-ELL}\equiv i[st]=OUT\land (\forall j\in Adj_i:j[st]=OUT)$.\\
        $\textsc{Removable-2DS-ELL}(i)\equiv i[st]=IN\land (\forall j\in Adj_i\cup\{i\}:((j\neq i\land j[st]=IN)$\\
        \quad\quad\quad\quad $\lor$ $(\exists k\in Adj_j, k\neq i: k[st]=IN)))$.\\
        $\textsc{Two-Addable-2DS-ELL}(i) \equiv i[st]=OUT\land(\forall j\in Adj^2_i\cup\{i\}:$\\
        \quad\quad\quad\quad $\lnot (\textsc{Addable-2DS-ELL}(j)\lor\textsc{Removable-2DS-ELL}(j))) \land$\\
        \quad\quad\quad\quad $(\exists j,k\in Adj^2_i, j[st]=IN, k[st]=IN:$\\
        \quad\quad\quad\quad $(\forall q\in Adj_j\cup Adj_k\cup\{j,k\}:(\exists r\in Adj_q:r[st]=IN\lor r=i)))$.\\
        $\textsc{Unsatisfied-2DS-ELL}(i)\equiv \textsc{Removable-2DS-ELL}(i)\lor \textsc{Two-Addable-2DS-ELL}(i)$.\\
        $\textsc{\Imped-2DS-ELL}(i) \equiv \textsc{Unsatisfied-2DS-ELL}(i) \land(\forall j\in Adj^4_i:$\\
        \quad\quad\quad\quad $(\lnot \textsc{Unsatisfied-2DS-ELL}(j) \lor i[id]>j[id]))$.\\
        \hline 
    \end{tabular}
    \caption{Macros used in the algorithm for \tds.}
    \label{table:macros-2ds-ell}
\end{table}

The algorithm for the 2-dominating set problem is as follows.
If a node $i$ is addable, then it turns itself in the DS, ensuring that $i$ and all it neighbouring nodes it stay dominated. 
As stated above, a node is \imped then it is either removable or 2-addable.
If a node is \imped and removable, then it turns itself out of the DS, ensuring that $i$ is not such a node that is not needed in the DS, but is still present in the DS. If $i$ is \imped and 2-addable, then there are two nodes $j$ and $k$ in the DS such that $j$ and $k$ can be removed, and $i$ can be added, and the resulting DS is still a valid DS. In this case, $i$ moves into the DS, and moves $j$ and $k$ out of the DS.

\begin{algorithm}\label{algorithm:rules-2ds-v1}Rules for node $i$.
$$
    \begin{array}{|l|}
        \hline 
        \textsc{Addable-2DS-ELL}(i)\longrightarrow i[st]= IN.\\
        \textsc{\Imped-2DS-ELL}(i)\longrightarrow\\
        \begin{cases}
            i[st]=OUT. & \text{if $i[st]=IN$.}\\
            {j[st]=OUT, k[st]=OUT, i[st]=IN.} & \text{if $i[st]=OUT$.}
        \end{cases}~\\
        // \text{The reference to $j$ and $k$ is from the definition of $\textsc{Two-Addable-2DS-ELL}(i)$}\\
        \hline 
    \end{array}
$$
\end{algorithm}

This is an ELLSS algorithm that works in three phases: first, every node $i$ checks if it is addable. If $i$ is not addable, then $i$ checks if it is \imped and removable, providing a minimal DS. And finally, $i$ checks if it is \imped and 2-addable, providing a 2DS.
Thus, this algorithm satisfies the conditions in \Cref{definition:ellssa}, where $F_1$ constitutes of the first action of \Cref{algorithm:rules-2ds-v1}, $F_2$ corresponds to its second action, and $S_f$ is the set of the states for which $\mathcal{P}_{d}^\prime$ holds true.
Thus, starting from any arbitrary state, the algorithm eventually reaches a state where $\mathcal{D}$ is 2-dominating set.

\begin{lemma}
    \Cref{algorithm:rules-mvc} is a silent eventually lattice-linear self-stabilizing algorithm for 2-dominating set.
\end{lemma}

\begin{proof}
    In an arbitrary non-feasible state (where the input graph $G$ does not manifest a valid \ds), there is at least one node that is addable. An addable node immediately executes the first instruction of \Cref{algorithm:rules-2ds-v1} and moves in the \ds. This implies that by the end of the first round, we obtain a valid (possibly non-minimal) \ds.
    
    If the input graph $G$ is in a feasible, but not optimal, state (where $G$ manifests a non-minimal \ds), then there is at least one node that is removable or 2-addable. This implies that there is at least one \imped node $i$ in that state (e.g., the node, which is removable or 2-addable, with the highest ID).
    Under \Cref{algorithm:rules-mvc}, any node in $Adj^4_i$ will not execute until $i$ changes its state. If $i$ is removable, then all its neighbours are being dominated by a node other than $i$. If $i$ is 2-addable, then there exists a pair of nodes $j$ and $k$ such that if $j$ and $k$ can move out and $i$ moves in, then all nodes in $Adj_i$, $Adj_j$ and $Adj_k$ will stay dominated, including $i$, $j$ and $k$. Thus $i$ must execute so that it becomes non-\imped. This shows that  the second rule in \Cref{algorithm:rules-2ds-v1} is lattice-linear.

    Notice that if an arbitrary node $j$ and $k$ can move out of the \ds given that all nodes stay dominated if $i$ moves in, then $j$ and $k$ must be the neighbours of $i$. This is assuming that $G$ is in a valid dominating set. Otherwise, it cannot be guaranteed that $i$ can dominate the nodes that only $j$ or $k$ are dominating.
    
    Since in a non-minimal, but valid, \ds, there is at least one node that is removable \imped, we have that with every move of a removable \imped node, the size of the \ds, manifested by $G$, reduces by 1.
    Now assume that $G$ manifests a \ds such that no node is addable or removable. Here, if $G$ does not manifest a 2-dominating set, then, from the discussion from the above paragraph there must exist at least one set of three nodes $i$, $j$ and $k$ such that $j$ and $k$ can move out and $i$ can move in guaranteeing that all nodes in $Adj_i$, $Adj_j$ and $Adj_k$ stay dominated, including $i$, $j$ and $k$. With every move of a 2-addable \imped node, the size of the \ds, manifested by $G$, reduces by 1.
    Also, notice that when an \imped node $i$ changes its state, no node in $Adj^4_i$ changes its state simultaneously. Thus, the validity of the \ds is not impacted when $i$ moves. Therefore, we have that \Cref{algorithm:rules-2ds-v1} is self-stabilizing.
    
    When $G$ manifests a 2-dominating set, no node is addable, removable or 2-addable. This shows that \Cref{algorithm:rules-2ds-v1} is silent.
\end{proof}

Note that in \Cref{algorithm:rules-2ds-v1}, the definition of \textsc{Removable} relies on the information about distance-2 neighbours, and consequently, the definition of \textsc{Two-Addable} relies on the information about distance-4 neighbours. Hence, because of the time complexity of evaluating if a node is \imped, the guards take $O(\Delta^8)$ time. 
This algorithm converges in $3n$ moves (or more precisely 1 round plus $2n$ moves).

In this algorithm, one of the actions is changing the states of 3 processes at once. 
However, it can be implemented in a way that a process changes its own state only. We sketch how this can be done as follows.
To require that a process only changes its own state, we will need additional variables so processes know that they are in the midst of an update where $i$ needs to add itself to $\mathcal{D}$ and $j$ and $k$ need to remove themselves from $\mathcal{D}$.
Intuitively, it will need a variable of the form $getout.i$ which will be set to $\{j, k\}$ to instruct $j$ and $k$ to leave the dominating set. 
When $j$ or $k$ are in the midst of leaving the dominating set, all the nodes in $Adj_i^6$ will have to wait until the operation is completed.
With this change, we note that the algorithm will not be able to tolerate incorrect initialization of $getout.i$ while preserving lattice-linearity. 

\section{Experiments}\label{section:experiments-ella}

In this section, we present the experimental results of time convergence of shared memory programs. We focus on the problem of maximal independent set (\Cref{algorithm:rules-mis}) as an example. 

\begin{figure}[ht]
    \centering
    \subfigure[]{
        \includegraphics[width=0.44\textwidth]{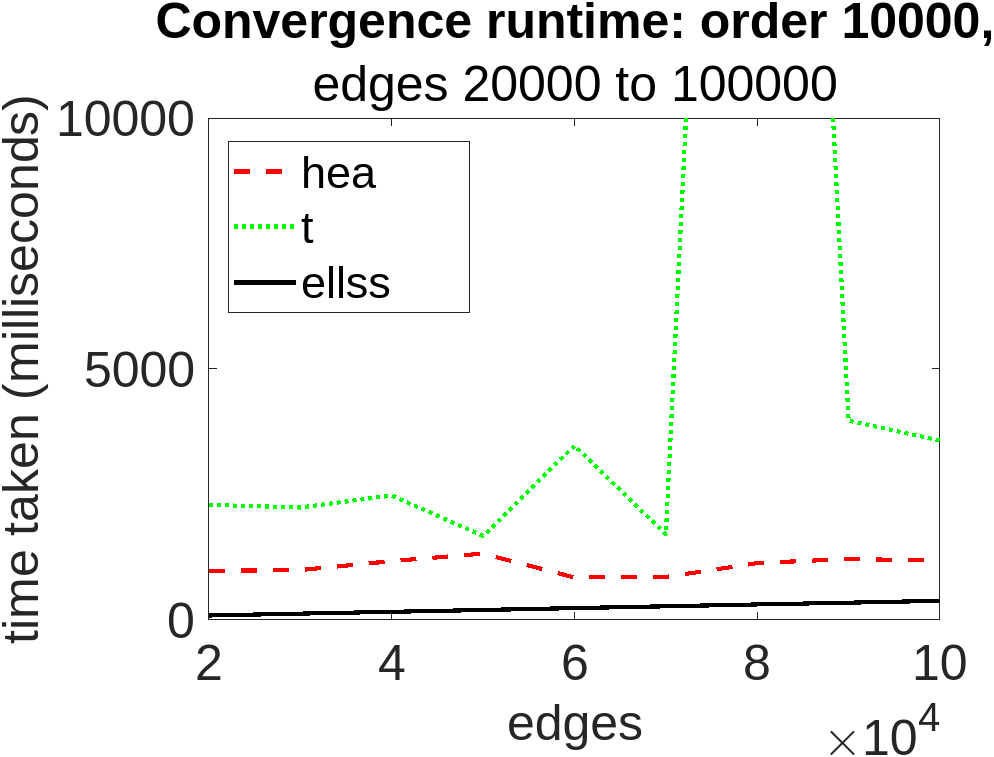}
    }
    \subfigure[]{
        \includegraphics[width=0.44\textwidth]{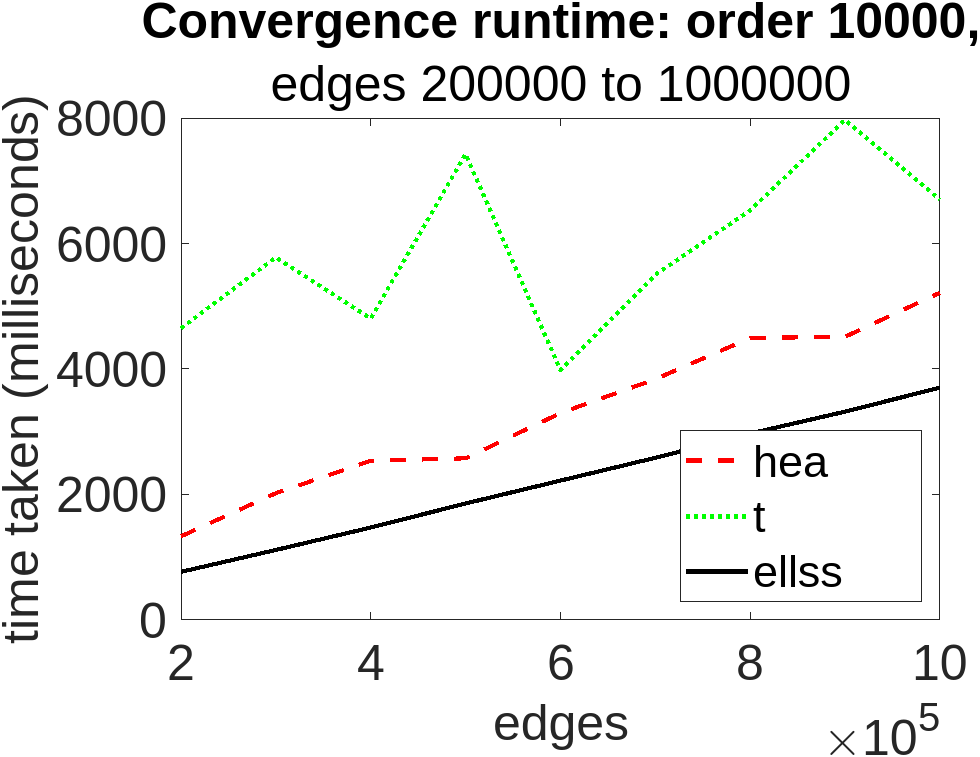}
    }
    \subfigure[]{
        \includegraphics[width=0.44\textwidth]{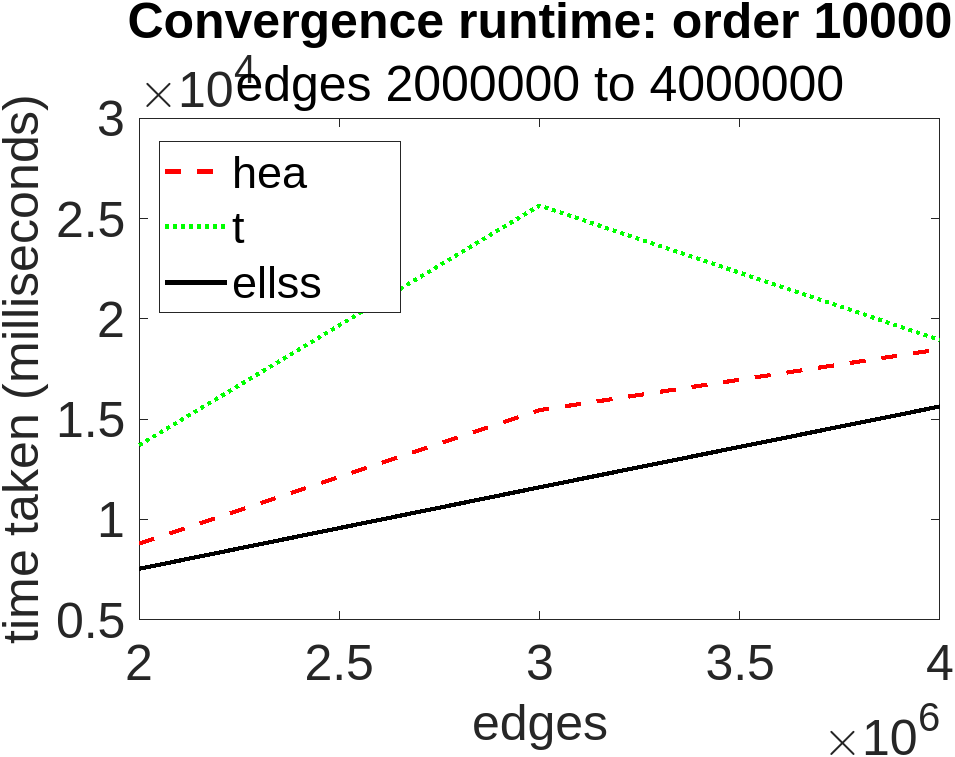}
    }
    \subfigure[]{
        \includegraphics[width=0.44\textwidth]{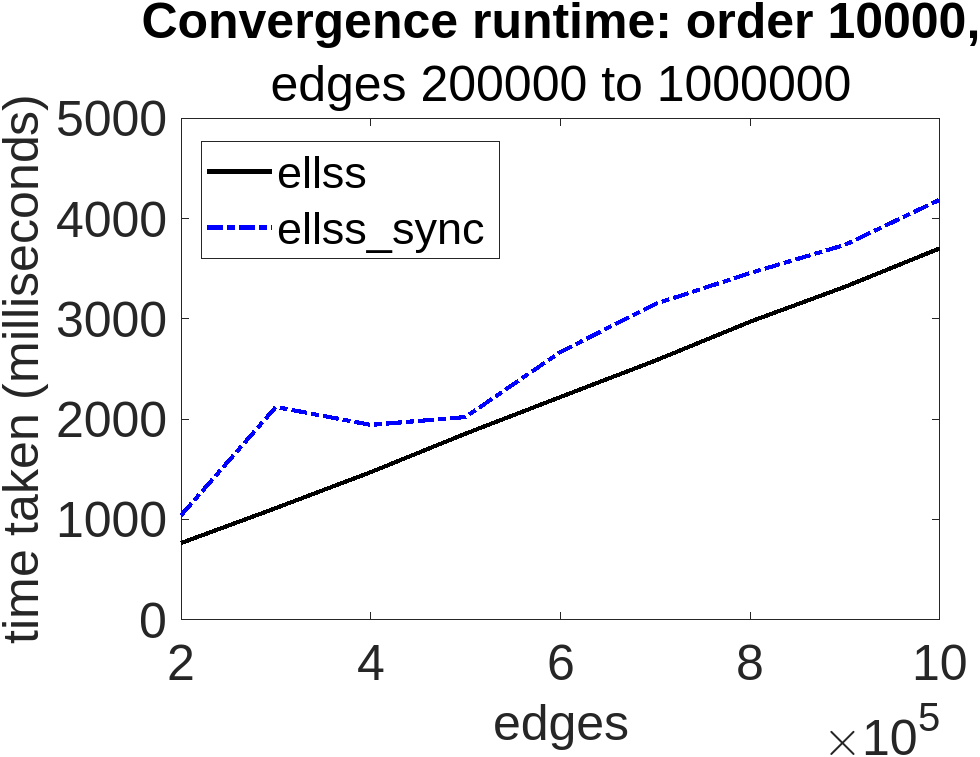}
    }
    \caption{Maximal Independent set algorithms convergence time on random graphs generated by \texttt{networkx} library of \texttt{python3}. All graphs are of 10,000 nodes. Comparision between runtime of \Cref{algorithm:rules-mis}, Hedetniemi et al. (2003) \cite{Hedetniemi2003} (marked as \texttt{hea}) and Turau (2007) \cite{Turau2007} (marked as \texttt{t}) and \Cref{algorithm:rules-mis}. (a) 20,000 to 100,000 edges, \Cref{algorithm:rules-mis}, \cite{Hedetniemi2003} and \cite{Turau2007}. (b) 200,000 to 1,000,000 edges, \Cref{algorithm:rules-mis}, \cite{Hedetniemi2003} and \cite{Turau2007}. (c) 2,000,000 to 4,000,000 edges, \Cref{algorithm:rules-mis}, \cite{Hedetniemi2003} and \cite{Turau2007}. (d) 20,000 to 100,000 edges, \Cref{algorithm:rules-mis} and \Cref{algorithm:rules-mis} lockstep synchronized.}
    \label{figure:figure-ella}
\end{figure}

We compare \Cref{algorithm:rules-mis} with the algorithms present in the literature for the maximal independent set problem. Specifically, we implemented the algorithms present in Hedetniemi et al. (2003) \cite{Hedetniemi2003} and Turau (2007) \cite{Turau2007}, and compare their convergence time. 
The input graphs were random graphs of order 10,000 nodes, generated by the \texttt{networkx} library of python. For comparing the performance results, all algorithms are run on the same set of graphs. 

The experiments are run on Cuda using the \texttt{gcccuda2019b} compiler. The program for \Cref{algorithm:rules-mis} was run asynchronously, and the algorithms in 
\cite{Hedetniemi2003} and \cite{Turau2007} are run under the required synchronization model.
The experiments are run on \texttt{Intel(R) Xeon(R) Platinum 8260 CPU @} \texttt{2.40} \texttt{GHz, cuda  V100S}. The programs are run using the command \texttt{nvcc $\langle$program$\rangle$.cu -G}. Here, each multiprocessor ran 256 threads. And, the system provided sufficient multiprocessors so that each node in the graph can have its own thread. All the observations are an average of 16 readings.

\Cref{figure:figure-ella} (a) (respectively, \Cref{figure:figure-ella} (b) and \Cref{figure:figure-ella} (c)) shows a line graph comparison of the convergence time for these algorithms with the number of edges varying from 20,000 to 100,000  (respectively, 200,000 to 1,000,000 and 2,000,000 to 4,000,000). So, the average degree is varying from 4 to 20 (respectively, 40 to 200 and 400 to 800).
Observe that
the convergence time taken by the program for \Cref{algorithm:rules-mis} is significantly lower than the other two algorithms.

Next, we considered how much of the benefit of \Cref{algorithm:rules-mis} can be allocated to asynchrony due to the property of lattice-linearity. For this, we compared the performance of \Cref{algorithm:rules-mis} running in asynchrony (to allow nodes to read old/inconsistent values) and running in lock-step (to ensure that they only reads the most recent values). \Cref{figure:figure-ella} (d) compares these results. We observe that the asynchronous implementation has lower convergence time.

We have performed the experiments on shared memory architecture that allows nodes to access all memory \textit{quickly}. This means that the overhead of synchronization is low. By contrast, if we had used a distributed system, where computing processors are far apart, the cost of synchronization will be even higher. Hence, the benefit of lattice-linearity (where synchronization is not needed) will be even higher.

\section{Summary of the Chapter}\label{section:ellss-summary}

We extended lattice-linearity from \cite{Garg2020} to the context of self-stabilizing algorithms. A key benefit of lattice-linear systems is that correctness is preserved even if nodes read old information about other nodes. However, the approach in \cite{Garg2020} relies on the assumption that the algorithm starts in  specific initial states, hence, it is not directly applicable in self-stabilizing algorithms. 

We began with the service demand based minimal dominating set (SDMDS) problem and designed a self-stabilizing algorithm for the same. Subsequently, we observed that it consists of two parts: One part makes sure that it gets the system to a state in $S_f$. The second part is a lattice-linear algorithm that constructs a minimal dominating set if it starts in some valid initial states, say a state in $S_f$. We showed that these parts have bounded interference, thus, they guarantee that the system stabilizes even if the nodes execute asynchronously. 

We defined the general structure of eventually lattice-linear self-stabilization to capture such algorithms. We demonstrated that it is possible to develop eventually lattice-linear self-stabilizing (ELLSS) algorithms for minimal vertex cover, maximal independent set, graph colouring and 2-dominating set problems. 

We also demonstrated that these algorithms substantially benefit from their ELLSS property. They outperform existing algorithms while they guarantee convergence without synchronization among processes.
\chapter{FULLY LATTICE-LINEAR ALGORITHMS}\label{chapter:flla}

In \Cref{chapter:ella}, we study algorithms that induce lattices only in a subset of the state space. The algorithms that we develop in \Cref{chapter:ella} are called eventually lattice-linear algorithms. Such algorithms are capable of inducing one or more lattices in a subset of the state space, and guarantee that the system will transition from an arbitrary global state to a state in one of the lattices, and then it transitions to an optimal state traversing through that lattice. 

In this chapter, we focus on studying whether lattices can be induced in the entire state space for non-lattice-linear problems.
Specifically, in this chapter, we alleviate the limitations of, and bridge the gap between, \cite{Garg2020} and \Cref{chapter:ella} by introducing fully lattice-linear algorithms (FLLAs). The former creates a single lattice in the state space and does not allow self-stabilization whereas the latter creates multiple lattices in a subset of the state space. FLLAs induce one or more lattices among the reachable states and can enable self-stabilization. This overcomes the limitations of \cite{Garg2020} and \Cref{chapter:ella}.
We present fully lattice-linear self-stabilizing algorithms for the minimal dominating set (\mds), graph colouring (\gc), minimal vertex cover (\mvc) and maximal independent set (\mis) problems, and a lattice-linear 2-approximation algorithm for vertex cover (\vc). 

We show that the algorithm developed by Goswami et al. \cite{Goswami2022} is lattice-linear.
Lattice-linearity follows that the moves of the robots are predictable. 
This allows us to show tighter bounds to the arena traversed by the robots under the algorithm.
As a consequence of tighter bounds on this arena, (1) we obtain a better convergence time bound for this algorithm, which is lower than that showed in \cite{Goswami2022}, and (2) we show that the gathering point of the robots can be uniquely determined from the initial, or any intermediate, global state. We show that this algorithm converges in $2n$ rounds, which is lower as compared to the time complexity bound ($2.5(n+1)$ rounds) shown in \cite{Goswami2022}.

The algorithms for \mds, \mvc and \mis converge in $n$ moves and the algorithm for \gc converges in $n+2m$ moves.  These algorithms are fully tolerant to consistency violations and asynchrony. 
The 2-approximation algorithm for \vc is the first lattice-linear approximation algorithm for an NP-Hard problem; it converges in $n$ moves. 

The algorithms present in this chapter tolerate asynchrony in AMR model (cf. \Cref{section:asynchrony}).

\subsubsection*{Organization of the Chapter}

\noindent This chapter is organized as follows.
In \Cref{section:ella-revisit}, we recap on eventually lattice-linear algorithms and discuss the motivation behind the theory present in this chapter. 
In \Cref{section:ds-andcolouring-algos}, we describe the general structure of a (fully) lattice-linear algorithm.
In \Cref{section:ds-ll}, we present a fully lattice linear algorithm for minimal dominating set.
We present a fully lattice linear algorithm for graph colouring in \Cref{section:gc-ll}.

We discuss why the design used to develop algorithms for minimal dominating set and graph colouring cannot be extended to develop algorithms for minimal vertex cover and maximal independent set in \Cref{section:vc-no-ll}. We present algorithms for minimal vertex cover and maximal independent set problems in \Cref{subsection:mvc-ll} and \Cref{subsection:mis-ll} respectively.

In \Cref{section:experiments}, we compare the convergence speed of the algorithm presented in \Cref{section:ds-ll} with other algorithms (for the minimal dominating set problem) in the literature.

We present a lattice-linear 2-approximation algorithm for vertex cover in \Cref{section:vc-approx-algo}. In \Cref{section:gsgs}, 
we study the lattice-linearity properties of the algorithm developed by Goswami et al. \cite{Goswami2022}.
Finally, we summarize the chapter in \Cref{section:flla-summary}. 

\section{Revisiting Eventually Lattice-Linear \textsc{Algorithms}}\label{section:ella-revisit}

Unlike the lattice-linear problems where the problem description creates a lattice among the states in $S$, there are problems where the states do not form a lattice naturally, i.e., in those problems, given a suboptimal global state, the problem does not specify a specific set of nodes to change their state. As a result, in such problems, there are instances in which the \imped nodes cannot be determined naturally, i.e., in those instances
$\exists s :\lnot\mathcal{P}(s) \wedge   (\forall i : \exists s' : \mathcal{P}(s')\land s[i]=s'[i]$). 

However, lattices can be induced in the state space algorithmically in these cases. In \Cref{chapter:ella}, we presented algorithms for some of such problems.
Specifically, the algorithms presented in \Cref{chapter:ella} partition the state space into two parts: feasible and infeasible states, and induce multiple lattices among the feasible states. These algorithms work in two phases. The first phase takes the system from an infeasible state to a feasible state (where the system starts to exhibit the desired property), which is an element of a lattice. In the second phase, only an \imped node can change its state. This phase takes the system from a feasible state to an optimal state. These algorithms converge starting from an arbitrary state; they are called \textit{eventually lattice-linear self-stabilizing algorithms.}

\begin{example}\textbf{\mds}.\label{example:dominating-set-definition}
    In the minimal dominating set problem, the task is to choose a minimal set of nodes $\mathcal{D}$ in a given graph $G$ such that for every node in $V(G)$, either it is in $\mathcal{D}$, or at least one of its neighbours is in $\mathcal{D}$. Each node $i$ stores a variable $i[st]$ with domain $\{IN, OUT\}$; $i\in\mathcal{D}$ iff $i[st]=IN$.
    \qed
\end{example}
    
\noindent\textit{Remark}: The minimal dominating set (MDS) problem is not a lattice-linear problem. This is because, for any given node $i$, an optimal state can be reached if $i$ does or does not change its state. Thus $i$ cannot be deemed as \imped or not \imped under the natural constraints of \mds.
    
\begin{exampledscont}
    Even though \mds is not a lattice-linear problem, lattice-linearity can be imposed on it algorithmically. \Cref{algorithm:ds-ellss} (present in the following) is based on the algorithm in \Cref{chapter:ella} for a more generalized version of the problem, the service demand based minimal dominating set problem. \Cref{algorithm:ds-ellss} consists of two phases. In the first phase, if node $i$ is \textit{addable}, i.e., if $i$ and all its neighbours are not in dominating set (\ds) $\mathcal{D}$, then $i$ enters $\mathcal{D}$. 
    This phase does not satisfy the constraints of lattice-linearity from \Cref{definition:impedensable-node}.
    However, once the algorithm reaches a state where nodes in $\mathcal{D}$ form a (possibly non-minimal) \ds, phase 2 imposes lattice-linearity. Specifically, in phase 2, a node $i$ leaves $\mathcal{D}$ iff it is \textit{\imped}, i.e., $i$ along with all neighbours of $i$ stay dominated even if $i$ moves out of $\mathcal{D}$, and $i$ is of the highest ID among all the removable nodes within its distance-2 neighbourhood. 
    \qed 
\end{exampledscont}

\begin{algorithm}\label{algorithm:ds-ellss} Eventually lattice-linear algorithm for \mds.
    \begin{center}
        \begin{tabular}{|l|}
            \hline
            \textsc{Addable-MDS-ELL}$(i)$ $\equiv i[st]=OUT\land (\forall j\in Adj_i:j[st]=OUT)$.\\
            \textsc{Removable-MDS-ELL}$(i)$ $\equiv i[st]=IN\land (\forall j\in Adj_i\cup\{i\}: ((j\neq i\land j[st]=IN)\lor$\\\quad\quad\quad\quad $(\exists k\in Adj_j, k\neq i: k[st]=IN)))$.\\
            \scriptsize{
            // Node $i$ can be removed without violating dominating set
            }\\
            \textsc{\Imped-MDS-ELL}$(i) \equiv \textsc{Removable-MDS-ELL}(i) \land$\\\quad\quad\quad\quad $(\forall j\in Adj^2_i:\lnot \textsc{Removable-MDS-ELL}(j) \lor i[id]>j[id])$.\\~\\
            Rules for node $i$:\\
            \textsc{Addable-MDS-ELL}$(i)$ $\longrightarrow i[st]=IN$. (phase 1)\\
            \textsc{\Imped-MDS-ELL}$(i)$ $\longrightarrow i[st] = OUT$. (phase 2)\\
            \hline
        \end{tabular}
    \end{center}
\end{algorithm}
    
\begin{exampledscont}\label{example:4-nodes}
    To illustrate the lattice imposed by phase 2 of \Cref{algorithm:ds-ellss}, consider an example graph $G_4$ with four nodes such that they form two disjoint edges, i.e., $V(G)=\{v_1,v_2,v_3,v_4\}$ and $E(G)=\{\{v_1,v_2\},\{v_3,v_4\}\}$. Assume that $G$ is initialized in a feasible state.
    
    The lattices formed in this case are shown in \Cref{figure:half-lattices-from-ds-example}. We write a state $s$ of this graph as $\langle v_1[st],$ $v_2[st],$ $v_3[st],$ $v_4[st]\rangle$. We assume that $v_i[id]>v_j[id]$ iff $i>j$.
    Due to phase 1, the nodes not being dominated move in the DS, which makes the system traverse to a feasible state. Now, due to phase 2, only \imped nodes move out, thus, lattices are induced among the feasible global states as shown in the figure.
    \qed
\end{exampledscont}
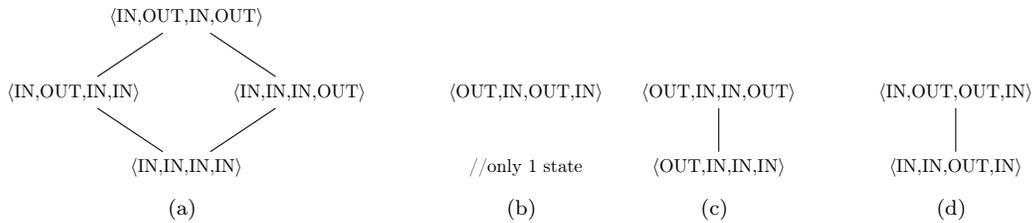
\begin{figure}[ht]
    \centering
    \subfigure[]{
        \begin{tikzpicture}[every node/.style={scale=.7}]
            \node at (0,0) (a1) {$\langle$IN,OUT,IN,OUT$\rangle$};
            \node at (-1.5,-1) (a2) {$\langle$IN,OUT,IN,IN$\rangle$};
            \node at (1.5,-1) (a3) {$\langle$IN,IN,IN,OUT$\rangle$};
            \node at (0,-2) (a4) {$\langle$IN,IN,IN,IN$\rangle$};
            \draw (a1) -- (a2);
            \draw (a1) -- (a3);
            \draw (a2) -- (a4);
            \draw (a3) -- (a4);
        \end{tikzpicture}
    }\quad\quad 
    \subfigure[]{
        \begin{tikzpicture}[every node/.style={scale=.7}]
            \node at (0,0) (a1) {$\langle$OUT,IN,OUT,IN$\rangle$};
            \node at (0,-1) (a2) {//only 1 state};
        \end{tikzpicture}
    }
    \subfigure[]{
        \begin{tikzpicture}[every node/.style={scale=.7}]
            \node at (0,0) (a1) {$\langle$OUT,IN,IN,OUT$\rangle$};
            \node at (0,-1) (a2) {$\langle$OUT,IN,IN,IN$\rangle$};
            \draw (a1) -- (a2);
        \end{tikzpicture}
    }\quad\quad 
    \subfigure[]{
        \begin{tikzpicture}[every node/.style={scale=.7}]
            \node at (0,0) (a1) {$\langle$IN,OUT,OUT,IN$\rangle$};
            \node at (0,-1) (a2) {$\langle$IN,IN,OUT,IN$\rangle$};
            \draw (a1) -- (a2);
        \end{tikzpicture}
    }
    \caption{The lattices induced in the problem instance in Example MDS continuation \ref{example:4-nodes}. Transitive edges are not shown for brevity.}
    \label{figure:half-lattices-from-ds-example}
\end{figure}

In \Cref{algorithm:ds-ellss}, lattices are induced among only some of the global states. After the execution of phase 1, the algorithm locks into one of these lattices.
Thereafter in phase 2, the algorithm executes lattice-linearly to reach the supremum of that lattice. Since the supremum of every lattice represents an \mds, this algorithm always converges to an optimal state. 

\section{Overcoming Limitations of \cite{Garg2020} and \Cref{chapter:ella}}

\label{section:ds-andcolouring-algos}

In this section, we introduce fully lattice-linear algorithms that induce a lattice structure among all reachable states. While defining these algorithms, we also distinguish them from the closely related work in \cite{Garg2020} and \Cref{chapter:ella}. 
Specifically, we discuss why developing algorithms for non-lattice-linear problems (such that the algorithms are lattice-linear, i.e., they induce lattices in the reachable state space) requires the innovation presented in this chapter.

In \cite{Garg2020}, authors consider lattice-linear problems. Here, the state space is induced under a predicate and forms one lattice. Such problems, for a given suboptimal state, specify a set of nodes that must change their state, in order for the system to reach an optimal state. Such problems possess only one optimal state, and hence a violating node must change its state. The acting algorithm simply follows that lattice to reach the optimal state. 

Certain problems, e.g., dominating set are not lattice-linear 
(cf. the remark below \Cref{example:dominating-set-definition}) and thus they cannot be modelled under the constraints of \cite{Garg2020}. That is, the problem cannot specify for an arbitrary suboptimal state, a specific set of nodes that must change their state.
Such problems are studied in \Cref{chapter:ella}. An interesting observation on the algorithms studied in \Cref{chapter:ella} is that they induce multiple lattices in a subset of the state space (c.f. \Cref{figure:half-lattices-from-ds-example}). 

\subsubsection*{Limitations of \cite{Garg2020}}

From the above discussion, we note that the general approach presented in \Cref{chapter:ella} is applicable to a wider class of problems. Additionally, many lattice-linear problems do not allow self-stabilization. In such cases, e.g., in SMP, if the algorithm starts in, e.g., the supremum of the lattice, then it may terminate declaring that no solution is available. Unless the supremum is the optimal state, the acting algorithm cannot be self-stabilizing.

\subsubsection*{Limitations of \Cref{chapter:ella}}
In eventually lattice-linear algorithms (e.g., \Cref{algorithm:ds-ellss} for \mds), the lattice structure is imposed only on a subset of states. Thus, by design, the algorithm has a set of rules, say $\mathcal{A}_1$, that operate in the part of the state space where the lattice structure does not exist, and another set of rules, say $\mathcal{A}_2$, that operate in the part of the state space where the lattice is induced. 
Since actions of $\mathcal{A}_1$ operate outside the lattice structure, a developer must guarantee that if the system is initialized outside the lattice structure, then $\mathcal{A}_1$ converges the system to one of the states participating in the lattice (from where $\mathcal{A}_2$ will be responsible for the traversal of the system through the lattice) and thus the developer faces an extra proof obligation. 
In addition, it also must be proven that 
actions of $\mathcal{A}_2$ and the actions of $\mathcal{A}_1$ do not interfere with each other. E.g., the developer (in the context of \Cref{algorithm:ds-ellss}) has to make sure that actions of $\mathcal{A}_2$ do not perturb the node to a state where the selected nodes do not form a dominating set.

\subsubsection*{Alleviating the Limitations of \cite{Garg2020} and \Cref{chapter:ella}} 
In this chapter, we investigate if we can benefit from the advantages of both \cite{Garg2020} and \Cref{chapter:ella}. We study if there exist fully lattice-linear algorithms where lattices can be imposed on all reachable states, forming single or multiple lattices. In the case that there are multiple optimal states and the problem requires self-stabilization, it would be necessary that multiple disjoint lattices are formed where the supremum of each lattice is an optimal state.
Self-stabilization also requires that these lattices are exhaustive, i.e., they collectively contain all states in the state space. 

Incorporating the property of self-stabilization ensures that the system can be allowed to initialize in any state, and asynchrony can be permitted. The initial state locks into one of the lattices, and due to the induction of $\prec$-lattices, such algorithms ensure a deterministic output (all local states visited by individual nodes form a total order, so an \imped node has only one choice of action, and thus, the global state of convergence can be predicted deterministically from the initial state or any intermediate state; the following sections contain examples of such algorithms).
Such algorithms would also permit multiple optimal states.
In addition, there will be no need to deal with interference between actions. 

\begin{definition}\label{definition:ll-algos}\textbf{Lattice-linear algorithms (LLA)}.
    Algorithm $A$ is an LLA for a problem $P$, iff there exists a predicate $\mathcal{P}$ and $A$ induces a $\prec$-lattice among the states of $S_1, ..., S_w \subseteq S (w\geq 1)$, such that
    \begin{itemize}
        \item State space $S$ of $P$ contains mutually disjoint lattices, i.e.
        \begin{itemize}
            \item $S_1, S_2, \cdots, S_w\subseteq S$ are pairwise disjoint.
            \item $S_1 \cup \cdots \cup S_w$ contains all the reachable states (starting from a set of initial states, if specified; if an arbitrary state can be an initial state, then $S_1 \cup \cdots \cup S_w=S$).
        \end{itemize}
        \item Lattice-linearity is satisfied in each subset under $\mathcal{P}$, i.e., 
        \begin{itemize}
            \item $P$ is deemed solved iff the system reaches a state where $\mathcal{P}$ is true
            \item $\forall k: 1 \leq k \leq w$, 
            $\mathcal{P}$ is lattice-linear with respect to the $\prec$-lattice induced in $S_k$ by $A$, i.e., $\forall s\in S_k: \lnot\mathcal{P}(s) \Rightarrow \exists i:
            \textsc{\Imped}(i,s,\mathcal{P})$.
        \end{itemize}
    \end{itemize}
\end{definition}

\noindent\textit{Remark}: Any algorithm that traverses a $\prec$-lattice of global states is a lattice-linear algorithm. An algorithm that solves a lattice-linear problem, under the constraints of lattice-linearity, e.g. the algorithm described in \Cref{example:mom-definition}, is also a lattice-linear algorithm.

\begin{definition}\textit{Self-stabilizing LLA}.
    Continuing from \Cref{definition:ll-algos}, $A$ is self-stabilizing only if $S_1 \cup S_2 \cup \cdots \cup S_w=S$ and $\forall k:1\leq k\leq w$, the supremum of the lattice induced among the states in $S_k$ is optimal.
\end{definition}

\section{Fully Lattice-Linear Algorithm for Minimal Dominating Set (\mds)}\label{section:ds-ll}

In this section, we present a lattice-linear self-stabilizing algorithm for \mds. \mds has been defined in \Cref{example:dominating-set-definition}.

We describe the algorithm as \Cref{algorithm:ds-ll}.
The first two macros are the same as \Cref{example:dominating-set-definition}. The definition of a node being \imped is changed to make the algorithm fully lattice-linear. Specifically, even allowing a node to enter into the dominating set (DS) is restricted such that only the nodes with the highest ID in their distance-2 neighbourhood can enter the \ds. 
Any node $i$ which is addable or removable will toggle its state iff it is \imped, i.e., iff any other node $j\in Adj^2_i:j[id]>i[id]$ is neither addable nor removable. In the case that $i$ is \imped, if $i$ is addable, then we call it \textit{addable-\imped}, otherwise, if it is removable, then we call it \textit{removable-\imped}.

\newpage 
\begin{algorithm}\label{algorithm:ds-ll}
    Algorithm for \mds.
\end{algorithm}
\begin{center}
    \begin{tabular}{|l|}
        \hline 
        \textsc{Removable-MDS-FLL}$(i)$ $\equiv i[st]=IN\land (\forall j\in Adj_i\cup\{i\}: ((j\neq i\land j[st]=IN)\lor$\\\quad\quad $(\exists k\in Adj_j, k\neq i: k[st]=IN)))$.\\
        \textsc{Addable-MDS-FLL}$(i)$ $\equiv i[st]=OUT\land (\forall j\in Adj_i:j[st]=OUT)$.\\
        \textsc{Unsatisfied-MDS-FLL}$(i)$ $\equiv \textsc{Removable-MDS-FLL}(i)\lor \textsc{Addable-MDS-FLL}(i)$.\\
        $\textsc{\Imped-MDS-FLL}(i)\equiv\textsc{Unsatisfied-MDS-FLL}(i)\land$\\ \quad\quad $(\forall j\in Adj^2_i:\lnot\textsc{Unsatisfied-MDS-FLL}(j)\lor$ $ i[id]>j[id])$.\\~\\
        Rules for node $i$.\\
        $\textsc{\Imped-MDS-FLL}(i) \longrightarrow i[st] = \lnot i[st]$.\\
        \hline 
    \end{tabular}
\end{center}
\begin{lemma}\label{lemma:ds-no-step-back}
    Any node in an input graph does not revisit its older state while executing under \Cref{algorithm:ds-ll}.
\end{lemma}

\begin{proof}
    Let $s$ be the global state at time $t$ while \Cref{algorithm:ds-ll} is executing. We have from \Cref{algorithm:ds-ll} that if a node $i$ is addable-\imped or removable-\imped, then no other node in $Adj^2_i$ changes its state.
    
    If $i$ is addable-\imped at $t$, then any node in $Adj_i$ is out of the \ds. After when $i$ moves in, then any other node in $Adj_i$ is no longer addable, so they do not move in after $t$. As a result, $i$ does not have to move out after moving in.
    
    If otherwise $i$ is removable-\imped at $t$, then all the nodes in $Adj_i\cup\{i\}$ are being dominated by some node other than $i$. So after when $i$ moves out, then none of the nodes in $Adj_i$, including $i$, becomes unsatisfied. 
    
    Let that $i$ is dominated and out, and some $j\in Adj_i$ is removable \imped. $j$ will change its state to $OUT$ only if $i$ is being covered by another node. Also, while $j$ turns out of the \ds, no other node in $Adj^2_j$, and consequently in $Adj_i$, changes its state. As a result, $i$ does not have to turn itself in because of the action of $j$.
    
    From the above cases, we have that $i$ does not change its state to $i[st]$ after changing its state from $i[st]$ to $i[st']$.
    throughout the execution of \Cref{algorithm:ds-ll}.
\end{proof}

To demonstrate that \Cref{algorithm:ds-ll} is lattice-linear, we define state value and rank, the auxiliary variables associated with nodes and global states, as follows:
$$
    \begin{array}{l}
        \textsc{State-Value-MDS}(i,s)=
        \begin{cases}
            1 & \text{if $\textsc{Unsatisfied-MDS}(i)$ in state $s$} \\
            0 & \text{otherwise}
        \end{cases}
    \end{array}
$$
$$
    \textsc{Rank-MDS}(s)=\sum\limits_{i\in V(G)}\textsc{State-Value-MDS}(i,s).
$$

\begin{theorem}\label{theorem:ds-ll}
    \Cref{algorithm:ds-ll} is a silent self-stabilizing and lattice-linear algorithm executed by $n$ nodes running asynchronously.
\end{theorem}

\begin{proof}
    We have from the proof of \Cref{lemma:ds-no-step-back} that if $G$ is in state 
    $s$ and $\textsc{Rank-MDS}(s)$ is non-zero, then at least one node will be \imped, e.g., the unsatisfied node in $V(G)$ with the highest ID.
    For any node $i$, we have that $\textsc{State-Value-MDS}(i)$ decreases whenever $i$ is \imped and never increases.
    As a result, $\textsc{Rank-MDS}$ monotonously decreases throughout the execution of the algorithm until it becomes zero. This shows that \Cref{algorithm:ds-ll} is self-stabilizing. Once \textsc{Rank-MDS} is zero, no node is \imped, so no node makes a move. This shows that \Cref{algorithm:ds-ll} is silent.
    
    Next, we show that \Cref{algorithm:ds-ll} is fully lattice-linear. We claim that there is one lattice corresponding to each optimal state. It follows that if there are $w$ optimal states for a given instance, then there are $w$ disjoint lattices $S_1, S_2, \cdots, S_w$ formed in the state space $S$. We show this as follows.
    
    We observe that an optimal state (manifesting a minimal dominating set) is at the supremum of its respective lattice, as there are no outgoing transitions from an optimal state. 
    
    Furthermore, given a state $s$, we can uniquely determine the optimal state that would be reached from $s$.  
    This is because in any given non-optimal state, the \imped nodes, that must change their state in order to reduce the ranks of the global state of the system, can be uniquely identified.  Additionally, the value, that these \imped nodes will update their local state to, is also unique. Thus, the optimal state reached from a given state $s$ can be uniquely identified. 
    
    This implies that starting from a state $s$ in $S_k (1\leq k\leq w)$, the algorithm cannot converge to any state other than the supremum of $S_k$. Thus, the state space of the problem is partitioned into $S_1, S_2\cdots, S_w$ where each subset $S_k$ contains one optimal state, say $s_{k_{opt}}$, and from all states in $S_k$, the algorithm converges to $s_{k_{opt}}$. 
    
    Each subset, $S_1, S_2, \cdots, S_w$, forms a $\prec$-lattice where $s[i]\prec s'[i]$ if and only if $\textsc{State-Value-MDS}(i,s)>\textsc{State-Value-MDS}(i,s')$ and $s\prec s'$ iff $\textsc{Rank-MDS}(s)>\textsc{Rank-MDS}(s')$. 
    This 
    shows that \Cref{algorithm:ds-ll} is lattice-linear.
\end{proof}

\begin{figure}[ht]
    \centering
    \subfigure[]{
        \begin{tikzpicture}[every node/.style={scale=.7}]
            \node at (0,0) (a1) {$\langle$IN,OUT,IN,OUT$\rangle$};
            \node at (-1.5,-1) (a2) {$\langle$IN,OUT,IN,IN$\rangle$};
            \node at (1.5,-1) (a3) {$\langle$IN,IN,IN,OUT$\rangle$};
            \node at (0,-2) (a4) {$\langle$IN,IN,IN,IN$\rangle$};
            \draw (a1) -- (a2);
            \draw (a1) -- (a3);
            \draw (a2) -- (a4);
            \draw (a3) -- (a4);
        \end{tikzpicture}
    }
    \subfigure[]{
        \begin{tikzpicture}[every node/.style={scale=.7}]
            \node at (0,0) (a1) {$\langle$OUT,IN,OUT,IN$\rangle$};
            \node at (-1.5,-1) (a2) {$\langle$OUT,IN,OUT,OUT$\rangle$};
            \node at (1.5,-1) (a3) {$\langle$OUT,OUT,OUT,IN$\rangle$};
            \node at (0,-2) (a4) {$\langle$OUT,OUT,OUT,OUT$\rangle$};
            \draw (a1) -- (a2);
            \draw (a1) -- (a3);
            \draw (a2) -- (a4);
            \draw (a3) -- (a4);
        \end{tikzpicture}
    }\\
    \subfigure[]{
        \begin{tikzpicture}[every node/.style={scale=.7}]
            \node at (0,0) (a1) {$\langle$OUT,IN,IN,OUT$\rangle$};
            \node at (-1.5,-1) (a2) {$\langle$OUT,IN,IN,IN$\rangle$};
            \node at (1.5,-1) (a3) {$\langle$OUT,OUT,IN,OUT$\rangle$};
            \node at (0,-2) (a4) {$\langle$OUT,OUT,IN,IN$\rangle$};
            \draw (a1) -- (a2);
            \draw (a1) -- (a3);
            \draw (a2) -- (a4);
            \draw (a3) -- (a4);
        \end{tikzpicture}
    }
    \subfigure[]{
        \begin{tikzpicture}[every node/.style={scale=.7}]
            \node at (0,0) (a1) {$\langle$IN,OUT,OUT,IN$\rangle$};
            \node at (-1.5,-1) (a2) {$\langle$IN,IN,OUT,IN$\rangle$};
            \node at (1.5,-1) (a3) {$\langle$IN,OUT,OUT,OUT$\rangle$};
            \node at (0,-2) (a4) {$\langle$IN,IN,OUT,OUT$\rangle$};
            \draw (a1) -- (a2);
            \draw (a1) -- (a3);
            \draw (a2) -- (a4);
            \draw (a3) -- (a4);
        \end{tikzpicture}
    }
    \caption{The lattices induced by \Cref{algorithm:ds-ll} on the graph $G_4$ described in \Cref{example:dominating-set-definition}. Transitive edges are not shown for brevity.}
    \label{figure:full-lattices-from-ds-example}
\end{figure}
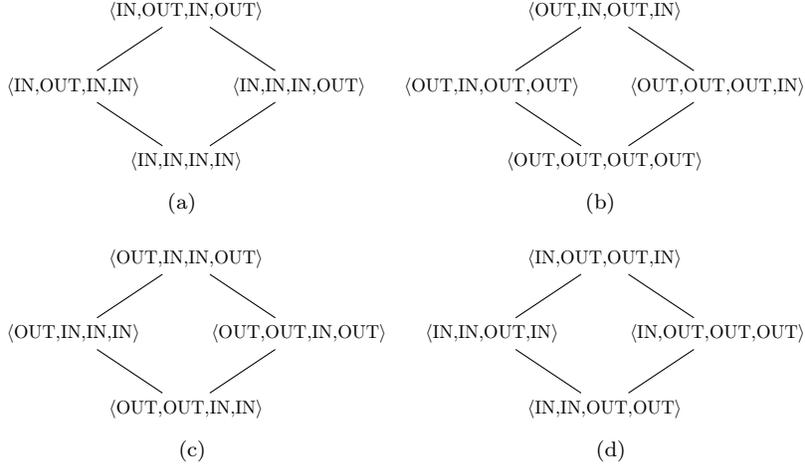

\begin{exampledscont}\label{example:fullylatticelinear}
    For $G_4$ the lattices induced under \Cref{algorithm:ds-ll} are shown in \Cref{figure:full-lattices-from-ds-example}; each vector represents a global state $\langle v_1[st]$, $v_2[st]$, $v_3[st]$, $v_4[st]\rangle$. 
    \qed 
\end{exampledscont}

\section{Fully Lattice-Linear Algorithm for Graph Colouring (\gc)}\label{section:gc-ll}

In this section, we describe a fully lattice-linear algorithm for \gc. We first define the \gc problem, and then we describe an algorithm for \gc.
Graph colouring is defined in \Cref{definition:gc} (\Cref{section:gc}).

We describe the algorithm as \Cref{algorithm:gc-ll}. Any node $i$, which has a conflicting colour with any of its neighbours, or if its colour value is reducible, is an \textit{unsatisfied} node. A node having a conflicting or reducible colour changes its colour to the lowest non-conflicting value iff it is \imped, i.e., any node $j$ in $Adj_i$ with ID more than $i$ is not unsatisfied. In the case that $i$ is \imped, if $i$ has a conflict with any of its neighbours, then we call it \textit{conflict-\imped}, if, otherwise, its colour is reducible, then we call it \textit{reducible-\imped}.

\begin{algorithm}\label{algorithm:gc-ll}Algorithm for \gc.
    \begin{center}
        \begin{tabular}{|l|}
            \hline 
            $\textsc{Conflicted-GC-FLL}(i)\equiv \exists j\in Adj_i:j[colour]=i[colour]$.\\
            $\textsc{Reducible-GC-FLL}(i)\equiv \exists c\in\mathbb{N},c<i[colour]: (\forall j\in Adj_i:c\neq j[colour])$.\\
            \textsc{Unsatisfied-GC-FLL}$(i)$ $\equiv \textsc{Conflicted-GC-FLL}(i)\lor \textsc{Reducible-GC-FLL}(i)$.\\

            $\textsc{\Imped-GC-FLL}(i)\equiv\textsc{Unsatisfied-GC-FLL}(i)\land $\\ \quad\quad $(\forall j\in Adj_i:\lnot\textsc{Unsatisfied-GC-FLL}(j)\lor$ $ i[id]>j[id])$.
            ~\\~\\
            Rules for node $i$.\\
            $\textsc{\Imped-GC-FLL}(i) \longrightarrow i[colour]=\min\{c\in \mathbb{N}: \forall j\in Adj_i, c\neq j[colour]\}$.\\
            \hline 
        \end{tabular}
    \end{center}
\end{algorithm}

\begin{lemma}\label{lemma:gc-decrease}
     Under \Cref{algorithm:gc-ll}, the colour value may increase or decrease at its first move, after which, its colour value monotonously decreases.
\end{lemma}

\begin{proof}
    
    When some node $i$ is deemed \imped for the first time, it may be conflicted or reducible. In either case, it obtains a colour value that is not conflicting with the colour value of its neighbours. The updated colour of $i$ will be a value from 1 to $|Adj_i|+1$. At this time, no neighbour of $i$ can change its colour. 
        
    Now, we show that $i$ will not become conflicted again, after becoming \imped for the first time. Under \Cref{algorithm:gc-ll}, any node $j$ in $Adj_i$ will not change its colour until it obtains the updated colour value of $i$. (If $j$ reads old information about $i[colour]$ then it will continue to wait for $i$ to execute as required by the guard \textsc{\Imped-GC-II}.) If in case some node $j$ in $Adj_i$ becomes impedensable, then it must obtain a colour value that is not equal to the copy of $i[colour]$ that it reads/stores. Thus $i$ does not become conflicted by the action of $j$.

    Thus, after $i$ becomes \imped for the first time, it only reduces its colour in every subsequent move.
\end{proof}

To demonstrate that \Cref{algorithm:gc-ll} is lattice-linear, we define the state value and rank, auxiliary variables associated with nodes and global states, as follows.
    $$
    \begin{array}{l}
        \textsc{State-Value-GC}(i,s)=
        \begin{cases}
            deg(i)+2 & \text{if $\textsc{Conflicted-GC-FLL}(i)$ in state $s$} \\
            i[colour] & \text{otherwise}
        \end{cases}
    \end{array}
    $$
$$\textsc{Rank-GC}(s)=\sum\limits_{i\in V(G)}\textsc{State-Value-GC}(i,s).$$

\begin{theorem}\label{theorem:gc-ll}
    \Cref{algorithm:gc-ll} is a silent self-stabilizing and lattice-linear algorithm executed by $n$ nodes running asynchronously.
\end{theorem}

\begin{proof}
    From the proof of \Cref{lemma:gc-decrease}, we have that for any node $i$, $\textsc{State-Value-GC}(i)$ decreases when $i$ is \imped and never increases. This is because $i$ can increase its colour only once, after which it obtains a colour that is not in conflict with any of its neighbours, so any move that $i$ makes after that will reduce its colour. Therefore, $\textsc{Rank-GC}$ monotonously decreases until no node is \imped. This shows that \Cref{algorithm:gc-ll} is self-stabilizing. 
    
    In any suboptimal global state, at least one node is \imped, e.g., the highest ID node that is unsatisfied. Thus, a suboptimal global state will transition to a global state with a lesser rank. Since there are only a bounded number of colour values from 1 to $deg(i)+1$, a node can become reducible only a bounded number of times. If no node is conflicted and no node is reducible, and no node is \imped, no node makes a move. This shows that \Cref{algorithm:gc-ll} is silent.
    
    \Cref{algorithm:gc-ll} exhibits properties similar to \Cref{algorithm:ds-ll} which are elaborated in the proof for its lattice-linearity in \Cref{theorem:ds-ll}. Thus, \Cref{algorithm:gc-ll} is also lattice-linear.
\end{proof}

\section{Limitations of using Simple Actions and Tiebreakers for Developing FLLA}
\label{section:vc-no-ll}

We studied that lattice-linear algorithms for \mds and \gc can be designed by simply using tie-breakers.
Hence, a natural question arises if a lattice-linear algorithm can be designed for other graph theoretic problems by using some tie-breaker. The answer is no. Specifically, we cannot extend this design to develop algorithms for all graph theoretic problems -- we study minimal vertex cover (\mvc) and maximal independent set (\mis) problems in this context.

We first show (\Cref{subsection:vc-no-tie-breaker}) the issues involved in an algorithm that simply uses a tie-breaker to decide which node enters or leaves the vertex cover. Specifically, we show that this design results in cyclic behaviour.
Such behaviour is observed when we use \textit{simple actions}, where a node only changes the state of itself when it evaluates that its guards are true, with arbitrary-distance tie-breaker.
Similar results can be derived for the \mis problem.
Subsequently, in \Cref{subsection:mvc-ll} (respectively, in \Cref{subsection:mis-ll}), we show that a lattice-linear algorithm can be developed for \mvc (respectively, \mis) with \textit{complex actions}, where a node is allowed to make changes to the variables of other nodes. Then, in \Cref{subsection:complex-actions-vc-is}, we elaborate on the properties of algorithms, that we present, for \mvc and \mis.

\subsection{Issues in Using Only a Tie-Breaker in Algorithm for Minimal Vertex Cover (\mvc)}\label{subsection:vc-no-tie-breaker}


Minimal vertex cover is defined in \Cref{definition:mvc} (\Cref{section:mvc}).

We could use the macros $addable$ (some edge of a subject node is not covered) and $removable$ (removing the node preserves the vertex cover) to design an algorithm for \mvc. 
However, this design results in a cyclic behaviour, with respect to the local state transition of a node, even with a tie-breaker with all other nodes in the graph.
To illustrate this, consider the execution of an algorithm with such a tie-breaker on a line graph of 4 nodes (ID'd 1-4, sequentially) where all nodes are initialized to $OUT$.
Here, node 4 can change its state to $IN$. Other nodes cannot change their state because there is a node with a higher ID that can enter the vertex cover. After node 4 enters the vertex cover, node 3 enters the vertex cover, as edge $\{2, 3\}$ is not covered. However, this requires node 4 to leave the vertex cover to keep it minimal. 

Observe, above, that node $4$ was initialized such that $4[st]=OUT$, then it changed to $4[st]=IN$ and subsequently changed again to $4[st]=OUT$. 
Thus, we see a cyclic behaviour which is not desired in a lattice-linear algorithm. 
%
%
%
%
This analysis also shows that the use of simple actions results in the system exhibiting a cyclic behaviour. However, we have that complex actions can be utilized to move around this issue, which we study in the following.

\subsection{Fully Lattice-Linear Algorithm for Minimal Vertex Cover (\mvc)}\label{subsection:mvc-ll}

In \Cref{subsection:vc-no-tie-breaker}, we discussed the issues that arise in using (1) only a tie-breaker, and (2) simple actions. Based on these limitations, in this section, we describe a lattice-linear algorithm that utilizes complex actions to solve the \mvc problem. 

We use the macros listed in \Cref{table:macros-mvc-fll}. 
A node $i$ is \textit{removable} iff $i$ is in the vertex cover, and all the neighbours of $i$ are also in the vertex cover.
$i$ is \textit{addable} iff $i$ is out of the vertex cover and there is some edge $\{i,j\}$ incident on $i$ such that $j$ is not in the vertex cover.
$i$ is \textit{unsatisfied} iff $i$ is removable or $i$ is addable.
$i$ is \textit{impedendable} iff $i$ is unsatisfied and there is no node $j$ in distance-3 of $i$, with ID greater than $i$, such that $j$ is unsatisfied.
\begin{table}[ht]
    \centering 
    \doublespacing 
    \begin{tabular}{|l|}
        \hline 
        \textsc{Removable-MVC-FLL}$(i)$ $\equiv i[st]=IN\land (\forall j\in Adj_i: j[st]=IN)$.\\
        \textsc{Addable-MVC-FLL}$(i)$ $\equiv i[st]=OUT\land (\exists j\in Adj_i:j[st]=OUT)$.\\
        \textsc{Unsatisfied-MVC-FLL}$(i)$ $\equiv \textsc{Removable-MVC-FLL}(i)\lor \textsc{Addable-MVC-FLL}(i)$.\\
        $\textsc{\Imped-MVC-FLL}(i)\equiv\textsc{Unsatisfied-MVC-FLL}(i)\land$\\ \quad\quad $ (\forall j\in Adj^3_i:\lnot\textsc{Unsatisfied-MVC-FLL}(j)\lor i[id]>j[id])$.\\
        \hline 
    \end{tabular}
    \caption{Macros used in the algorithm for \mvc.}
    \label{table:macros-mvc-fll}
\end{table}
The algorithm is defined as follows. An \textit{addable-\imped} node $i$ turns itself in and forces all its removable neighbours out. 
(after accounting for the fact that $i$ has already turned in). 
In this version of the algorithm, this complex action is assumed to be atomic. 
A \textit{removable-\imped} node $i$ will move out of the vertex cover. 

\begin{algorithm}\label{algorithm:vc-ll}Rules for node $i$.
\end{algorithm}
\centerline{
$\begin{array}{|l|}
    \hline 
    \textsc{\Imped-MVC-FLL}(i) \longrightarrow\\
    \begin{cases}
        i[st]=IN.~\forall j\in Adj_i:j[st]=OUT,\text{if $\textsc{Removable-MVC-FLL}(j)$}. & \text{if $\textsc{Addable-MVC-FLL}(i)$}.\\
        i[st] = OUT. & \text{otherwise}
    \end{cases}~\\
    \hline 
\end{array}$
}


\begin{lemma}\label{lemma:vc-no-step-back}
An addable (respectively, removable) node that enters (respectively, leaves) the vertex cover does not become removable (respectively addable) in any future time.

\end{lemma}

\begin{proof}
    Let $s$ be the state at time $t$ while \Cref{algorithm:vc-ll} is executing. We have from \Cref{algorithm:vc-ll} that if a node $i$ is addable-\imped or removable-\imped, then no other node in $Adj^3_i$ changes its state.

    If $i$ is removable-\imped at $t$, then all the nodes in $Adj_i$ are in the vertex cover. 
    Any node $j\in Adj_i$ cannot move out of the vertex cover until $i$ moves out. Hence, $i$ does not become addable after being removed at time $t$.

    If $i$ is addable-\imped at $t$, then some node in $Adj_i$ is out of the vertex cover. 
    After $i$ moves in and forces its removable neighbours out, $i$ is no longer addable and all nodes in $Adj_i$ are no longer removable.
    Now, $i$ can become removable only if some neighbour of $i$ enters the vertex cover. 
    Let $j$ be a neighbour of $i$. We consider two cases (1) $j[st]=OUT$ (2) $j[st]=IN$ and $j$ was forced out by $i$ (3) $j[st]=IN$ and $j$ was not forced out by $i$. 
    
    In the first case, as long as $j$ never enters the vertex cover, $i$ cannot become removable, thereby ensuring that edge $\{i, j\}$ remains covered. In the event $j$ enters the vertex cover, $i$ may be forced out. However, (as a result) $i$ does not become removable-\imped. 
    
    In the second case, $j$ is forced out of the vertex cover. This implies that all neighbours of $j$ are already in the vertex cover. Hence, $j$ does not become addable again. (This argument is the same as the above argument that showed that if a node is removed from the vertex cover, it does not become addable.) Since $j$ is never added back to the vertex cover, $i$ cannot become removable because of the action of $j$.
    
    In the third case, even if $j$ moves out afterwards, it cannot make $i$ removable, 
    so after time $t$, nodes in $Adj_i$ do not move in; as a result, $i$ does not have to move out after moving in. In addition, the nodes that $i$ turned out of the vertex cover do not have to move in after $t$, because the neighbours of those nodes are not changing their states simultaneously when their states are being changed.
    

    
    From the above cases, we have that $i$ does not change its state to $i[st]$ after changing its state from $i[st]$ to $i[st']$.
    throughout the execution of \Cref{algorithm:vc-ll}.
\end{proof}

To demonstrate that \Cref{algorithm:vc-ll} is lattice-linear, we define state value and rank, auxiliary variables associated with nodes and global states, as follows:
$$
    \begin{array}{l}
        \textsc{State-Value-MVC}(i,s)=
        \begin{cases}
            1 & \text{if $\textsc{Unsatisfied-MVC-FLL}(i)$ in state $s$} \\
            0 & \text{otherwise}
        \end{cases}
    \end{array}
$$
$$
    \textsc{Rank-MVC}(s)=\sum\limits_{i\in V(G)}\textsc{State-Value-MVC}(i,s).
$$



\begin{theorem}\label{theorem:vc-ll}
    \Cref{algorithm:vc-ll} is a silent self-stabilizing and lattice-linear algorithm executed by $n$ nodes running asynchronously.
\end{theorem}

\begin{proof}
    We have from the proof of \Cref{lemma:vc-no-step-back} that if $G$ is in state 
    $s$ and $\textsc{Rank-MVC}(s)$ is non-zero, then at least one node will be \imped, e.g., an unsatisfied node with the highest ID.
    For any node $i$, we have that $\textsc{State-Value-MVC}(i)$ decreases whenever $i$ is \imped and never increases. In addition, by the action of $i$, the state value of any other node in $G$ does not increase. (Note that adding node $i$ to the vertex cover may have caused a neighbour $j$ of $i$ to be removable. If this happens, $j$ is removed from the vertex cover in the same action atomically. Hence, $j$ does not become unsatisfied.)
    In effect, $\textsc{Rank-MVC}$ monotonously decreases throughout the execution of the algorithm until it becomes zero. This shows that \Cref{algorithm:vc-ll} is self-stabilizing. Once \textsc{Rank-MVC} is zero, no node is \imped, so no node makes a move. This shows that \Cref{algorithm:vc-ll} is silent.
    
    \Cref{algorithm:vc-ll} exhibits properties similar to \Cref{algorithm:ds-ll} (as well as \Cref{algorithm:gc-ll}) which are elaborated in the proof for its lattice-linearity in \Cref{theorem:ds-ll}. From there, we obtain that \Cref{algorithm:vc-ll} also is lattice-linear.
\end{proof}

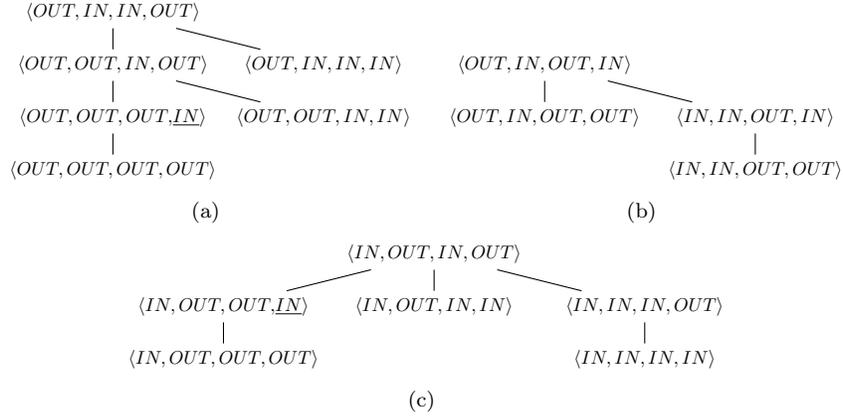
\begin{figure}[ht]
    \centering
    
    \subfigure[]{
        \begin{tikzpicture}[scale=.7,every node/.style={scale=.7}]
            \node at (0,3) (a6) {$\langle OUT,IN,IN,OUT\rangle$};
            \node at (0,2) (a2) {$\langle OUT,OUT,IN,OUT\rangle$};
            \node at (0,1) (a1) {$\langle OUT,OUT,OUT,$\underline{$IN$}$\rangle$};
            \node at (0,0) (a0) {$\langle OUT,OUT,OUT,OUT\rangle$};
            
            \node at (4,1) (a3) {$\langle OUT,OUT,IN,IN\rangle$};
            
            \node at (4,2) (a7) {$\langle OUT,IN,IN,IN\rangle$};

            \draw (a0) -- (a1) -- (a2) -- (a6);
            \draw (a3) -- (a2);
            \draw (a7) -- (a6);
        \end{tikzpicture}
    }
    \subfigure[]{
        \begin{tikzpicture}[scale=.7,every node/.style={scale=.7}]
            \node at (0,5) (a5) {$\langle OUT,IN,OUT,IN\rangle$};
            \node at (0,4) (a4) {$\langle OUT,IN,OUT,OUT\rangle$};
            
            \node at (4,4) (a13) {$\langle IN,IN,OUT,IN\rangle$};
            \node at (4,3) (a12) {$\langle IN,IN,OUT,OUT\rangle$};

            \draw (a4) -- (a5);
            \draw (a12) -- (a13) -- (a5);
        \end{tikzpicture}
    }
    \subfigure[]{
        \begin{tikzpicture}[scale=.7,every node/.style={scale=.7}]
            \node at (4,10) (a10) {$\langle IN,OUT,IN,OUT\rangle$};
            \node at (0,9) (a9) {$\langle IN,OUT,OUT,$\underline{$IN$}$\rangle$};
            \node at (0,8) (a8) {$\langle IN,OUT,OUT,OUT\rangle$};
            
            \node at (4,9) (a11) {$\langle IN,OUT,IN,IN\rangle$};
            
            \node at (8,9) (a14) {$\langle IN,IN,IN,OUT\rangle$};
            \node at (8,8) (a15) {$\langle IN,IN,IN,IN\rangle$};
            
            \draw (a8) -- (a9) -- (a10);
            \draw (a11) -- (a10);
            \draw (a15) -- (a14) -- (a10);
        \end{tikzpicture}
    }
    \caption{$\prec$-lattices formed by the global states of $G^p_4$ of 4 nodes forming a straight path $\langle v_1,v_2,v_3,v_4\rangle$ under \Cref{algorithm:vc-ll}. The nodes that are kicked out, by a node that decides to move into the vertex cover, are underlined.}
    \label{figure:vc-semilattice-example}
\end{figure}

\begin{example}\label{example:vc-semilattice}
    Let $G^p_4$ be a graph of four vertices forming a path $\langle v_1,v_2,v_3,v_4\rangle$ such that $v_i[id]>v_j[id]$ iff $i>j$. In \Cref{figure:vc-semilattice-example}, we show all possible state transitions that $G^p_4$ can go through under \Cref{algorithm:vc-ll}. The global states in the figure are of the form $\langle v_1[st],v_2[st],v_3[st],v_4[st]\rangle$.
    \qed 
\end{example}

\subsection{Fully Lattice-Linear Algorithm for Maximal Independent Set (\mis)}\label{subsection:mis-ll}

In this subsection, we describe an algorithm for the \mis. The issues, similar to the issues that we discussed in \Cref{subsection:vc-no-tie-breaker}, can be observed for \mis problem as well. For example, a similar behaviour can be observed on a path of 4 nodes, all initialized to $IN$. However, we can follow the general design of \Cref{algorithm:vc-ll}, that we described in \Cref{subsection:mvc-ll}, to develop an algorithm for \mis. \mis problem is defined in \Cref{definition:mis} (\Cref{section:mis}).

The macros that we use here are similar to the macros we used for \mvc, but with opposite polarity. We use the macros listed in \Cref{table:macros-mis-fll}. 
A node $i$ is \textit{addable} iff $i$ is out of the independent set (\is), and all the neighbours of $i$ are also out of the \is.
$i$ is \textit{removable} iff $i$ is in the \is and there is some edge $\{i,j\}$ incident on $i$ such that $j$ is also in the \is.
$i$ is \textit{unsatisfied} iff $i$ is removable or $i$ is addable.
$i$ is \textit{\imped} iff $i$ is unsatisfied and there is no node $j$ in distance-3 of $i$, with ID greater than $i$, such that $j$ is unsatisfied.

\begin{table}[ht]
    \centering 
    \doublespacing 
    \begin{tabular}{|l|}
        \hline 
        \textsc{Addable-MIS-FLL}$(i)$ $\equiv i[st]=OUT\land (\forall j\in Adj_i j[st]=OUT)$.\\
        \textsc{Removable-MIS-FLL}$(i)$ $\equiv i[st]=IN\land (\exists j\in Adj_i:j[st]=IN)$.\\
        \textsc{Unsatisfied-MIS-FLL}$(i)$ $\equiv \textsc{Removable-MIS-FLL}(i)\lor \textsc{Addable-MIS-FLL}(i)$.\\
        $\textsc{\Imped-MIS-FLL}(i)\equiv\textsc{Unsatisfied-MIS-FLL}(i)\land$\\ \quad\quad $(\forall j\in Adj^2_i:\lnot\textsc{Unsatisfied-MIS-FLL}(j)\lor i[id]>j[id])$.\\
        \hline 
    \end{tabular}
    \caption{Macros used in the algorithm for \mis.}
    \label{table:macros-mis-fll}
\end{table}

Given the similarities in the \mvc and \mis problems, the algorithm we develop here is similar to the algorithm we developed for \mvc. The algorithm is defined as follows. A \textit{removable-\imped} node $i$ turns itself out and moves all its addable neighbours into the independent set (after accounting for the fact that $i$ has already turned out). An \textit{addable-\imped} node $i$ will turn itself into the independent set. In the present design of the algorithm, the execution while $i$ is \imped is assumed to be executed atomically.

\begin{algorithm}\label{algorithm:is-ll}Rules for node $i$.
\end{algorithm}
\begin{center}
    $\begin{array}{|l|}
        \hline 
        \textsc{\Imped-MIS-FLL}(i) \longrightarrow\\
        \begin{cases}
            i[st]=OUT.~\forall j\in Adj_i: j[st]=IN, \text{if $\textsc{Addable-MIS-FLL}(j)$}. & \text{if $\textsc{Removable-MIS-FLL}(i)$}.\\
            i[st] = IN. & \text{otherwise}
        \end{cases}~\\
        \hline 
    \end{array}$
\end{center}

Since the behaviour of \Cref{algorithm:is-ll} is similar to the behaviour of \Cref{algorithm:vc-ll}, we briefly cover the description of the behaviour of \Cref{algorithm:is-ll} in the following: \Cref{lemma:is-no-step-back} and \Cref{theorem:is-ll}.

\begin{lemma}\label{lemma:is-no-step-back}
    Any node in an input graph does not revisit its older state while executing under \Cref{algorithm:is-ll}.
\end{lemma}

\begin{proof}
    Let $s$ be the state at time $t$ while \Cref{algorithm:is-ll} is executing. We have from \Cref{algorithm:is-ll} that if a node $i$ is addable-\imped or removable-\imped, then no other node in $Adj^3_i$ changes its state.
    
    If $i$ is removable-\imped at $t$, then some node in $Adj_i$ is in the independent set. After when $i$ moves out, and turns its addable neighbours in, then we have that $i$ is no longer removable and all nodes in $Adj_i$ are no longer addable, so after time $t$, nodes in $Adj_i$ do not move out; as a result, $i$ does not have to move in after moving out. In addition, the nodes that $i$ turned into the independent set do not have to move out after $t$, because the neighbours of those nodes are not changing their states simultaneously when their states are being changed.
    
    Let that at some instance of time, two nodes $i$ and $j$ simultaneously evaluate that they are removable-\imped. Since no nodes in $Adj_i^3$ (respectively, $Adj_{j}^3$) change their state under \Cref{algorithm:is-ll} until $i$ (respectively, $j$) changes its state, we have that $i$ and the nodes that $i$ would move into the independent set are not adjacent to $j$ or the nodes that $j$ would move into the independent set. Thus, no node by the action of $i$ becomes removable.
    
    If otherwise $i$ is addable-\imped at $t$, then all the nodes in $Adj_i$ are out of \is. So after when $i$ moves in, none of the non-unsatisfied nodes in $Adj_i$ (non-addable nodes in $Adj_i$), including $i$, becomes unsatisfied (addable). As a result, $i$ does not have to move out after moving in.
    
    From the above cases, we have that $i$ does not change its state to $i[st]$ after changing its state from $i[st]$ to $i[st']$.
    throughout the execution of \Cref{algorithm:is-ll}.
\end{proof}

To demonstrate that \Cref{algorithm:is-ll} is lattice-linear, we define state value and rank, auxiliary variables associated with nodes and global states, as follows:
$$
    \begin{array}{l}
        \textsc{State-Value-MIS}(i,s)= 
        \begin{cases}
            1 & \text{if $\textsc{Unsatisfied-IS}(i)$ in state $s$} \\
            0 & \text{otherwise}
        \end{cases}
    \end{array}
$$
$$
    \textsc{Rank-MIS}(s)=\sum\limits_{i\in V(G)}\textsc{State-Value-MIS}(i,s).
$$



\begin{theorem}\label{theorem:is-ll}
    \Cref{algorithm:is-ll} is a silent self-stabilizing and lattice-linear algorithm executed by $n$ nodes running asynchronously.
\end{theorem}

\begin{proof}
    We have from the proof of \Cref{lemma:is-no-step-back} that if $G$ is in state 
    $s$ and $\textsc{Rank-MIS}(s)$ is non-zero, then at least one node will be \imped, e.g., an unsatisfied node with the highest ID.
    For any node $i$, we have that $\textsc{State-Value-MIS}(i)$ decreases whenever $i$ is \imped and never increases. In addition, by the action of $i$, the state value of any other node in $G$ does not increase.
    In effect, $\textsc{Rank-MIS}$ monotonously decreases throughout the execution of the algorithm until it becomes zero. This shows that \Cref{algorithm:is-ll} is self-stabilizing.
    Once \textsc{Rank-MIS} is zero, no node is \imped, so no node makes a move. This shows that \Cref{algorithm:is-ll} is silent.
    
    \Cref{algorithm:is-ll} exhibits properties similar to \Cref{algorithm:ds-ll} (as well as \Cref{algorithm:gc-ll} and \Cref{algorithm:vc-ll}) which are elaborated in the proof for its lattice-linearity in \Cref{theorem:ds-ll}. From there, we obtain that \Cref{algorithm:is-ll} also is lattice-linear.
\end{proof}

\subsection{Complex Actions: Properties Shared by Algorithms for \mvc and \mis}\label{subsection:complex-actions-vc-is}

In this subsection, we study some behavioural aspects   of the algorithm for \mvc present in \Cref{subsection:mvc-ll}. Consequently, similar arguments for the algorithm for \mis will follow.

\Cref{algorithm:vc-ll} can be transformed into a simple action algorithm as follows. To accommodate that, we use the variable $i[addable]$ to set to be $true$ so that the surrounding nodes can then evaluate if they are removable. We use the following additional guards.
For a node $i$, \textit{else-pointed} is true iff a node $j$ in $Adj^4_i$ moved into the \vc (and set $j[addable]$ to $true$), and there is a node $k$ in $Adj_j$ that is removable.

\begin{center}
        $\textsc{Else-Pointed-MVC-FLL}(i)\equiv \exists j\in Adj^4_i:(j[addable]=true\land \exists k\in Adj_j: \textsc{Removable-MVC-FLL}(k))$
\end{center}

Consequently, the algorithm can be modified as follows. A node $i$ will not be enabled (\imped) if else-pointed is true for $i$. The modified algorithm that allows simple actions, thus, is defined as follows.

\begin{algorithm}\label{algorithm:vc-ll-simple-actions}
    Transformed \Cref{algorithm:vc-ll}, where nodes only execute simple actions.
\end{algorithm}
\begin{center}
    $\begin{array}{|l|}
        \hline
        \textsc{\Imped-MVC-FLL-II}(i)\equiv\\
        \quad\quad (\exists j\in Adj_i:j[addable]=true \land \textsc{Removable-MVC-FLL}(i))\lor\\
        \quad\quad\quad\quad (\lnot\textsc{Else-Pointed-MVC-FLL}(i)\land (\textsc{Unsatisfied-MVC-FLL}(i)\land\\ 
        \quad\quad\quad\quad (\forall j\in Adj^3_i:\lnot\textsc{Unsatisfied-MVC-FLL}(j)\lor\\ \quad\quad\quad\quad i[id]>j[id]))).\\~\\
        \text{Rules for node $i$}:\\
        \textsc{\Imped-MVC-FLL-II}(i)\longrightarrow
        \begin{cases}
            i[addable]=true & \text{if $i[st]=OUT$.}\\
            i[addable]=false & \text{if $i[st]=IN$.}\\
            i[st]=\lnot i[st] & \text{unconditionally.}\\
        \end{cases}
        ~\\
        \hline 
    \end{array}$
\end{center}

Next, we identify why \Cref{algorithm:vc-ll-simple-actions} can be reconciled with the inability to design an algorithm with simple actions from \Cref{subsection:vc-no-tie-breaker}. This analysis also helps us to obtain the correctness proof of \Cref{algorithm:vc-ll-simple-actions}.

\subsubsection*{Reconciling \Cref{subsection:vc-no-tie-breaker} and \Cref{algorithm:vc-ll-simple-actions}}

As discussed in \Cref{subsection:vc-no-tie-breaker}, if a tie-breaker in conjunction with simple actions is deployed, then the system would exhibit cyclic behaviour. On the other hand, Algorithm \ref{algorithm:vc-ll-simple-actions} uses only simple actions. We identify the subtlety, involved in both these results, that make this possible, in the following.


To explain how these results can coexist together, we describe the behaviour of \Cref{algorithm:vc-ll-simple-actions} with an example. Assume that this algorithm is deployed on a graph of 4 nodes forming a path, with node IDs being in the sequence $\langle 1,4,3,2\rangle$. In a simple algorithm that uses a tiebreaker with all nodes (which is considered in \Cref{subsection:vc-no-tie-breaker}), node 4 would execute first then node 3, then node 2 and finally node 1. This execution order is not preserved in \Cref{algorithm:vc-ll-simple-actions}, which we discuss as follows.

Specifically, let the initial global state in this graph be $\langle IN,OUT,OUT,OUT\rangle$. First, node 4 will move into the \vc. Now, the node that is unsatisfied and has the highest ID is node 3. However, \textsc{Else-Pointed}(3) is true, because $4[addable]$ is set to $true$ and node $1$
needs to move out of the vertex cover as part of the action that allowed node $4$ to enter the vertex cover.
In other words, node $3$ can execute only after node $1$ leaves the vertex cover. This is not permitted in a algorithm that uses simple actions with tie-breaker on node IDs. This effect is similar to that of priority inheritance, where node $1$ inherits the priority of node $4$ because it has to be forced out of the vertex cover by node $4$. Hence, in this specific case, 
node 1 has a higher priority of movement as compared to node 3, despite the fact that node 3 is unsatisfied and is of a higher ID, because of a recent action committed by node 4.

After node 1 moves out, node 3 moves in, and thence, the system reaches an optimal state. 

\subsubsection*{Correctness of \Cref{algorithm:vc-ll-simple-actions}}

\Cref{algorithm:is-ll} is lattice-linear with respect to state value and rank, defined as follows.
$$
    \begin{array}{l}
        \textsc{State-Value-MVC-II}(i,s)=
        \begin{cases}
            |Adj_i+1|\\ \quad\quad \text{if $\textsc{Unsatisfied-MVC-FLL}(i)$ in state $s$} \\
            |\{j\in Adj_i: \textsc{Unsatisfied-MVC-FLL}(j)\}|\\ \quad\quad \text{if in state $s$, $\lnot\textsc{Unsatisfied-MVC-FLL}(i)\land$}\\
            \quad\quad\quad\quad \text{$(\exists j\in Adj_i: \textsc{Unsatisfied-MVC-FLL}(j))$} \\
            0\\ \quad\quad \text{otherwise}
        \end{cases}
    \end{array}
$$
$$
    \textsc{Rank-MVC-II}(s)=\sum\limits_{i\in V(G)}\textsc{State-Value-MVC-II}(i,s).
$$

Since the working of this algorithm straightforwardly follows from the working of \Cref{algorithm:vc-ll}, we omit the proof of correctness of this algorithm. A similar algorithm can be developed for \mis, that deploys only simple actions.




\section{Experiments}\label{section:experiments}

In this section, we present the experimental results of convergence times from implementations run on real-time shared memory model. We implement the algorithm for minimal dominating set (\Cref{algorithm:ds-ll}), and compare it to algorithms by Hedetniemi et al. (2003) \cite{Hedetniemi2003} and Turau (2007) \cite{Turau2007}. We also present the runtime of a distance-1 transformation of \Cref{algorithm:ds-ll}. First, we present the transformation of \Cref{algorithm:ds-ll} in the following subsection.

\subsection{Transforming \Cref{algorithm:ds-ll} to Distance-1}\label{subsection:mds-d1-transformation}

In \Cref{algorithm:ds-ll}, we observe that the guards of a node $i$ are distance-4. First, we transform this algorithm to a distance-1 algorithm. To accomplish this, the nodes maintain additional variables, that provide them information about other nodes, as required.
We use additional variables and guards to propagate this information.
Due to the constraint of reading only distance-1 neighbours, the nodes may end up reading old information about the other nodes. However, due to lattice-linearity, such executions stay to be correct.

A straightforward transformation would require each node $i$ to maintain copies of all the variables of its distance-4 neighbours. However, we use only 4 additional variables. Note that the requirement of these four variables is independent of the number of nodes in $Adj_i^4$. 

We use $i[ldom]$ and $i[hdom]$ to assist in propagating the information about the macro $\textsc{Removable-DS}(i)$.
$i[hdom]$ stores the highest ID dominator of $i$: it is a node in $N_i$ of highest ID such that its state is $IN$. $i[ldom]$ stores the lowest ID dominator of $i$: it is a node in $N_i$ of lowest ID such that its state is $IN$. 
(If such a node does not exist, these variables are set to $\top$ (\textit{null}).)

Once Removable-DS is transformed to a distance-1 macro, unsatisfied-ds will also be a distance-1 macro, as $\textsc{Addadble-DS}$  is already a distance-1 macro. 

We use $i[uflag]$ and $i[hud1]$ to assist in propagating the information about $\textsc{\Imped-II-DS}$.
Node $i$ sets $i[uflag]$ to $true$ to indicate that $i$ is unsatisfied.
$i[hud1]$ stores the highest ID node in distance-1, i.e., in $N_i$, of $i$ that is unsatisfied.


Now, we describe the actions of the transformed distance-1 algorithm. 
We use the macros listed in \Cref{table:macros-transformed-mds-fll}. $i$ is \textit{hdom-outdated} iff $i[hdom]$ is not equal to the highest ID dominator of $i$. $i$ is \textit{ldom-outdated} iff $i[ldom]$ is not equal to the lowest ID dominator of $i$. 
$i$ is \textit{removable} iff every node $j \in N_i$ will stay dominated even if $i$ moves out of the dominating set. This will happen if either $j[st]=IN$, or, either $j[hdom]$ or $j[ldom]$ differs from $i$. 
Node $i$ is \textit{addable} if all nodes in $Adj_i$, along with $i$, are out of the \ds. 
$i$ is \textit{unsatisfied} if $i$ is removable or addable. $i$ is \textit{unsatisfied-flag-outdated} iff $i[uflag]$ is not equivalent to $i$ being unsatisfied. $i$ is \textit{hud1-outdated} iff $i[hud1]$ is not qual to the highest ID node in $N_i$ that is unsatisfied. $i$ is \textit{unsatisfied-\imped} if $i$ is the highest ID node in the distance-2 neighbourhood that is unsatisfied. $i$ is \textit{\imped} iff $i$ is hdom-outdated, ldom-outdated, unsatisfied-flag-outdated, hud1-outdated or unsatisfied-\imped.

\begin{table}[ht]
    \centering 
    \doublespacing 
    \begin{tabular}{|l|}
        \hline 
        variables of $i$: $st,ldom,hdom,uflag,hud1$\\
        $\textsc{HDom-Outdated}(i)\equiv i[hdom]\neq \text{arg} \max\{x[id]: x\in N_i\land x[st]=IN\}$.\\
        $\textsc{LDom-Outdated}(i)\equiv i[ldom]\neq \text{arg} \min\{x[id]: x\in N_i\land x[st]=IN\}$.\\
        $\textsc{Removable-\mds-D}\equiv i[st]=IN\land (\forall j\in N_i: ((j\neq i\land j[st]=IN)\lor ((j[ldom]\neq i\land$\\ \quad\quad$j[ldom]\neq \top)\lor(j[hdom]\neq i\land j[hdom]\neq \top))))$.\\
        $\textsc{Addable-\mds-D}(i) \equiv i[st]=OUT\land (\forall j\in Adj_i:j[st]=OUT)$.\\
        $\textsc{Unsatisfied-\mds-D}(i)\equiv \textsc{Removable-\mds-D}(i)\lor \textsc{Addable-MDS-D}(i)$.\\
        $\textsc{Unsatisfied-Flag-Outdated}(i)\equiv i[uflag]\not= \textsc{Unsatisfied-\mds-D}(i)$.\\
        $\textsc{HUD1-Outdated}(i)\equiv i[hud1]\neq \text{arg} \max\{x[id]: x\in N_i\land x[uflag]=true\}$.\\
        $\textsc{Unsatisfied-\Imped-MDS-D}(i)\equiv i[uflag]\land (\forall j\in Adj_i:(j[uflag]\Rightarrow j[id]<i[id])\land$\\ \quad\quad$(j[hud1]\neq \top\Rightarrow j[hud1]<i[id]))$.\\
        $\textsc{\Imped-\mds-D}(i)\equiv \textsc{HDom-Outdated}(i)\lor \textsc{LDom-Outdated}(i)\lor$\\ \quad\quad $\textsc{Unsatisfied-Flag-Outdated}(i)\lor \textsc{HUD1-Outdated}(i)\lor$\\ \quad\quad $\textsc{Unsatisfied-\Imped-MDS-D}(i)$.\\
        \hline 
    \end{tabular}
    \caption{Macros used in transformed algorithm for \mds.}
    \label{table:macros-transformed-mds-fll}
\end{table}

We describe the algorithm as follows. If $i$ is hdom-outdated, then it updates $i[hdom]$ with ID of the highest ID node in $N_i$ that is in the dominating set. If $i$ is ldom-outdated, then it updated $i[ldom]$ with ID of the lowest ID node in $N_i$ that is in the dominating set. If $i$ is unsatisfied-flag-outdated, then it updates $i[uflag]$ to correctly denote whether $i$ is unsatisfied. If $i$ is hud1-outdated, then $i$ updates $i[hud1]$ with the ID of the highest ID node in $N_i$ that is unsatisfied. If $i$ is unsatisfied-impedensable, then $i$ toggles $i[st]$.

\newpage 
\begin{algorithm}\label{algorithm:ds-ll-d1}
    Rules for node $i$.
\end{algorithm}
\begin{center}
    $\begin{array}{|l|}
        \hline 
        \textsc{\Imped-\ds-M}(i)\longrightarrow\\
        \begin{cases}
            hdom=\text{arg} \max\{x[id]:x\in N_i\land x[st]=IN\} & \text{if $\textsc{HDom-Outdated}(i)$}.\\
            ldom=\text{arg} \min\{x[id]:x\in N_i\land x[st]=IN\} & \text{if $\textsc{LDom-Outdated}(i)$}.\\
            uflag\not= \textsc{Unsatisfied-\mds-D}(i) & \text{if $\textsc{Unsatisfied-Flag-\Imped}(i)$}.\\
            i[hud1]=\text{arg} \max\{x[id]: x\in N_i\land x[uflag]=true\} & \text{if $\textsc{HUD1-Outdated}(i)$}.\\
            i[st]=\lnot i[st] & \text{if $\textsc{Unsatisfied-\Imped}(i)$}.\\
            i[uflag]=false & \text{unconditionally}.
        \end{cases}~\\
        \hline 
    \end{array}$
\end{center}

Deploying the above algorithm reduces the work complexity of evaluating the guards for a node to $O(\Delta)$, which was originally $O(\Delta^4)$ in \Cref{algorithm:ds-ll}. In principle, all lattice-linear distance-$x$ (where $x>1$) algorithms can be transformed into distance-1 algorithms by 
keeping a copy of all variables in $Adj^x$ \cite{Afek2002}. However, it will increase the space complexity of every node by $|Adj^x|$, without decreasing the time complexity of the evaluation of guards.
By contrast, the above algorithm increases the space complexity by only $O(1)$, while decreasing the time complexity from $O(\Delta^4)$ to $O(\Delta)$.

\subsection{Runtime Comparison}

While we see a significant reduction of the time complexity, of the evaluation of guards by a node, from $O(\Delta^4)$ in \Cref{algorithm:ds-ll} to $O(\Delta)$ in \Cref{algorithm:ds-ll-d1}, it is also worthwhile to compare the convergence time of these algorithms when they are implemented on real-time systems.
In this subsection, we compare the runtime of \Cref{algorithm:ds-ll-d1} with \Cref{algorithm:ds-ll} and other algorithms.

\begin{figure}[ht]
    \centering
    \subfigure[]{
        \includegraphics[width=.44\textwidth]{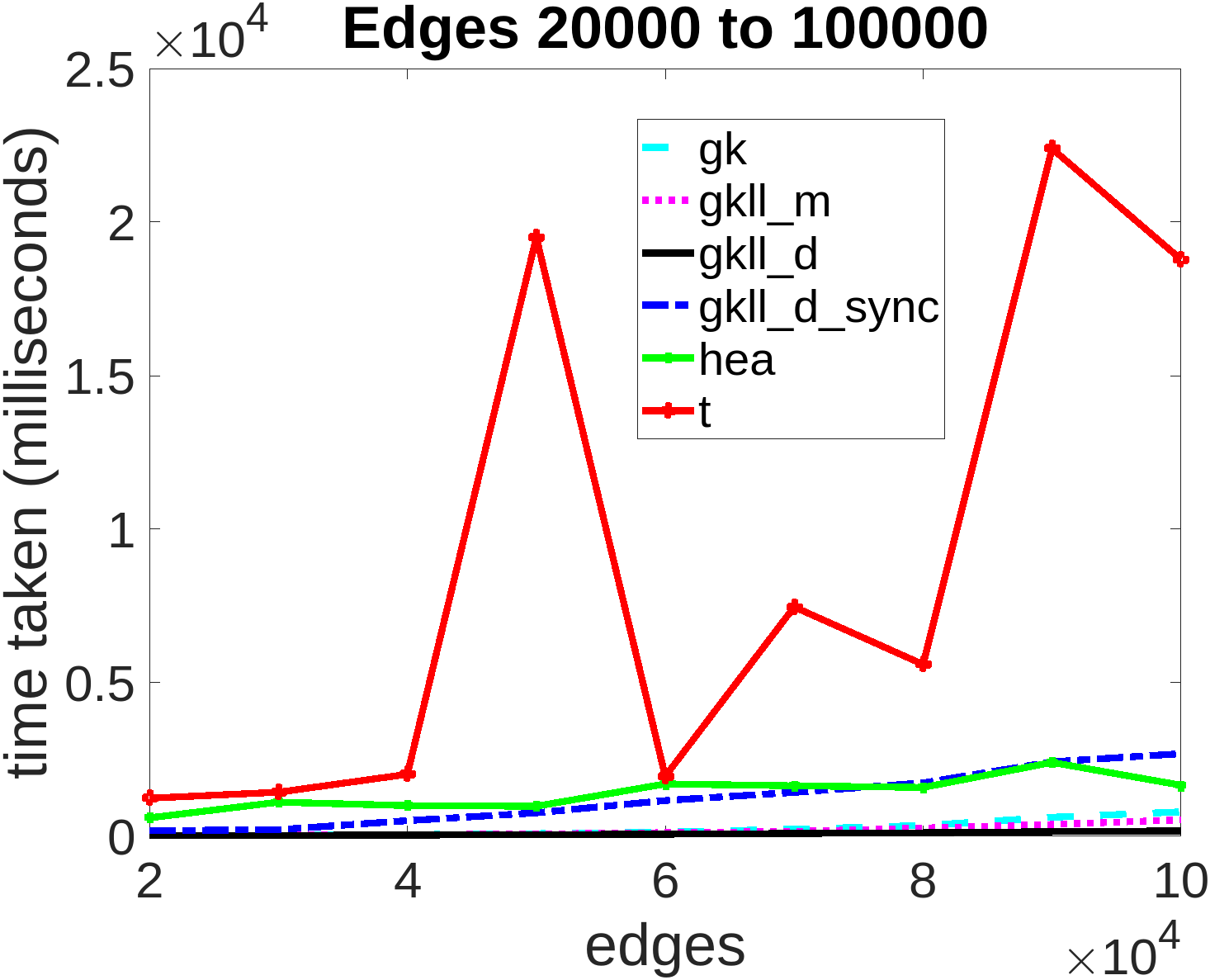}
    }
    \subfigure[]{
        \includegraphics[width=.44\textwidth]{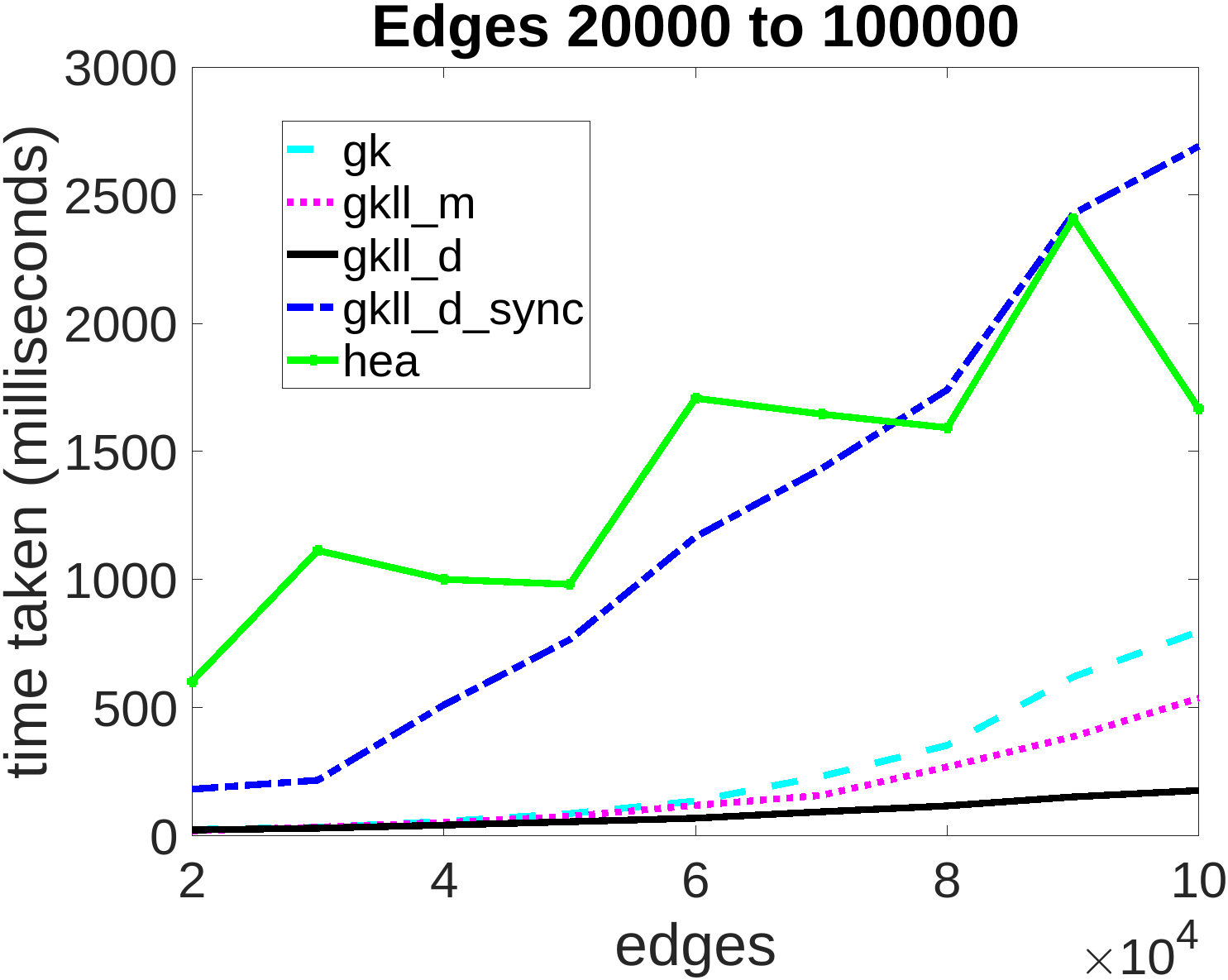}
    }
    \subfigure[]{
        \includegraphics[width=.44\textwidth]{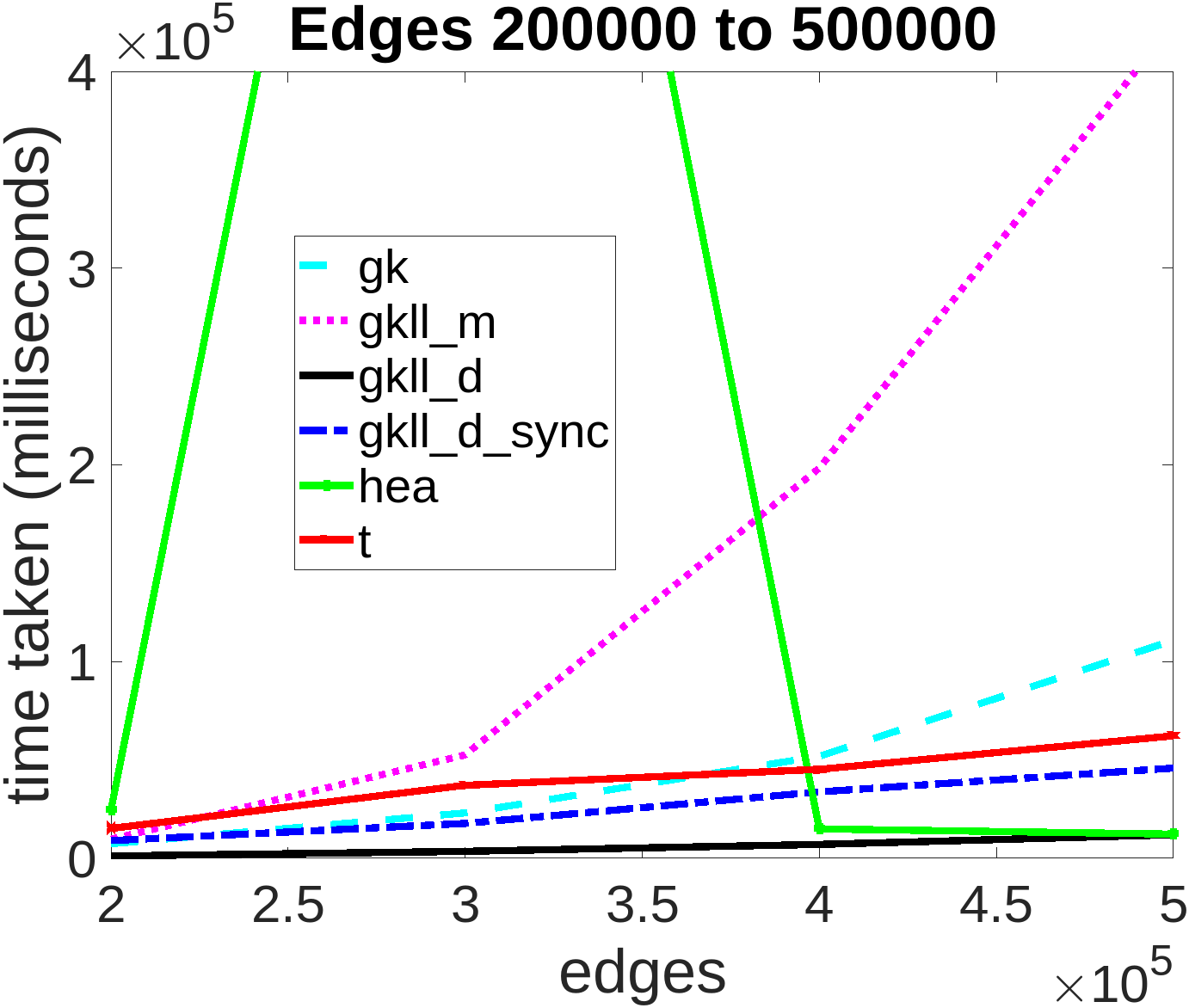}
    }
    \subfigure[]{
        \includegraphics[width=.44\textwidth]{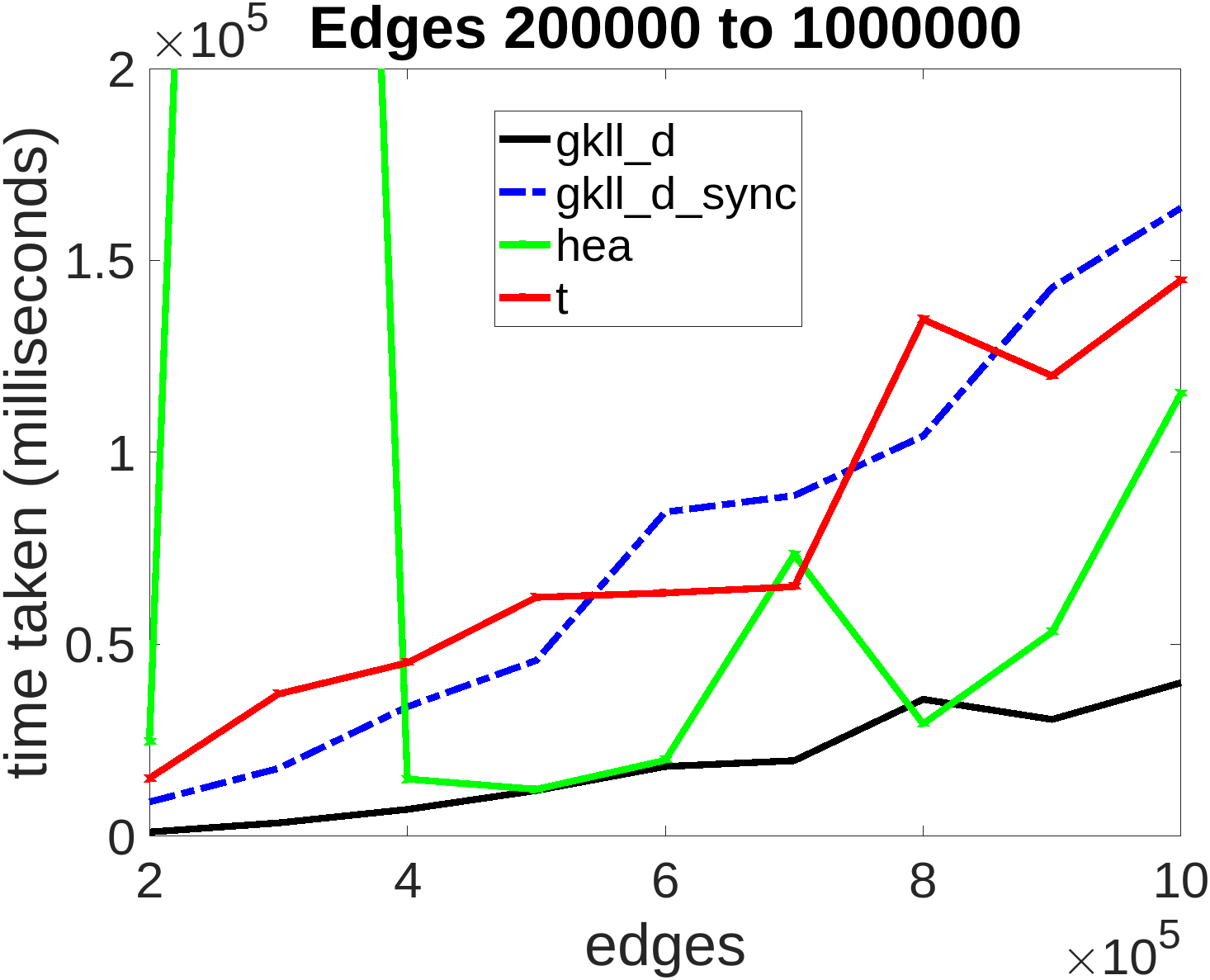}
    }
    \caption{Runtime comparison of \Cref{algorithm:ds-ll-d1}, \Cref{algorithm:ds-ll}, \Cref{algorithm:ds-ellss} and other algorithms for minimal dominating set in the literature. All graphs are of 10,000 nodes.}
    \label{figure:experiments-flla}
\end{figure}

We implemented \Cref{algorithm:ds-ll} (\texttt{gkll\_m}), 
\Cref{algorithm:ds-ll-d1} (\texttt{gkll\_d}), lockstep synchronized \Cref{algorithm:ds-ll-d1} (\texttt{gkll\_d\_sync}), \Cref{algorithm:ds-ellss} (\texttt{gk}), algorithms for minimal dominating set present in Hedetniemi et al. (2003) \cite{Hedetniemi2003} (\texttt{hea}) and Turau (2007) \cite{Turau2007} (\texttt{t}), and compare their convergence time. 
The input graphs were random graphs of order 10,000 nodes, generated by the \texttt{networkx} library of python. For comparing the performance results, all algorithms are run on the same set of graphs. 

The experiments are run on Cuda using the \texttt{gcccuda2019b} compiler. \texttt{gkll\_m}, \texttt{gkll\_d} and \texttt{gk} were run asynchronously, and the algorithms in \texttt{gkll\_d\_sync}, \texttt{hea} and \texttt{t} were run under the required synchronization model.
The experiments are run on \texttt{Intel(R) Xeon(R) Platinum 8260 CPU @} \texttt{2.40} \texttt{GHz, cuda  V100S}. The programs are run using the command \texttt{nvcc $\langle$program$\rangle$.cu}. Here, each multiprocessor ran 256 threads. And, the system provided sufficient multiprocessors so that each node in the graph can have its own thread. All the observations are an average of 3 readings.

\Cref{figure:experiments-flla} (a) (respectively, \Cref{figure:experiments-flla} (c) and \Cref{figure:experiments-flla} (d)) shows a line graph comparision of the convergence time for these algorithms with the number of edges varying from 20,000 to 100,000  (respectively, 200,000 to 500,000 and 200,000 to 1,000,000).
So, the average degree is varying from 4 to 20 (respectively, 40 to 100 and 40 to 200).
\Cref{figure:experiments-flla} (b) is same as \Cref{figure:experiments-flla} (a), except that the curve for \texttt{hea} is removed so that the other curves can be analyzed closely.
Similarly, \Cref{figure:experiments-flla} (c) and \Cref{figure:experiments-flla} (d) are similar, however, (1) \Cref{figure:experiments-flla} (c) shows curves for convergence time of graphs of average degree 40 to 100, whereas \Cref{figure:experiments-flla} (d) shows curves for convergence time of graphs of average degree 40 to 200, and (2) \Cref{figure:experiments-flla} (c) contains all 6 curves, whereas \Cref{figure:experiments-flla} (d) does not contain curves for \texttt{gkll\_m} and \texttt{gk}.
Observe that
the convergence time taken by the program for \texttt{gkll\_d} is consistently lower than the other algorithms.

In \Cref{figure:experiments-flla} (b), it can be observed that the runtime of \texttt{gkll\_m} is lower than the other algorithms (except \texttt{gkll\_d}, which is not surprising). However, in \Cref{figure:experiments-flla} (c), it can be observed that the runtime of \texttt{gkll\_m} increases more rapidly and overtakes the runtime of other algorithms (that is why we omitted \texttt{gkll\_m} from \Cref{figure:experiments-flla} (d)). This happens mainly because the nodes under \texttt{gkll\_m} are reading values of nodes at distance-4 from themselves. \texttt{gk} also converges comparatively quicker than (\texttt{gkll\_m}) (but not quicker than other algorithms) because its first phase is quicker: the addable nodes move in the dominating set ``carelessly'', whereas in \texttt{gkll\_m} the nodes moving in are ``careful'' as well as the nodes that move out of the dominating set, which adds to the convergence time in the case of \texttt{gkll\_m}.

Next, we discuss how much of the benefit of \texttt{gkll\_d} can be allocated to asynchrony due to the property of lattice-linearity. For this, observe the performance of \texttt{gkll\_d} running in asynchrony (to allow nodes to read old/inconsistent values)  against \texttt{gkll\_d\_sync} (which is the same algorithm as \texttt{gkll\_d} but running in lock-step, to ensure that the nodes only read the most recent values). We observe that the asynchronous implementation has a lower convergence time. This happens mainly because both the asynchronous and the synchronized algorithms have the same convergence time complexity, however, the cost of synchronization (time spent in synchronization, plus the requirement of at least one scheduling thread) is eliminated.

We have performed the experiments on shared memory architecture that allows the nodes to access memory \textit{quickly}. This means that the overhead of synchronization is low. By contrast, if we had implemented these algorithms on a distributed system instead, where computing processors are placed remotely, the cost of synchronization would be even higher. Hence, we anticipate the benefit of lattice-linearity (where synchronization is not needed) to be even higher.

\section{Lattice-Linear 
2-Approximation Algorithm for 
Vertex Cover (\vc)
}\label{section:vc-approx-algo}

It is highly alluring to develop parallel processing approximation algorithms for NP-Hard problems under the paradigm of lattice-linearity. In fact, it has been an open question if this is possible \cite{Garg2020}. We observe that this is possible. In this section, we present a lattice-linear 2-approximation algorithm for \vc. 

The following algorithm is the classic 2-approximation algorithm for \vc.
\textit{
    Choose an uncovered edge $\{A, B\}$, select both $A$ and $B$, repeat until all edges are covered.
}
Since the minimum \vc must contain either $A$ or $B$, the selected \vc is at most twice the size of the minimum \vc. 

While the above algorithm is sequential in nature, we demonstrate that we can transform it into a distributed algorithm under the paradigm of lattice-linearity as shown in \Cref{algorithm:vc-2-approx} (we note that this algorithm is \textit{not} self-stabilizing).

In \Cref{algorithm:vc-2-approx}, all nodes are initially out of the VC, and are not \textit{done}. In the algorithm, each node will check if all edges incident on it are covered. A node is called \textit{done} if it has already evaluated if all the edges incident on it are covered. A node is \imped if it is not done yet and it is the highest ID node in its distance-3 neighbourhood which is not done.
If an \imped node $i$ has an uncovered edge, and assume that $\{i,j\}$ is an uncovered edge with $j$ being the highest ID node in $Adj_i$ which is out (note that $i$ is out), then $i$ turns both $i$ and $j$ into the VC. If otherwise $i$ evaluates that all its edges are covered, then it declares that it is done ($i$ sets $i[done]$ to true), while staying out of the VC.

This is straightforward from the 2-approximation algorithm for VC. We have chosen 3-neighbourhood to evaluate \imped to ensure that no conflicts arise while execution from the perspective of the 2-approximation algorithm for VC.

\begin{algorithm}\label{algorithm:vc-2-approx}
    2-approximation lattice-linear algorithm for \vc.
\end{algorithm}
\begin{center}
    \begin{tabular}{|l|}
        \hline
        Init: $\forall i\in V(G), i[st]=OUT, i[done]=false$.\\
        \textsc{\Imped-VC2A}$(i)$ $\equiv i[done]=false\land (\forall j\in Adj^3_i: j[id]<i[id] \lor j[done]=true)$.\\~\\
        Rules for node $i$.\\
        \textsc{\Imped-VC2A}$(i)$ $\longrightarrow$\\
        \quad \textsc{if} $(\forall k\in Adj_i, k[st]=IN)$, then $i[done]=true$.\\
        \quad \textsc{else}, then\\
        \quad\quad\quad $j=\text{arg} \max\{x[id]:x\in Adj_i\land x[done]=false\}$.\\
        \quad\quad\quad $i[st]=IN$.\\
        \quad\quad\quad $j[st]=IN$, $j[done]=true$. 
        \\
        \quad\quad\quad $i[done]=true$.\\
        \hline
    \end{tabular}
\end{center}

Observe that the action of node $i$ is selecting an edge $\{i, j\}$ and adding $i$ and $j$ to the \vc. This follows straightforwardly from the classic 2-approximation algorithm. 

\subsection{Lattice-Linearity of \Cref{algorithm:vc-2-approx}}

To demonstrate that \Cref{algorithm:vc-2-approx} is lattice-linear, we define the state value and rank, auxiliary variables associated with nodes and global states, as follows.
$$
    \begin{array}{l}
        \textsc{State-Value-VC2A}(i,s)= 
        \begin{cases}
            |\{j\in Adj_i:s[j[st]]=OUT\}| & \text{if $s[i[st]]=OUT$} \\
            0 & \text{otherwise}
        \end{cases}
    \end{array}
$$

$$\textsc{Rank-VC2A}(s)=\sum\limits_{i\in V(G)}\textsc{State-Value-VC2A}(i,s).$$

\begin{theorem}\label{theorem:vertex-cover-2-approx}
    \Cref{algorithm:vc-2-approx} is a lattice-linear 2-approximation algorithm for \vc.
\end{theorem}

\begin{proof}
    \noindent\textit{Lattice-linearity}: In the initial state, every node $i$ has $i[done]=false$ and $i[st]=OUT$. Let $s$ be an arbitrary state at the beginning of some time step while the algorithm is under execution such that $s$ does not manifest a vertex cover. Let $i$ be some node such that $i$ is of the highest ID in its distance 3 neighbourhood such that some of its edges are not covered. Also, let $j$ be the node of the highest ID in $Adj_i$ for which $j[done]=false$, if one such node exists. We have that $i$ is the only \imped node in its distance-3 neighbourhood, and $j$ is the specific additional node, which $i$ turns in. Thus the states form a partial order where state $s$ transitions to another state $s'$ where $s\prec s'$ and for any such $i$, $s'[i[st]]=IN\land s'[i[done]]=true$ and $s'[j[st]]=IN\land s'[j[done]]=true$.
    
    If $s$ manifests a vertex cover, then no additional nodes will be turned in, and atmost one additional node (node $i$, as described in the paragraph above) will have $s'[i[done]]=true$.
    
    From the above observations, we have that for every node $i$, $\textsc{State-Value-VC2A}(i)$ is initially $deg(i)$; this value decreases monotonously and never increases; it becomes 0 after when $i$ is \imped. As a result, $\textsc{Rank}$ decreases monotonously until it becomes zero, because if $\textsc{Rank}$ is not zero, then at least one node is \imped (e.g., node of highest ID which has at least one uncovered edge).
    Thus, we have that \Cref{algorithm:vc-2-approx} also is lattice-linear. However, it induces only one lattice among the global states since the initial state is predetermined, thus $w=1$.
    
    \noindent\textit{2-approximability}: If a node $i$ is \imped and if one of its edges is uncovered, then it selects an edge (it points to the other node in that edge) and both the nodes in that edge turn in; thus it straightforwardly follows the standard sequential 2-approximation algorithm. If some node $k$ (at a distance farther than 3 from $i$) executes and selects $k'\in Adj_k$ to turn in, then and neither $i$ nor $j$ can be a neighbour of $k$ or $k'$.
    Thus any race condition is prevented.
    This shows that \Cref{algorithm:vc-2-approx} preserves the 2-approximability of the standard 2-approximation algorithm for \vc.
\end{proof}

\begin{figure}[ht]
    \centering
    \subfigure[]{
        \begin{tikzpicture}
            \node [circle, inner sep=1 pt, fill=black, label=right:$v_2$] (a2) at (0,0) {};
            \node [circle, inner sep=1 pt, fill=black, label=above:$v_1$] (a1) at (0,1) {};
            \node [circle, inner sep=1 pt, fill=black, label=below:$v_3$] (a3) at (0,-1) {};
            \node [circle, inner sep=1 pt, fill=black, label=left:$v_4$] (a4) at (-1,0) {};
            
            \node [circle, inner sep=1 pt, fill=black, label=right:$v_8$] (a8) at (3,0) {};
            \node [circle, inner sep=1 pt, fill=black, label=above:$v_7$] (a7) at (3,1) {};
            \node [circle, inner sep=1 pt, fill=black, label=below:$v_6$] (a6) at (3,-1) {};
            \node [circle, inner sep=1 pt, fill=black, label=left:$v_5$] (a5) at (2,0) {};
            
            \draw (a2) -- (a1); \draw (a2) -- (a3); \draw (a2) -- (a4);
            \draw (a8) -- (a5); \draw (a8) -- (a6); \draw (a8) -- (a7);
        \end{tikzpicture}
    }
    \subfigure[]{
        \begin{tikzpicture}[every node/.style={scale=.7}]
            \node[] (a1) at (0,0) {\begin{tabular}{l} $(OUT,OUT,OUT,OUT$\\ $OUT,OUT,OUT,OUT)$ \end{tabular}};
            \node[] (a2) at (-2,2) {\begin{tabular}{l} $(OUT,IN,OUT,IN$\\ $OUT,OUT,OUT,OUT)$ \end{tabular}};
            \node[] (a3) at (2,2) {\begin{tabular}{l} $(OUT,OUT,OUT,OUT$\\ $OUT,OUT,IN,IN)$ \end{tabular}};
            \node[] (a4) at (0,4) {\begin{tabular}{l} $(OUT,IN,OUT,IN$\\ $OUT,OUT,IN,IN)$ \end{tabular}};
            
            \draw (a4) -- (a3); \draw (a4) -- (a2);
            \draw (a3) -- (a1); \draw (a2) -- (a1);
        \end{tikzpicture}
    }
    \caption{Execution of \Cref{algorithm:vc-2-approx}: (a) input graph, and (b) lattice induced in the input graph. Transitive edges are not shown for brevity.}
    \label{figure:vc-ll-example}
\end{figure}
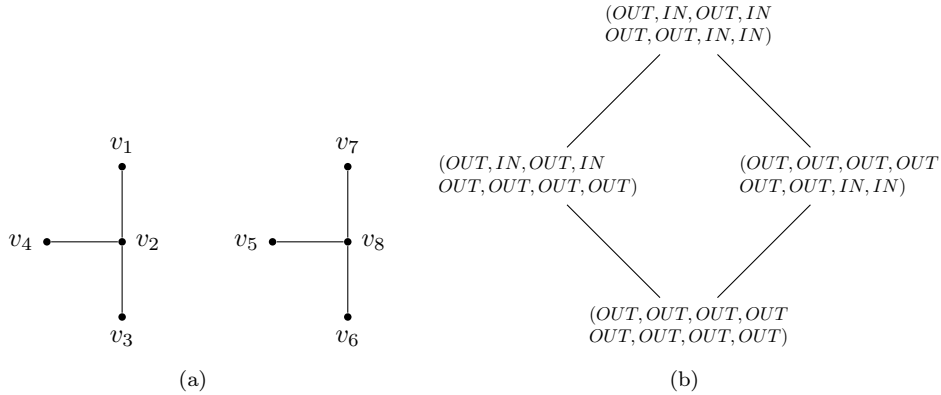

\begin{example}
    In \Cref{figure:vc-ll-example} (a), we show a graph (containing eight nodes $v_1,...,v_8$) and the lattice induced by \Cref{algorithm:vc-2-approx} in the state space of that graph. We are omitting how the value of $i[done]$ gets modified; we only show how the vertex cover is formed. Only the reachable states are shown. Each node in the lattice represents a tuple of states of all nodes $\langle v_1[st],v_2[st],\cdots,v_8[st]\rangle$.
    \qed 
\end{example}

In \Cref{algorithm:vc-2-approx}, local state of any node $i$ is represented by two variables $i[done]$ and $i[st]$. Observe that in this algorithm, the definition of a node being \imped depends on $i[done]$ and not $i[st]$. Therefore the transitions and consequently the structure of the lattice depends on $i[done]$ only, 
whose domain is of size 2. 

In this algorithm, a node $i$ makes changes to the variables of another node $j$, which is, in general, not allowed in a distributed system. We observe that this algorithm can be transformed into a lattice-linear distributed system algorithm where any node only makes changes only to its own variables. We describe the transformed algorithm, next.

\subsection{Distributed Version of \Cref{algorithm:vc-2-approx}}\label{appendix:distributed-vc-2-approx}

In \Cref{algorithm:vc-2-approx}, we presented a lattice-linear 2-approximation algorithm for \vc. In that algorithm, the states of two nodes $i$ and $j$ were changed in the same action.

Here, we present a mapping of that algorithm where $i$ and $j$ change their states separately. The key idea of this algorithm is when $i$ intends to add $j$ to the \vc, $i[point]$ is set to $j$.  When $i$ is pointing to $j$, $j$ has to execute and add itself to the \vc. Thus, the transformed algorithm is as shown in \Cref{algorithm:vc-2-approx-distributed}.

\newpage 
\begin{algorithm}\label{algorithm:vc-2-approx-distributed} \Cref{algorithm:vc-2-approx} transformed where every node modifies only its own variables.
\end{algorithm}
\begin{center}
    \small 
    \begin{tabular}{|l|}
        \hline
        Init: $\forall i\in V(G), i[st]=OUT, i[done]=false, i[point]=\top$.\\
        \textsc{Else-Pointed-VC2A}$(i)$ $\equiv \exists j\in Adj^4_i: \exists k\in Adj_j:$\\ \quad\quad\quad\quad $k[point]=j \land j[done]=false$.\\
        \textsc{\Imped-VC2A-II}$(i)$ $\equiv$\\ 
        \quad\quad\quad\quad $(i[done]=false\land(\exists j\in Adj_i: j[point]=i)\lor$\\
        \quad\quad\quad\quad $(\lnot \textsc{Else-Pointed-VC2A}(i)\land$\\
        \quad\quad\quad\quad $(i[done]=false\land (\forall j\in Adj^3_i: j[id]<i[id]\lor$\\ \quad\quad\quad\quad $j[done]=true))).$
        \\~\\
        Rules for node $i$:\\
        \textsc{\Imped-VC2A-II}$(i)$ $\longrightarrow$\\
        \quad \textbf{if} $(\exists j\in Adj_i: j[point]=i)$, then\\
        \quad\quad\quad \textbf{if} $(\forall k\in Adj_i: k[st]=IN)$, then $i[done]=true$\\
        \quad\quad\quad \textbf{else}, then $i[st]=IN, i[done]=true$.\\
        \quad \textbf{else}, then\\
        \quad\quad\quad \textbf{if} $(\forall k\in Adj_i, k[st]=IN)$, then $i[done]=true$.\\
        \quad\quad\quad \textbf{else}, then\\
        \quad\quad\quad\quad\quad $j=\text{arg} \max\{x[id]:x\in Adj_i\land x[done]=false\}$.\\
        \quad\quad\quad\quad\quad $i[st]=IN$.\\
        \quad\quad\quad\quad\quad $i[point]=j$.\\
        \quad\quad\quad\quad\quad $i[done]=true$.\\
        \hline
    \end{tabular}
\end{center}

\noindent\textit{Remark}: Observe that in this algorithm, any $j$ chosen throughout the algorithm by some \imped $i$ does not move in if it is already covered. Thus, we have that this algorithm computes a minimal as well as a 2-approximate vertex cover. On the other hand, \Cref{algorithm:vc-2-approx} is a faithful replication of the classic sequential 2-approximation algorithm for vertex cover.

\section{Gathering Robots on Triangular Grid}\label{section:gsgs}



In this 
section, we study the problem of gathering distance-1 myopic robots on an infinite triangular grid.
We show that the algorithm developed by Goswami et al. \cite{Goswami2022} is lattice-linear. Hence, this algorithm will run correctly even if the robots run in asynchrony where the robots can execute on old information about other robots.
Because of lattice-linearity, this algorithm works correctly even if the robots are equipped with a unidirectional \textit{camera} to see neighbouring robots.
Authors of \cite{Goswami2022} assumed a distributed scheduler, which would require an omnidirectional camera, capable to get fresh values from all neighbouring locations.

Lattice-linearity follows that the moves of the robots are predictable. 
This allows us to show tighter bounds to the arena traversed by the robots under the algorithm.
As a consequence of tighter bounds on this arena, (1) we obtain a better convergence time bound for this algorithm, which is lower than that showed in \cite{Goswami2022}, and (2) we show that the gathering point of the robots can be uniquely determined from the initial, or any intermediate, state. We show that this algorithm converges in $2n$ rounds, which is lower as compared to the time complexity bound ($2.5(n+1)$ rounds) shown in \cite{Goswami2022}.

\subsection{Gathering of Distance-1 Myopic Robots on Infinite Triangular Grid (\gmrit)}\label{subsection:gmrit}

This chapter focuses on the problem where the input is a swarm of robots
with minimal capabilities. 
Each robot is present at a vertex on an infinite triangular grid. 
In the initial global state, the robots form a connected graph on the underlying grid. The robots agree on an axis (i.e. a direction and its orientation).
The robots can move only across one edge at a time.
Each robot is myopic, i.e., it can only sense if another robot is present at an adjacent vertex.
Robots do not have an abiliy to \textit{communicate} with each other. 
Under these constraints, it is required that all robots gather at one point.

\newcounter{diags}
\newcounter{rdiag}
\newcounter{vert}
\begin{figure}[ht]
    \centering
    \subfigure{
    \begin{tikzpicture}[x=1.1cm,y=0.6351cm] 
        \foreach \diag in {0,...,4}{
            \draw (\diag,0) -- (5,-5+\diag);
        }
        \foreach \diag in {8,...,5}{
            \draw (0,-9+\diag) -- (5,\diag-14);
        }

        \foreach \diag in {4,...,1}{
            \draw (0,-9+\diag) -- (\diag,-9);
        }
        
        \foreach \rdiag in {0,...,4}{
            \draw (\rdiag,-9) -- (5,-\rdiag-4);
        }
        \foreach \rdiag in {8,...,5}{
            \draw (0,-\rdiag) -- (5,5-\rdiag);
        }
        \foreach \rdiag in {4,...,1}{
            \draw (0,-\rdiag) -- (\rdiag,0);
        }
        
        \foreach \vert in {0,...,10}{
            \draw (\vert*.5,0) -- (\vert*.5,-9);
        }

        \draw[dashed, red, line width=2pt] (0,0) -- (0,-8);
        \draw[dashed, red, line width=2pt] (5,0) -- (5,-10);

        \draw[dashed, red, line width=2pt] (-.5,-.5) -- (5.5,-.5);
        \draw[dashed, red, line width=2pt] (-.5,-6.5) -- (5.5,-6.5);
        
        \draw[dashed, red, line width=2pt] (-.5,-4.5) -- (3,-8) -- (5,-10);
        \draw[dashed, red, line width=2pt] (6,-5) -- (3,-8);
        \draw[dashed, red, line width=2pt] (-.5,-6) -- (2.5,-9);
        \draw[dashed, red, line width=2pt] (6,-5.5) -- (2.5,-9);
        
        \draw[dashed, red, line width=2pt] (0,-5) -- (5,-5);
        
        \node[font=\boldmath] at (-.25,-.25) {$A$};
        \node[font=\boldmath] at (-.25,-6.75) {$B$};
        \node[font=\boldmath] at (5.25,-6.75) {$C$};
        \node[font=\boldmath] at (5.25,-.25) {$D$};
        \node[font=\boldmath] at (2.5,-9.25) {$P$};
        
        \node[font=\boldmath] at (-.25,-5.25) {$B'$};
        \node[font=\boldmath] at (4.75,-5.75) {$C'$};
        \node[font=\boldmath] at (2.5,-8) {$Q$};

        \node[font=\boldmath] at (5.25,-10.25) {$E$};
        \node[font=\boldmath] at (5.25,-5) {$J$};
        
        \node [circle, inner sep=2pt, fill=black, draw=black] at (0,-1) {};
        \node [circle, inner sep=2pt, fill=black, draw=black] at (0,-2) {};
        \node [circle, inner sep=2pt, fill=black, draw=black] at (.5,-2.5) {};
        \node [circle, inner sep=2pt, fill=black, draw=black] at (.5,-3.5) {};
        \node [circle, inner sep=2pt, fill=black, draw=black] at (.5,-5.5) {};
        \node [circle, inner sep=2pt, fill=black, draw=black] at (1,-2) {};
        \node [circle, inner sep=2pt, fill=black, draw=black] at (1,-6) {};
        \node [circle, inner sep=2pt, fill=black, draw=black] at (1.5,-1.5) {};
        \node [circle, inner sep=2pt, fill=black, draw=black] at (1.5,-2.5) {};
        \node [circle, inner sep=2pt, fill=black, draw=black] at (1.5,-6.5) {};
        \node [circle, inner sep=2pt, fill=black, draw=black] at (2,-1) {};
        \node [circle, inner sep=2pt, fill=black, draw=black] at (2,-3) {};
        \node [circle, inner sep=2pt, fill=black, draw=black] at (2,-4) {};
        \node [circle, inner sep=2pt, fill=black, draw=black] at (2,-6) {};
        \node [circle, inner sep=2pt, fill=black, draw=black] at (2.5,-.5) {};
        \node [circle, inner sep=2pt, fill=black, draw=black] at (2.5,-4.5) {};
        \node [circle, inner sep=2pt, fill=black, draw=black] at (2.5,-5.5) {};
        \node [circle, inner sep=2pt, fill=black, draw=black] at (2.5,-6.5) {};
        \node [circle, inner sep=2pt, fill=black, draw=black] at (3,-1) {};
        \node [circle, inner sep=2pt, fill=black, draw=black] at (3,-5) {};
        \node [circle, inner sep=2pt, fill=black, draw=black] at (3.5,-1.5) {};
        \node [circle, inner sep=2pt, fill=black, draw=black] at (3.5,-5.5) {};
        \node [circle, inner sep=2pt, fill=black, draw=black] at (4,-1) {};
        \node [circle, inner sep=2pt, fill=black, draw=black] at (4,-5) {};
        \node [circle, inner sep=2pt, fill=black, draw=black] at (4,-6) {};
        \node [circle, inner sep=2pt, fill=black, draw=black] at (4.5,-.5) {};
        \node [circle, inner sep=2pt, fill=black, draw=black] at (4.5,-4.5) {};
        \node [circle, inner sep=2pt, fill=black, draw=black] at (4.5,-6.5) {};
        \node [circle, inner sep=2pt, fill=black, draw=black] at (5,-1) {};

        \node [rectangle, inner sep=4pt, fill=darkgreen, draw=darkgreen] at (3,-8) {};

    \end{tikzpicture}
    }
    \subfigure{
        \begin{tikzpicture}[scale=.5,every node/.style={scale=.9}]
            \draw[->, line width=2pt] (0,0) -- (0,4);
            \node at (.5,2) {$+y$};
        \end{tikzpicture}
    }
    \caption{Robots on an infinite triangular grid: one on every round highlighted vertex.}
    \label{figure:infinite-triangular-grid}
\end{figure}
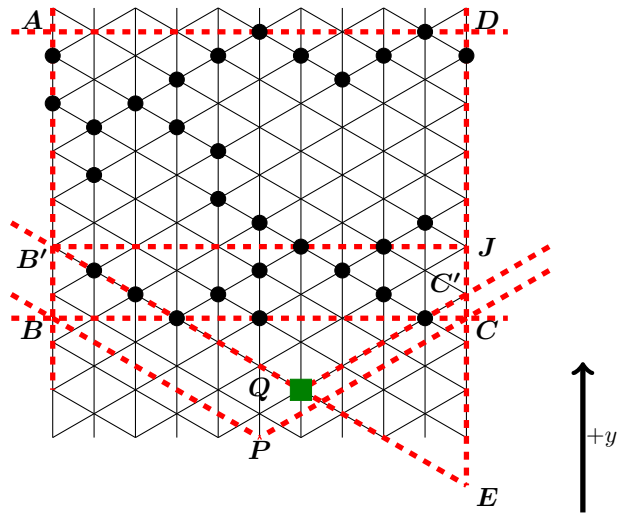

\subsubsection*{Problem Statement}
The input is a global state $s$ that describes the location of $n$ robots placed on the grid points of an infinite triangular grid $G$ such that the robots form a connected graph. 
The \gmrit problem requires that all robots gather at one vertex of $G$ and stay forever at that vertex subject to the following constraints:

\begin{itemize}
    \item \textit{Visibility:}
    A robot can only determine if another robot is present in a neighbouring location. It cannot exchange data with another robot.
    \item \textit{One Axis Agreement:}
    All robots agree on one axis and the orientation of that axis. (cf. $y$-axis as shown in \Cref{figure:infinite-triangular-grid}).
\end{itemize}

All robots are independent and identical from a physical and computational perspective and do not have an ID.
They are oblivious to the coordinates of their location on the infinite triangular grid $G$.
Observe that we can allow a global state $s$ to be a multiset of vertices, each of which is the location of a robot.
For $s$, its \textit{visibility graph} is the subgraph of $G$ induced by the set of vertices in $s$.

Notice that by definition of the problem statement, a solution to \gmrit will provide silent self-stabilization.
An instance of \gmrit
is shown in \Cref{figure:infinite-triangular-grid}. 
Here, each round highlighted vertex represents a robot. 
Observe that the visibility graph of this global state is a connected graph.

Next, we discuss how \gmrit problem is not a lattice-linear problem. This can be illustrated by a system containing two robots $x_1$ and $x_2$ present at locations $l_1$ and $l_2$ ($l_1$ and $l_2$ are different vertices on the same edge) on $G$. In such a system $x_1$ can move to $l_2$, in which case, $x_2$ is not \imped, or otherwise, $x_2$ can move to $l_1$, in which case, $x_1$ is not \imped. Hence, no specific robot can be deemed \imped, though, the global state is suboptimal.

\subsubsection*{Problem Specific Definitions}\label{subsection:gsgs-definitions}

Some of the definitions that we discuss in this subsection are from \cite{Goswami2022}. 
A \textit{horizontal layer} is a line perpendicular to the $y$-axis that passes through at least one robot.
The \textit{top layer} of a global state $s$ is a horizontal layer such that there is no horizontal layer above it (e.g., $AD$ in \Cref{figure:infinite-triangular-grid}).
\textit{Bottom layer} of $s$ is a horizontal layer such that there is no horizontal layer below it (e.g., $BC$ in \Cref{figure:infinite-triangular-grid}).

A \textit{vertical layer} is parallel to the $y$-axis such that it passes through at least one robot.
The \textit{left layer} of $s$ is the vertical layer such that there is no vertical layer on its left (e.g., $AB$ in \Cref{figure:infinite-triangular-grid}).
The \textit{right layer} of $s$ is the vertical layer such that there is no vertical layer on its right (e.g., $CD$ in \Cref{figure:infinite-triangular-grid}).

As seen in \Cref{figure:infinite-triangular-grid}, vertices in $G$ are intersections of three groups of parallel lines; one of these groups are lines parallel to the $y$-axis. We use $p$-axis (positive slope) and $n$ axis (negative slope) to denote the other group of parallel lines.
The \textit{positive slant} is a line
parallel to $p$-axis (e.g., $BP$ in \Cref{figure:infinite-triangular-grid}) and \textit{negative slant} is a line
parallel to $n$-axis (e.g., $CP$ in \Cref{figure:infinite-triangular-grid}).
The \textit{bottom $l2r$ slant} of $s$ is a negative slant that passes through a robot such that there is no negative slant on its left passing through a robot (e.g., $B'Q$ in \Cref{figure:infinite-triangular-grid}).
The \textit{bottom $r2l$ slant} of $s$ is a positive slant that passes through a robot such that there is no positive slant on its right passing through a robot (e.g., $C'Q$ in \Cref{figure:infinite-triangular-grid}).
Note that a negative slant and a positive slant can be imaginary, or a line in $G$.

The \textit{depth} of $s$ is the distance between its top layer and its bottom layer.
The \textit{width} of $s$ is defined as the distance between its left layer and right layer.

As shown in \Cref{figure:infinite-triangular-grid}, a polygon $ABPCD$ is a \textit{bounding polygon} of a global state $s$ if
(1) $AB$ and $CD$ are line segments of the left layer and the right layer of $s$ respectively, (2) $AD$ and $BC$ are line segments of the top layer and the bottom layer of $s$ respectively, and (3) $P$ is the point of intersection between the negative slant passing through $B$ and the positive slant passing through $C$.

Note that these definitions (top/bottom layer, etc.) are only used for discussion of the protocol and proofs. The robots are not aware of them. Similarly, the robots can distinguish between $up$ and $down$, but not $left$ and $right$.

\subsection{General Idea of the Algorithm}

\begin{figure}[ht]
    \centering
    \subfigure[]{
    \begin{minipage}{.25\textwidth}
        \centering 
        \begin{tikzpicture}[x=1.5cm,y=.8660cm] 
        \node [circle, fill = black, inner sep = 2pt, label=left:$i$] (i) at (0,0) {};
        \node [circle, fill = black, inner sep = 1pt, label=below:$v_1$] (v1) at (0,-1) {};
        \node [circle, fill = black, inner sep = 1pt, label=left:$v_2^\ell$] (v2l) at (-.5,-.5) {};
        \node [circle, fill = black, inner sep = 1pt, label=right:$v_2^r$] (v2r) at (.5,-.5) {};
        \node [circle, fill = black, inner sep = 1pt, label=left:$v_3^\ell$] (v3l) at (-.5,.5) {};
        \node [circle, fill = black, inner sep = 1pt, label=right:$v_3^r$] (v3r) at (.5,.5) {};
        \node [circle, fill = black, inner sep = 1pt, label=above:$v_4$] (v4) at (0,1) {};

        \node [circle, draw = black, inner sep = 2pt] (v1) at (0,-1) {};
        \node [circle, draw = black, inner sep = 2pt] (v2l) at (-.5,-.5) {};
        \node [circle, draw = black, inner sep = 2pt] (v3l) at (-.5,.5) {};

        \draw (i) -- (v1); \draw (i) -- (v2l); \draw (i) -- (v2r); \draw (i) -- (v3l); \draw (i) -- (v3r);
        \draw (i) -- (v4);
        
        \draw (v1) -- (v2l) -- (v3l) -- (v4) -- (v3r) -- (v2r) -- (v1);
    \end{tikzpicture}
    \end{minipage}
    }
    \renewcommand{\thesubfigure}{(b)}
    \subfigure[]{
    \begin{minipage}{.35\textwidth}
        \subfigure{
        \begin{tikzpicture}[x=1cm,y=.5773cm] 
            \node [circle, fill = black, inner sep = 2pt] (i) at (0,0) {};
            \node [circle, draw = black, inner sep = 2pt] (v1) at (0,-1) {};
            \node [circle, fill = black, inner sep = .5pt] (v2l) at (-.5,-.5) {};
            \node [circle, fill = black, inner sep = .5pt] (v2r) at (.5,-.5) {};
            \node [circle, fill = black, inner sep = .5pt] (v3l) at (-.5,.5) {};
            \node [circle, fill = black, inner sep = .5pt] (v3r) at (.5,.5) {};
            \node [circle, fill = black, inner sep = .5pt] (v4) at (0,1) {};
    
            \draw (i) -- (v1); \draw (i) -- (v2l); \draw (i) -- (v2r); \draw (i) -- (v3l); \draw (i) -- (v3r);
            \draw (i) -- (v4);
            
            \draw (v1) -- (v2l) -- (v3l) -- (v4) -- (v3r) -- (v2r) -- (v1);
            
            \node at (0,-1.5) {$\ast$};
        \end{tikzpicture}
    }
    \subfigure{
        \begin{tikzpicture}[x=1cm,y=.5773cm] 
            \node [circle, fill = black, inner sep = 2pt] (i) at (0,0) {};
            \node [circle, draw = black, inner sep = 2pt] (v1) at (0,-1) {};
            \node [circle, draw = black, inner sep = 2pt] (v2l) at (-.5,-.5) {};
            \node [circle, fill = black, inner sep = .5pt] (v2r) at (.5,-.5) {};
            \node [circle, fill = black, inner sep = .5pt] (v3l) at (-.5,.5) {};
            \node [circle, fill = black, inner sep = .5pt] (v3r) at (.5,.5) {};
            \node [circle, fill = black, inner sep = .5pt] (v4) at (0,1) {};
    
            \draw (i) -- (v1); \draw (i) -- (v2l); \draw (i) -- (v2r); \draw (i) -- (v3l); \draw (i) -- (v3r);
            \draw (i) -- (v4);
            
            \draw (v1) -- (v2l) -- (v3l) -- (v4) -- (v3r) -- (v2r) -- (v1);
            
            \node at (0,-1.5) {$\ast$};
        \end{tikzpicture}
    }
    \subfigure{
        \begin{tikzpicture}[x=1cm,y=.5773cm] 
            \node [circle, fill = black, inner sep = 2pt] (i) at (0,0) {};
            \node [circle, fill = black, inner sep = .5pt] (v1) at (0,-1) {};
            \node [circle, draw = black, inner sep = 2pt] (v2l) at (-.5,-.5) {};
            \node [circle, draw = black, inner sep = 2pt] (v2r) at (.5,-.5) {};
            \node [circle, fill = black, inner sep = .5pt] (v3l) at (-.5,.5) {};
            \node [circle, fill = black, inner sep = .5pt] (v3r) at (.5,.5) {};
            \node [circle, fill = black, inner sep = .5pt] (v4) at (0,1) {};
    
            \draw (i) -- (v1); \draw (i) -- (v2l); \draw (i) -- (v2r); \draw (i) -- (v3l); \draw (i) -- (v3r);
            \draw (i) -- (v4);
            
            \draw (v1) -- (v2l) -- (v3l) -- (v4) -- (v3r) -- (v2r) -- (v1);
            
            \node at (0,-1.5) {$\ast$};
        \end{tikzpicture}
    }
    \subfigure{
        \begin{tikzpicture}[x=1cm,y=.5773cm] 
            \node [circle, fill = black, inner sep = 2pt] (i) at (0,0) {};
            \node [circle, draw = black, inner sep = 2pt] (v1) at (0,-1) {};
            \node [circle, draw = black, inner sep = 2pt] (v2l) at (-.5,-.5) {};
            \node [circle, draw = black, inner sep = 2pt] (v2r) at (.5,-.5) {};
            \node [circle, fill = black, inner sep = .5pt] (v3l) at (-.5,.5) {};
            \node [circle, fill = black, inner sep = .5pt] (v3r) at (.5,.5) {};
            \node [circle, fill = black, inner sep = .5pt] (v4) at (0,1) {};
    
            \draw (i) -- (v1); \draw (i) -- (v2l); \draw (i) -- (v2r); \draw (i) -- (v3l); \draw (i) -- (v3r);
            \draw (i) -- (v4);
            
            \draw (v1) -- (v2l) -- (v3l) -- (v4) -- (v3r) -- (v2r) -- (v1);
            
            \node at (0,-1.5) {$\ast$};
        \end{tikzpicture}
    }\\
    \subfigure{
        \begin{tikzpicture}
            \node[] at (0,0) {case 1};
            \node[] at (1.5,0) {case 2};
            \node[] at (3,0) {case 3};
            \node[] at (4.5,0) {case 4};
        \end{tikzpicture}
    }\\
    \subfigure{
        \begin{tikzpicture}[x=1cm,y=.5773cm] 
            \node [circle, fill = black, inner sep = 2pt] (i) at (0,0) {};
            \node [circle, fill = black, inner sep = .5pt] (v1) at (0,-1) {};
            \node [circle, draw = black, inner sep = 2pt] (v2l) at (-.5,-.5) {};
            \node [circle, fill = black, inner sep = .5pt] (v2r) at (.5,-.5) {};
            \node [circle, draw = black, inner sep = 2pt] (v3l) at (-.5,.5) {};
            \node [circle, fill = black, inner sep = .5pt] (v3r) at (.5,.5) {};
            \node [circle, fill = black, inner sep = .5pt] (v4) at (0,1) {};
    
            \draw (i) -- (v1); \draw (i) -- (v2l); \draw (i) -- (v2r); \draw (i) -- (v3l); \draw (i) -- (v3r);
            \draw (i) -- (v4);
            
            \draw (v1) -- (v2l) -- (v3l) -- (v4) -- (v3r) -- (v2r) -- (v1);
            
            \node at (-.75,-.75) {$\ast$};
        \end{tikzpicture}
    }
    \subfigure{
        \begin{tikzpicture}[x=1cm,y=.5773cm] 
            \node [circle, fill = black, inner sep = 2pt] (i) at (0,0) {};
            \node [circle, draw = black, inner sep = 2pt] (v1) at (0,-1) {};
            \node [circle, draw = black, inner sep = 2pt] (v2l) at (-.5,-.5) {};
            \node [circle, fill = black, inner sep = .5pt] (v2r) at (.5,-.5) {};
            \node [circle, draw = black, inner sep = 2pt] (v3l) at (-.5,.5) {};
            \node [circle, fill = black, inner sep = .5pt] (v3r) at (.5,.5) {};
            \node [circle, fill = black, inner sep = .5pt] (v4) at (0,1) {};
    
            \draw (i) -- (v1); \draw (i) -- (v2l); \draw (i) -- (v2r); \draw (i) -- (v3l); \draw (i) -- (v3r);
            \draw (i) -- (v4);
            
            \draw (v1) -- (v2l) -- (v3l) -- (v4) -- (v3r) -- (v2r) -- (v1);
            
            \node at (-.75,-.75) {$\ast$};
        \end{tikzpicture}
    }
    \subfigure{
        \begin{tikzpicture}[x=1cm,y=.5773cm] 
            \node [circle, fill = black, inner sep = 2pt] (i) at (0,0) {};
            \node [circle, fill = black, inner sep = .5pt] (v1) at (0,-1) {};
            \node [circle, draw = black, inner sep = 2pt] (v2l) at (-.5,-.5) {};
            \node [circle, fill = black, inner sep = .5pt] (v2r) at (.5,-.5) {};
            \node [circle, fill = black, inner sep = .5pt] (v3l) at (-.5,.5) {};
            \node [circle, fill = black, inner sep = .5pt] (v3r) at (.5,.5) {};
            \node [circle, fill = black, inner sep = .5pt] (v4) at (0,1) {};
    
            \draw (i) -- (v1); \draw (i) -- (v2l); \draw (i) -- (v2r); \draw (i) -- (v3l); \draw (i) -- (v3r);
            \draw (i) -- (v4);
            
            \draw (v1) -- (v2l) -- (v3l) -- (v4) -- (v3r) -- (v2r) -- (v1);
            
            \node at (-.75,-.75) {$\ast$};
        \end{tikzpicture}
    }\\
    \subfigure{
        \begin{tikzpicture}
            \node[] at (5,0) {};
            \node[] at (6,0) {case 5};
            \node[] at (7.75,0) {case 6};
            \node[] at (9.5,0) {case 7};
            \node[] at (11,0) {};
        \end{tikzpicture}
    }
    \end{minipage}
    }
    \begin{minipage}{.1\textwidth}
        \subfigure{
        \begin{tikzpicture}
            \draw[->, line width=2pt] (0,0) -- (0,2);
            \node at (.5,1) {$+y$};
        \end{tikzpicture}
    }
    \end{minipage}
    \caption{(a) Naming conventions for neighbourhood of a robot. (b) Cases where a node is \imped. Note that the mirror images of these local states are also \imped.}
    \label{figure:local-states}
\end{figure}
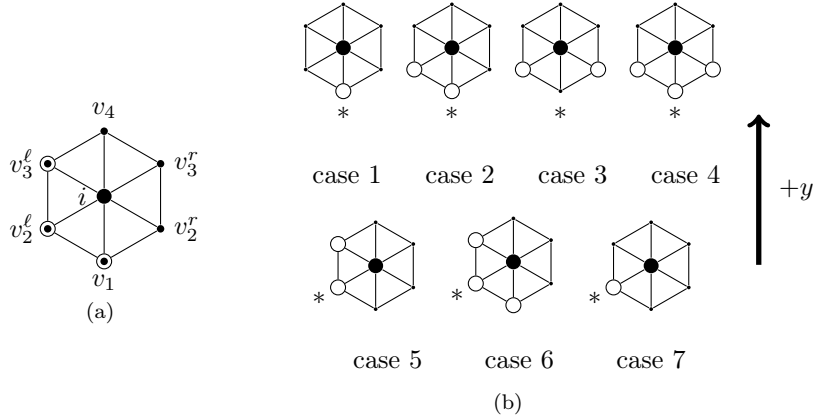

A robot has six possible neighbouring locations. The naming convention for these locations is as shown in \Cref{figure:local-states} (a) \cite{Goswami2022}. Since each robot has 6 neighbouring locations, it can be in one of $2^6$ possible local states. Of these, the robot can move in only 11 states. Of these, 7 states are shown in \Cref{figure:local-states} (b) \cite{Goswami2022}. The other 4 states are mirror images of those shown in cases 2, 5, 6 and 7 in \Cref{figure:local-states} (b).

Authors of \cite{Goswami2022} show that the robots do not move out of the bounding polygon. They also show that the visibility graph induced among the robots stay connected, and the dimensions of the bounding polygon reduce with every round. 

\subsection{GSGS Algorithm \cite{Goswami2022} for \gmrit problem}\label{subsection:algorithm-gsgs}

In this section, we reword the algorithm in \cite{Goswami2022} to demonstrate its lattice-linearity. 
We define the macros listed in \Cref{table:macros-gmrit-fll}. 
For a set $L$ of locations around a node $i$, $\textsc{At}(i,L)$ is true iff if there is at least one robot at each location in $L$. $\textsc{Only-At}(i,L)$ is true iff $\textsc{At}(i,L)$, and there is no other robot at locations other than locations in $L$.
A robot $i$ is \textit{extreme} if (1) there is no robot on \textit{top} ($v_4(i)$) of $i$ and (2) if there is a robot on the left ($v_2^\ell$ or $v_3^\ell$) of $i$, then there is no robot on the right ($v_2^r$ or $v_3^r$) of $i$.
If a robot $i$ is extreme, and there is no robot around it, then $i$ is a \textit{terminating} robot.
If $i$ is extreme, and there is a robot only on $v_3(i)$ or there are robots only on both $v_1(i)$ and $v_3(i)$, then $i$ is a \textit{staying} robot.

If $i$ is extreme, and there is a robot on $v_1(i)$ and no robot on $v_3(i)$, then $i$ is a \textit{downward \imped} robot.
If $i$ is not a downward \imped robot, not a staying robot, and not a terminating robot, then it is a \textit{downslant \imped} robot.
If $i$ is not an extreme robot, and there is a robot on both $v_2(i)$ and no robot at its y-coordinate $>$ 0, then $i$ is a \textit{non-extreme \imped} robot.
\begin{table}[ht]
    \centering 
    \doublespacing 
    $\begin{array}{|l|}
        \hline 
        \textsc{Extreme}(i)\equiv \lnot\textsc{At}(i,\{v_4\})\land((\textsc{At}(i,\{v_2^r\})\lor\\
        \quad\quad \textsc{At}(i,\{v_3^r\}))\Rightarrow(\lnot \textsc{At}(i,\{v_2^\ell\})\land \lnot\textsc{At}(i,\{v_3^\ell\}))) \\
        \textsc{Terminating}(r)\equiv \textsc{Extreme}(i)\land (\forall q\in \{v_1,v_2^r,v_3^r,v_4,v_3^\ell,v_2^\ell\}:\lnot\textsc{At}(i,\{q\})).\\
        \textsc{Staying}(i)\equiv \textsc{Extreme}(i)\land (\textsc{Only-At}(i,\{v_3^r\}) \lor\\
        \quad\quad \textsc{Only-At}(i,\{v_3^\ell\}) \lor \textsc{Only-At}(i,\{v_1,v_3^r\}) \lor \textsc{Only-At}(i,\{v_1,v_3^\ell\})).\\
        \textsc{Downward}(i)\equiv \textsc{Extreme}(i)\land \textsc{At}(i,\{v_1\})\land \lnot(\textsc{At}(i,\{v_3^r\})\lor \textsc{At}(i,\{v_3^\ell\})).\\
        \textsc{Downslant-Right}(i)\equiv \textsc{Extreme}(i)\land \lnot \textsc{Downward}(i)\land \lnot\textsc{Staying}(i) \land\\
        \quad\quad \lnot\textsc{Terminating}(r) \land \textsc{At}(i,\{v_2^r\}).\\
        \textsc{Downslant-Left}(i)\equiv \textsc{Extreme}(i)\land \lnot \textsc{Downward}(i)\land \lnot\textsc{Staying}(i) \land\\
        \quad\quad \lnot\textsc{Terminating}(r) \land \textsc{At}(i,\{v_2^\ell\}).\\
        \textsc{Non-Extreme}(i)\equiv \lnot\textsc{Extreme}(i) \land \textsc{At}(i,\{v_2^r,v_2^\ell\}) \land  \lnot(\textsc{At}(i,\{v_3^r\}) \lor \textsc{At}(i,\{v_3^\ell\}) \lor \textsc{At}(i,\{v_4\})).\\
        \textsc{\Imped-GSGS}(i)\equiv \textsc{Downward}(i)\lor \textsc{Downslant-Right}(i) \lor\\
        \quad\quad \textsc{Downslant-Left}(i) \lor \textsc{Non-Extreme}(i).\\
        \hline 
    \end{array}$
    \caption{Macros used in the algorithm for \gmrit problem.}
    \label{table:macros-gmrit-fll}
\end{table}
The algorithm is described as follows.
If a robot $i$ is \textit{downward \imped}, then $i$ moves downwards to $v_1(i)$.
If $i$ is \textit{downslant \imped}, then $i$ moves to $v_2(i)$.
If $i$ is a \textit{non-extreme \imped} robot, then $i$ moves to $v_1(i)$.

\newpage 
\begin{algorithm}\label{algorithm:gsgs-algo}
    Rules for robot $i$.
\end{algorithm}
\begin{center}
    $\begin{array}{|l|}
        \hline 
        \textsc{\Imped-GSGS}(i)\longrightarrow
        \begin{cases}
            move(i,v_1(i)) & \text{if $\textsc{Downward}(i)$}\\
            move(i,v_2^r(i)) & \text{if $\textsc{Downslant-Right}(i)$}\\
            move(i,v_2^\ell(i)) & \text{if $\textsc{Downslant-Left}(i)$}\\
            move(i,v_1(i)) & \text{if $\textsc{Non-Extreme}(i)$}
        \end{cases}~\\
        \hline 
    \end{array}$
\end{center}

In \cite{Goswami2022}, authors assume a distributed scheduler. Next, we show that \Cref{algorithm:gsgs-algo} is lattice-linear, Thus, it will be correct even in asynchrony. 

\subsubsection*{Lattice-Linearity}\label{subsection:gsgs-lattice-linearity}

In this subsection, we show lattice-linearity of \Cref{algorithm:gsgs-algo}. Among the lemmas and theorems presented here, \Cref{lemma:lr-layers} and \Cref{lemma:horizontal-layer-moves-down} are adopted from \cite{Goswami2022}. We use them to help prove some properties of \Cref{algorithm:gsgs-algo}. 
All other results show or arise from the lattice-linearity of \Cref{algorithm:gsgs-algo}. 

\begin{lemma}\label{lemma:gmrit-llp}
    The predicate $\forall i:\lnot\textsc{\Imped-GSGS}(i)$ is a lattice-linear predicate on $n$ robots, and the visibility graph does not get disconnected by the actions under \Cref{algorithm:gsgs-algo}.
\end{lemma}

\begin{proof}
    In this proof, we consider the 7 cases as shown in \Cref{figure:local-states} (b) and show that if robot $i$ is \imped, it must execute to reach the goal state.
    We show that if a robot $i$ is \imped, then there exists at least one robot $j$ around $i$ which does not move until $i$ moves. Specifically, \Cref{algorithm:gsgs-algo} imposes that $j$ `waits' for $i$ to move. It means that if $i$ does not move, then the robots cannot find the gathering point.
    
    \textit{Case 1}: This robot $i$ is a downward \imped robot. The other robot that is present below it is not extreme and is also not a non-extreme \imped robot because $i$ is present above it, so it will not move until $i$ changes its location.

    \textit{Case 2}: This robot $i$ is a downward \imped robot. There are two robots, $x_1$ and $x_2$, present at locations $v_1(i)$ and $v_2(i)$ respectively. $x_1$ is not extreme and is also not non-extreme \imped because $i$ is present above it. So $x_1$ will not move until $i$ changes its location. $x_2$ may be \imped.
    $x_2$ can only move to the location of $x_1$ thereby resulting in case 1.
    In this possibility, the robot $i$ remains \imped and its required action does not change.

    \textit{Case 3}: This robot $i$ is a non-extreme \imped robot. There are two robots, $x_1$ and $x_2$, present at locations $v_2(i)$-left and $v_2(i)$-right respectively. For $x_1$ or $x_2$ to be impedensable, there must be some robot at the $v_1(i)$ location, which is not the case, thus, they are not \imped. Hence, $x_1$ and $x_2$ will not move until $i$ changes its location.

    \textit{Case 4}: This robot $i$ is a non-extreme \imped robot. There are three robots, $x_1$, $x_2$ and $x_3$, present at locations $v_2(i)$-left, $v_1(i)$ and $v_2(i)$-right respectively. $x_2$ is not \imped. $x_1$ and $x_3$ may be \imped, based on their local states. If one or both of them move, they will move to the location of $x_2$, resulting in case 2 or case 1.
    In these possibilities, the robot $i$ remains \imped and its required action does not change.

    \textit{Case 5}: This robot $i$ is a downslant \imped robot. There are two robots, $x_1$ and $x_2$, present at locations $v_2(i)$ and $v_3(i)$ respectively. $x_1$ is not extreme and is also not non-extreme \imped. So $x_1$ will not move until $i$ changes its location. $x_2$ may be \imped, based on its local state.
    $x_2$ can move to the location of $x_1$ or $i$ thereby resulting in case 7.
    In this possibility, the robot $i$ remains \imped and its required action does not change.
    
    \textit{Case 6}: This robot $i$ is a downslant \imped robot. There are three robots, $x_1$, $x_2$ and $x_3$, present at locations $v_1(i)$, $v_2(i)$ and $v_3(i)$ respectively. $x_1$ and $x_2$ are not extreme and are also not non-extreme \imped. Initially, $x_1$ and $x_2$ cannot move. $x_3$ may be downward \imped, based on its local state.
    $x_3$ can only move to the location of $x_2$ thereby resulting in case 2. After this, one or both of $x_1$ and $x_2$ can move to the location of $x_1$, resulting in case 1.
    In these possibilities, $i$ remains \imped, its required action may change, but the graph does not get disconnected if it executes under case 6 (using old information, if it assumes, despite the movement of other robots, that it falls in case 6).

    \textit{Case 7}: This robot $i$ is a downslant \imped robot. The other robot $x_1$ that is present at $v_2(i)$ is not extreme and is also not a non-extreme \imped robot, so $x_1$ will not move until $i$ changes its location.

    From these cases, we also have that an \imped robot stays connected to the robots that were its neighbours before it moved. This implies that the visibility graph stays connected after any \imped robots move.
\end{proof}

The robots executing \Cref{algorithm:gsgs-algo}, as shown in \cite{Goswami2022}, stay in the bounding polygon $ABPCD$.
Next, we show, using the above proof, a tighter polygon bounding the robots. To define this polygon, we let $Q$ to be the point such that it is an intersection between the bottom $l2r$ slant of $s$ and the bottom $r2l$ slant of $s$ (cf. \Cref{figure:infinite-triangular-grid}). Let $B'$ be the point of intersection between left layer ($AB$) and the bottom $l2r$ slant of $s$ and let $C'$ be the point of intersection between the right layer ($CD$) and the bottom $r2l$ slant of $s$.
We show that the robots never step out of the polygon $AB'QC'D$, which is tighter than $ABPCD$. 

\begin{observation}\label{observation:imped-5-6}
    If the neighbouring robot, say $j$ of an \imped robot $i$ moves then $i$ or $j$ fall under case 5 or case 6.
\end{observation}

\begin{lemma}\label{lemma:lower-slants}
    Throughout the execution of \Cref{algorithm:gsgs-algo}, the bottom $r2l$ slant and the bottom $l2r$ slant will not change.
\end{lemma}

\begin{proof}
    In a global state $s$, a 
    robot present at the bottom $l2r$ slant of $s$ or a bottom $r2l$ slant of $s$ is represented in cases 5 and 7. From \Cref{algorithm:gsgs-algo} and the proof of \Cref{lemma:gmrit-llp}, if a robot is present at bottom $l2r$ slant (respectively, bottom $r2l$ slant), it will never move below ($v_1(i)$) or left ($v_2^\ell(i)$) of its location (respectively, below ($v_1(i)$) or right ($v_2^r(i)$) of its location).
\end{proof}

\begin{lemma}\label{lemma:lr-layers}\cite{Goswami2022}
    Throughout the execution of \Cref{algorithm:gsgs-algo}, left layer does not move leftwards and right layer does not move rightwards.
\end{lemma}

\begin{lemma}\cite{Goswami2022}\label{lemma:horizontal-layer-moves-down}
    In every round of \Cref{algorithm:gsgs-algo}, the top layer moves at least 1/2 unit in the negative direction of the $y$-axis.
\end{lemma}

\begin{corollary}\label{corollary:never-step-out}(From \Cref{lemma:lower-slants} and \Cref{lemma:lr-layers})
    The robots will never step out of the polygon $AB'QC'D$.
\end{corollary}

\begin{theorem}\label{theorem:gathering-point}
\Cref{algorithm:gsgs-algo} is a lattice-linear self-stabilizing algorithm for the \gmrit problem on $n$ robots executing asynchronously.
\end{theorem}

\begin{proof}
    From \Cref{lemma:lower-slants}, we have that bottom $l2r$ slant and bottom $r2l$ slant do not change. From \Cref{corollary:never-step-out}, we have that the robots will never step out of the polygon $AB'C'DQ$.
    From \Cref{lemma:horizontal-layer-moves-down}, we have that the top layer moves down by at least half a unit in the negative direction of $y$-axis. Thus we have that the robots converge at the point of intersection of the bottom $l2r$ slant and the bottom $r2l$ slant, and the robots will eventually gather at that point.
\end{proof}

\begin{corollary}\label{corollary:gathering-point}(From \Cref{lemma:horizontal-layer-moves-down}, \Cref{corollary:never-step-out} and \Cref{theorem:gathering-point})
    The point, $Q$, where the robots gather, can be uniquely determined from the initial global state.
\end{corollary}  

\subsubsection*{Time Complexity Properties}

In \cite{Goswami2022}, authors showed that \Cref{algorithm:gsgs-algo} converges in $2.5(n+1)$ rounds.
Based on \Cref{corollary:gathering-point} which identifies a predictable gathering point, 
we show that a maximum of $2n$ rounds is sufficient, which is a tighter bound. 

\begin{theorem}\label{theorem:time-gsgs}
    \Cref{algorithm:gsgs-algo} converges in $2n$ rounds.
\end{theorem}

\begin{proof}

We use \Cref{figure:infinite-triangular-grid} to discuss convergence of robots in \Cref{algorithm:gsgs-algo}. As discussed in \Cref{subsection:gsgs-definitions} and \Cref{subsection:gsgs-lattice-linearity}, let $A$, $B'$, $Q$, $C'$ and $D$ be the points obtained by pairwise intersection of the top layer, left layer, bottom $l2r$ slant, bottom $r2l$ slant and right layer. 
Let $h_\ell$ be the depth of the line segment $AB'$, $h_r$ be the depth of the line segment $C'D$, and $w$ be the width of the line segment $AD$. Note that a unit of length of the depth of $AB'$ or $C'D$ is $\sqrt{3}$ times a unit of length of the width of $AD$ due to the geometry of $G$.
Since the robots form a connected graph, $w \leq n$. And, if $w > 0$ then $h_\ell+h_r\leq n$. If $w=0$, we define $h_\ell=0$ and $h_r=n$.

In the case where $w=0$, it can be clearly observed that \Cref{algorithm:gsgs-algo} converges in $n$ rounds. 
Next, we consider if $w>0$. Without the loss of generality, let $h_r\geq h_\ell$. Thus, $h_\ell\leq n/2$. 

Let $E$ be the point of intersection between the bottom $l2r$ slant and the right layer
(cf. \Cref{figure:infinite-triangular-grid}).
We draw a horizontal line (perpendicular to the $y$-axis) through $B'$ and use $J$ to denote its intersection with $DC'$.
Thus, the depth of $AB'JD$ is $h_\ell$. 
Additionally, observe that the length of $B'E$ on the $n$-axis is $w$. Thus the height of $JE$ is $w/2$ units on the $y$-axis.
This means that the depth of $B'EJ$ is $w/2$. By construction of $E$, the depth of $B'QC'J$ is upper bounded by the depth of $B'EJ$. Thus, the depth of $B'QC'J$ cannot exceed $w/2\leq n/2$ units.

Thus, the total depth of $AB'QC'D$ is equal to the sum of the depth of $AB'JD$ and the depth of $B'QC'J$, which is 
upper bounded by $n/2+n/2=n$ units.
From \Cref{lemma:horizontal-layer-moves-down}, the total number of rounds required for the robots to gather is upper bounded by $2n$ moves.
\end{proof} 

\subsection{Revised Algorithm for \gmrit}\label{section:gsgs-new-algo}

In this section, we present a revised algorithm that simplifies the proof of lattice-linearity. This algorithm is based on the difficulties involved in the proof of \Cref{lemma:gmrit-llp} where we needed to consider the possible actions taken by the neighbours of an \imped robot. Our proof would have been simpler
if all the neighbours of an \imped robot $i$ would not be allowed to move until $i$ moves.
Additionally, from Observation \ref{observation:imped-5-6}, we have that if a robot $j$, neighbouring to an \imped node $i$, is \imped, then $i$ or $j$ fall in case $5$ or case $6$.

These issues can be alleviated by removing cases 5 and 6 from the algorithm.
The macros that we utilize are as follows. A robot is \textit{downward \imped} if its local state is one of those represented in cases 1, 2, 3 or 4 (and their mirror images; cf \Cref{figure:local-states} (b)). A robot is is \textit{downslant \imped} if its local state is that represented in case 7.
$$
\begin{array}{l}
    \textsc{Downward-II}(i)\equiv (\textsc{At}(i,\{v_1\})\land \lnot\textsc{At}(i,\{v_3^\ell, v_4, v_3^r\}))\lor \textsc{Only-At}(i,\{v_2^\ell, v_2^r\}).\\
    \textsc{Downslant-Left-II}(i)\equiv \textsc{Only-At}(i,\{v_2^\ell\}).\\
    \textsc{Downslant-Right-II}(i)\equiv \textsc{Only-At}(i,\{v_2^r\}).\\
\end{array}
$$

The revised algorithm is as follows. A downward impedensable robot moves to $v_1(i)$ location, and a downslant \imped robot moves to $v_2(i)$ location.

\begin{algorithm}\label{algorithm:gsgs-new-algo}
    Rules for robot $i$.
\end{algorithm}
\begin{center}
    $\begin{array}{|l|}
        \hline 
        \textsc{Downward-II}(i)\longrightarrow move(i,v_1(i)).\\
        \textsc{Downslant-Right-II}(i)\longrightarrow move(i,v_2^r(i)).\\
        \textsc{Downslant-Left-II}(i)\longrightarrow move(i,v_2^\ell(i)).\\
        \hline 
    \end{array}
    $
\end{center}

In \Cref{algorithm:gsgs-new-algo}, because of the removal of cases 5 and 6, any robot around an \imped robot does not move. Thus lattice-linearity of this algorithm can be visualized more intuitively. Consequently, we have the following lemma.

\begin{lemma}
    The predicate $\forall i: \lnot(\textsc{Downward-II}(i)$ $\lor$
        $\textsc{Downslant-Right-II}(i)$ $\lor$
        $\textsc{Downslant-Left-II}(i))$,
    is a lattice-linear predicate on $n$ robots, and the visibility graph does not get disconnected by the actions under \Cref{algorithm:gsgs-new-algo}.
\end{lemma}

\Cref{lemma:horizontal-layer-moves-down} shows the top-layer moves down in each round. This proof is not affected by the removal of cases 5 and 6, as the robot executing in cases 5 or 6 is \textit{not} a top-layer robot. 
Consequently, \Cref{algorithm:gsgs-new-algo} follows the properties as described in \Cref{lemma:horizontal-layer-moves-down}, \Cref{theorem:gathering-point} and hence \Cref{theorem:time-gsgs}. 
Thus, we obtain the following theorem.

\begin{theorem}
    \Cref{algorithm:gsgs-new-algo} is a lattice-linear self-stabilizing algorithm for \gmrit problem on $n$ robots executing asynchronously. It converges in $2n$ rounds, and the robots gather at $Q$.
\end{theorem}

\section{Summary of the chapter}\label{section:flla-summary}

In this chapter, we 
introduced \textit{fully lattice-linear algorithms} that are tolerant to asynchrony.
Such algorithms induce lattices in the state space even if the underlying problem does not specify, in a suboptimal global state, a set of nodes that must change their states. 

\subsection{Theoretical Achievements}



We bridge the gap between lattice-linear problems \cite{Garg2020} and eventually lattice-linear algorithms (\Cref{chapter:ella}).
%
Fully lattice-linear algorithms overcome the limitations of \cite{Garg2020} and \Cref{chapter:ella}. 
Additionally, such algorithms can be developed even for problems that are not lattice-linear. This overcomes a key limitation of \cite{Garg2020} where the system fails if nodes cannot be deemed \imped, or not \imped, naturally.
Since the lattice structures exist in the entire (reachable) state space, we overcome a limitation of \Cref{chapter:ella} where only in a subset of global states, lattice-linearity is observed. 

\subsection{New Fully Lattice-Linear Algorithms}

We present algorithms for minimal dominating set (\mds), graph colouring (\gc), minimal vertex cover (\mvc) and maximal independent set (\mis). 
Of these, \mds and \gc relied on tie-breakers on node IDs, a common approach for breaking ties in the literature.  
We observe that a similar design cannot be directly extended to develop an algorithm for  \mvc and \mis. However, the use of complex actions -- that permit a node to change the values of the variables of other node as well as its own -- enable the design of algorithms for \mvc and \mis. We also observe that complex actions can be revised into simple actions -- where a node can only change its own values -- without losing lattice-linearity. However, these revised algorithms utilize the phenomenon of priority inheritance to accomplish this.

We also provide a fully lattice-linear 2-approximation algorithm for vertex cover. This algorithm is the first lattice-linear approximation algorithm for an NP-Complete problem.


In \cite{Garg2020} lattice-linearity is studied in only those systems where the state space forms a distributive lattice where all pairs of global states have a join (supremum) and meet (infimum), and join and meet operations distribute over each other.
We observe that some of these requirements are not required to provide correctness under asynchrony. 
Specifically, we observe that a system allows asynchrony if the state space forms $\prec$-lattices, where the join between any two states is defined, but the definition of meet is not required. This aspect is more overtly observed in instances of \mvc. Specifically, \Cref{figure:vc-semilattice-example} shows that we have a $\prec$-lattice but not a distributive lattice. 

Fully lattice-linear algorithms considered in this chapter preserve an advantage of \cite{Garg2020} that was lost in the extension by \Cref{chapter:ella}. Specifically, in \cite{Garg2020}, the final configuration could be uniquely determined from the initial state, whereas in \Cref{chapter:ella}, all global states (specifically, the infeasible states) do not form a lattice, so starting from an arbitrary state, the state of convergence cannot be predicted. In fully lattice-linear algorithms that we introduce in this chapter, the state space is split into multiple lattices and the algorithm starts in one of them. Hence, the state of convergence can be uniquely determined by the initial state.

\subsection{Distance-1 Transformation and Experiments}

We have that a lattice-linear algorithm can be transformed to a distance-1 algorithm by having the nodes keep a copy of the variables, of the other nodes, that they want to evaluate their guards with. We transform \Cref{algorithm:ds-ll} to a distance-1 algorithm by using a minimal set of variables needed to evaluate said guards. 

We also demonstrate that these algorithms substantially benefit from using the fact that they satisfy the property of lattice-linearity. Specifically, they outperform existing algorithms when they utilize the fact that they are correct without synchronization among processes, i.e., they are correct even if a node is reading old/inconsistent values of its neighbours.

\subsection{Gathering Myopic Robots}

We also show lattice-linearity of the algorithm developed by Goswami et al \cite{Goswami2022} for gathering robots on an infinite triangular grid. This removes the assumptions of synchronization from the algorithm and thus makes a system running this algorithm fully tolerant to asynchrony. We also present a revised algorithm that simplifies the proof of lattice-linearity without losing any of the desired properties (e.g., convergence time, stabilization).

Lattice-linearity implies that the locations, possibly visited by a robot, form a total order. The total order is a result of the fact that we are able to determine all and the only robots in any global state that are \imped, and an \imped robot has only one choice of action. By making this observation, it can also be noticed that we can closely predict the executions that the robots would perform. As a result, we are able to (1) compute the exact arena traversed by the robots throughout the execution of the algorithm (\Cref{lemma:lower-slants} and \Cref{lemma:lr-layers}), and (2) deterministically predict the point of gathering of the robots (\Cref{corollary:gathering-point}).

We also provided a better upper bound on the time complexity of this algorithm. Specifically, we show that it converges in $2n$ rounds, whereas \cite{Goswami2022} showed that a maximum of $2.5(n+1)$ rounds are required.
This was possible due to the observations that followed from the proof of lattice-linearity of this algorithm.

\chapter{PARTIAL ORDER-INDUCING SYSTEMS}\label{chapter:dag}

In the previous chapters, we explored several special cases of algorithms that can converge without synchronization. Specifically, we studied how algorithms, that impose the nodes to visit their local states in a total order, tolerate asynchrony. Consequently, the global state transition graph forms a $\prec$-lattice, and thus, we call such systems lattice-linear systems. We also studied several example problems and algorithms that act as concrete evidence of our theory. However, as we will study in this chapter, an arbitrary system, that can converge in asynchrony, may not be lattice-linear. Lattice induction is a sufficient condition to allow asynchrony, but it is not a necessary condition. In this chapter, we complete the theory that explains the behaviour of an arbitrary system that can tolerate asynchrony.
To this end,  we introduce partial order-inducing problems and partial order-inducing algorithms.

We show that induction of a $\prec$-DAG (induced among the global states -- that forms as a result of a partial order induced among the local states visited by individual nodes) is a necessary and sufficient condition to allow an algorithm to run in asynchrony.

In the chapter, we first provide a comprehensive description of partial order-inducing problems and partial order-inducing algorithms, along with some simple examples. Then we show some properties of an algorithm that can converge under asynchrony, which include the condition that we discussed in the above paragraph.

An important conclusion from the above observation is that if we want to show that an algorithm can converge without synchronization, then we do not have to generate the entire global state transition system and check for the absence of cycles. Rather, we only need to show that the local state transition graph forms a partial order (PO). Thus, the complexity of determining the correctness of such systems is significantly reduced, and so this observation is fruitful in writing social and formal proofs that show tolerance of an algorithm to asynchrony.

In this chapter, we study problems such as the dominant clique (DC) problem, the shortest path (SP) problem and the maximal matching (MM) problem.
We show that DC and SP are \po-inducing problems. Among these, DC allows self-stabilization, whereas the algorithm that we present for the SP does not.
We demonstrate that MM is not a \po-inducing problem. We present a \po-inducing algorithm for it. This algorithm allows self-stabilization.
We study the upper bound to the convergence time of a \po-inducing algorithm.
We show how inducing a partial order among the local states of all individual nodes is crucial to allow asynchrony: it is necessary and sufficient condition to allow asynchrony.

A observation that immediately follows is that since a total order is a special case of a partial order, all lattice-linear problems and algorithms are, respectively, \po-inducing problems and algorithms.

\subsubsection*{Organization of the Chapter}

\noindent This chapter is organized as follows. 
In \Cref{section:dip}, we study the characteristics of \po-inducing problems, and
in \Cref{section:dia}, we study the characteristics of \po-inducing algorithms, with examples.
In \Cref{section:dag-properties}, we study the properties of \po-inducing algorithms. While the previous sections provide simple, but sufficient, examples of asynchrony tolerant systems, this section is crucial from the perspective of the theory that we establish in this chapter.
Finally, we summarize the chapter in \Cref{section:dag-summary}.

\section{Natural partial order induction: \po-inducing \textit{Problems}}
\label{section:dip}

In this section, we discuss properties of problems where a partial order (PO) among the nodes visited by individual nodes is 
induced \textit{naturally}.
It means that in any suboptimal state, the problem definition itself specifies the nodes that must change their state, in order for the system to reach an optimal state.


\subsection{Embedding a \ldag among global states}\label{subsection:<-dag}

To explain the embedding of a \ldag, 
first, we define a partial order $\prec_l$ among the local states of a node.
This partial order 
defines
all the possible transitions that a node is allowed to take.
The partial order $\prec_l$ is used to restrict how node $i$ can execute: $i$ can go from state $s[i]$ to $s'[i]$ only if $s[i] \prec_l s'[i]$.

Using 
$\prec_l$,
we define $\prec_g$ that orders the global states.
The predicate $s \prec_g s^\prime$ is true iff the predicate $(\forall i:$ $s[i]=s'[i]\lor s[i]\prec_l s'[i]) \land (\exists i:s[i]\prec_ls'[i])$ is true; $s=s'$ iff $\forall i:s[i] = s'[i]$. 
For brevity, we use $\prec$ to denote $\prec_l$ and $\prec_g$: $\prec$ corresponds to $\prec_l$ while comparing local states, and $\prec$ corresponds to $\prec_g$ while comparing global states. 
We also use the symbol `$\succ$' which is such that $s\succ s'$ iff $s' \prec s$.
Similarly, we use symbols `$\preceq$' and `$\succeq$'; e.g., $s\preceq s'$ iff  $s=s' \lor s \prec s'$.
We call the \dag, formed from such partial order, a \textit{\ldag}.

\begin{definition}\label{definition:<-dag}
    \textbf{\boldmath \ldag}. 
    Given a partial order $\prec_l$ that orders the local states visited by $i$ (for each $i$), the \ldag corresponding to $\prec_l$ is defined as follows:
    $s \prec s'$ iff $(\forall i: s[i] \preceq_l s'[i]) \wedge (\exists i: s[i] \prec_l s'[i])$.
\end{definition}

A \ldag constraints how global states can transition among one another: state $s$ can transition to state $s'$ iff $s\prec s'$.
By varying $\prec_l$ that identifies a partial order among the local states of a node, one can obtain different \ldag{s}. A \ldag, embedded in the state space, is useful for permitting the algorithm to execute asynchronously.
Under proper constraints on the structure of \ldag, convergence can be ensured. We elaborate on this in \Cref{subsection:properties-dip}. 


\subsection{General Properties of \po-Inducing Problems}\label{subsection:properties-dip}

Asynchrony can be allowed in any system that imposes a condition that any node $i$ changes its state only if it evaluates that an optimal global state cannot be reached with the current local state of $i$. We call such a node an \textit{\imped node} (\textit{indispensable} to change for progress, an \textit{impediment} to progress if it does not change its state). Let $\mathcal{P}$ be the predicate governing a system such that it determines the state transitions of that system, that is, it determines which nodes have to change their state. And, let $\mathcal{P}(s)$ be true only if no nodes in global state $s$ are impedensable.

\subsubsection*{\Imped Node and \Imped Global State}

In a multiprocessor system, a node can have single or multiple choices of action when it changes its state, but it can be in only one state at a given instance. A node $i$ is enabled iff it is \imped.
In this case, (1) the nodes are also capable of discarding more than one local state, and (2) consequently, a partial order is induced among the local states visited by individual nodes.

Let $i$ be \imped in a global state $s$, and let $st$ be the current local state of $i$. Node $i$ does not revisit $st$, and it also does not visit some other local states.
To explain this, let $\textsc{Equicorrupt}(st,i)$ be the set of local states of $i$ that are deemed to be violating iff state $st$ is is a violating local state in $i$. So if $st$ is discarded, all states in $\textsc{Equicorrupt}(st,i)$ are also not visited by $i$ throughout the execution.
This ensures that $i$ must move up in the partial order. We have the definition of an \imped node as follows.

\begin{definition}\label{definition:imped-node-updated}\textbf{Impedensable node (updated definition).} $\textsc{\Imped}(i,s,\mathcal{P})\equiv \lnot \mathcal{P}(s)\land(\forall s':(s'\succeq s)\Rightarrow(s'[i]\in \textsc{Equicorrupt}(s[i],i)\Rightarrow\lnot \mathcal{P}(s')))$.
\end{definition}

\begin{examplemaxcont}
    Consider the execution of the nodes in global state $\langle 2,2,3\rangle$ under the algorithm for the max problem presented in \Cref{subsection:colouring-and-max}. $\textsc{Equicorrupt}(2,1)$ is the set of local states of node 1 that are equally corrupt as local state 2. Due to the total order, there is only one node in each level, and so, e.g., the set $\textsc{Equicorrupt}(2,1)=\{2\}$ has only one value.
    \qed 
\end{examplemaxcont}

As we will study in \Cref{subsection:dc} and in the following parts of the chapter, the states in $\textsc{Equicorrupt}(st,i)$ are essentially the local states of $i$ that are at the same level in the partial order.
Thus, a global state $s$ is suboptimal, that is, $s$ is \imped, if and only if it contains at least one \imped node. Formally,

\begin{definition}\textbf{\textit{Impedensable global state.}}\label{definition:imped-state-updated}
    $\textsc{\Imped}(s,\mathcal{P})\equiv \exists i:\textsc{\Imped}(i,s,\mathcal{P})$.
\end{definition}

\subsubsection*{PO-inducing problems}

A \textit{\po-inducing problem}
$P$ can be represented by a predicate $\mathcal{P}$, where $P$ stipulates that any local state $st$ of $i$, that is deemed in violation by $i$, will make any global state $s$ suboptimal if $s[i]=st$. So $i$ never revisits $st$.
As a result,
%
predicate $\mathcal{P}$ induces a partial order among the local states visited by a node, for all nodes
(no cycles).
Consequently, the discrete structure $\mathcal{S}$ that gets induced among the global states is a \ldag, as described in \Cref{definition:<-dag}. 
We say that $\mathcal{P}$, satisfying \Cref{definition:imped-node-updated}, is \textit{\po-inducing} with respect to that \ldag.
$\mathcal{P}$ is used by the nodes to determine if they are \imped, using \Cref{definition:imped-node-updated} and \Cref{definition:imped-state-updated}.

\begin{definition}\textbf{\po-Inducing Predicate.}\label{definition:po-inducing-predicate}
    $\mathcal{P}$ is a \po-inducing predicate with respect to a \ldag induced among the global states iff $\forall s\in S: \lnot\mathcal{P}(s) \Rightarrow \exists i:\textsc{\Imped}(i,s,\mathcal{P})$.
\end{definition}

\noindent \textbf{\textit{Remark}}: Since a total order is a special case of a partial order, all lattice-linear problems are \po-inducing problems.

Now we complete the definition of \po-inducing problems. In a \po-inducing problem $P$, given any suboptimal global state, we can identify all nodes that should not retain their state.
$\mathcal{P}$ is thus designed conserving this nature of problem $P$.

\begin{definition}\label{definition:dip}
\textbf{\po-inducing problem (DIP)}.
A problem $P$ is \po-inducing
iff there exists a predicate $\mathcal{P}$ and a \ldag $\mathcal{S}$, induced among the global states, such that
(1) $P$ is solved iff the system reaches a state where $\mathcal{P}$ is true,
(2) $\mathcal{P}$ is \po-inducing with respect to $\mathcal{S}$, i.e., $\forall s: \neg \mathcal{P}(s) \Rightarrow \exists i:\textsc{\Imped}(i,s,\mathcal{P})$, and
(3) $\forall s:(\forall i:\textsc{\Imped}(i,s,\mathcal{P})\Rightarrow (\forall s':\mathcal{P}(s')\Rightarrow s'[i]\neq s[i]))$.
\end{definition}

\subsubsection*{Successors of Global States}

When a system allows asynchrony, then a global state $s$ can transition to one of multiple other global states. Let $S_s$ be a set of such states. $S_s$ will contain the states which $s$ can transition to if all nodes are reading fresh local states of other nodes. 

\begin{definition}\textbf{\textit{Successors of a global state}}.
    $\textsc{Successors}(s)\equiv \{s':s'\succ s\}$.
\end{definition}


There may be are some global states that do not have any successors. We call them \textit{terminal successors}.

\begin{definition}\label{definition:terminal-successors}\textbf{\textit{Terminal Successors}}.
    $s'\in \textsc{Terminal-Successors}(s)$, iff $\{s'|$ $s'\in \textsc{Successors}(s) \land$ $\textsc{Successors}(s')$ $=$ $\phi\}$.
\end{definition}

\begin{examplemaxcont}
    Going back to \Cref{example:max} (Chapter 2), the only terminal successor in the state transition graph shown in \Cref{figure:colouring-and-max} (b) is $\langle 3,3,3\rangle$.
    \qed 
\end{examplemaxcont}


\subsubsection*{Self-Stabilizing Predicates}

$\mathcal{P}$ satisfies \Cref{definition:ss-dip} only if starting from any arbitrary state, the system converges to an optimal state. This, in turn, is possible only if all terminal successors in $\mathcal{S}$ are optimal states.
$\mathcal{P}$ can be true in other states as well. 

\begin{definition}\label{definition:ss-dip}\textbf{Self-stabilizing \po-inducing predicate}.
    Continuing from \Cref{definition:dip}, $\mathcal{P}$ is a self-stabilizing \po-inducing predicate if and only if all terminal successors in the \ldag induced by $\mathcal{P}$ are optimal states, i.e. $\forall s,s'\in S: \textsc{Terminal-Successor}(s,s')\Rightarrow \mathcal{P}(s')=true$.
\end{definition}

\begin{examplemaxcont}
    We have shown an incomplete state transition graph in \Cref{figure:colouring-and-max} (b), however, it can be trivially noticed that the predicate for the max problem, as noted in Example Max: Continuation \ref{example:max-predicate} is a self-stabilizing \po-inducing predicate.
    \qed 
\end{examplemaxcont}

The main intent of this chapter is to establish some fundamental properties of \po-inducing systems. However, to understand these properties, we need to understand the behaviour of such systems. Thus, in the remaining part of this section, we study some \po-inducing problems, and explore how the induction of a partial order among the local states, under the acting algorithm, allows the nodes to execute without synchronization.

\subsection{Dominant Clique (\dc) Problem}\label{subsection:dc}

In this section, we describe an algorithm for the dominant clique problem, which is defined in the following paragraph. The algorithm that we describe in this subsection is an \textit{embarrassingly parallel algorithm}, i.e., in this algorithm, there is no transfer of data among the nodes. However, we use this trivial algorithm to build the intuition behind the induction of a partial order among the local states visited by individual nodes, and how that gives rise to a \ldag among the global states.

\begin{definition}\textbf{Dominant Clique}.
    In the dominant clique problem, the input is an arbitrary graph $G$ such that for the variable $i[cliq]$ of each node $i$, $i[cliq]\subseteq N_i$ and $\{i\}\subseteq i[cliq]$. The task is to (re-)evaluate $i[cliq]$ such that (1) all the nodes in $i[cliq]$ form a clique,
    and (2) there exists no clique $c$ in $G$ such that $i[cliq]$ is a proper subset of $c$.
\end{definition}

\noindent Thus, we define the DC problem by the following predicate.
\begin{center}
    $\mathcal{P}_{dc}\equiv (i\in i[cliq])\land(\forall j,k\in i[cliq]: (j\neq k)\Rightarrow(k\in Adj_j))\land(\not\exists j\in Adj_i: j\not\in i[cliq]\land (\forall k\in i[cliq]:k\in Adj_j))$
\end{center}
The local state of a node $i$ is defined by $\langle i[cliq]\rangle$. An \imped node $i$ in a state $s$ is a node for which (1) all the nodes in $i[cliq]$ do not form a clique, or otherwise (2) there exists some node $k$ in $Adj_i$ such that $i[cliq]\cup\{k\}$ is a valid clique, but $k$ is not in $i[cliq]$. Formally,
\begin{center}

$\textsc{\Imped-DC}(i)\equiv$ $(i\not\in i[cliq])\lor \lnot(\forall j,k\in i[cliq]:j\neq k\land j\in Adj_k))$ $\lor$\\ $(\exists j\in Adj_i:j\not\in i[cliq]\land (\forall k\in i[cliq]:k\in Adj_j))$.\\
\end{center}
The algorithm that we develop next is a self-stabilizing algorithm, which means that the nodes can be initialized arbitrarily. Thus, $i[cliq]$ may contain the nodes that are not connected to $i$ by an edge. The algorithm is defined as follows. If all the nodes in $i[cliq]$ do not form a clique, then $i[cliq]$ is reset to be $\{i\}$. If there exists some node $j$ in $Adj_i$ such that $i[cliq]\cup\{j\}$ is a clique, but $j$ is not in $i[cliq]$, then $j$ is added to $i[cliq]$. 

\begin{algorithm}\label{algorithm:dc-dip}
    Rules for node $i$ in state $s$.
\end{algorithm}
\begin{center}
    $\begin{array}{|l|}
            \hline
            \textsc{\Imped-DC}(i)\longrightarrow\\
            \begin{cases}
                i[cliq]=\{i\} & \text{if $(i\not\in i[cliq]\lor(\exists j,k\in$} \text{\quad $i[cliq]:j\neq k\land j\not\in Adj_k))$}\\
                i[cliq]= i[cliq]~\cup\{j\} & \text{otherwise}\\
                 & \text{\quad (where $j$ is such that $\forall i\in i[cliq]: j\in Adj_i$)}
            \end{cases}~\\
            \hline
        \end{array}$
\end{center}

\begin{lemma}\label{lemma:dc-dip}
    The dominant Clique problem is a \po-inducing problem.
\end{lemma}

\begin{proof}
    For a node $i$, $i[cliq]$ contains the nodes that $i$ is connected with, and the nodes in $i[cliq]$ should form a clique. A global state does not manifest a dominant clique if at least one node $i$ in $s$ does not store a set of nodes forming a maximal clique with itself, i.e. (1) $i[cliq]$ is not a maximal clique, that is, there exists a $j$ in $Adj_i\setminus i[cliq]$ such that $i[cliq]\cup\{j\}$ forms a valid clique, or (2) the nodes in $i[cliq]$ do not form a clique.
    
    Next, we need to show that if some node $i$ in state $s$ is violated, then for any global state $s'$ such that $s'\succeq s$, if $s'[i]=s[i]$, then $s'$ will not manifest a dominant clique. This is straightforward from the definition itself, that if a node $i$ is \imped, then $i$ does not store a set of nodes forming a maximal clique with itself. Thus, if $i$ is \imped in $s$, and $i$ has the same state in some $s'$ such that $s'\succ s$, then $s'$, as well, does not satisfy $\mathcal{P}_{dc}$.
\end{proof}

To present the abstraction of the partial order induced among the local states, we define the state value of a local state as follows.
\begin{center}
    $
    \begin{array}{l}
        \textsc{State-Value-DC}(i,s)=\\
            \begin{cases}
                |C|-|i[cliq]|:C = \text{largest superset of $i[cliq]$ that is a valid clique} & \text{if $i[cliq]$ is a clique}.\\
                deg(i)+1 & \text{otherwise}.
            \end{cases}
    \end{array}
    $
\end{center}

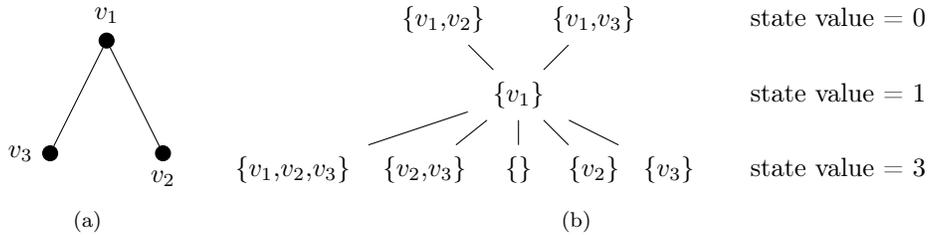
\begin{figure}[ht]
    \centering
    \subfigure[]{
        \begin{tikzpicture}[scale=1.5]
            \node [circle, draw=black,fill=black,inner sep=2pt,label=above:$v_1$] (a) at (0,0) {};
            \node [circle, draw=black,fill=black,inner sep=2pt,label=left:$v_3$] (b) at (-.5,-1) {};
            \node [circle, draw=black,fill=black,inner sep=2pt,label=below:$v_2$] (c) at (.5,-1) {};
            
            \draw (a) -- (b); \draw (a) -- (c);
        \end{tikzpicture}
    }
    \subfigure[]{
        \begin{tikzpicture}
            \node (a) at (0,0) {\begin{tabular}{c}\{$v_1$\}\end{tabular}};
            \node (b) at (-1,1) {\begin{tabular}{c}\{$v_1$,$v_2$\}\end{tabular}};
            \node (c) at (1,1) {\begin{tabular}{c}\{$v_1$,$v_3$\}\end{tabular}};
            
            \node (d) at (-3,-1) {\begin{tabular}{c}\{$v_1$,$v_2$,$v_3$\}\end{tabular}};
            \node (e) at (-1.25,-1) {\begin{tabular}{c}\{$v_2$,$v_3$\}\end{tabular}};
            \node (f) at (0,-1) {\begin{tabular}{c}\{\}\end{tabular}};
            \node (g) at (1,-1) {\begin{tabular}{c}\{$v_2$\}\end{tabular}};
            \node (h) at (2,-1) {\begin{tabular}{c}\{$v_3$\}\end{tabular}};
            \draw (a) -- (b); \draw (a) -- (c);\draw (a) -- (d); \draw (a) -- (e); \draw (a) -- (f); \draw (a) -- (g); \draw (a) -- (h);
            
            \node at (4.25,1) {\begin{tabular}{c}state value = 0\end{tabular}};
            \node at (4.25,0) {\begin{tabular}{c}state value = 1\end{tabular}};
            \node at (4.25,-1) {\begin{tabular}{c}state value = 3\end{tabular}};
        \end{tikzpicture}
    }
    \caption{(a) Input graph. (b) The acronym s.v. stands for state value. Partial order among local states of node 1 (all edges are directed upwards).}
    \label{figure:dc}
\end{figure}

$\mathcal{P}$ induces a partial order among the local states, which can be abstracted by state value as defined above: for a pair of global states $s$ and $s'$, $s[i]\prec s'[i]$ iff $\textsc{State-Value-DS}(i,s')<\textsc{State-Value-DS}(i,s)$. As an instance, the partial order induced among the local states of node $v_1$ (of the graph in \Cref{figure:dc} (a)) is shown in \Cref{figure:dc} (b).

To present the abstraction of the \ldag induced among the global states, we define the rank of a global state as follows.
\begin{center}
    $\textsc{Rank-DC}(s)=\sum\limits_{i\in V(G)}\textsc{State-Value-DC}(i,s).$
\end{center}

\begin{figure}[ht]
    \begin{tikzpicture}[scale=.9,every node/.style={scale=.9}]
        \node (a) at (0,-1) {\begin{tabular}{c}$\langle$\{1\},\{2\},\{3\}$\rangle$\end{tabular}};
        
        \node (b1) at (-5,.5) {\begin{tabular}{c}$\langle$\{1,2\},\{2\},\{3\}$\rangle$\end{tabular}};
        \node (b2) at (-1.5,.5) {\begin{tabular}{c}$\langle$\{1\},\{1,2\},\{3\}$\rangle$\end{tabular}};
        \node (b3) at (1.5,.5) {\begin{tabular}{c}$\langle$\{1\},\{2\},\{1,3\}$\rangle$\end{tabular}};
        \node (b4) at (5,.5) {\begin{tabular}{c}$\langle$\{1,3\},\{2\},\{3\}$\rangle$\end{tabular}};

        \node (c1) at (-7.5,2.5) {\begin{tabular}{c}$\langle$\{1,2\},\{1,2\},\{3\}$\rangle$\end{tabular}};
        \node (c2) at (-4.5,2.5) {\begin{tabular}{c}$\langle$\{1,2\},\{2\},\{1,3\}$\rangle$\end{tabular}};
        \node (c3) at (-1.5,2.5) {\begin{tabular}{c}$\langle$\{1\},\{1,2\},\{1,3\}$\rangle$\end{tabular}};

        \node (c6) at (1.5,2.5) {\begin{tabular}{c}$\langle$\{1\},\{1,2\},\{1,3\}$\rangle$\end{tabular}};
        \node (c4) at (4.5,2.5) {\begin{tabular}{c}$\langle$\{1,3\},\{1,2\},\{3\}$\rangle$\end{tabular}};
        \node (c5) at (7.5,2.5) {\begin{tabular}{c}$\langle$\{1,3\},\{2\},\{1,3\}$\rangle$\end{tabular}};
        
        \node (d1) at (-3,4) {\begin{tabular}{c}$\langle$\{1,2\},\{1,2\},\{1,3\}$\rangle$\end{tabular}};
        \node (d2) at (3,4) {\begin{tabular}{c}$\langle$\{1,3\},\{1,2\},\{1,3\}$\rangle$\end{tabular}};
        
        \draw (a) -- (b1); \draw (a) -- (b2);\draw (a) -- (b3);\draw (a) -- (b4);
        
        \draw (b1) -- (c1);\draw (b2) -- (c1);
        \draw (b1) -- (c2);\draw (b3) -- (c2);
        \draw (b2) -- (c3);\draw (b3) -- (c3);

        \draw (b2) -- (c4);\draw (b4) -- (c4);
        \draw (b3) -- (c5);\draw (b4) -- (c5);
        \draw (b2) -- (c6);\draw (b3) -- (c6);
        
        \draw (c1) -- (d1);\draw (c2) -- (d1);\draw (c3) -- (d1);
        \draw (c4) -- (d2);\draw (c5) -- (d2);\draw (c6) -- (d2);
    \end{tikzpicture}
    \caption{\ldag, assuming that initial state is $\langle\{1\},\{2\},\{3\}\rangle$; we replaced writing $v_i$ by $i$ for brevity. In all these states, the second guard of \Cref{algorithm:dc-dip} is false.
    Observe that any other state will converge to one of these states and then converge to one of the optimal states in this \ldag. (Transitive edges are not shown; all edges are directed upwards.)}
    \label{figure:dc-dag}
\end{figure}
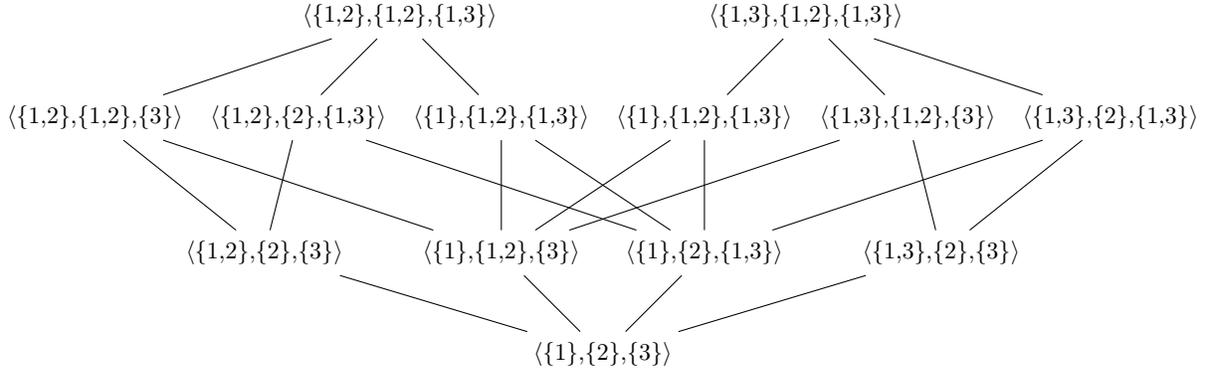

Under \Cref{algorithm:dc-dip}, the global states of the graph in \Cref{figure:dc} (a) form a \ldag that we show in \Cref{figure:dc-dag}. For a pair of global states $s$ and $s'$, $s\prec s'$ iff $\textsc{Rank-DC}(s')<\textsc{Rank-DC}(s)$.
In \Cref{figure:dc-dag}, a global state is represented as $\langle\langle v_1[cliq]\rangle,\langle v_2[cliq]\rangle,\langle v_3[cliq]\rangle\rangle$.
The state space for this instance has a total 512 states. In the figure, we only show the states where the second guard is false in all the nodes. All the global states where the second guard is true in some nodes will converge to one of the states present in this figure, and then it will converge to one of the terminal successors.

Notice that $\forall i\lnot \textsc{\Imped-DC}(i)$ is a self-stabilizing \po-inducing predicate and satisfies \Cref{definition:ss-dip}; \Cref{algorithm:dc-dip} that utilizes this predicate is a self-stabilizing algorithm.

\begin{theorem}\label{theorem:dc-dip}
    \Cref{algorithm:dc-dip} is a silent self-stabilizing algorithm for the dominant clique problem on $n$ nodes executing asynchronously.
\end{theorem}

\begin{proof}
    We need to show that (1) \Cref{algorithm:dc-dip} traverses a \ldag of global states, (2) for all suboptimal states, $\exists$ a terminal successor, and (3) all terminal global states are optimal states.
    
    Let the current state be $s$.
    If $s$ is suboptimal, then for at least for one of the nodes $i$: (1) $i[cliq]$ is not a maximal clique, that is, there exists a $j$ in $Adj_i\setminus i[cliq]$ such that $i[cliq]\cup\{j\}$ forms a valid clique, or (2) the nodes in $i[cliq]$ do not form a clique.
    
    In the case that $s$ is suboptimal and the first case holds true for some node $i$, then under \Cref{algorithm:dc-dip}, $i$ will include a node $j$ in $i[cliq]$ which forms a clique with the nodes already present in $i[cliq]$, which reduces the state value of $i$ by at least 1.

    In the case that $s$ is suboptimal and the second case holds true for some node $i$, then under \Cref{algorithm:dc-dip}, $i$ will change $i[cliq]$ to be $\{i\}$, which reduces the state value of $i$ from $deg(i)+1$ to some value less than or equal to $deg(i)$.

    This shows a partial order being induced among the local states visited by an arbitrary node $i$. Thus under \Cref{algorithm:dc-dip}, an arbitrary graph will follow a \ldag of states and if it transitions from a state $s$ to another state $s'$, then we have that $s'\succ s$ such that rank of $s'$ is less than the rank of $s$.

    If some node is \imped, then the rank of the corresponding global state is non-zero. When a \imped node $i$ makes an execution, then its state value reduces, until it becomes 0. Thus if there is a global state $s$ with rank greater than 0, then there exists at least one \imped node in it. When any node performs execution in $s$ then $s$ transitions to some state with rank less than $s$. This shows that for every suboptimal global state, there exists at least one terminal successor.

    Let that $s$ is a terminal successor. This implies that $\mathcal{P}(s)$ is true: no node is \imped in $s$, so any node will not change its state and $s$ manifests a dominant clique. Thus we have that all terminal states are optimal states, and \Cref{algorithm:dc-dip} is silent.
\end{proof}

\Cref{algorithm:dc-dip} is a distance-1 algorithm and guarantees converges in asynchrony. This is because in a given state $s$ some node $i$ is \imped iff $i$ does not store a dominant clique, thus, $s$ will never transition to an optimal state without $i$ changing its state.

\subsection{Shortest Path (SP) Problem}\label{subsection:sp}

\begin{definition}\textbf{Shortest path.}
    In the shortest path problem, the input is a weighted arbitrary connected graph $G$ (all edge weights are positive) and a destination node $v_{des}$. Every node $i$ stores $i[p]$ (initialized with $\top$) and $i[d]$ (initialized with $\infty$). The task is to compute, $\forall i\in V(G)$, the length $i[d]$ of a shortest path from $i$ to $v_{des}$, and the parent $i[p]$ through which an entity would reach $v_{des}$ starting from $i$.
\end{definition}

The positive weights assigned for every edge $\{i,j\}\in E(G)$ denote the cost that is required to move from node $i$ to node $j$. In this problem, if we would have considered the local state of a node $i$ to be represented only by the variable $i[d]$ then
the local states of the nodes would form a total order. Consequently, the resultant discrete structure formed among the global states will be a $\prec$-lattice. This was shown in \cite{Garg2020}. 
On the other hand, in applications such as source routing \cite{Medhi2017}
where the source node specifies the path that should be taken, the local states form a partial order: such a system cannot be simulated within a total order.
For brevity,
we only represent the next hop, in $i[p]$.
The SP problem can be represented by the following predicate, where, $w(i,j)$ is the weight of edge $\{i,j\}$.
\begin{center}
    $\mathcal{P}_{sp}\equiv \forall i: (i[d]=dis(i,v_{des})=\min\{dis(j,v_{des})+w(i,j):j\in Adj_i\})\land$\quad $(i[p]=\text{arg }\min\{dis(j,v_{des})+w(i,j):j\in Adj_i\})$.
\end{center}
The local state of a node $i$ is defined by $\langle i[p], i[d]\rangle$. An \imped node $i$ in a state $s$ is a node for which its current parent is not a direct connection to the shortest path from $i$ to $v_{des}$. Formally,

\begin{center}
    $\textsc{\Imped-SP}(i)\equiv (i[d]\neq 0\land i=v_{des})\lor (\exists j\in Adj_i: i[d]>j[d]+w(i,j))$.
\end{center}

The algorithm that we develop next is not a self-stabilizing algorithm. We require that each node $i$ has $i[d]$ initialized to $\infty$, however, $i[p]$ can be arbitrarily initialized. However, as we later prove in this subsection, this algorithm converges even without enforcing any synchronization mechanism. The algorithm is defined as follows. If an \imped node $i$ is $v_{des}$, then $i[d]$ is updated to 0 and $i[p]$ is updated to $v_{des}$. Otherwise, $i[p]$ is updated to the $j$ in $Adj_i$ for which $j[d]+w(i,j)$ is minimum.

\begin{algorithm}\label{algorithm:sp-dip}Rules for node $i$.
\end{algorithm}
\begin{center}
    $
    \begin{array}{|l|}
        \hline 
        \textsc{\Imped-SP}(i)\longrightarrow\\
        \begin{cases}
            i[d]=0, i[p]=i & \text{if $i=v_{des}$}\\
            \langle i[d],i[p]\rangle = \langle j[d]+w(i,j), j\rangle: j =  \text{arg} \min\{k[d]+w(i,k):k\in Adj_i\} & \text{otherwise}
        \end{cases}~\\
        \hline 
    \end{array}
    $
\end{center}

We show in the following
that \Cref{algorithm:sp-dip} is a \po-inducing algorithm, and the properties of this algorithm imply that the \sp problem is a \po-inducing problem.

\begin{lemma}
    The shortest path problem is a \po-inducing problem.
\end{lemma}

\begin{proof}
    For a node $i$, $i[d]$ contains the distance of $v_{des}$ from node $i$. A global state $s$ does not manifest all correct distances if for at least one node $i$ in $s$, (1) $dis(i,v_{des})\neq i[d]$, that is, $i$ does not store a shortest path from $i$ to $v_{des}$, or (2) the parent of $i$ is not a valid direct connection in a shortest path from $i$ to $v_{des}$.
    
    Next, we need to show that if some node $i$ in state $s$ violates $\mathcal{P}_{sp}$, then for each global state $s'$ such that $s'\succ s$, if $s'[i]=s[i]$, then $s'$ will not manifest all shortest paths. This is straightforward from the definition itself, that if a node $i$ is \imped, then either $i=v_{des}$ and it is not pointing to itself through $i[p]$, or there is at least one other node $j$ such that $i[d]>j[d]+w(i,j)$. If $i$ is \imped in $s$, and $i$ has the same state in some global state $s'$ such that $s'\succ s$, then $i$ stays \imped in $s'$ as well, and $s'$ does not satisfy $\mathcal{P}_{sp}$.
\end{proof}

To present the abstraction of the induction of an \ldag, we define the state value and rank as follows.

\begin{figure}[ht]
    \centering
    \subfigure[]{
        \begin{tikzpicture}[scale=2]
            \node [circle, inner sep=2pt, fill=black, draw=black, label=above:$v_1$] (v1) at (0,0) {};
            \node [circle, inner sep=2pt, fill=black, draw=black, label=left:$v_2$] (v2) at (-1,-1) {};
            \node [circle, inner sep=2pt, fill=black, draw=black, label=right:$v_3$] (v3) at (1,-1) {};
            \node [circle, inner sep=2pt, fill=black, draw=black, label=below:{$v_4=v_{des}$}] (v4) at (0,-2) {};
            
            \draw (v4) -- node[below] {2} (v2); \draw (v4) -- node[below] {1} (v3);
            \draw (v1) -- node[above] {2} (v2); \draw (v1) -- node[above] {3} (v3);
        \end{tikzpicture}
    }
    \subfigure[]{
        \begin{tikzpicture}
            \node (a1) at (0,0) {\begin{tabular}{l}$\langle\langle\top,\infty\rangle$,$\langle\top,\infty\rangle$,\\$\langle\top,\infty\rangle$,$\langle\top,\infty\rangle\rangle$\end{tabular}};
            \node (a2) at (0,1.5) {\begin{tabular}{l}$\langle\langle\top,\infty\rangle$,$\langle\top,\infty\rangle$,\\$\langle\top,\infty\rangle$,$\langle v_4,0\rangle\rangle$\end{tabular}};
            \node (a3) at (-2,3) {\begin{tabular}{l}$\langle\langle\top,\infty\rangle$,$\langle v_4,2\rangle$,\\$\langle\top,\infty\rangle$,$\langle v_4,0\rangle\rangle$\end{tabular}};
            \node (a4) at (2,3) {\begin{tabular}{l}$\langle\langle\top,\infty\rangle$,$\langle\top,\infty\rangle$,\\$\langle v_4,1\rangle$,$\langle v_4,0\rangle\rangle$\end{tabular}};
            \node (a5) at (0,4.5) {\begin{tabular}{l}$\langle\langle\top,\infty\rangle$,$\langle v_4,2\rangle$,\\$\langle v_4,1\rangle$,$\langle v_4,0\rangle\rangle$\end{tabular}};
            \node (a6) at (-2,6) {\begin{tabular}{l}$\langle\langle v_2,4\rangle$,$\langle v_4,2\rangle$,\\$\langle v_4,1\rangle$,$\langle v_4,0\rangle\rangle$\end{tabular}};
            \node (a7) at (2,6) {\begin{tabular}{l}$\langle\langle v_3,4\rangle$,$\langle v_4,2\rangle$,\\$\langle v_4,1\rangle$,$\langle v_4,0\rangle\rangle$\end{tabular}};
            
            \draw (a1) -- (a2);
            \draw (a2) -- (a3); \draw (a2) -- (a4);
            \draw (a3) -- (a5); \draw (a3) -- (a6);
            \draw (a4) -- (a5); \draw (a4) -- (a7);
            \draw (a5) -- (a6); \draw (a5) -- (a7);
        \end{tikzpicture}
    }
    \caption{(a) Input graph. (b) \ldag induced among the global states in evaluating for the shortest path parblem in the graph shown in (a); a global state is represented as $\langle\langle p.v_1,d.v_1\rangle,...,\langle p.v_4,d.v_4\rangle\rangle$. Transitive edges are not shown.
    }
    \label{figure:sp-global-states}
\end{figure}
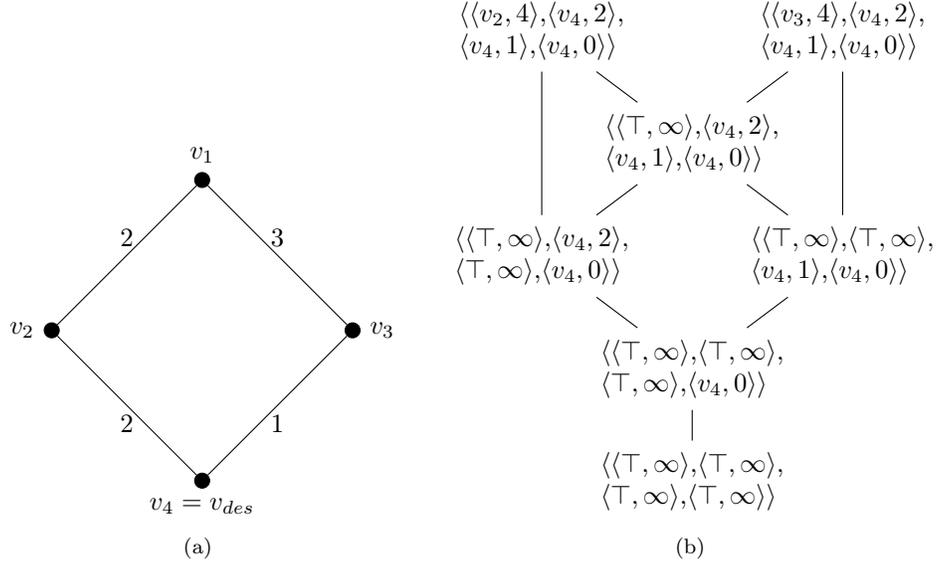

\begin{center}
    $
    \textsc{State-Value-SP}(i,s)=i[d]-dis(i,v_{des}).
    $

    $\textsc{Rank-SP}(s)=\sum\limits_{i\in V(G)}\textsc{State-Value-SP}(i,s).$
\end{center}
Under \Cref{algorithm:sp-dip}, the global states form a \ldag. We show an example in \Cref{figure:sp-global-states}.
\Cref{figure:sp-global-states} (a) is the input graph and \Cref{figure:sp-global-states} (b) is the \ldag induced among the global states. For a pair of global states $s$ and $s'$, $s\prec s'$ iff $\textsc{Rank-SP}(s')<\textsc{Rank-SP}(s)$. In \Cref{figure:sp-global-states}, a global state is represented as $\langle\langle v_1[p]$, $v_1[d]\rangle$, $...$, $\langle v_4[p]$, $v_4[d]\rangle\rangle$.

The above algorithm requires that all nodes are initialized where $i[d]=\infty,i[p]=\top$. If nodes were initialized arbitrarily (e.g., if for all nodes $i$, $i[d]=0$) then the algorithm does not compute shortest paths. Hence, $\forall i~\lnot\textsc{\Imped-SP}(i)$ is a \po-inducing predicate but is not self-stabilizing; in turn, \Cref{algorithm:sp-dip}, that utilizes this predicate, is not self-stabilizing. 

\begin{theorem}\label{theorem:sp-dip}
    \Cref{algorithm:sp-dip} solves the shortest path problem, on a connected positive weighted graph, on $n$ nodes executing asynchronously.
\end{theorem}

\begin{proof}
    We need to show that (1) \Cref{algorithm:sp-dip} traverses a \ldag of global states, (2) for all suboptimal states, $\exists$ a terminal successor, and (3) all terminal global states are optimal states.
    
    Let the current state be $s$.
    If $s$ is suboptimal, then for at least one of the nodes $i$: (1) $i[p]\neq i\land i=v_{des}$, that is, $i$ is the destination node and is not pointing to itself, or (2) $dis(i,v_{des})\neq i[d]$, that is, $i$ does not store a shortest path from $i$ to $v_{des}$.
    
    In the case that $s$ is suboptimal and the first case holds true for some node $i$, then under \Cref{algorithm:dc-dip}, $i$ updates $i[d]$ to 0 and $i[p]$ to $i$, which reduces the state value of $i$ to 0.

    In the case that $s$ is suboptimal and the second case holds true for some node $i$, then under \Cref{algorithm:sp-dip}, $i$ will reduce its $i[d]$ value and update $i[p]$, which reduces the state value of $i$ at least by 1.

    This shows that a partial order is induced among the local states visited by an arbitrary node $i$. Thus under \Cref{algorithm:dc-dip}, an arbitrary graph will follow a \ldag of global states and if it transitions from a state $s$ to another state $s'$, then we have that $s'\succ s$ such that rank of $s'$ is less than the rank of $s$.

    If no node is \imped, then this implies that all nodes have computed the shortest distance in their $i[d]$ variable, and thus the rank is 0. Thus if there is a global state $s$ with rank greater than 0, then there exists at least one \imped node in it. When any node performs execution in $s$ then $s$ transitions to some state $s'$ such that the rank of $s'$ is less than the rank of $s$. This shows that for every suboptimal global state, there exists at least one terminal successor.

    Let that $s$ is a terminal successor. Then, $\mathcal{P}(s)$ is true: no node is \imped in $s$, so any node will not execute and $s$ manifests correct shortest path evaluation for all nodes. Thus we have that all terminal states are optimal states, and \Cref{algorithm:sp-dip} is silent.
\end{proof}

\Cref{algorithm:sp-dip} is a distance-1 algorithm and converges even if the nodes run without synchronization in AA model. This is because in a given state $s$, some node $i$ is \imped iff there is a shorter path that $i$ can follow to reach $v_{des}$, thus, $s$ will never transition to an optimal state without $i$ changing its state.

\subsection{Limitations of Modelling Problems as \po-Inducing Problems}\label{subsection:dag-prob-limitations}

Unlike the \po-inducing problems where the problem description creates a \ldag among the states in $S$, there are problems where the states do not form a \ldag naturally. Such problems are non-\po-inducing problems. In such problems, there are instances in which the \imped nodes cannot be distinctly determined, i.e., in those instances
$\exists s :\lnot\mathcal{P}(s) \land (\forall i: \exists s':\mathcal{P}(s')\land s[i]=s'[i]$).

\begin{definition}\textbf{Maximal matching}.
    In the maximal matching problem, the input is an arbitrary graph $G$. For all $i$, $i[match]$ has the domain $Adj_i\cup\{\top\}$. The task is to compute the $i[match]$ (for ach node $i$) such that (1) $\forall i:i[match]\neq \top\Rightarrow (i[match])[match]=i$, and (2) if $i[match]=\top$, then there must not exist a $j$ in $Adj_i$ such that $j[match]=\top$.
\end{definition}

Maximal matching (MM) is a non-\po-inducing problem.
This is because, for any given node $i$, an optimal state can be reached if $i$ does or does not change its state. Thus $i$ cannot be deemed as \imped or not \imped under the natural constraints of \mm.
This can be illustrated through a simple instance of a 3 nodes network forming a simple path $\langle A,B,C\rangle$. Initially no node is paired with any other node. Here, \mm can be obtained by matching $A$ and $B$.
Thus, $C$ is not \imped. 
Another maximal matching can be obtained by matching $B$ and $C$, in which case $A$ is not \imped.
Thus the problem itself does not define which node is \imped.

We observe that it is possible to induce a \ldag in non-\po-inducing problems algorithmically. We call such algorithms non-\po-inducing algorithms, which we study in the following section.

\section{Imposed \po-Induction: \po-inducing \textit{Algorithms}}\label{section:dia}

In this section, we study algorithms that can be developed for problems that cannot be represented by a predicate under which the global states form a \ldag.
This is because, as described in \Cref{subsection:dag-prob-limitations}, in a suboptimal global state, the problem does not specify a specific set of nodes that must change their state.

\subsection{General Properties of \po-Inducing Algorithms}\label{subsection:properties-dia}

Non-\po-inducing problems do not naturally define which node is \imped. There may be multiple optimal states. However, \imped nodes can be defined algorithmically.

\begin{definition}\label{definition:dia}\textbf{\po-inducing algorithms (DIA)}.
$A$ is a DIA for a problem $P$, represented by predicate $\mathcal{P}$, iff
(1) $P$ is solved iff the system reaches a state where $\mathcal{P}$ is true, and
  (2) $\mathcal{P}$ is \po-inducing with respect to $\mathcal{S}$ induced in $S$ by $A$, i.e. $\forall s\in S: \lnot\mathcal{P}(s) \Rightarrow \exists i:
     \textsc{\Imped}(i,s,\mathcal{P})$.
\end{definition}

\noindent \textbf{\textit{Remark}}: An algorithm that traverses a \ldag $\mathcal{S}$ of global states is a DIA. Thus, an algorithm that solves a \po-inducing problem, under the constraints of \po-induction, e.g. Algorithm \ref{algorithm:dc-dip}, is a DIA.

\noindent \textbf{\textit{Remark}}: Since a total order is a special case of a partial order, all lattice-linear algorithms are \po-inducing algorithms.

\begin{definition}\label{definition:ss-dia}\textbf{Self-stabilizing DIA}.
    Continuing from \Cref{definition:dia}, $A$ is self-stabilizing only if in the \ldag $\mathcal{S}$ induced by $A$,
    $\forall s,s'\in S:\textsc{Terminal-Successor}(s,s')\Rightarrow\mathcal{P}(s')=true$.
\end{definition}

In the remaining part of this section, we study the maximal matching problem, a non-\po-inducing problem, and explore how a \po-inducing algorithm can be developed for such a problem. We will see, again, that the induction of a partial order among the local states, under the acting algorithm, allows the nodes to execute without synchronization.

\subsection{Maximal Matching (MM) Problem}\label{subsection:mm}

As discussed in \Cref{subsection:dag-prob-limitations}, MM is not a \po-inducing problem.
However, a \po-inducing algorithm can be developed for this problem, which we discuss in the following. 

The local state of a node $i$ is defined by $\langle i[match]\rangle$. We use the macros listed in \Cref{table:macros-mm-dag}. A node $i$ is \textit{wrongly matched} if $i$ is pointing to some node $j$, but $j$ is pointing to some node $k\neq i$.
A node $i$ is \textit{matchable} if $i$ is not pointing to any node, i.e. $i[match]=\top$, and there exists a node $j$ adjacent to $i$ which is also not pointing to any node.
A node $i$ is being \textit{pointed to}, or $i$ is \textit{$i$-pointed}, if $i$ is not pointing to any node,
and there exists a node $j$ adjacent to $i$ which is pointing to $i$.
A node sees that another node is being pointed, or $i$ ``sees'' \textit{else-pointed}, if some node $j$ around (in 2-hop neighbourhood of) $i$ is pointing to another node $k$ and $k$ is not pointing to anyone.
A node is \textit{unsatisfied} if it is wrongly matched or matchable. A node $i$ is \textit{\imped} if $i$ is i-pointed, or otherwise, given that $i$ does not see else-pointed, $i$ is the highest ID unsatisfied node in its distance-2 neighbourhood.
\begin{table}[ht]
    \centering 
    \doublespacing 
    \begin{tabular}{|l|}
        \hline
        $\textsc{Wrongly-Matched-MM}(i)\equiv i[match]\neq \top\land$ $(i[match])[match]\neq i\land (i[match])[match]\neq \top$.\\
        $\textsc{Matchable-MM}(i)\equiv i[match]=\top\land (\exists j\in Adj_i:$ $j[match]=\top)$.\\
        $\textsc{I-Pointed-MM}(i)\equiv i[match]=\top\land(\exists j\in Adj_i:$ $j[match]=i)$.\\
        $\textsc{Else-Pointed-MM}(i)\equiv \exists j\in Adj^2_i,\exists k\in Adj_j:$ $j[match]=k\land k[match]=\top$.\\
        $\textsc{Unsatisfied-MM}(i)\equiv\textsc{Wrongly-Matched-MM}(i)\lor$ $\textsc{Matchable-MM}(i)$.\\
        $\textsc{\Imped-MM}(i) \equiv\textsc{I-Pointed-MM}(i)\lor$ $(\lnot \textsc{Else-Pointed-MM}(i)\land(\textsc{Unsatisfied-MM}(i)\land$\\
        \quad \quad $(\forall j\in Adj^2_i:i[id]>j[id]\lor\lnot\textsc{Unsatisfied-MM}(j)))$.\\
        \hline
    \end{tabular}
    \caption{Macros used in the algorith for \mm.}
    \label{table:macros-mm-dag}
\end{table}

The algorithm that we develop next is a self-stabilizing algorithm, which means that the nodes can be initialized arbitrarily. Thus, $i[match]$ may store some node that is not connected to $i$ by an edge, or some node $j$ in $Adj_i$ but $j[match]$ may not store $i$. The algorithm for an arbitrary node $i$ can be defined as follows. If $i$ is \imped and i-pointed, then $i$ starts to point to the node which is pointing at $i$. If $i$ is wrongly matched and \imped, then $i$ takes back its pointer, i.e. $i$ starts pointing to $\top$. Otherwise (if $i$ is matchable and \imped), $i$ chooses a node $j$ which is not pointing to anyone, i.e. $j[match]=\top$, and $i$ starts pointing to $j$.

\begin{algorithm}\label{algorithm:mm-dia}
    Rules for node $i$.
\end{algorithm}
\begin{center}
    $
        \begin{array}{|l@{}|}
            \hline
            \textsc{\Imped-MM}(i)\longrightarrow\\
            \begin{cases}
                i[match]=j:j\in Adj_i: j[match]=i & \text{if $\textsc{I-Pointed-MM}(i)$}.\\
                i[match]=\top & \text{if $\textsc{Wrongly-Matched-MM}(i)$}.\\
                i[match]=j:j\in Adj_i: j[match]=\top & \text{otherwise}.
            \end{cases}~\\
            \hline
        \end{array}
    $
\end{center}

\begin{lemma}\label{lemma:mm-dag-structure}
    \Cref{algorithm:mm-dia} induces a \ldag in the global state space under AMR model.
\end{lemma}

\begin{proof}
    The \ldag is induced in the global state space with respect to the state values, which we prove in the following. Let $s$ be a suboptimal state that the input graph is in. A node $i$ is \imped in $s$ (1) if $i$ is wrongly matched, (2) if $i$ is matchable, or (3) if $i$ is being pointed at by another node $j$, but $i$ does not point back to $j$ or any other node. We elaborate on all these cases in the following paragraphs of this proof. In all these cases, we assume AMR model, that is, if a node reads a local state of another node, then it does not read an older local state from that node in a subsequent read operation.

    We show that if some node $i$ is \imped in some state $s$, then for any state $s':s'\succ s$, if $s'[i]=s[i]$, then $s'$ will not form a maximal matching under \Cref{algorithm:mm-dia}.
    
    In the case if $i$ is wrongly matched in $s$ and is \imped, and it is pointing to the same node in $s'$ as well, then $i$ stays to be \imped in $s'$. This is because since $i$ is \imped, $i$ is also the highest ID node that is unsatisfied, so all other nodes within distance-2 of $i$ will wait for $i$ to take back its pointer before taking any action. Thus $s'$ does not have a correct matching. 
    
    In the case if in $s$, $i$ is being pointed to by some node but $i$ does not point to any node, and $i$ stays in the same state in $s'$, then $i$ stays to be \imped in $s'$. This can be explained as follows. Let $j$ be the node that is pointing to $i$, i.e., $j[match]=i\land i[match]=\top$. Now since $\textsc{I-Pointed-MM}(i)$ is true, so for all unsatisfied nodes within distance-2 of $i$, $\textsc{Else-Pointed}$ is true. Thus, they will not take any action until $i$ does. Also since $i[match]=\top$, $j$ will not retreat its pointer as it does not fall under the constraints of $\textsc{Wrongly-Matched}$. Thus $s'$ does not form a correct matching.
    
    Finally, in the case if $i$ is matchable and \imped in $s'$, and it stays the same in $s'$, then it is still \imped as any other node in $Adj_i$ will not initiate matching with it. This is because all the unsatisfied nodes within distance-2 of $i$ have IDs less than that of $i$. Also, for the same reason, any node in $Adj^2_i$ will not take any action until $i$ does. Thus $s'$ does not manifest a maximal matching.
\end{proof}

To present the abstraction of the induction of an \ldag under \Cref{algorithm:mm-dia}, we define the state value and rank as follows.

$$
\begin{array}{l}
    \textsc{State-Value-MM}(i,s)=
    \begin{cases}
        3 & \text{if $\textsc{Wrongly-Matched-MM}(i)$}. \\
        2 & \text{if $\textsc{Matchable-MM}(i)\land\lnot\textsc{I-Pointed-MM}(i)$}.\\
        1 & \text{if $\textsc{I-Pointed-MM}$(i)}.\\
        0 & otherwise.
    \end{cases}
\end{array}
$$
$$\textsc{Rank-MM}(s)=\sum\limits_{i\in V(G)}\textsc{State-Value-MM}(i,s).$$

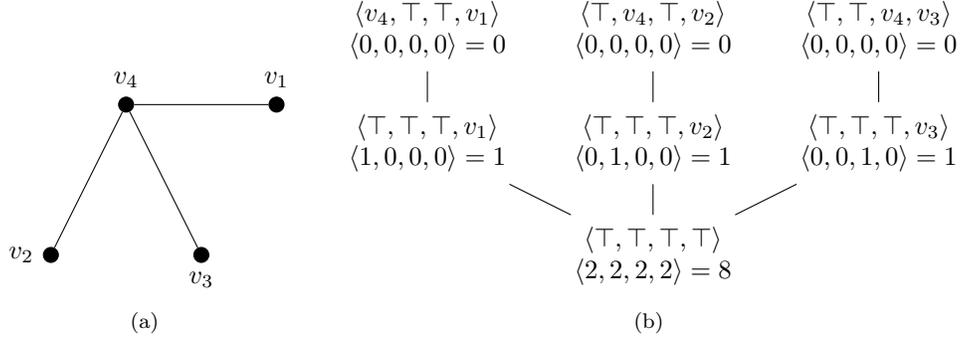
\begin{figure}[ht]
    \centering
    \subfigure[]{
        \begin{tikzpicture}[scale=2]
            \node [circle, draw=black,fill=black,inner sep=2pt,label=above:$v_4$] (a) at (0,0) {};
            \node [circle, draw=black,fill=black,inner sep=2pt,label=left:$v_2$] (b) at (-.5,-1) {};
            \node [circle, draw=black,fill=black,inner sep=2pt,label=below:$v_3$] (c) at (.5,-1) {};
            \node [circle, draw=black,fill=black,inner sep=2pt,label=above:$v_1$] (d) at (1,0) {};
            
            \draw (a) -- (b); \draw (a) -- (c);\draw (a) -- (d);
        \end{tikzpicture}
    }
    \subfigure[]{
        \begin{tikzpicture}
            \node (a) at (0,0) {\begin{tabular}{c}$\langle \top,\top,\top,\top\rangle$\\$\langle 2,2,2,2\rangle=8$\end{tabular}};
            \node (b) at (-3,1.5) {\begin{tabular}{c}$\langle \top,\top,\top,v_1\rangle$\\$\langle 1,0,0,0\rangle=1$\end{tabular}};
            \node (c) at (0,1.5) {\begin{tabular}{c}$\langle \top,\top,\top,v_2\rangle$\\$\langle 0,1,0,0\rangle=1$\end{tabular}};
            \node (d) at (3,1.5) {\begin{tabular}{c}$\langle \top,\top,\top,v_3\rangle$\\$\langle 0,0,1,0\rangle=1$\end{tabular}};
            
            \node (b2) at (-3,3) {\begin{tabular}{c}$\langle v_4,\top,\top,v_1\rangle$\\$\langle 0,0,0,0\rangle=0$\end{tabular}};
            \node (c2) at (0,3) {\begin{tabular}{c}$\langle \top,v_4,\top,v_2\rangle$\\$\langle 0,0,0,0\rangle=0$\end{tabular}};
            \node (d2) at (3,3) {\begin{tabular}{c}$\langle \top,\top,v_4,v_3\rangle$\\$\langle 0,0,0,0\rangle=0$\end{tabular}};
            \draw (a) -- (b); \draw (a) -- (c);\draw (a) -- (d);
            \draw (b2) -- (b); \draw (c2) -- (c);\draw (d2) -- (d);
        \end{tikzpicture}
    }
    \caption{(a) Input graph. (b) The state transition diagram, a \ldag, assuming that the initial global state is $\langle \top,\top,\top,\top\rangle$. In every state, the first row shows the global state, the second row shows the respective local state values of nodes and the rank of the global state. Observe that any other state will converge to one of these states and then converge to one of the optimal states in this \ldag. (All edges are directed upwards; transitive edges are not shown for brevity).}
    \label{figure:mm-global-states}
\end{figure}

Under \Cref{algorithm:mm-dia}, the global states form a \ldag. We show an example in \Cref{figure:mm-global-states}: \Cref{figure:mm-global-states} (a) is the input graph and \Cref{figure:mm-global-states} (b) is the induced \ldag. For a pair of global states $s$ and $s'$, $s\prec s'$ iff $\textsc{Rank-MM}(s')<\textsc{Rank-MM}(s)$. In \Cref{figure:mm-global-states}, a global state is represented as $\langle\langle v_1[match]\rangle$, $...$, $\langle v_4[match]\rangle\rangle$.

Since the solution presented for this problem is self-stabilizing, $\forall i\lnot \textsc{\Imped-MM}(i)$ forms a self-stabilizing predicate with respect to the \ldag induced by \Cref{algorithm:mm-dia}. Thus, \Cref{algorithm:mm-dia} is a \po-inducing self-stabilizing algorithm and satisfies \Cref{definition:ss-dia}. 

\begin{theorem}\label{theorem:mm-dia}
    \Cref{algorithm:mm-dia} is a \po-inducing algorithm for the maximal matching problem on $n$ nodes executing asynchronously in AMR model.
\end{theorem}

\begin{proof}[Proof of \Cref{theorem:mm-dia}]
    We show that (1) \Cref{algorithm:mm-dia} traverses a \ldag of global states, under AMR model, that has the properties as mentioned in the above lemma, (2) for all suboptimal states $\exists$ a terminal successor, and (3) all terminal global states are optimal states.

    If $s$ is suboptimal, then for at least one of the nodes $i$: (1) $i$ is wrongly matched, (2) $i$ is matchable, or (3) $i$ is being pointed at by another node $j$, but $i$ does not point back to $j$ or any other node.

    If $s$ is suboptimal and some node $i$ is being pointed to by some node $j$ and $i$ does not point to any node, then under \Cref{algorithm:mm-dia}, $i$ will point back to $j$, and thus the state value of $i$ will get reduced from $1$ to $0$.
    
    In the case that $s$ is suboptimal and some node $j$ is wrongly matched or matchable with $\lnot$\textsc{Else-Pointed-MM}$(j)$, then at least one node (e.g., a node with highest ID which is wrongly matched or matchable) will be unsatisfied and \imped. Let that $i$ is unsatisfied and \imped. Here $i$ is either wrongly matched or matchable. If $i$ is wrongly matched, then $i$ will change its pointer and start pointing to $\top$, in which case its state value will change from 3 to 2, 1, or 0. If $i$ is matchable then, $i$ will start pointing to some node $j$ in $Adj_i$, in which case, the state value of $i$ will change from 2 to 0 and the state value of $j$ will change from 2 to 1.

    In all the above cases, we have that under \Cref{algorithm:mm-dia}, if $s$ is a suboptimal state, then its rank will be of some value greater than zero because at least one of the nodes will be \imped. $s$ will transition to some state $s'$ whose rank is less than that of $s$. Thus, we have that \Cref{algorithm:mm-dia} transitions $s$ to $s'$ and thus decreases the rank of the system. This shows that (1) \Cref{algorithm:mm-dia} traverses a \ldag that has the properties as mentioned in \Cref{lemma:mm-dag-structure}, (2) for all suboptimal states $\exists$ a terminal successor.

    In the case that $s$ is a terminal successor, then none of the nodes will be enabled. So no node will change its state. This state will manifest a maximal matching. This shows that all terminal successors are optimal states, and \Cref{algorithm:mm-dia} is silent.
\end{proof}

\Cref{algorithm:mm-dia} is a distance-4 algorithm. From \Cref{theorem:mm-dia}, we have that \Cref{algorithm:mm-dia} converges even if the nodes run withoutout synchronization in AMR model. 
However, \Cref{algorithm:mm-dia} cannot tolerate asynchrony in AA model. This is because a node $i$ in its current state may wrongly evaluate itself to be i-pointed or unsatisfied-\imped if it gets information, from other nodes, out of order. This can result in $i$ changing its state incorrectly. As a consequence of execution in $AA$ model, $i$ can keep repeating such execution and as a result, the system may not obtain an optimal state.

\section{Properties of \po-Induction}\label{section:dag-properties}

In the previous sections, we discussed example algorithms that converge even without synchronization. 
In this section, we use that intuition and describe some characteristics of general \po-inducing systems. We study how the induction of a partial order among the local states visited by individual nodes is a necessary and sufficient condition to allow asynchrony (\Cref{subsection:dag<->asynch}). 
We also study some time-complexity properties of an algorithm that induces a \ldag among the global states (\Cref{subsection:dag-tc}).

\subsection{\po-Induction to Obtain Asynchrony}\label{subsection:dag<->asynch}

In this subsection, we study whether \ldag induction is necessary and sufficient for asynchronous execution.

\begin{theorem}\label{theorem:sufficiency}
Let $P$ be a problem that requires an algorithm to converge to a state where $\mathcal{P}$ is true. 
Let A be an algorithm for $P$ that is correct under 
a central scheduler.
Let $\mathcal{S}$ be the transition graph that is formed under $A$, where nodes are allowed to read old values in some communication model $M$.
\\~\\
\noindent If $\mathcal{S}$ forms a \ldag, then A guarantees convergence in asynchrony in the model $M$.
\end{theorem}

\begin{proof}
    \Cref{definition:dia} follows that the local states form a partial order if every node, under Algorithm $A$, rejects each violating local state permanently. Consequently, a \ldag is induced among the global states.
    
    A \ldag, induced under $\mathcal{P}$, allows asynchrony because if a node, reading old values, reads the current state $s$ as $s'$, then $s'\prec s$. So $\lnot\mathcal{P}(s')\Rightarrow \lnot\mathcal{P}(s)$ because $\textsc{\Imped}(i,s',\mathcal{P})$ and $s'[i]=s[i]$.
\end{proof}

The limitation of 
the above theorem is as follows. An algorithm $A$ guarantees convergence in asynchrony in some model only if it induces a \ldag in that model. For instance, a \Cref{algorithm:mm-dia} developed for maximal matching induces a \ldag in AMR model. As we discussed in \Cref{subsection:mm}, such an algorithm does not necessarily induce a \ldag in AA model, so it may not guarantee convergence in AA model.

In the above theorem, we showed that a \ldag is sufficient for allowing asynchronous executions. 
Next, we study if a \ldag is guaranteed to be induced among the global states given that an algorithm is correct under asynchrony. In other words, we study if a \ldag is necessary for asynchrony.
To study that, we first examine the representation of a global state, which is a mathematical abstraction of a multiprocessor system.

Let $C$ be an algorithm that runs correctly under a central scheduler. 
Let that $R^{cent}_s$ be the set of states, that $s$ can transition to, under $C$, i.e. $\forall s'\in R^{cent}_s$, $\langle s,s'\rangle$ is a valid transition under $C$.
If $C$ and $s$ are given, $R^{cent}_s$ can be correctly computed. 
In synchronous systems, $s$ is an abstraction of the global state that only contains the \textit{current} local states of the nodes.
However, if some algorithm $A$ were to be executed in asynchrony, then a set $R^{asyn}_s$ of resulting states cannot be computed correctly for a given state $s$. This is because some nodes would be reading the old values of other nodes. Thus, $s$ must also contain the details of the information that nodes have about other nodes.

Let $\mathcal{S}$ be the original transition system, and $\mathcal{S}_{ext}$ be its extended version, where a given state $s_{ext}$ identifies local states of individual nodes and the information that the nodes maintain about other nodes. 
Similarly, $\mathcal{S}_{ext}$ can be constructed back to $\mathcal{S}$, by removing the information that the nodes have about other nodes, and then merging the resulting global states
(along with the transition edges) that are the same.

Observe that in all the algorithms presented in this chapter, the \textsc{State-Value} of a node $i$
provides information about how \textit{bad} the current local state of $i$ is, with respect to an optimal global state farthest from its current global state $s$. Thus, the evaluation of the state value of $i$ can utilize the information about the local states of other nodes in $s$. In the following theorem, we use this observation as leverage to show some interesting properties of \po-inducing systems. Herein, instead of $\mathcal{S}$ we elaborate on the necessity of $\mathcal{S}_{ext}$ being a \ldag to allow asynchrony.

\begin{theorem}\label{theorem:necessity}
    Let $P$ be a problem that requires an algorithm to converge to a state where $\mathcal{P}$ is true. 
Let A be an algorithm for $P$ that is correct under 
a central scheduler.
Let $\mathcal{S}$ be the transition graph that is formed under $A$, where nodes are allowed to execute asynchronously in some communication model $M$.
\\~\\
    \noindent If algorithm $A$ guarantees convergence in asynchrony under the communication model $M$, then $\mathcal{S}_{ext}$ (the extended transition system) forms a \ldag in the model $M$. 
\end{theorem}

\begin{proof}
    Since $A$ guarantees to terminate under asynchrony and the extended state space $S_{ext}$ captures the effect of asynchrony (where a node may read old values of the variables of other nodes), there will be no cycles present among the global states in $\mathcal{S}_{ext}$, i.e., $\mathcal{S}_{ext}$ forms a \dag. Next, we transform $\mathcal{S}_{ext}$ to a \ldag.

    Let $S^{ext}_o$ be the set of optimal states.
    For each optimal global state $o\in S^{ext}_o$, for each node $i$ in $o$, assign 
    the state value of $i$ to be 0. Thus, the rank of $o$ is 0 (sum of state values of all nodes).
    For every non-optimal global state $s$, we assign 
    the state value of every node $i$ to be $\top$.
    Subsequently, if all successors of state $s$ have a non-null (non-$\top$) rank (i.e., $\forall s': \langle s, s'\rangle \in E(\mathcal{S}_{ext}) \Rightarrow (\forall i: \textsc{State-Value}(s'[i]) \neq \top$)) then we set the state value of each node $i$ in $s$ to be $(\max\{\textsc{Rank}(s') | \langle s, s'\rangle \in E(\mathcal{S}_{ext})\})/n+1 $.
    We do this recursively until no more updates happen to any state value of any global state in $\mathcal{S}_{ext}$.
    
    Since there are no cycles, this procedure will terminate in finite time. 
    Observe that using the above procedure, we obtain a valid \ldag from $\mathcal{S}_{ext}$: in this \ldag, for all reachable global states $s'$ and $s''$, $s'\prec s''$ (i.e., $\langle s',s''\rangle$ is in $E(\mathcal{S}_{ext})$) iff $s'[i]\preceq s''[i]$, $\forall i:[1:n]$.
\end{proof}

If $\mathcal{S}_{ext}$ forms a \ldag then so does $\mathcal{S}$, Hence, from \Cref{theorem:sufficiency} and \Cref{theorem:necessity}, we have

\begin{corollary}\label{corollary:dag<->partial-order}
    Let $P$ be a problem that requires an algorithm to converge to a state where $\mathcal{P}$ is true. 

    \noindent An Algorithm $A$ guarantees convergence in asynchrony iff it induces a \ldag among the (extended) global states.
\end{corollary}

The above corollary shows that a \ldag is a necessary and sufficient condition for a parallel processing system to guarantee convergence in asynchrony.
For reasons of space, we have moved the subsection on the time complexity properties of \po-inducing algorithms to Appendix \ref{subsection:dag-tc}.

\subsection{Time Complexity Properties of an Algorithm Traversing a \ldag}\label{subsection:dag-tc}

\begin{theorem}\label{theorem:general-convergence-time}
    Given a system of $n$ processes, with the domain of (state values having) size not more than $m$ for each process, the acting algorithm will converge in $n\times (m-1)$ moves.
\end{theorem}

\begin{proof}
    Assume for contradiction that the underlying algorithm converges in $x\geq n\times (m-1)+1$ moves. This implies, by pigeonhole principle, that at least one of the nodes $i$ is revisiting their state $st$ after changing to $st'$. If $st$ to $st'$ is a step ahead transition for $i$, then $st'$ to $st$ is a step back transition for $i$ and vice versa. For a system where the global states form a \ldag, we obtain a contradiction since step-back actions are absent in such systems.
\end{proof}

\begin{exampledscont}\label{example:number-of-moves}
    Consider phase 2 of \Cref{algorithm:ds-ellss}. 
    As discussed earlier, this phase is lattice-linear. The domain of each process $\{IN$, $OUT\}$ is of size 2. Hence, phase 2 of \Cref{algorithm:ds-ellss} requires at most $n\times (2-1)=n$ moves. (Phase 1 also requires atmost $n$ moves. But this fact is not relevant with respect to \Cref{theorem:general-convergence-time}.)
    \qed 
\end{exampledscont}

\begin{examplesmpcont}
    Observe from \Cref{figure:smplattice} that any system of 3 men and 3 women with arbitrary preference lists will converge in $3\times (3-1)=6$ moves. This comes from 3 men (resulting in 3 processes) and 3 women (domain size of each man (process) is 3).
    \qed 
\end{examplesmpcont}

\begin{corollary}\label{corollary:convergence-time-dag-multivariable}
    Let that each node $i$ stores atmost $r$ variables,  $i[var_1],...,i[var_r]$ (with domain sizes $z_1,...z_r$ respectively) contribute independently to the formation of the partial order. 
    Then an algorithm traversing the resultant \ldag will converge in $n\times ((\prod_{j=1}^r z_j)-1)$ moves.
\end{corollary}

\subsubsection*{Corollaries from \Cref{chapter:flla}}

\begin{corollary}
    (From \Cref{theorem:ds-ll} and \Cref{theorem:general-convergence-time}) \Cref{algorithm:ds-ll} converges in $n$ moves.
\end{corollary}

\begin{corollary}
    (From \Cref{theorem:gc-ll} and \Cref{theorem:general-convergence-time}) \Cref{algorithm:gc-ll} converges in $\sum\limits_{i\in V(G)}deg(i)+1=n+2m$ moves.
\end{corollary}

\begin{proof}
    This can be reasoned as follows: first, a node may increase its colour, once, to resolve its colour conflict with a neighbouring node. Then it will decrease its colour, whenever it moves. Depending on the colour value of its neighbours and when they decide to move, a node $i$ can decrease its colour almost $deg(i)$ times.
\end{proof}

\begin{corollary}
    (From \Cref{theorem:vc-ll} and \Cref{theorem:general-convergence-time}) \Cref{algorithm:vc-ll} converges in $n$ moves.
\end{corollary}

\begin{corollary}
    (From \Cref{theorem:is-ll} and \Cref{theorem:general-convergence-time}) \Cref{algorithm:is-ll} converges in $n$ moves.
\end{corollary}

\begin{corollary}\label{corollary:vc-2-approx-time-complexity}
    (From \Cref{theorem:vertex-cover-2-approx} and \Cref{corollary:convergence-time-dag-multivariable}) \Cref{algorithm:vc-2-approx} converges in $n$ moves.
\end{corollary}

\subsubsection*{Corollaries from this chapter}
\begin{corollary}(From \Cref{theorem:dc-dip} and \Cref{corollary:convergence-time-dag-multivariable})
    \Cref{algorithm:dc-dip} converges in $\sum\limits_{i\in V(G)}deg(i)=2m$ moves. In terms of rounds, it converges in $\Delta$ rounds, where $\Delta$ is the maximum degree of the input graph.
\end{corollary}

\begin{corollary}(From \Cref{theorem:sp-dip} and \Cref{corollary:convergence-time-dag-multivariable})
    \Cref{algorithm:sp-dip} converges converges in $\mathcal{D}$ rounds, where $\mathcal{D}$ is the diameter of the input graph.
\end{corollary}

\begin{corollary}\label{corollary:tc-mm}(From \Cref{theorem:mm-dia} and \Cref{corollary:convergence-time-dag-multivariable})
    \Cref{algorithm:mm-dia} converges in $2n$ moves.
\end{corollary}
\begin{proof}
    This is because, as explained in \Cref{lemma:mm-dag-structure}, any node $i$ goes from state value 3 to 2 or 3 to 1, and then 2 to 0 or 1 to 0. Hence, there are atmost two transitions that $i$ goes through, with respect to its state value, which can happen due to the movement of $i$ or some node in $Adj_i$.
\end{proof}

\section{Summary of the Chapter}\label{section:dag-summary}

In this chapter, we focused on the problem of finding necessary and sufficient conditions for an algorithm to execute correctly without synchronization. We observe that the induction of a partial order among the local states is necessary and sufficient for multiprocessor algorithms to allow execution without any synchronization.

In \po-inducing problems, all unsatisfied nodes are enabled and can (and must) therefore evaluate their guards and take a corresponding action at any time. In a non-\po-inducing problem, however, all unsatisfied nodes are not enabled. Only the \imped nodes are enabled; these nodes satisfy some additional constraints, in addition to being unsatisfied. In the algorithm for maximal matching, for example, we note that a tie-breaker is the key to deciding which nodes are \imped; in algorithms for dominant clique and shortest path, on the other hand, do not require any tie-breaking strategy.

The sufficiency nature of \ldag implies that such algorithms can be executed asynchronously, thereby eliminating the cost of synchronization. This is especially important in today's multiprocessor architecture where synchronization overhead is the Achilles heel of parallel algorithms.
The necessity of this result means that 
an \ldag exists in these algorithms even if the algorithms were designed without any prior assumption of an \ldag. 

Finally, we have that since a total order is a special case of a partial order, all lattice-linear systems are \po-inducing systems.
\chapter{RELATED WORK}\label{chapter:literature}

In this chapter, we discuss works from several areas of computer science that we find related to our work.

\section{Asynchronous Circuits}\label{section:asynchronous-circuits}

In this dissertation, we studied algorithms that are tolerant to asynchrony. There has also been recent development of hardware architectures that allow programs to run asynchronously, i.e., such architectures run programs without synchronization among their components. 
\textit{Asynchronous circuits} are the circuits that do not synchronize their components centrally. Such circuits are also called \textit{clockless} or \textit{self-timed circuits}.
We discuss some literature in this area, however, we do not go into the details of the hardware architecture of such circuits. The reader is directed to \cite{Brzozowski1995} for a comprehensive discussion on asynchronous circuits.

Asynchronous circuits contain communicating components, and the input and output ports of these components control the computation. On the other hand, in synchronous circuits, the computation is controlled by a global clock, that triggers the transition of the circuit from one state to another.
Authors of \cite{Renaudin2000} opine that asynchronous logic is the key technology for telecommunication applications.

Asynchronous circuits, as compared to synchronous circuits, have (1) low power consumption, (2) high performance, and (3) low noise and electromagnetic emissions \cite{VanBerkel1999}. This is mainly because, respectively, (1) asynchronous circuits are clock-driven whereas asynchronous circuits are data-driven, (2) asynchronous circuits are self-timed, and (3) asynchronous circuits implement a distributed control which results in low current peaks, as compared to synchronous circuits which implement a central control \cite{Renaudin2000}. In addition, with the increase of the number and circuit size of the components on a chip, it is an increasingly complex problem to time all the components using a global clock, however, this problem is eliminated in asynchronous circuits.

The University of Manchester designed the AMULET2e, an embedded chip that incorporates a 32-bit ARM-compatible asynchronous core, a cache, and several other system functions \cite{Furber1999,Furber1995,Garside1996}.
The 80C51 microcontroller Philips Semiconductors and Philips Research, and then later redesigned to its asynchronous version \cite{Gageldonk1998}. 80C51, along with its successors -- other integrated chips designed based on 80C51 -- is the first asynchronous integrated chip that was commercially available.
Cogency designed the Digital Signal Processor, and then later redesigned to its asynchronous version \cite{Paver1998}.
University of Osaka, University of Kochi and Sharp Corporation designed a self-timed data-driven multimedia processor \cite{Komori1989,Terada1995,Terada1999}. An application for which it can be used is digital television receivers. Its peak performance is 8600 memory operations per second, and where it consumes below 1 watt.

Some other asynchronous chips are MiniMIPS designed by Caltech \cite{Abrial2001}, AMULET3i designed by University of Manchester \cite{Garside1999}, TITAC2 designed by Tokyo University \cite{Martin1997}, and MICA designed by TIMA Laboratory \cite{Renaudin1994}. The details of these chips are summarized in \cite{Renaudin2000}.

Asynchronous circuits have not received much attention in industry and academia. One of the contributing factors is the unavailability of software tools that can run on these circuits -- programs that can tolerate asynchrony. This dissertation studies the necessary and sufficient conditions that make an algorithm tolerant to asynchrony. In addition, we not only develop new algorithms but also show that many existing algorithms in the literature are tolerant to asynchrony. This dissertation lays a theoretical foundation for asynchronous algorithms -- that can run on asynchronous circuits -- and provides ways to determine if an algorithm being developed is tolerant to asynchrony.

\section{Other Abstractions in Concurrent Computing}

In this section, we discuss some existing models that guarantee the progress or convergence of multiprocessor algorithms in the presence of node failures or the absence of synchronization at different levels.

\subsection{Lock-Free and Wait-Free Algorithms}

An algorithm is \textit{non-blocking} if in a system running such algorithm, if a node fails or is suspended, then it does not result in failure or suspension of another node. A non-blocking algorithm is \textit{lock-free} if system-wide progress can be guaranteed, and it is \textit{wait-free} if progress can be guaranteed per node. A lock-free algorithm completes a given operation in a finite number of system steps, whereas a wait-free algorithm completes a given operation in a finite number of its own steps. 

Non-blocking algorithms are very useful in designing memory transaction and input-output protocols due to the fact that such algorithms guarantee global (system-wide) or local (with respect to one computational node) progress. They allow processes that fail, in performing an operation due to contention, to continue processing other tasks and not continue to wait.

There is a vast literature on non-blocking algorithms. We note some of them as follows.
\begin{itemize}
    \item Authors of \cite{Michael2002} present algorithms for dynamic lock-free hash tables and list-based sets.
    \item A lock-free stack algorithm is presented in \cite{Hendler2004}.
    \item Algorithms for implementing lock-free singly-linked lists are presented in \cite{Valois1995}.
    \item A lock-free algorithm to implement a binary search tree is present in \cite{Natarajan2014}.
    \item A wait-free sorting algorithm is studied in \cite{Shavit1997}, which sorts an array of size $N$ using $n\leq N$ processors.
    \item An $O(n)$ time wait-free approximate agreement algorithm is presented in \cite{Attiya1994}.
\end{itemize}

A large class of lock-free algorithms, as shown in \cite{Alistarh2016}, under the scheduling conditions that are close to those implemented in commercially available hardware, stochastically behave as wait-free algorithms.

In the context of non-blocking algorithms, \textit{contention} is the race condition that arises when multiple processes try to access the same resource (e.g., a variable, an array index or a memory location) simultaneously.
Lock-free algorithms are fast when contention is low, however, they do not provide an upper bound on the time complexity of individual operations when contention is high \cite{Afek1995} (this paper uses the term `non-blocking' to refer to lock-free algorithms).
Authors of \cite{Dwork1997} showed that an adversary can cause $O(n)$ contention in a wait-free algorithm that is being executed by $n$ processes in asynchrony. 

There are some subtle differences between non-blocking and asynchronous algorithms. Non-blocking algorithms allow the nodes to return without waiting for an operation to complete, whereas asynchronous algorithms allow multiple nodes to perform operations concurrently such that the computing nodes do not block the progress of each other. Due possibility of contention, the scheduler is required to continuously check for the failed processes and completed tasks so that it can assign tasks to idle processes and guarantee progress, whereas asynchronous algorithms, the class of algorithms that this dissertation studies, eliminate the requirement to schedule tasks to processes.

In this dissertation, we are interested in asynchronous algorithms. Asynchronous algorithms are non-blocking, but not vice-versa.
In addition, we do not assume a scenario where a node fails or turns byzantine. We assume that all nodes run correctly, however, their speeds can differ.

A key characteristic of \po-inducing algorithms is that they  permit the algorithm to execute asynchronously. And, a key difference between non-blocking and asynchronous algorithms is the \textit{system-perspective} for which they are designed.
To understand this, observe that from a perspective, the  asynchronous algorithms considered in this dissertation are wait-free.
Each node reads the values of other nodes. Then, it executes an action, if it is enabled, without synchronization. More generally, in an asynchronous algorithm, each node reads the state of its relevant neighbours to check if the guard evaluates to true. It can, then, update its state without coordination with other nodes. 

That said, the goal of asynchronous algorithms is not the progress / blocking of individual nodes 
(e.g., success of insert request in a linked list and a binary search tree, respectively, in \cite{Valois1995} and \cite{Natarajan2014}).
Rather it focuses on the progress from the perspective of the system, i.e., the goal is not about the progress of an action by a node but rather that of the entire system.  
For example, in the algorithm for minimal dominating set present in this dissertation, if one of the nodes is slow or does not move, the system will not converge. 
However, the nodes can run without any coordination and they can execute on old values, instead of requiring a synchronization primitive to ensure convergence.
In fact, the notion of \imped (recall that in the algorithms that we present in this dissertation, in any global state, all enabled nodes are \imped) captures this. An \imped node has to make progress in order for the system to make progress.

\subsection{Starvation-Free Algorithms}

\textit{Starvation} happens when requests of a higher priority prevent a request of lower priority from entering the critical section indefinitely. To prevent starvation, algorithms are designed such that the priority of pending requests are increased dynamically. Consequently, a low-priority request eventually obtains the highest priority. Such algorithms are called \textit{starvation-free} algorithms.

Starvation-free algorithms have been developed to solve several problems in multiprocessor systems. Some of the interesting works that deploy the starvation-free protocols are listed as follows.

\begin{itemize}
    \item Authors of \cite{Kim2005} present a starvation-free algorithm to schedule queued traffic in a network switch. This algorithm is based on the dynamic priority increment of the waiting requests.
    \item Authors of \cite{Attiya2010} implement a starvation-free distributed directory algorithm for shared objects. They show that this algorithm can serve concurrent requests and works correctly even in asynchrony.
    \item Authors of \cite{Hesselink2013} introduce the notions of buffered semaphore and polite semaphore that facilitate starvation-free mutual exclusion. They showed, for three existing algorithms (present in, respectively, \cite{Morris1979,Martin1997,Udding1986}), that they are implementations of one abstract algorithm and operate on one or the other of these semaphores.
    \item Authors of \cite{Lejeune2015}, on the other hand, view priority increment as a priority violation. They present an algorithm, for mutual exclusion in distributed systems, that postpones the priority increment of pending requests, and consequently the number of priority violations. Their algorithm is an extension of Karnar-Chaki algorithm \cite{Kanrar2010}. They show that their methods (1) have a low message overhead (in comparison to Kanrar–Chaki algorithm and Chang’s priority-based algorithm  \cite{Chang1994}), (2) keep the same waiting time, and (3) tolerate the peaks of request load well.
\end{itemize}

In asynchronous algorithms, priority modification is not the key. For example, in the algorithms that we presented in this dissertation, we do not modify priority so as to ensure that some nodes can execute. Rather, the algorithms that are designed under the asynchronous model have the property that the nodes can execute independently given that there is at least one guard that holds true.

\subsection{Serializability}

\textit{Serializability} in a distributed system allows only those executions to be executed concurrently which can be modelled to some permutation of a sequence of those executions. In other words, serializability does not allow nodes to read and execute on old information of each other: only those executions are allowed in concurrency such that reading fresh information, as if the nodes were executing in an interleaving fashion, would give the same result. Serializability is heavily utilized in database systems, and thus, the executions performed in such systems are called \textit{transactions}.

\begin{itemize}
    \item A \textit{multidatabase system} is a system of multiple autonomous and heterogeneous (local) databases. Authors of \cite{Georgakopoulos1991} addressed the problem of how global serializability can be ensured in such a system. In their model, if some local database commits a pair of global transactions $\mathcal{T}_1$ and $\mathcal{T}_2$ in that sequence, but the local serialization is reversed, then such a schedule is aborted.
    \item Authors of \cite{Fle1982} consider the problem in which the sequence of operations performed by a transaction may be repeated infinitely often. They describe a synchronization algorithm allowing only those schedules that are serializable in the order of commitment.
    \item Authors of \cite{Papadimitriou1979} show that corresponding to several transactions, determining whether a sequence of read and write operations is serializable is an NP-Complete problem. They also present some polynomial time algorithms that approximate such serializability.
    \item  JavaSpecs is a distributed data management tool produced by Sun Microsystem. Authors of \cite{Busi2001} showed that serializability is satisfied in JavaSpaces only if we restrict to output, input, and read operations. Serializability, on the other hand, is not satisfied in the presence of test for absence or event notification.
\end{itemize}

In serializability protocols, schedules that cannot be serialized are aborted. However, in asynchronous systems, no process is aborted: all processes freely read from each other and perform executions independently. The asynchronous execution considered in this dissertation is not \textit{serializable}, especially, since the reads can be from an old global state. Even so, the algorithm converges, and does not suffer from the overhead of synchronization required for serializability.

\subsection{RedBlue Systems}

In \textit{redblue} systems (e.g., \cite{Li2012}), the rules can be divided into two non-empty sets: red rules, which must be synchronized, and blue rules, which can run in a lazy manner and do not have to be synchronized. Lattice-linear and asynchronous systems in general are the systems in which red rules are absent as an enabled node can execute independently regardless of which action is to be executed.

\subsection{Local Mutual Exclusion}

In \textit{local mutual exclusion}, at a given time, some nodes block other nodes while entering to critical section. This can be done, e.g., by deploying semaphores. 

\begin{itemize}
    \item Authors of \cite{Keane2001} study the group mutual exclusion problem, where nodes request for various ``sessions'' repeatedly, and it is required that (1) individual processes cannot be in different sessions concurrently, (2) multiple processes can be in the same session concurrently, and (3) is a process tries to enter a session, it is eventually able to do so.
    \item Authors of \cite{Khanna2020} propose a leader-based algorithm that deploys local mutual exclusion to solve resource allocation problem in Flying Ad hoc Networks.
    \item Authors of \cite{Raymond1989} presented an algorithm for distributed mutual exclusion in computer networks, that uses a spanning tree of the subject network. In this algorithm, the number of messages exchanged per critical section depends on the topology of this tree, typically this value is $O(n)$.
    \item Authors of \cite{Yang1995} an algorithm with $O(\lg n)$ time complexity for mutual exclusion among $n$ nodes. Specifically, this algorithm requires atomic reads and writes and in which all spins are local (here a spin means a busy wait in which a node, in this case, waits on locally accessible shared variables).
\end{itemize}

We see, in algorithms based on local mutual exclusion, that they require additional data structures/variables to ensure that access is provided to (and blocking is deployed on) a certain set of processes. In asynchronous algorithms, nodes do not block each other. In non-lattice-linear problems, we see that usually a tie-breaker is required to ensure the correctness of the executions, however, if a problem is naturally lattice-linear, then it is not required. This is because in the case of non-lattice-linear problems, it may be desired that all unsatisfied nodes do not become enabled, however, in the case of lattice-linear problems, as we see in \cite{Garg2020}, all unsatisfied nodes can be enabled. And, all enabled nodes can read values and perform executions asynchronously, where they are allowed to read old values, which is not allowed in algorithms that deploy mutual exclusion.

\section{Fixed Point Theorem}

Fixed point theorems are extensively studied concepts in Mathematics. We discuss some closely related results from fixed point theory in this section, along with their applications. We include this discussion here because our work can be seen as a application of the fixed point theory -- in our work, we develop systems, and theory of such systems, which converge at a global state, meaning that once an optimal global state is reached, then the system continues to be in that state for the rest of the execution. This is precisely the definition of a fixed point; we discuss the formal definition of this term in the following.

Let $f$ be a function with the same domain and codomain. If for some input $x$, $f(x)=x$, then $x$ is a \textit{fixed point} of $f$. A function can have multiple fixed points. For example, let $f_{ds}$ be a function that realizes the functionality of \Cref{algorithm:ds-ll} for the minimum dominating set: given a graph $G$, for an input global state $s$, $f_{ds}$ will return the set of global states that $G$ will transition to under \Cref{algorithm:ds-ll}, if $G$ is initialized in $s$. Observe, for example, in \Cref{figure:full-lattices-from-ds-example}, that all the supremum of each lattice is a fixed point for $f_{ds}$. In general, an optimal state, a state in which $G$ manifests a minimal dominating set, is a fixed point for $f_{ds}$. Similarly, a function $f_{smp}$ can be simulated for the algorithm for the stable marriage problem as presented in \Cref{example:mom-definition}. It can be observed that for the instance of this problem as presented in Example \smp continuation \ref{examplesmpcont:smp-preferences} in \Cref{section:lattice-linearity}, there are multiple global states that act as a fixed point for $f_{smp}$ (a set of all states for this instance is shown in \Cref{figure:smplattice}). However, starting from the infimum of the lattice, i.e., $\langle 1,1,1\rangle$ we only reach the state $\langle 1,2,2\rangle$, which is one of the fixed points for this instance of the stable marriage problem. Note that all algorithms studied in this dissertation can be simulated, each, as a function with the same domain and codomain.

A \textit{complete lattice} is a lattice for which there exists a unique infimum and a unique supremum. The lattices present in \Cref{figure:full-lattices-from-ds-example} are, each, a complete lattice. The lattice present in \Cref{figure:smplattice} is also a complete lattice.

A fixed-point theorem provides the conditions under which a fixed point exists in a system. We only study the fixed-point theorems that are most relevant to this dissertation. 

There are some works that study fixed-point theorems in discrete systems, whereas some other works study fixed-point theorem in continuous systems. 

It was shown in \cite{Knaster1928} that a function $f$ whose domain and codomain are subsets of a set, which is increasing under set-theoretical inclusion, has at least one fixed point. This result was generalized in \cite{Tarski1955}, which provides fixed point theorems in lattices. We discuss how the results in \cite{Tarski1955} relate to our work, next.


The lattices that we study in this dissertation are induced under the `$\preceq$' operation. Let $(L,\preceq)$ be a lattice, and let $f$ be an order-preserving function with respect to the $\preceq$ operation (e.g., $f_{ds}$ and $f_{smp}$) that traverses through $L$. Alfred Tarski (1955) \cite{Tarski1955} showed that the fixed points of $f$ in $L$ form a complete lattice under $\preceq$. He further showed that if $F$ is a set of order-preserving commutative functions, then the fixed points of all the functions in $F$ form a complete lattice. He also presented its applications in set theory, Boolean algebra, topology and real functions.

Brouwer (1911) \cite{Brouwer1911} showed that if $f$ is a continuous function in a multidimensional closed simplex onto itself, then there exists a point $x$ such that $f(x)=x$.
Kakutani \cite{Kakutani1941}, further, provided the generalization of this theorem and studies a multidimensional closed simplex, where $f$ is a point to set function $f$. He showed that if $f$ is upper semi-continuous, then there exists a point $x_0$ such that $x\in f(x)$.

The fixed-point theorems have many applications. One of the applications is in the development of parallel processing algorithms that are tolerant to asynchrony, as we describe in \Cref{chapter:mulmod}, \Cref{chapter:ella} and \Cref{chapter:flla}. Fixed-point theorem implies that all the optimal states, that belong to the same lattice $L$, form a lattice $L'$ on their own. The infimum of $L'$ is the optimal state that a system will converge to, if it is initialized in the infimum of $L$.
A lattice-linear system that initializes in the infimum of a lattice is required to have at least one fixed point. A self-stabilizing system requires that the supremum is a fixed point. A silent self-stabilizing system requires that only the supremum is a fixed point.

In the next subsections, we present the applications of fixed point computation and how our results apply to them. 

\subsection{Conflict-free Replicated Datatypes}

Introduced in \cite{Shapiro2011}, a \textit{conflict-free replicated datatype} (CRDT) is a replicated data structure which can be accessed and modified by multiple processes. Each process has access to a distinct replica and the data structure is guaranteed to converge in s self-stabilizing manner even when these processes execute in asynchrony.

An example of such data structure is a vector for which the only allowed operation is to monotonically increase or decrease the values stored in it. The different states that this vector traverses through form a $\prec$-lattice ($\prec$-lattice is called \textit{semilattice} in \cite{Shapiro2011}). 

The theory of CRDTs has been extended to many other works. We note some of these works. The authors of \cite{Kleppmann2017} present an algorithm, as well as formal semantics, for a JSON data structure. This data structure automatically resolves concurrent modifications such that no updates are lost -- all replicas converge towards the same state. A replicated set datatype is presented in \cite{Deftu2013}. The authors of \cite{Nasirifard2019} study integration of CRDTs with blockchain technologies to alleviate the additional latency between executing and committing transactions.

\subsection{Fixed point iteration}

In numerical analysis a \textit{fixed point iteration} is the method of computing a fixed point. Essentially, given a function $f$ with the same domain and codomain, and an element $x_0$ in the codomain of $f$, we input $x_0$ in $f$, and recursively provide the output of $f$ as an input to it.
Mathematically, we perform the recursion $$x_{i+1}=f(x_i)\text{, for }i=0,1,2,...$$
This iterative computation is performed until $f$ stutters on a point, i.e., until a point $x_l$ is found such that $f(x_l)=x_l$. In such a case, $x$ is said to be a fixed point of $f$.

To find the least fixed point of a function $f$ with respect to a lattice $L$, we start the above iterative computation from the infimum $x_{inf}$ of $L$.
Notice that this is how, e.g., the algorithm for stable marriage problem functions, to reach the optimal state. 

The \textit{least fixed-point} of a function $f$ in a lattice $L$ is the least element $x$ in $L$ for which $f(x)=x$. Similarly, the \textit{greatest fixed-point} is the greatest element which is a fixed point for $f$. For example, in the lattice present in \Cref{figure:smplattice}, $\langle 1,2,2\rangle$ is the least fixed-point of $f_{smp}$. To find the set of all fixed points of $f$ with respect to a lattice, we initialize the above iterative computation starting from every point in a lattice.

Fixed point iteration is used by compilers for code optimization, e.g., through abstract interpretation \cite{Cousot1977} (where a compiler gains information about the semantics of a program -- e.g., control flow, data flow -- without performing all the computations). Apart from this, the fixed-point theorem has been used in several other fields such as economics, e.g., where John Nash introduced the Nash equilibrium \cite{Nash1951} by exploiting the Kakutani fixed-point theorem (this theorem provides sufficient conditions for a function, whose domain and codomain are set-valued, that is defined on a convex, compact subset of Euclidean space to have a fixed point).

The vector in the PageRank algorithm \cite{Page1999} is a fixed point with respect to linear transformation, and by extension, all Markov chain models also search for a fixed point where they stabilize with respect to transition probability.

\subsection{Modal $\mu$-Calculus}

$\mu$-calculus is a logic that describes the properties of a transition system $\mathcal{S}$ (cf. \Cref{section:preliminaries-distributed-systems} for the definition of a transition system). A transition system can be an infinite graph. However, most transition systems, that we study in this dissertation, have finite size. We do not discuss the $\mu$-calculus in detail, the reader is directed to \cite{Bradfield2007} for an introduction on the topic. $\mu$-calculus provides second-order expressive power which makes it extremely powerful logic in model checking. It utilizes the recursive computation of fixed-point iteration to express the operators of temporal logic.

\subsection{Extending fixed-point logics to \dag-inducing systems}

Observe that lattices and \dags are special cases of simplices, where a global state, an $n$-dimensional tuple, can have one or more outgoing edges pointing to other global states.
Thus, we have that a nondecreasing function whose domain is a set of $n$-dimensional tuples, forming a \dag, is a special case of the fixed point theorem by Kakutani \cite{Kakutani1941}. Thus we have a corollary from Kakutani's theorem, which we state as follows.

\begin{corollary}
    (Of \cite{Kakutani1941}) Let $f$ be a nondecreasing point-to-set function whose domain and codomain is a finite set of $n$-dimensional tuples forming a \dag. Then, there is at least one point $x$ such that $x\in f(x)$.
\end{corollary}

\section{Lattice-Linearity}\label{section:literature-lattice-linearity}

In \cite{Garg2020}, the authors have studied lattice-linear problems which possess a predicate under which the states naturally form a lattice among all states. Problems like the stable marriage problem, job scheduling
and others are studied in \cite{Garg2020}. We study the theory established in \cite{Garg2020} in detail in \Cref{section:lattice-linearity}.

In \cite{Garg2022}, the authors have studied lattice-linearity in several dynamic programming problems.

The key idea of lattice-linearity is that a process/node determines that its local state is not feasible in any reachable optimal global state. In other words, it has to change its state to reach an optimal state.
Thus, if node $i$ changes its state from $st.i$ to $st'.i$ it never revisits state $st.i$ again. Consequently, the local states visited by a node form a total order. 
Hence, it can change its state even if it is relying on the old values of its neighbours. 
As a result, the nodes can run without synchronization and the system is guaranteed to reach an optimal state.

Garg, in \cite{Garg2020}, studied problems in which a distributive lattice is formed among the global state, where a meet and join can be found for any given pair of states, and meet and join distribute over each-other. However, we find that to allow asynchrony, a more relaxed data structure can be allowed. Specifically, in a $\prec$-lattice, for a pair of global states, their join can be found, however, their meet may not be found. For instance, in the instance that we study in \Cref{figure:full-lattices-from-ds-example}, both meet and join can be found for a pair of global states in a $\prec$-lattice, however, in the instance that we study in \Cref{figure:vc-semilattice-example}, a join can be found for a pair of global states in a $\prec$-lattice but a meet is not always found.

\subsubsection*{Showing Property of Asynchrony for Existing Algorithms}
Johnson's algorithm for computing shortest paths \cite{Johnson1977} and Gale-Shapey algorithm for stable marriage \cite{Gale1962} have been shown to be tolerant to asynchrony in \cite{Garg2020}.
The top trading cycle algorithm for housing market problem by Shapley and Scarf \cite{Shapley1974} (attributed to Gale) was shown to be lattice-linear in \cite{Garg2021}.




\section{Problems Studied in this Dissertation}

In this section, we discuss the related work on specific problems that we study in this dissertation. While doing so, we compare the properties of existing algorithms for these problems against the algorithms that we study in this dissertation. Note that in several cases, we do not develop new algorithms for problems, rather we show that an existing algorithm has better properties than originally proven, e.g., the guarantee of convergence in asynchrony.

\subsubsection*{Multiplication}

\noindent 
In \cite{Cesari1996}, the authors presented three parallel implementations of the Karatsuba algorithm for long integer multiplication on a distributed memory architecture. Two of the implementations have time complexity of $O(n)$ on $n^{\lg 3}$ processors. The third algorithm has complexity $O(n \lg n)$ on $n$ processors.

We show that the Cesari-Maeder parallelization of the Karatsuba's algorithm for multiplication is tolerant to asynchrony. This algorithm converges in $O(n)$ time. We also study a parallel processing algorithm for modulo operation, which is tolerant to asynchrony.

\subsubsection*{Modulo}

\noindent 
In \cite{Zeugmann1992}, the authors have presented parallel processing algorithms for inverse, discrete roots, or a large power modulo a number that has only small prime factors. A hardware circuit implementation for mod is presented in \cite{Butler2011}.

\subsubsection*{Dominating Set}

\noindent 
Self-stabilizing algorithms for the minimal dominating set problem have been proposed in several works in the literature, for example, in \cite{Xu2003,GODDARD2008,Chiu2014}. Apart from these, the algorithm in \cite{Hedetniemi2003} converges in $O(n^2)$ moves, and the algorithm in \cite{Turau2007} converges in $9n$ moves under an unfair distributed scheduler. The best convergence time among these works is $4n$ moves.

We study a generalized version of minimal dominating set, the service demand based minimal dominating set problem, which is a more practical generalization of \mds than other algorithms present in the literature. We present an eventually lattice-linear self-stabilizing algorithm, converges in 1 round plus $n$ moves (within $2n$ moves), and does not require a synchronous environment.
In addition, evaluation of guards takes only $O(\Delta^4)$ time, which is better than the algorithm presented in \cite{Kobayashi2022}.
We also study a fully lattice-linear self-stabilizing algorithm for minimal dominating set that converges in $n$ moves and is fully tolerant to consistency violations. These results preset an improvement over the other algorithms present in the literature.

\subsubsection*{Vertex Cover}

\noindent 
Self-stabilizing algorithms for the vertex cover problem
has been studied in Astrand and Suomela (2010) \cite{Astrand2010} that converges in $O(\Delta)$ rounds, and Turau (2010) \cite{Turau2010} that converges in $O(\min\{n, \Delta^2, \Delta \log_3 n\})$ rounds.

We present an eventually lattice-linear self-stabilizing algorithm for minimal vertex cover; it converges in 1 round plus $n$ moves (within $2n$ moves).
We also present a fully lattice-linear algorithm for minimal vertex cover that converges in $n$ moves. These algorithms guarantee convergence in asynchrony.

\subsubsection*{Independent Set}

\noindent 
Self-stabilizing algorithm for maximal independent set has been presented in \cite{Turau2007}, that converges in $\max\{3n-5, 2n\}$ moves under an unfair distributed scheduler, \cite{GODDARD2008} that converges in $n$ rounds under a distributed or synchronous scheduler, \cite{Hedetniemi2003} that converges in $2n$ moves.

We present an eventually lattice-linear self-stabilizing algorithm for maximal independent set; it converges in 1 round plus $n$ moves (within $2n$ moves). We also present a fully lattice-linear algorithm for maximal independent set that converges in $n$ moves. These algorithms guarantee convergence in asynchrony.

\subsubsection*{Graph Colouring}

\noindent 
Self-stabilizing algorithms for graph colouring have been presented in several works, including  \cite{Bhartia2016,Checco2017,Duffy2013,Duffy2008,Galan2017,Leith2006,Motskin2009,Chakrabarty2020}. The best convergence time among these algorithms is $n\times \Delta$ moves, where $\Delta$ is the maximum degree of the input graph.

We study an eventually lattice-linear self-stabilizing algorithm for graph colouring; it converges in $n+4m$ moves.
We also study a fully lattice-linear self-stabilizing algorithm for graph colouring; it converges in $n+2m$ moves. These algorithms guarantee convergence in asynchrony.

\subsubsection*{2-Dominating Set}

\noindent 
The 2-dominating set is not an extensively studied problem. The problem was introduced in \cite{Bollobas1990}. A self-stabilizing algorithm for the 2-dominating set problem has been studied in \cite{Maruyama2022}. This algorithm converges in $O(nD)$ rounds under a distributed scheduler, where $D$ is the diameter of $G$.

We study an eventually lattice-linear self-stabilizing algorithm for 2-dominating set; it converges in 1 round plus $2n$ moves (within $3n$ moves), and is tolerant to asynchrony.

\subsubsection*{Robot Gathering on Discrete Grids}

\noindent 
In a general case, it is impossible to gather a system of robots if their visibility graph is not a connected graph. One-axis agreement and distance-1 myopia are the minimal capabilities that robots need to converge on a triangular grid \cite{Goswami2022}.

A system of robots with minimal capabilities has been studied with several output requirements, including gathering \cite{Goswami2022,DAngelo2016,Flocchini2005}, dispersion \cite{Augustine2018}, arbitrary pattern formation \cite{Bose2020}. Gathering of robots has been studied more recently in \cite{Bhagat2015,Klasing2010}. We focus on systems of robots on grids, mainly the papers that study gathering.

Robots placed on an infinite rectangular grid were studied in \cite{Poudel2021}, where the authors presented two algorithms for gathering. A synchronous scheduler is assumed and the robots require, respectively, distance-2 and distance-3 visibility.
Moreover, under the latter algorithm, robots may not gather at one point but will gather a horizontal line segment of unit length.

Robots placed on an infinite triangular grid were studied in \cite{Cicerone2021}, where the authors provided an algorithm to form any arbitrary pattern.
They require full visibility.
Their algorithm works only when the the initial global state is asymmetric.
Authors of \cite{Shibata2021} have studied gathering problem of 7 robots -- initially, 6 of them form a hexagon and one robot is present at the centre of that hexagon. They require the initial state to form a connected visibility graph; the system finally reaches a global state where the maximum distance between two robots is minimized. A synchronous scheduler is assumed.
In \cite{DAngelo2016}, authors  characterized the problem of gathering on a tree and finite grid.

We study the algorithm presented by \cite{Goswami2022} and show that it can converge in asynchrony. This algorithm converges assuming that the robots are myopic, and can use a unidirectional camera, that sees one neighbour at a time. The robots form an arbitrary connected graph initially.
Apart from the property of being able to converge in asynchrony, we also show that this algorithm converges in $2n$ rounds; the authors of the original algorithm showed that it converges in $2.5(n+1)$ rounds.

\subsubsection*{Maximal Matching}

\noindent 
A distributed self-stabilizing algorithm for the maximal matching problem is presented in \cite{Hsu1992}; this algorithm converges in $O(n^3)$ moves. The algorithm in \cite{Hanckowiak2001} converges in $O(\log^4n)$ moves under a synchronous scheduler. The algoritrithm for maximal matching presented in \cite{Goddard2003} converges in $n+1$ rounds. Hedetniemi et al. (2001) \cite{Hedetniemi2001} showed that the algorithm presented in \cite{Hsu1992} converges in $2m+n$ moves.

The \po-inducing algorithm for maximal matching, present in this dissertation, converges in $2n$ moves and is tolerant to asynchrony. This is an improvement to the algorithms present in the literature.
\chapter{CONCLUSION}\label{chapter:conclusion}

Synchronization constraints affect how and when the nodes read data from each other. These constraints permit the user to design the algorithm at a high level and omit some of the details of the underlying system. Hence, they make the design of algorithms easier.
However, enforcing synchronization creates an overhead that can lead to suboptimal use of computational resources. This is especially problematic, as we see a rise in the development and usage of large multiprocessor systems. 

In this dissertation, we focused on developing algorithms that work correctly even without synchronization. 
We explore the necessary and sufficient properties of algorithms that allow them to execute without synchronization.

We showed that 
local state transitions being abstracted as a partial order
is both necessary and sufficient for an algorithm to allow asynchrony. Focusing on the \textit{necessary} condition, we find that if there is any existing algorithm that allows to be executed under asynchrony, then it induces a $\prec$-\dag in the global state space. 

\subsubsection*{Organization of the Chapter}

\noindent In \Cref{section:summary-contributions}, we discuss the specific contributions of this dissertation.
In \Cref{section:transformation}, we discuss how a \po-inducing algorithm can be transformed to other models.
In \Cref{label:section:impact}, we discuss the practical applications and theoretical impact of this dissertation in multiprocessor systems technology. Finally, in \Cref{section:future-work}, we discuss future work directions that arise from this dissertation.

\section{Summary of Contributions}\label{section:summary-contributions}

In this section, we summarise some key objective contributions of this dissertation.

\subsubsection*{Self-Stabilizing Lattice-Linear Problems}

\noindent 
We observed the existence of lattice-linear problems that allow self-stabilization.
We show that the parallel processing version developed by \cite{Cesari1996} for Karatsuba's multiplication algorithm (cf. \cite{Karatsuba1962}) has properties of asynchrony; we present one other algorithm for multiplication and two algorithms for the modulo operation that are lattice-linear, and thus, guarantee convergence in asynchrony. These algorithms are self-stabilizing, and in any given global state, all \imped nodes can be identified.

\subsubsection*{Eventually Lattice-Linear Algorithms}

\noindent 
We observed that eventually lattice-linear algorithms can be developed for non-lattice-linear problems. These algorithms induce one or more lattices only in a subset of the state space. They guarantee that starting from an arbitrary global state, the system traverses to a state in a lattice, and then traverses to an optimal state through that lattice.
We develop eventually lattice-linear self-stabilizing algorithms for minimal dominating set, minimal vertex cover, maximal independent set and graph colouring problems.

\subsubsection*{Fully Lattice-Linear Algorithms}

\noindent 
We observed that fully lattice-linear algorithms can be developed for non-lattice-linear problems. Such algorithms induce lattices in the entire state space. 
We develop fully lattice-linear self-stabilizing algorithms for minimal dominating set, graph colouring, minimal vertex cover and maximal independent set problems. We also develop a lattice-linear 2-approximation algorithm for vertex cover; this algorithm is not self-stabilizing.

We showed for an algorithm developed in \cite{Goswami2022}, that solves the gathering problem of myopic robots on an infinite triangular grid, that it is lattice-linear.

\subsubsection*{\po-Inducing Systems}

\noindent 
We observed that the induction of a partial order in the local state transition graph is a necessary and sufficient condition to allow asynchrony.
An observation that immediately follows is that since a total order is a special case of a partial order, all lattice-linear problems and algorithms are, respectively, \po-inducing problems and algorithms.

We showed that the dominant clique problem and the shortest path problem (where the path is required to be generated) are \po-inducing problems, and maximal matching is a non-\po-inducing problem. We present algorithms for these problems; the local state transition graph induced by these algorithms forms a partial order -- it cannot be modelled within the constraints of a discrete structure such as the total order.

We derive time-complexity properties of a \po-inducing algorithm. As direct corollaries, we obtain the time-complexity properties of all the fully lattice-linear algorithms that we present in \Cref{chapter:flla} and all the \po-inducing algorithms that we present in \Cref{chapter:dag}.

\section{Transforming Our Algorithms to Other Models}\label{section:transformation}

In this dissertation, for the sake of simplicity, we presented algorithms that often require a node to read the values of its distance-$x$ neighbours $x \geq 1$. 
For similar reasons, we present algorithms where nodes easily read data from each other as if they had direct access to the memory of each other.
We note that these can be easily extended to other models of practical use while preserving the properties of interest. We identify some of these extensions, next.

\subsubsection*{Transforming to Distance-1}

\noindent 
Some algorithms that we study in this dissertation require to read information about the nodes at distance-$x$ from themselves, where $x>1$. An algorithm, under which the nodes require to read information from other nodes at a high distance, can be costly to execute.

Since \po-inducing algorithms are tolerant to asynchrony and the nodes executing based on old information, a trivial transformation to distance-1 is to keep a copy of all variables of nodes in distance-$x$ (cf. \cite{Afek2002}). However, as we see in \Cref{subsection:mds-d1-transformation}, a transformation that we present for \Cref{algorithm:ds-ll}, a distance-4 algorithm for minimal dominating set, we may not need to keep a copy of all such variables. As we see in \Cref{figure:experiments-flla}, the transformed algorithm has a substantial benefit in runtime as compared to the original algorithm.

\subsubsection*{Transforming to Message Passing Model}

\noindent 
The algorithms that we study in this dissertation appear to execute such that the nodes can seamlessly read data from each other. As a result, these algorithms appear to be executing in the shared memory model. If the nodes, however, are placed remotely from each other, then the algorithm will need to be executed in the message passing model. We can very simply transform a shared memory \po-inducing algorithm into an algorithm that is executable in the message passing model. We discuss this in the following paragraph.

Since the algorithms we study are tolerant to asynchrony, we can assume, for message passing model, that every time a node changes its state, it announces its new state to other nodes. The target nodes may continue to perform executions based on old values until they receive this information. However, since the algorithm that they are executing is tolerant to asynchrony, their execution will not worsen the rank of the system.

\section{Application and Impact of this Dissertation}\label{label:section:impact}

In addition to applications in improving computing technology, this dissertation has an impact on the theory of multiprocessor systems and the way they are designed. Among such topics, we discuss some key areas in the following.

\subsubsection*{Designing a \po-inducing algorithm}

\noindent 
We note that the techniques for \po-inducing algorithms are often different. 
However, a \po-inducing algorithm is closely related to the properties of the problem at hand.
Additionally, the partial order imposed among states may involve auxiliary variables.



If while developing an algorithm, it is analyzed if it is \po-inducing, then a lot of assumptions about the properties of the system implementing that algorithm.

Being able to design systems that fully tolerate asynchrony has been a desired, albeit unattainable, objective. \po-inducing systems provide with a deterministic, discrete, guarantee to attain such fault tolerance. 

\subsubsection*{Writing Proofs}

\noindent 
One immediate implication of our theory is in its application in writing proofs. Developing systems that can tolerate asynchrony has been difficult, and one of the reasons that adds to the intricacy is writing proofs of correctness of such systems. Our theory not only insists on the possibility of simplifying such proofs, but also provides an upper bound to the time complexity of the runtime of asynchronous algorithms. \Cref{corollary:dag<->partial-order} implies that to show that an algorithm is tolerant to asynchrony, we only need to show that the local states visited by individual nodes can be abstracted as a partial order, rather than to generate the entire state space and checking for the existence of a cycle. Thus, our theory simplifies the detail that a proof would require in order to show tolerance to asynchrony of a multiprocessing system. In addition, \Cref{corollary:convergence-time-dag-multivariable} provides the upper bound of the time complexity of a \po-inducing algorithm.


\subsubsection*{Showing Property of Asynchrony for Existing Algorithms}

Studying whether existing algorithms exploit lattice-linearity is extremely interesting and beneficial. Specifically, it allows us to eliminate costly, sophisticated, assumptions of synchronization from existing systems, instead of redesigning them from scratch.

A large number of problems can be solved by algorithms that may converge without requiring any synchronization. A high fraction of such algorithms may already be published, but it has not been proven so.
Recently, there has been work that shows for some existing algorithms that they do not require synchronization. Such work includes the results present in this dissertation, where we show for existing algorithms that they do not require synchronization. 
We discuss this in the following paragraph.

We show tolerance to asynchrony in Cesari-Maeder parallelization \cite{Cesari1996} of Karatsuba's multiplication and GSGS algorithm for converging myopic robots on an infinite triangular grid \cite{Goswami2022} in, respectively, \Cref{chapter:mulmod} (\Cref{subsection:karatsuba}) and \Cref{chapter:flla} (\Cref{section:gsgs}).


There has been work by other authors who have proved for existing algorithms that they can be executed without synchronization. This work falls under the umbrella of \po-inducing algorithms. These results are listed in Related Work (\Cref{section:literature-lattice-linearity}).

The above results 
are a consequence of the observation that these algorithms stipulate that all \imped nodes must update their local states, and that as a consequence, the local state transition graph forms a partial order. 

\section{Future Work}\label{section:future-work}


In the following, we discuss some interesting future work that can be extended from this dissertation. 

\subsubsection*{Algorithm Design}

\noindent 
One promising direction is to continue studying what other existing algorithms tolerate asynchrony. This will have an impact on alleviating the assumptions of synchronization from existing systems, and will thus, avoid having to design systems from scratch. Such work may have immediate applications in industrial computing technology. 

Tolerating asynchrony does remove all requirements of synchronization, however, we have not seen any considerable improvement in the termination detection of such algorithms. Even these algorithms use the tools to detect termination, same as the algorithms that require synchronization. Despite numerous and ongoing efforts, developing a better termination detection protocol seems to be, as of yet, an unsolved problem. It would be very useful to bring about an improvement in the area of termination detection of algorithms that execute without synchronization.

\subsubsection*{Systems Design}

\noindent 
The existing algorithms that we have found results for, present a potential impact in discrete software engineering systems and operating systems. For example, we show for an existing parallel processing algorithm for multiplication (cf. \Cref{chapter:mulmod}) that it is tolerant to asynchrony. Most computational machines contain a circuit that computes the multiplication of a pair of bitstrings, which makes our result vital from the perspective of operating systems. Thus, along with theoretical problems and algorithms, a demanding and promising direction is to study algorithms that have applications in operating systems and firmware design. This will have an impact on improving the robustness of multicore, distributed and GPU systems. If a critical system is not straightforwardly tolerant to asynchrony, then it is worthwhile to investigate what minimal changes we can make to make it tolerant to asynchrony.

Another related research area is timeless circuits. Such circuits do not use a clock to organise the steps of its components, rather, the output of a component is fully dependent on the input fed to it by its predecessor component in the circuit (cf. \Cref{section:asynchronous-circuits}). Such circuits have not received sufficient attention due to the scarcity of asynchronous algorithms, and the lack of any improvement in termination detection. Our research has a potential to motivate more substantial research in these areas; we expect our models and algorithms to perform even better when implemented on timeless circuits. 

\subsubsection*{Machine Intelligence}

\noindent 
Another application of our work is in machine learning, deep learning, and AI systems. We have recently started investigating such systems and found that many such systems can be divided into phases, where each individual phase can run asynchronously. However, when we merge those phases into one system, then it is required to put barriers between them. Specifically, transitioning between phases requires synchronization, and the current phase needs to be completed on all computing nodes before starting the next phase. Given the applications of such systems, a direction that is worth pursuing is to study if such systems can be allowed to run without synchronization between phases, and if not, then what minimal changes can be made to allow asynchrony.

\clearpage
\singlespacing 
\renewcommand*{\bibname}{Bibliography\vspace{0.175in}}
\addcontentsline{toc}{chapter}{BIBLIOGRAPHY}
\bibliography{t.bib}
\bibliographystyle{ieeetr}

\newpage

\doublespacing

\appendix
\addtocontents{toc}{\protect\renewcommand{\protect\cftchappresnum}{APPENDIX }}

\chapter{OVERVIEW OF GRAPH THEORY AND DISTRIBUTED ALGORITHMS}\label{chapter:introductory-concepts}

The intention behind adding this appendix is to assist a reader, who is new to distributed systems and graph theory, in understanding the concepts on which this dissertation is based. However, we assume that a reader is acquainted with Boolean algebra. Part of \Cref{section:preface-algo} and \Cref{section:preface-graphs} is taken verbatim from Burning Geometric Graphs (A T Gupta, Master's Thesis, 2020) \cite{Gupta2020}.

\section{Preface to Algorithms}\label{section:preface-algo}

Lionardo Pisano, more popularly known as Fibonacci, was an Italian mathematician, and he introduced the conventional Indian mathematical methods to Europe in the $13^{th}$ century \cite{Sigler2002}. In his book, \textit{Liber Abaci} (liberation from abacuses/abaci), he introduces \textit{modus Indorum} (method of the Indians). Until then, the abacus was used across Europe to perform mathematical calculations. Pisano introduced a mathematics which was more efficient: computations could be performed on numbers without bounds on their digit length.

Earlier than Pisano, Muhammad ibn Musa al-Khwarizmi, a Persian mathematician, in $9^{th}$ century, wrote \textit{kit\=ab al-his\=ab al-hind\=i} (book of Indian arithmetic) and \textit{kitab al-jam' wa'l-tafriq al-ḥis\=ab al-hind\=i} (book of addition and subtraction in Indian arithmetic). 
A few centuries later, Al-Khwarizmi's texts were translated to Latin \cite{Blair2021}.

It is due to the work of Al-Khwarizmi and Pisano that the methods of Indian arithmetic spread across Europe, and a person who could perform computations without the use of abacus was called \textit{Ma\~estro-de-abaci}. 
After a series of nomenclatural adaptations, the Europeans started to call this new form of mathematics, which could be performed without abacus and without bounds on the input size, \textit{algorithms}\index{algorithms: etymology}. 

Since then, numerous efforts have been made to translate human intelligence and computing ability into artificial machinery. Blaise Pascal \cite{Dasgupta2014} built a machine in the $17^{th}$ century which could perform addition and subtraction. Gottfried Wilhelm Leibniz built a machine, during the same time, which could perform multiplication and division as well. Charles Babbage built the famous \textit{Difference Engine}\index{difference engine} which could do similar computations \textit{automatically}, that is, once the input numbers are supplied to it, it was able to do the computation without any human intervention. This machine was able to prepare tables: it was able to compute polynomials of degree $2$ for consecutive integers; this was called the \textit{method of differences}. Babbage built the first prototype of this machine in 1822.

Luigi Frederico Menabrea explained with reference to the Difference Engine that it was limited only to one type of computations, it could not be applied to solve numerous other problems in which mathematicians might be interested. This led Babbage to design the \textit{Analytical Engine}\index{analytical engine}, which could solve the full range of algebraic problems. The generality of the Analytical Engine is discussed in Menabrea’s Italian article \textit{Sketch of the Analytical Engine} (1842), which was translated into English by Augustus Ada \cite{Menabrea1843}.

Augustus Ada, countess of Lovelace, proposed that Babbage's design could be used to compute function of any number of functions. On Babbage's request, she wrote some additional notes to her \textit{memoir}, most famous one of them is the \textit{Note G}\index{note G - Augustus Ada}, in which, firstly, she anticipated an issue: whether computers can exhibit \textit{intelligence}, or, \textit{original thought}, and secondly, in this note she wrote a sequence of operations (an \textit{algorithm}) to compute Bernoulli numbers on the Analytical Engine.

Modern definition of the word algorithm is as follows. An \textit{algorithm}\index{algorithms: definition} is a step-by-step procedure used to solve a problem given that it halts in finite time for any given input.

After some decades, Alan Mathison Turing worked on decision functions (or a decision problem, as he presents in \cite{Turing1937}) and their representations. A decision function is a sequence of mathematical instructions whose outputs are \textit{accept} or \textit{reject} on an arbitrary input string. A decision function solves a decision problem. The set of strings which a function accepts is the \textit{language} of that function. This set of strings defines the problem which that function solves.

Turing initiated the design of what we call the Turing Machine \index{turing machine: invention} which works on these formal languages to compute for any decision problem. We do not discuss the Turing Machine; the reader is advised to refer \cite{Turing1937,Garey1979} to study the Turing Machine in detail. We start with a brief discussion on graphs, on which the chapters in this dissertation are majorly based.

\section{Preface to Graphs}\label{section:preface-graphs}

A \textit{graph}\index{graphs} is a representation of entities and their relations: generally, a graph tells which entities are related (unweighted graph); sometimes the relations may have some associated cost or weightage (weighted graphs). Formally, a graph is a mathematical object which represents entities as vertices, and relations as edges between those vertices: if two entities are related, then there will be an edge between their corresponding vertices in the graph. The number of vertices in a graph is its \textit{order}, and the number of edges in a graph is its \textit{size}. Generally, given a graph, all its edges represent the same type of relation.

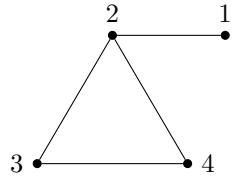
\begin{figure}[ht]
    \centering
    \begin{tikzpicture}
        \node [circle, draw=black, fill=black, inner sep=1pt, minimum size=3pt, label=above:{$1$}] (A) at (1.5,1) {};
        \node [circle, draw=black, fill=black, inner sep=1pt, minimum size=3pt, label=above:{$2$}] (B) at (0,1) {};
        \node [circle, draw=black, fill=black, inner sep=1pt, minimum size=3pt, label=left:{$3$}] (C) at (-1,-.707) {};
        \node [circle, draw=black, fill=black, inner sep=1pt, minimum size=3pt, label=right:{$4$}] (D) at (1,-.707) {};
        
        \draw (A) -- (B);
        \draw (B) -- (C); \draw (B) -- (D);
        \draw (C) -- (D);
    \end{tikzpicture}
    \caption{A sample graph represented by the sequence $G_4^1=\langle 4$, $7$, $8$, $11$, $13$, $15$, $2$, $1$, $3$, $4$, $2$, $4$, $2$, $3\rangle$.}
    \label{figure:graph-example}
\end{figure}

An example of an unweighted graph is presented in \Cref{figure:graph-example}. In this graph, $1$, $2$, $3$ and $4$ are the vertices and, for example, there is an edge between vertices 1 and 2. This graph can be represented by the sequence $G_4^1=\langle 4$, $7$, $8$, $11$, $13$, $15$, $2$, $1$, $3$, $4$, $2$, $4$, $2$, $3\rangle$. The first element in $G_4^1$, 4, represents the total number of vertices. The second element of $G_4^1$, 7, represents that the vertices connected to the first vertex start at position 7 in $G_4^1$. Similarly, the third element of $G_4^1$, 8, represents that the vertices connected to the second vertex start at position 8 in $G_4^1$. So the first vertex, vertex 1, is only connected to vertex 2, and vertex 2 is connected to vertices 1, 3 and 4. Similarly, the fourth and fifth elements of $G_4^1$, 11 and 13, represent that the vertices connected to the third and fourth vertices start at position 11 and 13 in $G_4^1$ respectively. The sixth element, 15, represents that $G_4^1$ contains only 14 elements. 

A graph in which all the edges are of equal weight is also viewed as an unweighted graph. Generally, in the graphs that are deemed as unweighted graphs, all edges are represented with the weight value 1. In the graph represented in \Cref{figure:graph-example} also, we consider the weight on all edges to be $1$, which we do not show explicitly for brevity.

Some of the graphs that we study in this dissertation are weighted graphs. A \textit{weighted graph} is a graph that has a weight associated to the edges that connect the vertices. A weighted edge represents a cost or weightage of a relation between a pair of entities.
Such cost-associated, or weighted, edges signify some real-world properties.
For example, if two computers are connected in a network, we can model them as a pair of vertices, and there will be an edge connecting those vertices. The frequency of communication between them can be modelled by assigning that much weight to that edge. A graph constructed in this way will provide information about the frequency of communication among computers in a network. 
For another example, a weighted edge joining a pair of vertices (where these vertices represent two cities) can represent that there is a direct road connecting those cities, and its weight would represent the distance between their corresponding cities.
In the above computer network example, the edges represent weightage (priority) of the connections, whereas in the city example, the edges would represent the cost of the connections.

We also study directed graphs. In an undirected graph, we have edges \textit{directed} both to and from the vertices that they connect. So, an undirected edge joining two vertices $a$ and $b$ has directions both from $a$ to $b$ and from $b$ to $a$. Thus, an undirected edge is effectively a bidirectional edge, as it can be depicted from the description of  $G^1_4$ and its representation in \Cref{figure:graph-example}. A \textit{directed graph} is a graph in which the edges are unidirectional. A \textit{mixed graph} is a graph that contains both unidirectional and bidirectional edges.

\begin{figure}[ht]
    \centering
    \begin{tikzpicture}
        \node [circle, draw=black, fill=black, inner sep=1pt, minimum size=3pt, label=above:{$1$}] (A) at (1.5,1) {};
        \node [circle, draw=black, fill=black, inner sep=1pt, minimum size=3pt, label=above:{$2$}] (B) at (0,1) {};
        \node [circle, draw=black, fill=black, inner sep=1pt, minimum size=3pt, label=left:{$3$}] (C) at (-1,-.707) {};
        \node [circle, draw=black, fill=black, inner sep=1pt, minimum size=3pt, label=right:{$4$}] (D) at (1,-.707) {};
        
        \draw[thick][->] (A) -- (B);
        \draw[thick][<-] (B) -- (C); \draw[thick][<-] (B) -- (D);
        \draw (C) -- (D);
        
        \node at (.75,1.25) {\textbf{\texttt{1}}};
        \node at (-.75,.15) {\textbf{\texttt{3}}};
        \node at (.75,.15) {\textbf{\texttt{4}}};
        \node at (-0,-.957) {\textbf{\texttt{5}}};
    \end{tikzpicture}
    \caption{A sample graph represented by the sequence $G_4^2=\langle 4$, $7$, $8$, $8$, $10$, $12$, $\langle 2,1\rangle$, $\langle 2, 3\rangle$, $\langle 4, 5\rangle$, $\langle 2,4\rangle$, $\langle 3,5\rangle\rangle$.}
    \label{figure:directed-graph-example}
\end{figure}
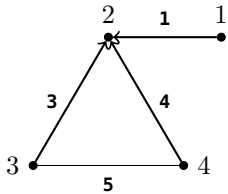

An example of a mixed weighted graph is presented in \Cref{figure:directed-graph-example}. In this graph, $1$, $2$, $3$ and $4$ are the vertices and, for example, there is an edge from vertex 1 to vertex 2. This graph can be represented by the sequence $G_4^2=\langle 4$, $7$, $8$, $8$, $10$, $12$, $\langle 2,1\rangle$, $\langle 2, 3\rangle$, $\langle 4, 5\rangle$, $\langle 2,4\rangle$, $\langle 3,5\rangle\rangle$. The first element in $G_4^2$, 4, represents the total number of vertices. The second element of $G_4^2$, 7, represents that the vertices that have edges from vertex 1 start at position 7 in $G_4^2$. Similarly, the third element of $G_4^2$, 8, represents that the vertices that have edges from vertex 2 start at position 8 in $G_4^2$. So vertex 1 has an edge to vertex 2 only; the weight of this edge is 1. Similarly, the fourth and fifth elements of $G_4^2$, 8 and 10, represent that the vertices that have edges from vertex 3 and vertex 4 start at position 8 and 10 in $G_4^2$ respectively. The sixth element, 12, represents that $G_4^2$ contains only 11 elements. Since the start index is 8 corresponding to both vertex 2 and vertex 3, we have that there is no edge that goes from vertex 2 to any other vertex. The edge between vertex 3 and vertex 4 is a bidirectional edge, which can be represented as having no arrows, as in \Cref{figure:directed-graph-example}, or having arrows on both its ends.

Note that we do not need to represent the edges of a weighted graph as a two-value tuple. We can represent such a graph as a single dimension tuple, where each edge will occupy two consecutive indices in that tuple, where the first of those indices represents the direction of the edge and the second index represents its weight. For example, $G_4^2$ can also be represented as $G_4^3=\langle 4$, $7$, $9$, $9$, $13$, $17$, $2$, $1$, $2$, $3$, $4$, $5$, $2$, $4$, $3$, $5\rangle$

In a directed graph, a \textit{supremum} of a pair vertices $a$ and $b$ is the closest vertex from $a$ and $b$ that has a path from both $a$ and $b$. Similarly, an \textit{infimum} of $a$ and $b$ is the closest vertex to $a$ and $b$ that has a path to both $a$ and $b$.

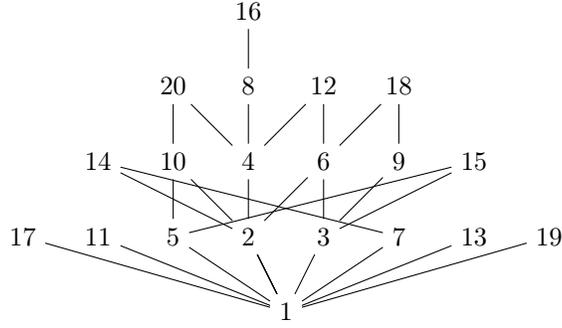
\begin{figure}
    \centering
    \begin{tikzpicture}
        \node (n1) at (0,0) {1};
        
        \node (n17) at (-3.5,1) {17};
        \node (n11) at (-2.5,1) {11};
        \node (n5) at (-1.5,1) {5};
        \node (n2) at (-.5,1) {2};
        \node (n3) at (.5,1) {3};
        \node (n7) at (1.5,1) {7};
        \node (n13) at (2.5,1) {13};
        \node (n19) at (3.5,1) {19};
        
        \node (n15) at (2.5,2) {15};
        \node (n10) at (-1.5,2) {10};
        \node (n4) at (-.5,2) {4};
        \node (n6) at (.5,2) {6};
        \node (n9) at (1.5,2) {9};
        \node (n14) at (-2.5,2) {14};
        
        \node (n20) at (-1.5,3) {20};
        \node (n8) at (-.5,3) {8};
        \node (n12) at (.5,3) {12};
        \node (n18) at (1.5,3) {18};
        
        \node (n16) at (-.5,4) {16};
        
        \draw (n1) -- (n2); \draw (n1) -- (n3); \draw (n1) -- (n5); \draw (n1) -- (n7); \draw (n1) -- (n11); \draw (n1) -- (n13); \draw (n1) -- (n17); \draw (n1) -- (n19); 
        
        \draw (n2) -- (n4); 
        \draw (n2) -- (n6); \draw (n3) -- (n6); 
        \draw (n1) -- (n2); \draw (n1) -- (n2); 
        \draw (n3) -- (n9); 
        \draw (n2) -- (n10); \draw (n5) -- (n10); 
        \draw (n2) -- (n14); \draw (n7) -- (n14); 
        \draw (n3) -- (n15); \draw (n5) -- (n15); 
        \draw (n4) -- (n8); 
        \draw (n6) -- (n12); \draw (n4) -- (n12); 
        \draw (n6) -- (n18); \draw (n9) -- (n18); 
        \draw (n8) -- (n16); 
        \draw (n4) -- (n20); \draw (n10) -- (n20); 
    \end{tikzpicture}
    \caption{A subgraph of the prime-factorizability lattice for natural numbers from 1 to 20. All edges are directed upwards; arrows are not shown for brevity.}
    \label{figure:prime-factorizability-lattice}
\end{figure}

A \textit{complete lattice} is a directed graph in which, for every pair of vertices $a$ and $b$, there exists a unique infimum and supremum of $a$ and $b$. As an example, consider a directed graph in which the vertices are natural numbers, and there is an edge from $a$ to $b$ iff $b$ is equal to $a$, multiplied by a prime number. Notice that this directed graph is a complete lattice: for every pair of numbers $a$ and $b$, there is a unique supremum (which is the lowest common multiple (LCM) of $a$ and $b$) and there is a unique infimum (which is the highest common factor (HCF) of $a$ and $b$). We call this lattice a \textit{prime-factorizability lattice}. We have shown a subgraph of the prime-factorizability lattice for numbers from 1 to 20 in \Cref{figure:prime-factorizability-lattice}. In a prime-factorizability lattice, a number $a$ is a factor of $b$ iff there is a path from $a$ to $b$. Note that the prime-factorizability lattice is of infinite order and size. The supremum of, for example, 16 and 20 is 80, which is not shown in \Cref{figure:prime-factorizability-lattice}.

In this dissertation, the lattices that we study have a supremum defined for any given pair of vertices, however, an infimum may not be found. Such lattices come within the class of \textit{incomplete lattices}.

The prime-factorizability lattice can be used to study several properties of numbers, for example: (1) iff there is a path from $a$ to $b$, then $a$ is a factor of $b$, (2) iff $b$ is a multiple of $a$, then the supremum of $a$ and $b$ is $b$, and the infimum of $a$ and $b$ is $a$, (3) iff the infimum of a pair of numbers $a$ and $b$ is 1, then $a$ and $b$ are coprime, and (4) if $p$ is a prime number, then $p$ is connected to 1 by a direct edge.

\begin{figure}[ht]
    \centering 
    \begin{tikzpicture}[scale=.8,every node/.style={scale=.8}]
        \node at (0,0) (a) {$\langle$1,1,1$\rangle$};
        
        \node at (-3,1) (b) {$\langle$2,1,1$\rangle$};
        \node at (0,1) (c) {$\langle$1,2,1$\rangle$};
        \node at (3,1) (d) {$\langle$1,1,2$\rangle$};
        
        \node at (4,2) (e) {$\langle$1,1,3$\rangle$};
        \node at (2,2) (f) {$\langle$1,2,2$\rangle$};
        \node at (4,3) (g) {$\langle$1,2,3$\rangle$};
        
        \node at (0,2) (h) {$\langle$1,3,1$\rangle$};
        \node at (2,3) (i) {$\langle$1,3,2$\rangle$};
        \node at (4,4) (j) {$\langle$1,3,3$\rangle$};
        
        \node at (-4,2) (k) {$\langle$2,1,2$\rangle$};
        \node at (-2,4) (l) {$\langle$2,1,3$\rangle$};
        
        \node at (-2,2) (m) {$\langle$2,2,1$\rangle$};
        \node at (0,3) (n) {$\langle$2,2,2$\rangle$};
        \node at (2,5) (o) {$\langle$2,2,3$\rangle$};
        
        \node at (-2,3) (p) {$\langle$2,3,1$\rangle$};
        \node at (0,5) (q) {$\langle$2,3,2$\rangle$};
        \node at (3,6) (r) {$\langle$2,3,3$\rangle$};
        
        \node at (-4,3) (s) {$\langle$3,1,1$\rangle$};
        \node at (2,4) (t) {$\langle$3,1,2$\rangle$};
        \node at (-2,5) (u) {$\langle$3,1,3$\rangle$};
        
        \node at (-4,4) (v) {$\langle$3,2,1$\rangle$};
        \node at (4,5) (w) {$\langle$3,2,2$\rangle$};
        \node at (0,6) (x) {$\langle$3,2,3$\rangle$};
        
        \node at (-4,5) (y) {$\langle$3,3,1$\rangle$};
        \node at (-3,6) (z) {$\langle$3,3,2$\rangle$};
        \node at (0,7) (parent) {$\langle$3,3,3$\rangle$};
        
        \draw (a) -- (b);
        \draw (a) -- (c);
        \draw (a) -- (d);
        \draw (d) -- (e);
        \draw (d) -- (f);
        \draw (c) -- (f);
        \draw (e) -- (g);
        \draw (f) -- (g);
        \draw (c) -- (h);
        \draw (f) -- (i);
        \draw (h) -- (i);
        \draw (g) -- (j);
        \draw (i) -- (j);
        \draw (b) -- (k);
        \draw (d) -- (k);
        \draw (e) -- (l);
        \draw (k) -- (l);
        \draw (b) -- (m);
        \draw (c) -- (m);
        \draw (m) -- (n);
        \draw (n) -- (o);
        \draw (g) -- (o);
        \draw (l) -- (o);
        \draw (f) -- (n);
        \draw (k) -- (n);
        \draw (h) -- (p);
        \draw (m) -- (p);
        \draw (p) -- (q);
        \draw (i) -- (q);
        \draw (n) -- (q);
        \draw (o) -- (r);
        \draw (j) -- (r);
        \draw (q) -- (r);
        \draw (b) -- (s);
        \draw (k) -- (t);
        \draw (s) -- (t);
        \draw (t) -- (u);
        \draw (l) -- (u);
        \draw (s) -- (v);
        \draw (m) -- (v);
        \draw (n) -- (w);
        \draw (t) -- (w);
        \draw (v) -- (w);
        \draw (w) -- (x);
        \draw (u) -- (x);
        \draw (o) -- (x);
        \draw (p) -- (y);
        \draw (v) -- (y);
        \draw (y) -- (z);
        \draw (w) -- (z);
        \draw (q) -- (z);
        \draw (z) -- (parent);
        \draw (r) -- (parent);
        \draw (x) -- (parent);
    \end{tikzpicture}
    \caption{An example lattice where each vertex is a tuple. All edges are pointing upwards. We have not shown the arrows for brevity.}
    \label{figure:lattice-of-tuples}
\end{figure}
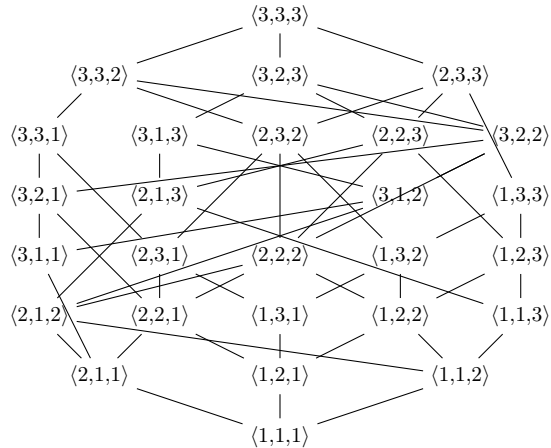

Another example lattice is shown in \Cref{figure:lattice-of-tuples}. In this lattice, each vertex is a tuple of three numbers. There is an edge from tuple $a$ to tuple $b$ iff (1) all elements of $b$ (sequentially) are equal to or greater than the elements of $a$, that is, $\forall i, 1\leq i\leq 3: b[i]\geq a[i]$, (2) all but one elements (sequentially) of $a$ and $b$ are different, that is, $\exists i, 1\leq i\leq 3: (a[i]\neq b[i] \land (\forall j\neq i: a[i]=b[i]))$, and (3) at the index $i$ where $a$ and $b$ are different, $b[i]-a[i]=1$. Notice that the graph shown in \Cref{figure:lattice-of-tuples} is a complete lattice.

\section{Transitive Edges}

Let that $R$ be a relation and for a pair of objects $a$ and $b$, let $a R b$ be true if and only if $b$ is related to $a$ under $R$. We say that $R$ is a \textit{transitive relation} only if the following is true: for any three objects, $a$, $b$ and $c$, if $a R b$ is true and $b R c$ is true, then $a R c$ is also true.

As an illustration of a relation that a graph can represent, consider the lattice presented in \Cref{figure:lattice-of-tuples}. Let that each tuple in this lattice represents some state of a system. Let this lattice be $\mathcal{L}$. We know that all the edges in this lattice are pointing upwards. An algorithm $A$ that follows this lattice can take a system from a state $s_1$ to another state $s_2$ in one step only if there is an edge from $s_1$ to $s_2$. If we consider a relation represented by a graph such as $\mathcal{L}$, then we have that $s_2$ is related to $s_1$ if and only if there is an edge from $s_1$ to $s_2$. It also means that $s_2$ is related to $s_1$ if and only if $A$ is able to take the system from $s_1$ to $s_2$ in one step.

In this dissertation, at several places, we discuss transitive edges. If we impose transitivity in $\mathcal{L}$, then it would mean we are considering $\mathcal{L}'$ instead of lattice $\mathcal{L}$, such that there are edges from a state $s_1$ to another state $s_2$ in $\mathcal{L}'$ if and only if there is a path from $s_1$ to $s_2$ in $\mathcal{L}$. An algorithm $A'$, that follows the lattice $\mathcal{L}'$, will be able to transition the system from $s_1$ to $s_2$ in $\mathcal{L}'$ in one step only if there is a path from $s_1$ to $s_2$ in $\mathcal{L}$. We discuss more on transitivity, and some applications transitivity in directed graphs, in \Cref{chapter:preliminaries} and the chapters that follow.

\section{Preface to Parallel Processing}

The soul of this dissertation lies in methods of parallel processing. Contemporary examples of parallel processing systems are a graphical processing unit (GPU) or a cluster of computers connected to each other via Ethernet, or computers even more remotely connected to each other, where, between a pair of computers, several channels can be involved, e.g., one or more of an Ethernet cable, WiFi, optical cable.

When a single processor solves a problem, then it may solve that problem in say, $k$ time units. But when we use a system comprising several processors, then we can distribute the work among those processors. We already have a single processor powerful enough, now one would expect that a bunch of, say, $n$ processors would be $n$ times as powerful. However, there are other costs that a parallel processing system must pay in order to utilize such increased power due to parallelization. Let us consider the following example, an instance of job distribution in a kitchen rather than a multiprocessor system.

Let us assume that a cook takes 2 hours to prepare a certain dish. Let us assume that this dish does not require any cooking, and only mixing certain vegetables and spices, but the vegetables have to be prepared and sliced, each, in a certain way which takes this much time in total. Now let us assume that we have 3 cooks at our disposal. Here, the cooks are the job-doers. In the same way, in a computer system, a processor is the job-doer. 3 cooks might take about 40-45 minutes to prepare this dish, as they can prepare all the ingredients consequently and then mix them at the end. Taking 40-45 minutes is reasonable.
Now assume that there are 100 cooks. Ideally, they should take less than 2 minutes altogether to prepare the dish. However, all those cooks should communicate their states in terms of progress and the part of the preparation they want to take up. Now on average, if I am one of those cooks, I will take about a minute and a half to talk to one other cook, and about 150 minutes to talk to all other cooks. Similarly, all cooks would want to talk to all other cooks, maybe more than once. This is going to take time even more than what a single cook would take to prepare the dish. So most of the power of the job-doers in our kitchen is consumed in communication for synchronization.

As we assumed above, a single processor would take $k$ time units to compute for a problem. But if we use $n$ processors, it may take more than $\dfrac{k}{n}$ time units to solve the same problem, even when the work is evenly divided. The extra time that they take is invested in communication for synchronization. Similar to cooks in our kitchen example, processors need to communicate with each other and they need to be synchronized with each other.

This dissertation investigates the behaviour of the systems where we eliminate the need for synchronization among the processors, and thereby, the costs incurred to enforce it. Some problems naturally allow asynchrony, while in other problems, we have to add details to the executions algorithmically.

For example, we could develop an algorithm for these 100 cooks and associate their names with the part of preparation that they have to take care of. Then they would simply work asynchronously and get their parts done, and then mix the ingredients at the end. Depending on the vegetables, it would take roughly two minutes to cut them precisely and mix them all with the required spices. Now any cook could choose any part of the preparation, other than what he is assigned, so this \textit{problem} does not allow asynchrony naturally, and therefore we must \textit{impose} on them their respective jobs \textit{algorithmically}. We do come across such problems in computer science as well.

There are, however, problems that naturally allow asynchrony. For example, consider that an aquarium company of pre-telephonic times, established near a freshwater lake, requires 100 pounds of live fish to put in its aquariums. Assume that a ship can collect 10 pounds of fish in a day. So it would take 10 days for a single ship to collect that amount of fish. If that aquarium company owns 5 ships, each ship can be sent off with the target of 20 pounds of fish each. This job can be done in roughly 2 days. There is no job association with, e.g., the name of the ship; all ships have to catch a specific, same, amount of fish, so such a system \textit{naturally} allows asynchrony.

Now assume that we have not assigned the amount of fish to all boats and require them to return, all, at the same time. One might think that this requirement will take the least possible time for all ships to come back, but it is not so. In the above system, the ships might have returned to port at different times, because they may be working at different speeds in catching the fish. However, in this latter case, we will have to utilize synchronization, so that we can ensure that the ships stay in communication with each other regarding the amount of fish each of them has to catch, and return to port at the same time. But since these are pre-telephonic times, exchanging information through longboats, will take considerable time in itself and much time will be consumed only in waiting for information to arrive. 

In this dissertation, we study the characteristics of problems that allow asynchrony naturally, and the structure of executions of algorithms that impose asynchrony in solving problems that do or do not allow asynchrony naturally.

\section{Preface to Acyclic State Transitions}

Above, we discussed the amount of resources and time used to enforce synchronization in a kitchen and a fish-catching example. In a multiprocessor system, where multiple processors perform execution to solve a problem, resources and time are invested, with a similar proportion, to enforce synchronization. In this dissertation, by the word \textit{system} we refer to a multiprocessor system.

The progress made by a system can be evaluated by analyzing its \textit{global state}, which is represented by the values stored in variables throughout all the processes. For now, we will call the state of convergence to be a global state where the system is deemed to have solved the problem at hand, and the solution is represented by the state that the system is in (by the values stored in variables throughout all the processes when the system is in the state of convergence). If a system is not synchronized, then it can potentially make mistakes, which may take it \textit{farther away} from the state of convergence. Synchronization prevents these mistakes from happening. Furthermore, the assumption of synchronization restricts when the nodes will read from other nodes and take action, and consequently, it makes the design of algorithms easy.

There are different types of synchronization, which we describe at various places throughout this dissertation. The type of synchronization that a system needs to enforce depends on the nature of the problem and the structure of the system. An absence of an apt synchronization primitive, where it would be otherwise required, can make the system make mistakes. These mistakes are caused because of executions that the processes make based on old and inconsistent information. We call such faults \textit{consistency violation faults}. Such mistakes result in the system to commit cyclic state transitions, where a system makes some progress, but then moves further away from the state of convergence because of a consistency violation.
This can repeat continually, and everytime the system makes progress, it will again move farther away from the state of convergence; this happens due to the absence of an apt synchronization primitive, even if the algorithm is otherwise correct on a uniprocessor system.

Proper synchronization ensures that an acyclic structure is induced among the state transitions, as it ensures that the data flows among the processes in a consistent fashion.

In this dissertation, we study the properties of systems that guarantee convergence without synchronization. The essential property for an algorithm to allow asynchrony is that it should be able to enforce acyclic state transitions in the system. There are multiple processors in a system running at least one process each. Acyclic state transitions can be enforced even in asynchrony as follows. Let us assume that every process performs execution such that a local state once discarded is never revisited again (a \textit{local state} of a process is represented by the values of only its own variables).
However, the processes need to make such transitions carefully: they must ensure that if they are discarding their local state, then that local state is infeasible for any possible desired state of convergence. Enforcing such guarantees is problem-dependent; this, we explore throughout this dissertation. If such guarantees can be made, then even if a node is reading old values, it can safely make a transition if it decides so, because its current local state would be infeasible for any desired state of convergence.
Consequently, the system exhibits acyclic state transitions, which is the main subject of exploration for this dissertation.

\section{An Example Graph Theoretic Problem}

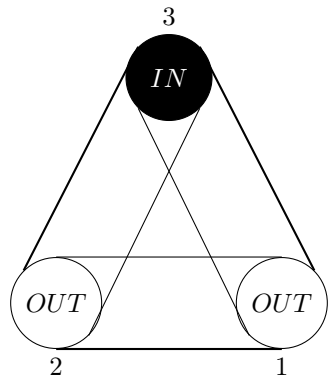
\begin{figure}[ht]
    \centering
    \begin{tikzpicture}
        \node [circle, draw=black, fill=contrastcolour, inner sep=4 pt, label=above:3] (a) at (0,0) {\color{backgroundcolour}$~IN~$};
        \node [circle, draw=black, fill=backgroundcolour, inner sep=4 pt, label=below:1] (b) at (1.5,-3) {\color{contrastcolour}$OUT$};
        \node [circle, draw=black, fill=backgroundcolour, inner sep=4 pt, label=below:2] (c) at (-1.5,-3) {\color{contrastcolour}$OUT$};
        
        \draw (a.south west) -- (b.south west);
        \draw[thick] (a.north east) -- (b.north east);

        \draw[thick] (a.north west) -- (c.north west);
        \draw (a.south east) -- (c.south east);

        \draw (b.north) -- (c.north);
        \draw[thick] (b.south) -- (c.south);
    \end{tikzpicture}
    \caption{Fort under attack: archers positioned at the tower marked $IN$.}
    \label{figure:fort}
\end{figure}

Consider, e.g., a fort containing three \textit{indexed} towers; each pair of towers has a straight path connecting them. This is demonstrated in \Cref{figure:fort}. Consider that the fort is under attack. All three paths are well protected with high and strong walls. The only entry point to enter the fort are the towers. The problem, thus, is to place archers \textit{minimally}; this problem is not of how many archers to place, the problem is \textit{where} to place them. We can place archers on all three towers, any two towers or just one tower. Notice that if we place archers in only one of the towers, they can protect the tower that they are placed in, as well as the other two towers. Therefore, for this problem, we have that placing one or more archers in only one tower is sufficient to protect all three towers, and hence the entire fort.

The threads in a multiprocessing system have a distinct ID associated to them. An algorithm for this problem computes a \textit{minimal} set of towers such that all towers can be protected. An example algorithm for such a problem makes each processor simulate a distinct tower. We call a processor, that performs computation in a multiprocessor system, a computation \textit{node}. A node can be in states $IN$ or $OUT$; if a node is $IN$, then it means that the algorithm decided to place archers in its corresponding tower, or otherwise, if a node is $OUT$, then it means that the algorithm decided \textit{not} to place any archers in its corresponding tower.

Consider an algorithm in which the nodes check if one of their neighbours is $IN$; if none of their neighbours are $IN$, then they will \textit{move} $IN$.
Suppose that initially, all nodes are $OUT$.
Now let that node 2 and node 3 execute together. They will \textit{see} that their neighbouring nodes are $OUT$, so both node 2 and node 3 will change their state to $IN$. In such an output, notice that the fort is protected, but not minimally. Thus, to allow only a minimal set of nodes to move in, we need some synchronization primitive to be deployed among the nodes. Most algorithms in the literature, that are developed for multiprocessor systems, assume some synchronization primitive to be deployed among the nodes.

In the paragraph above, we discussed an algorithm for minimal dominating set problem where the computing nodes should not be allowed to run in asynchrony. Now consider another algorithm for this problem, in which, if a set of nodes want to move in, they will allow the node with a higher ID in their neighbourhood to move first, and wait until then. In such an algorithm, node 3 will move (change its state to) $IN$. Node 1 and node 2 will wait for node 3 to move first. After when node 3 moves $IN$, when node 1 and node 2 evaluate their required \textit{action}, they will see that they are already being protected by node 3 (tower 3), so they will not perform any move (i.e., they will not change their state). Thus, we have a minimal set of towers that are able to protect the entire fort. Notice that this algorithm, unlike the algorithm described in the above paragraph, can be allowed to run asynchronously for this problem. This is because we used a tie-breaker based on node IDs, so \textit{race conditions} like the previous algorithm do not evolve. Thus, \Cref{figure:fort} shows the output of this algorithm, where we obtain a minimal set of locations to place the archers. In this dissertation, we study the properties of such algorithms: algorithms that can be allowed to run without synchronization.

The problem described above is the dominating set problem. In the \textit{dominating set} problem, the input is a graph $G$, and the task is to compute a dominating set $\mathcal{D}$ such that every node $i$ should be \textit{dominated}: (1) either $i$ is \textit{in} the dominating set, or (2) at least one of the neighbours of $i$ is \textit{in} the dominating set. We consider all and only the nodes that are $IN$ to be in $\mathcal{D}$.
In the \textit{minimal dominating set} problem, the input is a graph $G$, and the task is to compute the dominating set problem on $G$ and compute $\mathcal{D}$ such that if any node is removed from $\mathcal{D}$, then some node in $G$ becomes not dominated. The dominating set problem solves the problem of \textit{domination}, whereas the minimal dominating set problem solves the problem of \textit{minimality} along with domination.

Assume that in the initial state, all the nodes that represent the three towers are $OUT$. With such an input setting, notice that in the two algorithms that we described in the above paragraphs, the former algorithm solves the dominating set problem, even if it is run in asynchrony. However, it can solve the minimal dominating set problem only if it uses synchronization (in this case, local mutual exclusion or a central scheduler). The latter algorithm solves the minimal dominating set problem with or without synchronization.

In this dissertation, we study the properties that enable an algorithm to guarantee convergence in asynchrony. We study several problems for which such algorithms can be developed, and present example algorithms. We also study example problems where some specific algorithm designs do not work, and what algorithm design approaches would work. In such cases, we also study how the properties of the subject problems are responsible for certain algorithm design approaches to not work, and how these properties are correlated to the design approaches that work.
\chapter{PUBLICATIONS FROM THIS DISSERTATION}\label{chapter:publications}

We wrote the following peer-reviewed papers, published until the date of the defence, that emerged from this dissertation.

\begin{enumerate}
    \item [{[6]}] Gupta, A. T. and Kulkarni, S. S. (2024) Tolerance to Asynchrony of an Algorithm for Gathering Myopic Robots on an Infinite Triangular Grid. In Proceedings of the 19th European Dependable Computing Conference. EDCC 2024.
    \item [{[5]}] Gupta, A. T. and Kulkarni S. S. (2024) Eventually Lattice-Linear Algorithms. Journal of Parallel and Distributed Computing. (extended version of conference paper [1])
    \item[{[4]}] Gupta, A. T. and Kulkarni, S. S. (2023) Inducing Lattices in Non-Lattice Linear Problems. In Proceedings of the 42nd International Symposium on Reliable Distributed Systems. SRDS 2023. (extended version of brief announcement [2])
    \item[{[3]}] Gupta, A. T. and Kulkarni, S. S. (2023) Lattice Linearity of Multiplication and Modulo. In: Dolev, S., Schieber, B. (eds) Stabilization, Safety, and Security of Distributed Systems. SSS 2023. 
    \item[{[2]}] Gupta, A. T. and Kulkarni S. S. (2022) Brief Announcement: Fully Lattice Linear Algorithms. In: Devismes, S., Petit, F., Altisen, K., Di Luna, G.A., Fernandez Anta, A. (eds) Stabilization, Safety, and Security of Distributed Systems. SSS 2022. (full conference paper: [4])
    \item[{[1]}] Gupta, A. T., and Kulkarni, S. S. (2021) Extending Lattice Linearity for Self-stabilizing Algorithms. In: Johnen C., Schiller E.M., Schmid S. (eds) Stabilization, Safety, and Security of Distributed Systems. SSS 2021. (journal version: [5])
\end{enumerate}

\nocite{wiki}

\end{document}